\documentclass[useAMS,usenatbib]{mn2e}
\usepackage{graphicx}
\usepackage{subfigure}
\usepackage{lscape}
\usepackage{longtable}
\usepackage{threeparttable}
\usepackage{pdflscape}
\usepackage[usenames,dvipsnames,svgnames,table]{xcolor}

\title[A Test of Seven Widely-Used Spectral Synthesis Models Against Multi-Band Photometry of YMCs.]{A Comprehensive Comparative Test of Seven Widely-Used Spectral Synthesis Models Against Multi-Band Photometry of Young Massive Star Clusters}
\author[A. Wofford]{A. Wofford$^{1}$\thanks{E-mail:
wofford@iap.fr}, S. Charlot$^{1}$, G. Bruzual$^{2}$, J.J Eldridge$^{3}$, D. Calzetti$^{4}$, A. Adamo$^{5}$, 
 \newauthor M. Cignoni$^{6}$, S. E. de Mink$^{7}$, D.A. Gouliermis$^{8,9}$, K. Grasha$^{4}$, E.~K.~Grebel$^{10}$,
\newauthor  J. Lee$^{6}$, G. \"Ostlin$^{5}$, L.J. Smith$^{6}$, L. Ubeda$^{6}$, E. Zackrisson$^{11}$\\
$^{1}$Sorbonne Universit\'es, UPMC-CNRS, UMR7095, Institut d'Astrophysique de Paris, F-75014 Paris, France\\
$^{2}$Instituto de Radioastronomía y Astrofísica, UNAM, Campus Morelia, M{\'e}xico\\
$^{3}$Dept. of Physics, University of Auckland, Auckland, New Zealand\\
$^{4}$Dept. of Astronomy, University of Massachusetts -- Amherst, Amherst, MA, USA\\
$^{5}$Dept. of Astronomy, The Oskar Klein Centre, Stockholm University, AlbaNova University Centre, SE-106 91 Stockholm, Sweden\\
$^{6}$Space Telescope Science Institute, Baltimore, MD, USA\\
$^{7}$Astronomical Institute Anton Pannekoek, Amsterdam University, Amsterdam, The Netherlands\\
$^{8}$University of Heidelberg, Centre for Astronomy, Institute for Theoretical Astrophysics, Albert-Ueberle-Str.\,2, 69120 Heidelberg, Germany\\
$^{9}$Max Planck Institute for Astronomy,  K\"{o}nigstuhl\,17, 69117 Heidelberg, Germany\\
$^{10}$Astronomisches Rechen-Institut, Zentrum f\"ur Astronomie der Universit\"at Heidelberg, M\"onchhofstr, Heidelberg, Germany\\
$^{11}$Dept. of Physics and Astronomy, Uppsala University, Box 515, SE-751 20 Uppsala, Sweden 
}

\begin{document}

%\date{Accepted 1988 December 15. Received 1988 December 14; in original form 1988 October 11}

\pagerange{\pageref{firstpage}--\pageref{lastpage}} \pubyear{2002}

\maketitle

\label{firstpage}

\begin{abstract}
We test the predictions of spectral synthesis models based on seven different massive-star prescriptions against Legacy ExtraGalactic UV Survey (LEGUS) observations of eight young massive clusters in two local galaxies, NGC~1566 and NGC~5253, chosen because predictions of all seven models are available at the published galactic metallicities. The high angular resolution, extensive cluster inventory and full near-ultraviolet to near-infrared photometric coverage make the LEGUS dataset excellent for this study. We account for both stellar and nebular emission in the models and try two different prescriptions for attenuation by dust. From Bayesian fits of model libraries to the observations, we find remarkably low dispersion in the median $E(B-V)$ ($\sim0.03\,$mag), stellar masses ($\sim10^4M_\odot$) and ages ($\sim1\,$Myr) derived for individual clusters using different models, although maximum discrepancies in these quantities can reach 0.09~mag and factors of 2.8 and 2.5, respectively. This is for ranges in median properties of 0.05--0.54~mag, 1.8--10$\times10^4M_\odot$ and 1.6--40\,Myr spanned by the clusters in our sample. In terms of best fit, the observations are slightly better reproduced by models with interacting binaries and least well reproduced by models with single rotating stars. Our study provides a first quantitative estimate of the accuracies and uncertainties of the most recent spectral synthesis models of young stellar populations, demonstrates the good progress of models in fitting high-quality observations, and highlights the needs for a larger cluster sample and more extensive tests of the model parameter space.
\end{abstract}

\begin{keywords}
(ISM:) H{\,\sc{ii}} regions -- galaxies: star clusters: general -- galaxies: star formation -- ultraviolet: galaxies -- stars: early-type.
\end{keywords}

%%%%%%%%%%%%%%%%%%%%%%%%%%%%%%
% Introduction
%%%%%%%%%%%%%%%%%%%%%%%%%%%%%%

\section{Introduction}

Young massive clusters (YMCs) are dense aggregates of young stars that are considered to be fundamental building blocks of galaxies \citep{por10}. Determining accurate stellar masses and ages for large samples of individual YMCs in a wide range of galaxy environments is essential for studies of star cluster populations and their evolution \citep{ada10, bas12, fall12, bau13, cha14}, the star formation rates and spatially-resolved star formation histories of galaxies \citep{gla10, wof11, cha15, cal15}, and the effects on the evolution of galaxies of the radiative, chemical, and mechanical feedback of YMCs  (e.g., \citealt{cal15b}, hereafter C15b). 

The task of observing large samples of star clusters in galaxies with a wide range of properties was recently completed by the  \textit{Hubble Space Telescope} (\textit{HST}) Treasury program, LEGUS: Legacy Extragalactic Ultraviolet Survey (GO-13364; \citealt{cal15}). LEGUS consists of high spatial resolution ($\sim0.07"$) images of portions of 50 nearby ($\le13$ Mpc) galaxies taken with the UVIS channel of the Wide Field Camera Three (WFC3) in broad band filters F275W (2704 \AA), F336W (3355 \AA), F438W (4325 \AA), F555W (5308 \AA), and F814W (8024 \AA). The survey includes galaxies of different morphological types and spans factors of $\sim10^3$ in both star formation rate (SFR) and specific star formation rate (sSFR), $\sim10^4$ in stellar mass ($\sim10^7-10^{11}\,\rmn{M_\odot}$), and $\sim10^2$ in oxygen abundance ($12+\rmn{log\,O/H}=7.2-9.2$). Some of the targets in the survey have high quality archival images in some filters covering similar bandpasses required by LEGUS, from the Wide Field Channel of \textit{HST}'s Advanced Camera for Surveys (ACS), or in fewer cases, from ACS's High Resolution Channel (HRC). For such targets, LEGUS observed in filters that complete the five band coverage. 

At the distances of  LEGUS galaxies ($3-13\,$Mpc), star clusters cannot be resolved into individual stars, and they usually appear as light over-densities, slightly more extended than the stellar PSF. In such cases, a standard technique for deriving cluster masses and ages is the comparison of observed and computed predictions of the integrated light of clusters in various photometric bands. Such predictions are obtained by convolving spectral synthesis models with filter system throughputs. With regards to populations of massive stars, at fixed star formation history and stellar initial mass function (IMF), a major contributor to uncertainties in spectral synthesis models of young populations are uncertainties in massive-star evolutionary tracks \citep{lei14}. For a given initial metallicity, such tracks provide the mass, temperature, and luminosity of a star as a function of age. Significant uncertainties still remain in massive-star evolution because of our poor knowledge of some complex physical processes, which still require an empirical calibration, such as the efficiency of convective heat transport and interior mixing, and the effects of close-binary interactions. 

In recent years, independent groups working on massive star evolutionary tracks have attempted to reproduce three key observational constraints. First, blue loop stars are observed in the color-magnitude diagrams of nearby metal-poor dwarf irregular star-forming galaxies. The most recent Padova tracks for single non-rotating massive stars computed with the code {\tt{PARSEC}} (Bressan et al. 2012)  are able to reproduce these loops at metallicities of $Z=0.001$ and $Z=0.004$. This is accomplished by enhancing the overshooting at the base of the convective envelope during the first dredge-up, and invoking large mixing lengths of two and four times the pressure scale height \citep{tan14}. New {\tt{PARSEC}} massive-star tracks at other metallicities including solar are presented in \cite{che15}.

Second, nitrogen enhancements are observed on the surfaces of main sequence stars of typically $15\,\rmn{M_\odot}$ \citep{hun09}. In such stars, this product of nuclear burning is not expected at the surface, since these stars do not develop strong stellar winds capable of exposing the central nitrogen. The Geneva tracks for single rotating stars  (\citealt{geo13}; \citealt{eks12}) favor surface nitrogen enhancements. This is because rotation effectively mixes inner and outer stellar layers. 

Finally, it is now well established that massive stars are in binary systems with close to 70 per cent of them interacting over the course of their evolution (e.g. \citealt{mas09};  \citealt{san12,san13};  \citealt{chi12}). Processes that occur during the evolution of binaries include envelope removal from the binary, accretion of mass by the secondary, or even complete mergers (e.g. \citealt{pos92}; \citealt{lan12};  \citealt{dem13,dem14}). Several independent groups have developed evolutionary tracks that account for interacting binaries and are useful for spectral population synthesis models (e.g., \citealt{van98};  \citealt{zha04,zha05}; and references therein). In this work, we use models computed with the Binary Population and Spectral Synthesis code, {\tt{bpass}}, which is last described in Eldridge et al. (in preparation). Hereafter, we will refer to single star and binary tracks that are implemented in {\tt{bpass}} as the Auckland tracks. Compared to other types of evolutionary tracks, those including rotation and interacting binaries require the inclusion of a large number of uncertain physical parameters. However in this work, we use standard rotating and binary models and do not vary any of these uncertain parameters to achieve a better fit.

Another contributor to the uncertainties of spectral synthesis models of massive-star populations are uncertainties in models of massive-star atmospheres, which give the individual spectra of the stars as a function of time. We delay the discussion on this topic to further down in the paper. 
 
In this work, we use near-ultraviolet (NUV) to near-infrared photometry of two available LEGUS galaxies, spectral synthesis models based on seven different flavors of massive star evolution (older and state-of-the-art), and two different prescriptions for attenuation by dust, in order to address five questions: 1) how well do the different spectral synthesis models fit the observations, 2) is there a preferred flavor of massive star evolution, 3) is there a preferred prescription for attenuation by dust, 4) how well do the observations and models constrain the YMC properties, and 5) how different are the YMC properties obtained with the different models. The outline of our work is as follows. In the first four sections we describe the sample (Section 2), observations (Section 3), models (Section 4), and method for comparing models to observations (Section 5). Section 6 presents our results, which are discussed in Section 7. Finally, we provide a summary and conclude in section 8.  

%%%%%%%%%%%%%%%%%%%%%%%%%%%%%%
% Sample
%%%%%%%%%%%%%%%%%%%%%%%%%%%%%%

\section{Sample}\label{sec:sample} 

We select our sample of YMCs using four main criteria and cluster masses and ages determined by Adamo et al. (in preparation, hereafter A16, NGC 1566) and C15b (NGC 5253). The first criterion is that all clusters are detected in the five LEGUS bands. The second criterion is that the ages of clusters are $\le50\,\rmn{Myr}$, which ensures that there are massive stars in the YMCs. For reference, 50 Myr is the approximate main sequence life time of a  single non-rotating 10 M$_\odot$ star \citep{mey94}. The third criterion is that the clusters have masses of $\ge5\times10^4\,$ M$_\odot$, which mitigates the effect of the stochastic sampling of the stellar initial mass function (IMF, \citealt{cer04}). A16 and C15b derive cluster properties using {\tt{yggdrasil}} spectral synthesis models (\citealt{zac11}, http://ttt.astro.su.se/projects/yggdrasil/yggdrasil.html).  The version of {\tt{yggdrasil}} used in the latter papers take model spectra of simple stellar populations (SSPs) computed with {\tt{starburst99}} \citep{lei99, vaz05}, as input to photoionization models computed with {\tt{cloudy}} (last described in \citealt{fer13}). An SSP corresponds to a system where the stars are born instantaneously and are thus coeval. C15b and A16 obtain their cluster properties by using two different sets of older massive-star evolutionary tracks, which are described in detail in section 4, Padova (Po) and Geneva (Go) tracks. In our sample, we include all clusters with Po- or Go-based properties satisfying our mass and age criteria. The last selection criterion is that the average H{\sc\,ii}-region gas-phase oxygen abundance of the host galaxy must correspond closely to a metallicity for which massive-star tracks from the Padova, Geneva, and Auckland collaborations exist. The above collaborations recently released tracks which account for different non-standard physics that are mentioned in the introduction. Since at the time of writing Geneva tracks for rotating stars are only available at $Z=0.002$ \citep{geo13} and $Z=0.014$ \citep{eks12}, we focus on these two metallicity regimes. 

We encountered three main difficulties in assembling a sample satisfying our requirements. Firstly, stellar populations with masses $\ge5\times10^4\,$M$_\odot$ and ages $\le50\,\rmn{Myr}$ are significantly less numerous than less massive clusters of similar ages. Cluster formation is a stochastic process undergoing size-of-sample effects (Larsen 2002). More massive clusters are more likely to form in galaxies with higher SFRs (Whitmore 2000, Larsen 2002) or within longer time scales (Hunter et al 2003). Secondly, for the SFR range covered by the LEGUS sample, we do not expect a large population of YMCs within the mass and age limits imposed in this work. Indeed, YMCs satisfying our mass and age criteria are more frequently found in starburst galaxies, which are a minority in the LEGUS sample \citep{cal15}. Finally, at the time of writing, the LEGUS method for finding true clusters relies on visual inspection in order to remove observational artifacts, contaminants, and false positives. Only a handful of galaxies had been visually inspected by the LEGUS team at the time of this work. Our final sample is composed of six clusters in galaxy NGC 1566 and two clusters in galaxy NGC 5253. Next we provide a brief description of these galaxies, whose main properties, including morphology, redshift, distance, distance modulus, oxygen abundance, SFR, stellar mass, and color excess associated with attenuation within the Milky Way are summarized in Table~\ref{tab1}.

%%%%%%%%%%%%%%%%%%%%%%%%%%%%%%
% Table 1
%%%%%%%%%%%%%%%%%%%%%%%%%%%%%%

\begin{table*} 
\caption{Galaxy properties.}
\label{tab1}
\begin{threeparttable}[b]
\begin{tabular}{lcccccccc}
\hline
NGC & Morph.\tnote{a} & $z$\tnote{b} & Dist.\tnote{c} & $\mu$\tnote{d} & 12+log(O/H)\tnote{e} & SFR(UV)\tnote{f} & M$_*$\tnote{g} & $E(B-V)_{MW}$\tnote{h}\\
\# & \hfill & \hfill & (Mpc) & mag & (PT) / (KK) & ($M_\odot/yr$) &  ($M_\odot$) & mag \\
\hline
1566 &  SABbc & 0.005017 &  13.2 & 30.60 & 8.63 / 9.64 & 5.67 & 2.7E10 & 0.005\\
5253 & Im & 0.001358 &  3.15 & 27.78 & 8.25  & 0.10 & 2.2E8 & 0.049\\
\hline
\end{tabular}
\begin{tablenotes}
\item [a]  Morphological type given by the NASA Extragalactic Database, NED.
\item [b] Redshift from NED.
\item [c] Distance, as listed in table 1 of \cite{cal15} and used in this work. We became aware of a revised distance of NGC 1566 ($\sim18$ Mpc, corresponding distance modulus of 31.28 mag) too late for including it in the present work.
\item [d] Distance modulus.
\item [e] For NGC 1566, globally averaged abundance \citep{mou10}. The two values, (PT) and (KK), are the oxygen abundances on two calibration scales: the PT value, in the left-hand-side column, is from the empirical calibration of \cite{pil05}, the KK value, in the right-hand-side column, is from the theoretical calibration of \cite{kob04}. For NGC 5253, value from \cite{mon12}.
\item [f] Average star formation rate calculated from the GALEX far-UV, corrected for dust attenuation, as described in \cite{lee09}. We note that SFR indicators calibrated for the youth of the starburst in NGC 5253 yield SFR$\sim0.4\,M_\odot$ \citep{cal15b}.
\item [g] Stellar masses obtained from the extinction-corrected B-band luminosity, and color information, using the method described in \cite{bot09} and based on the mass-to-light ratio models of \cite{bel01}.
\item [h] Color excess associated with extinction due to dust in Milky Way. Uses \cite{sch11} extinction maps and \cite{fit99} reddening law with Rv=3.1.
\end{tablenotes}
\end{threeparttable} 
\end{table*} 

\subsection{NGC 1566}

NGC 1566 is a nearly face-on grand-design spiral galaxy that is the brightest member of the Dorado group \citep{agu04}. It has an intermediate-strength bar type (SAB) and hosts a low-luminosity AGN. Its Seyfert classification varies between 1 and 2, depending on the activity phase \citep{com14}. Its globally-averaged gas-phase oxygen abundance is 12+log(O/H)=8.63 or 9.64, depending on the calibration that is given in Table~\ref{tab1}. We select six clusters in this galaxy, excluding the nuclear region because of the presence of the AGN. 

\subsection{NGC 5253}

NGC 5253 is a dwarf starburst galaxy of morphological type Im. It has a fairly flat \citep{wes13} galactocentric profile of the gas-phase oxygen abundance  with a mean value of 12+Log(O/H)=8.25 \citep{mon12}. This value of the oxygen abundance is similar to the one reported by \cite{bre11}, 12+Log(O/H)=8.20$\pm$0.03. Calzetti et al. (2015b) recently studied the brightest clusters in this galaxy. Using extraordinarily well-sampled UV-to-near-IR spectral energy distributions and models which are described in section~\ref{sec:models} below, they obtain unprecedented constraints on dust attenuations, ages, and masses of 11 clusters. In particular they find two clusters with ages and masses satisfying our selection criteria and for which their fits are excellent, i.e., $\chi^2\sim1$. Using their notation, the selected clusters are \#5, which is located within the so-called ``radio nebula"; and \#9, which is  located outside of the ``radio nebula" but still within the starburst region of the galaxy. 

%%%%%%%%%%%%%%%%%%%%%%%%%%%%%%
% Observations
%%%%%%%%%%%%%%%%%%%%%%%%%%%%%%

\section{Observations}

 NGC 1566 was observed in the five LEGUS broad band filters, F275W, F336W, F438W, F555W, and F814W. NGC 5253 was observed in WFC3/UVIS filters F275W and F336W and ACS/HRC filters F435W (4311 \AA), F550M (5578 \AA), and F814W (8115 \AA), where observations in the latter three filters are archival (PID 10609, PI Vacca). The pixel scales of WFC3/UVIS and ACS/HRC are 0.039 and 0.025 arcsec/pixel, respectively. In order to preserve the highest angular resolution, the aligned UVIS images of NGC 5253 were all re-sampled to the pixel scale of HRC. For NGC 1566, photometry is performed with a circular aperture of 0.156" in radius, with the background measured within an annulus of 0.273" in inner radius and 0.039" in width; while for NGC 5253, the aperture is 0.125" in radius and the annulus has an inner radius of 0.5" and width of 0.075". For NGC 5253, C15b provide luminosities, which we first convert to the AB system, and then to the Vega system using conversion factors 1.49, 1.18, -0.10, 0.03, and 0.43 for filters F275W, F336W, F435W, F550M, and F814W, respectively. The conversion factors were computed from the spectrum of Vega that is described in \cite{boh07}. For NGC 1566, the observational errors in the different bands were obtained by summing in quadrature the photometric error produced by IRAF's task PHOT and the standard deviation derived from the aperture correction. For NGC 5253, the observational errors are driven by crowding and uncertainties in the aperture correction. The latter are at the level of $\pm15$\% for the UV--optical filters used here. When combined with other smaller contributions, the total uncertainty amounts to $\pm0.175$ mag. For further details on the data reduction of NGC 1566 and NGC 5253, see A16 and C15b, respectively. The coordinates and photometry of the YMCs are given for each galaxy in Tables~\ref{tab2} and~\ref{tab3}, respectively. 
 
%%%%%%%%%%%%%%%%%%%%%%%%%%%%%%
% Table 2
%%%%%%%%%%%%%%%%%%%%%%%%%%%%%%

\begin{table*}
\centering
\caption{Location \& photometry of clusters in NGC 1566.}
 \label{tab2}
\begin{threeparttable}
 \begin{tabular}{lllllllll}
  \hline
ID  &RA\tnote{a} & DEC\tnote{a}     & F275W\tnote{b}   & F336W\tnote{b}    & F438W\tnote{b}    & F555W\tnote{b}  & F814W\tnote{b} \\
\hfill & \hfill & \hfill & mag & mag & mag & mag & mag \\
\hline
1&64.98079821&-54.93067859&17.465$\pm$0.069&17.816$\pm$0.090&19.298$\pm$0.062&19.356$\pm$0.045&19.248$\pm$0.061\\
2&64.98077227&-54.93850353&17.541$\pm$0.070&17.875$\pm$0.092&19.307$\pm$0.064&19.325$\pm$0.047&19.088$\pm$0.060\\
3&65.01218101&-54.94103612&16.946$\pm$0.067&17.360$\pm$0.088&18.774$\pm$0.059&18.848$\pm$0.041&18.608$\pm$0.054\\
4&65.02378180&-54.94377250&18.410$\pm$0.069&18.682$\pm$0.090&20.054$\pm$0.061&20.069$\pm$0.043&19.800$\pm$0.058\\
5&65.00041201&-54.94429422&20.966$\pm$0.084&20.575$\pm$0.092&21.448$\pm$0.067&21.051$\pm$0.042&20.376$\pm$0.057\\
6&65.02121621&-54.95053748&17.697$\pm$0.069&18.047$\pm$0.089&19.506$\pm$0.060&19.554$\pm$0.042&19.406$\pm$0.055\\ 
\hline
\end{tabular}
\begin{tablenotes}
\item [a] Right ascension (RA) and declination (DEC) in decimal notation (J2000). The values were obtained from the F555W frame (aligned and registered).
\item [b] Apparent Vega magnitudes and photometric errors. The listed photometry is from a 4-pixel radius aperture and is corrected to infinite aperture and for foreground Milky Way extinction using E(B-V)=0.008. We use an average aperture correction. 
\end{tablenotes}
\end{threeparttable} 
\end{table*}

%%%%%%%%%%%%%%%%%%%%%%%%%%%%%%
% Table 3
%%%%%%%%%%%%%%%%%%%%%%%%%%%%%%

\begin{table*}
\centering
\caption{Location \& photometry of clusters in NGC 5253.}
 \label{tab3}
\begin{threeparttable}
 \begin{tabular}{llllllll}
  \hline
  ID  &RA\tnote{a} & DEC\tnote{a}     & F275W\tnote{b}   & F336W\tnote{b}    & F435W\tnote{b}    & F550M\tnote{b}  & F814W\tnote{b} \\
\hfill & \hfill & \hfill & mag & mag & mag & mag & mag \\
\hline
   5 & 204.98328 & -31.64015 & 17.048	$\pm$0.175 &	17.072	$\pm$0.175 &	18.013	$\pm$0.175 &	17.938	$\pm$0.175 &	16.905	$\pm$0.175\\
   9 & 204.9813 & -31.64149 & 16.877	$\pm$0.175 &	17.112	$\pm$0.175 &	17.851	$\pm$0.175 &	17.517	$\pm$0.175 &	16.783	$\pm$0.175\\
\hline    \end{tabular}
\begin{tablenotes}
\item [a] Right ascension (RA) and declination (DEC) in decimal notation (J2000). The values were obtained from the F336W image.
\item [b] Apparent Vega magnitude and photometric error based on luminosities reported in Calzetti et al. (2015b). The listed photometry is from a 5-pixel radius aperture and is corrected to infinite aperture and for foreground Milky Way extinction using E(B-V)=0.049. 
\end{tablenotes}
\end{threeparttable} 
\end{table*}

\begin{figure*}
\centering
\begin{subfigure}
  \centering
  \includegraphics[width=.99\columnwidth]{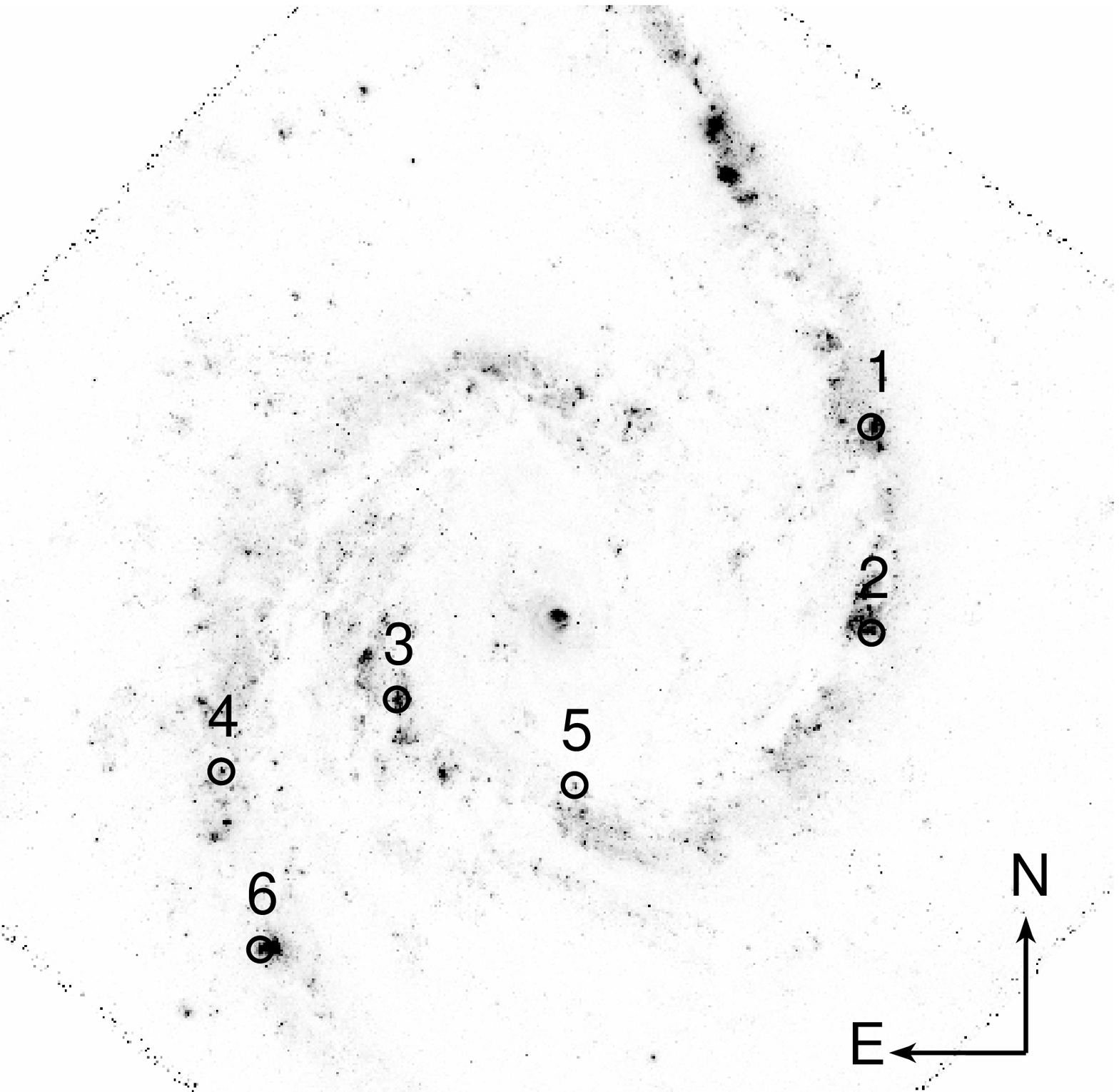}
\end{subfigure}%
\begin{subfigure}
  \centering
  \includegraphics[width=.99\columnwidth]{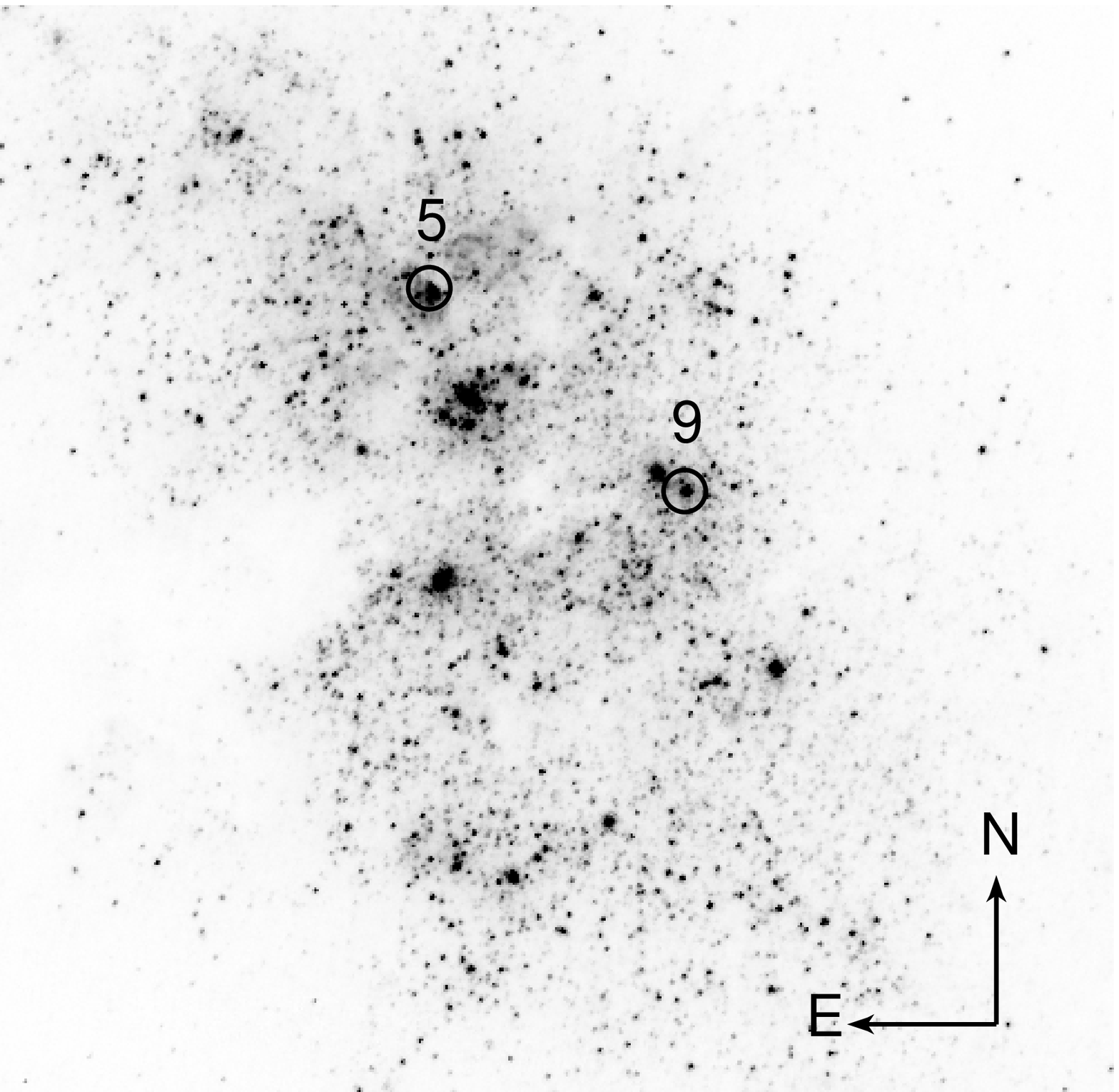}
\end{subfigure}
\caption{WFC3 UVIS F275W images of galaxies NGC 1566 (left) and NGC 5253 (right). We mark the locations of clusters with circles of radii 1.6" ($\sim$100 pc, NGC 1566) and 1.25" ($\sim$19 pc, NGC 5253). Photometry was extracted from circles of radii 10 times smaller. North is up and east is to the left.}
\label{fig1}
\end{figure*}

Figure~\ref{fig1} shows LEGUS NUV images of the two galaxies, where we have marked the locations of clusters in our sample. For NGC 1566, the final LEGUS IDs of clusters were not available at the time of writing and thus we assign our own IDs, while for NGC 5253, we use the IDs of C15b, for easier comparison with their work. 

%%%%%%%%%%%%%%%%%%%%%%%%%%%%%%
% Figure 1
%%%%%%%%%%%%%%%%%%%%%%%%%%%%%%

%%%%%%%%%%%%%%%%%%%%%%%%%%%%%%
%%% Models
%%%%%%%%%%%%%%%%%%%%%%%%%%%%%%

\section{Models}\label{sec:models}

The spectral synthesis models computed in this paper account for the contributions of stars, the ionized gas, and dust. For a more meaningful  comparison of the results obtained in this paper with the models used to select the clusters in our sample, we adopt the same stellar and nebular parameters as in the latter models, which are given in \cite{zac11}. In particular, we use simple stellar populations (SSPs) with initial masses of $10^6\,M_\odot$ and an IMF such that the number of stars in the mass range [m, $m+dm$] is given by $N(m)\,dm\propto m^{-\alpha}$, where $\alpha=1.3$ and $\alpha=2.35$ in the mass ranges 0.1--0.5\,M$_\odot$ and 0.5--100\,M$_\odot$, respectively. We compute models for ages in the range $6\leq\log(t/{\rm yr})\leq9$ in steps of 0.1. In our approach, the stars and ionized gas have the same metallicity. Since some of the stellar evolutionary tracks that we try are only available at two metallicities ($Z=0.002$ and $Z=0.014$), in our Bayesian fitting approach, metallicity is not a free parameter. Depending on the massive-star evolution flavor, we use tracks corresponding to metallicities of $Z=0.002$ or $Z=0.004$ (NGC 5253) and $Z=0.014$ or $Z=0.020$ (NGC 1566). These metallicities roughly correspond to the average gas-phase metallicities of the galaxies in our sample, as gauged by their oxygen abundances (see Table~\ref{tab1}).\footnote{The highest O/H value of NGC 1566 given in Table~\ref{tab1} yields a metallicity much larger than $Z=0.02$ when adopting the \cite{asp09} solar abundance scale. We exclude this O/H value and adopt the lowest value corresponding to $Z=0.014$, because some of the evolutionary tracks used in this work are not available at $Z>0.014$. However, if available, we use $Z=0.020$ instead, because it is in between the high/low O/H values quoted for this galaxy.} The elemental abundances in the ionized nebula are scaled solar abundances. We use the reference solar abundance set of Asplund et al. (2009), which corresponds to $Z_\odot=0.014$ and $12+{\rmn{log(O/H)}}=8.69$. In the next subsections, we expand on the different model components, explain our procedure for computing medium- and broad-band magnitudes, and discuss the impact of the ionized gas on the predicted magnitudes and colors. 

\subsection{Stellar component}

In this work, we focus on how updates in models of massive-star evolution affect the derived properties of YMCs. For this purpose, we test different generations and flavors of massive-star evolution (seven in total), where the different flavors account for different astrophysics. The main differences between these models lie in the stellar evolutionary tracks and stellar atmospheres employed. 

\subsubsection{Massive-star evolutionary tracks}

We compute models using the following massive-star evolutionary tracks: 1) older Padova for single non-rotating stars (\citealt{bre93}, $Z=0.020$; \citealt{fag94}, $Z=0.004$); 2) older Geneva for single non-rotating stars with high-mass loss (\citealt{mey94}, $Z=0.020$ and $Z=0.004$); 3) new Padova for single non-rotating stars; 4) new Geneva for single non-rotating stars; 5) new Geneva for single rotating stars; 6) new Auckland for single non-rotating stars; and 7) new Auckland for interacting binaries. Hereafter, we will use the IDs provided in the first column of Table~\ref{tab4} to refer to these tracks and to spectral synthesis models which are based on these tracks. Table~\ref{tab4} also gives the name of the city associated with the tracks (column 2), a description of the type of stellar evolution (column 3), the name of the spectral population synthesis code where each flavor of tracks is implemented (column 4), the metallicity of the tracks (column 5), the reference for each set of tracks (column 6), and the references for the population synthesis codes where the different sets of tracks are implemented (column 7). In this work, we do not account for pre-main sequence stars. The implications of this approach are discussed in section~\ref{sec:discussion}.

%%%%%%%%%%%%%%%%%%%%%%%%%%%%%%
% Table 4
%%%%%%%%%%%%%%%%%%%%%%%%%%%%%%

\begin{table*} 
\centering
\caption{Massive-star evolutionary tracks and spectral synthesis codes.}
 \label{tab4}
\begin{threeparttable}
\begin{tabular}{lllllllll}
\hline
ID\tnote{a} &city of&type of track\tnote{c}& spectral\tnote{d} &  metallicity\tnote{e}  & Ref.\tnote{f} & Ref.\tnote{g} \\
\hfill&tracks\tnote{b}&\hfill&synthesis&tracks&pop. syn.\\
(1)&(2)&(3)&(4)&(5)&(6)&(7)\\
\hline
Po &Padova & single non-rotating&{\tt{starburst99}}&0.004 / 0.020&1&1\\
Go    &Geneva & single non-rotating&{\tt{starburst99}}&0.002 / 0.020&2&1\\
Pn&Padova & single non-rotating&{\tt{galaxev}}&0.004 / 0.014&3&2\\
Gn&Geneva & single non-rotating&{\tt{starburst99}}&0.002 / 0.014&4&1\\
Gr&Geneva& single rotating\tnote{k}&{\tt{starburst99}}&0.002 / 0.014&4&1\\
An&Auckland & single non-rotating&{\tt{bpass}}&0.004 / 0.014&4&3\\
Ab&Auckland & interacting binaries&{\tt{bpass}}&0.004 / 0.014&4&3\\
\hline
\end{tabular}
\begin{tablenotes}
\item [a] ID of set of tracks and models based on corresponding set of tracks. The first letter is the letter of the city of the tracks. We use "o"/"n" to designate older/newer versions of the tracks. 
\item [b] City where tracks were computed.
\item [c] Type of evolution of massive stars.
\item [d] Spectral synthesis code where tracks are implemented.
\item [e] Metallicity of tracks. We use the low value for NGC 5253 [12+log(O/H)=8.25] and the high value for NGC 1566 [12+log(O/H) = 8.63 - 9.64, depending on calibration]. Using the solar photospheric abundances of \cite{asp09}, the correspondence between metallicity and oxygen abundance is: $Z=0.002$ [12+log(O/H)=7.84], $Z=0.004$ [12+log(O/H)=8.15], $Z=0.014$ [12+log(O/H)=8.69], and $Z=0.020$ [12+log(O/H)=8.84]. 
\item [f] Reference for the massive-star tracks. 1.  \citealt{bre93} ($Z=0.020$), \citealt{fag94} ($Z=0.004$). 2. \citealt{mey94} (high-mass loss). 3. \citealt{tan14} ($Z=0.004$); Chen et al ($Z=0.014$, 2014b in prep.; high metallicity). 4. \citealt{eks12} ($Z=0.014$), \citealt{geo13} ($Z=0.002$). 5. Eldridge et al. 2008, in prep. ($Z=0.004$, $Z=0.014$).
\item [g] References and websites of population synthesis codes: 1. \citep{lei14}; http://www.stsci.edu/science/starburst99/docs/default.htm. 2. (\citealt{bc03}; Charlot \& Bruzual in prep.; www.iap.fr/~charlot/bc2003).  3. Eldridge et al. (2009, 2012, in prep.); http://bpass.auckland.ac.nz/index.html
\item[h] Rotation velocity is 40 per cent of the break-up velocity on the zero-age main sequence.
\end{tablenotes}
\end{threeparttable} 
\end{table*}  

As previously mentioned, we use cluster ages and masses obtained by A16 (NGC 1566) and C15b (NGC 5253) with Po and/or Go tracks for selecting the YMCs in our sample. For a more homogeneous comparison with the rest of models in the present paper, we re-determined Po and Go based cluster properties. This is because A16 and C15b use different approaches for fitting the observations to the Bayesian approach used in the present work. In Sections 6 and 7 we present our results and compare them to values reported in A16 and C15b.

\subsubsection{Binary Tracks}

The Auckland binary tracks used in this work are described in detail in Eldridge et al. (in preparation). In summary, the {\tt{bpass}} models are synthetic stellar populations that include the evolutionary pathways from interacting binary stars that rotate. The stellar models are based on a version of the Cambridge {\tt{stars}} code, originally created by \cite{egg71} and the version employed here is described in \cite{eld08}. The combination of the evolution models into a synthetic population are described in \cite{eld09} and \cite{eld12}. However the version used here has been significantly improved since these works and are v2.0 to be described in detail in Eldridge et al. (in preparation). The improvements include increasing the number of stellar evolution models that can be calculated and the number of stellar models at each metallicity.

The models cover a range of masses and include mass-loss rates by stellar winds from \cite{vin01}, \cite{dej88}, and \cite{nug00}. The stellar wind mass-loss rates are scaled from Z=0.020. This is because in nearby massive stars the mass-loss rates are more likely to have a similar composition to other nearby massive stars, as deduced by \cite{nie12}, which is closer to a metal mass fraction of $Z=0.020$ rather than $Z=0.014$. However the mass loss of the population is dominated by binary interactions and so the stellar winds only have a secondary effect.

The range of initial parameter distributions for the binary stars are similar to those inferred from binary populations by \cite{san12}. Other enhancements include using the full grid from the Potsdam Wolf-Rayet (WR) model atmosphere grid. Classical Wolf-Rayet stars have lost their hydrogen envelope and can be the evolved descendants of: a) single stars with initial mass above a threshold ($\sim25\,M_\odot$ at $Z=0.020$) which increases as metallicity decreases; b) secondary stars of interacting binary systems which are initially less massive than this limit and have accreted enough mass to reach it; and c) primary stars of interacting binary systems initially less massive than this limit, whose hydrogen envelope has been transferred to the secondary. The model atmosphere grid includes lower metallicities and does not rely on  an extrapolation of the Solar metallicity grid to these lower metallicities, as in the past. This has greatly improved the accuracy of the predicted spectra at young ages when WR stars are a significant contribution to the observed light. 

\subsubsection{Stellar Atmospheres}

In addition to stellar evolutionary tracks, spectral population synthesis requires spectra of individual stars, which in our case come from various empirical and/or theoretical libraries. Theoretical libraries use stellar atmospheres that are characterized by parameters: luminosity, effective temperature, mass loss rate, surface gravity and chemical composition of the atmosphere. Differences in stellar atmospheres can affect predictions of stellar-population fluxes in broad- and medium-band filters via i) differences in stellar opacities, which at fixed stellar parameters (effective temperature, surface gravity, and metallicity) slightly affect the global shape of the spectrum; and ii) differences in fluxes of stellar spectral lines, which in our case is not an issue because, a) LEGUS bands do not contain strong stellar lines, and b) we use low-resolution spectra to convolve with filters. For each population synthesis code used in this work, Table~\ref{tab5} provides references for the empirical and theoretical libraries and codes used for the massive star spectra in this work. We include this information for B main-sequence stars, O main-sequence stars, WR stars, and cooler red supergiants (RSG). 

%%%%%%%%%%%%%%%%%%%%%%%%%%%%%%
% Table 5
%%%%%%%%%%%%%%%%%%%%%%%%%%%%%%

\begin{table*} 
\caption{Stellar Atmospheres.}
\label{tab5}
\begin{threeparttable}[b]
\begin{tabular}{lllll}
\hline
spectral & B MS & O MS & WR & RSG \\
synthesis &  \hfill & \hfill & \hfill \\
\hline
{\tt{galaxev}} & {\tt{Tlusty}}\tnote{a} & {\tt{Tlusty}} + {\tt{WM-Basic}}\tnote{b} & {\tt{PoWR}}\tnote{c} & {\tt{Miles}}\tnote{d} + {\tt{UVBLUE}}\tnote{e} \\
{\tt{starburst99}} & {\tt{ATLAS9}}\tnote{f} & {\tt{WM-Basic}} &  {\tt{CMFGEN}}\tnote{g} &  {\tt{BaSeL v3.1}}\tnote{h} \\
{\tt{bpass}} & {\tt{BaSeL v3.1}} & {\tt{WM-Basic}} & {\tt{PoWR}} & {\tt{BaSeL v3.1}} \\
\hline
\end{tabular}
\begin{tablenotes}
\item [a]  {\tt{Tlusty}} (\citealt{hub88}; \citealt{hub93}; \citealt{hub94}).
\item [b] {\tt{WM-Basic}} (\citealt{pau01}; \citealt{smi02}).
\item [c] {\tt{PoWR}} (\citealt{gra02}; \citealt{ham03}; \citealt{ham04}). 
\item [d] {\tt{Miles}} (RSG cooler than $10^4$ K; optical; \citealt{san06}; \citealt{fal11}) 
\item [e] {\tt{UVBLUE}} (RSG cooler than $10^4$ K; UV; \citealt{rod05}; http://www.inaoep.mx/~modelos/uvblue/uvblue.html).
\item [f] {\tt{ATLAS9}} \citep{kur92}
\item [g] {\tt{CMFGEN}} (\citealt{hil98}; \citealt{smi02}).
\item [h] {\tt{BaSeL v3.1}} (\citealt{lej97}; \citealt{wes99}).
\end{tablenotes}
\end{threeparttable} 
\end{table*} 

The PoWR atmospheres used in this work include WR stars that show Balmer absorption lines as the ones seen by \cite{dri95} in NGC 3603, and which are believed to be main sequence core hydrogen burning stars, rather than evolved stars \citep{con95, dek97}. Such stars have been found in clusters with ages $1-3$ Myr. In models, these stars have luminosities of log $L>6$ under Galactic metallicity. In {\tt{galaxev}}, for the 100 $M_\odot$ tracks for instance, all WR phases (WNL, WNE, and WC) occur at log $L>6$. In {\tt{bpass}}, for a hydrogen mass fraction of $X>0.4$, H-burning WR stars appear mostly as O stars but contribute to He{\sc\,ii} wind emission lines. For $X<0.4$, H-burning stars are included but there are not many of them. The evolution of these stars might be very different at low metallicities (cf. Hainich et al. 2015).

With regards to RSGs, an important effect of binary evolution is to reduce the number of RSGs to about a third of the number predicted by single star evolution models. In addition, the atmosphere models for RSGs are less important than how cool the stellar models become, which is linked to the details of the assumed mixing length. Currently, Auckland binary models evolve to much cooler temperatures than single-star models, as the stars approach core-collapse. Work is in progress trying to understand the temperature of resolved nearby RSGs as well as the temperatures of  RSG progenitors.
 
\subsubsection{Ionizing Fluxes of Stellar Populations}

\begin{figure*}
\begin{subfigure}
\centering
\includegraphics[width=0.99\columnwidth]{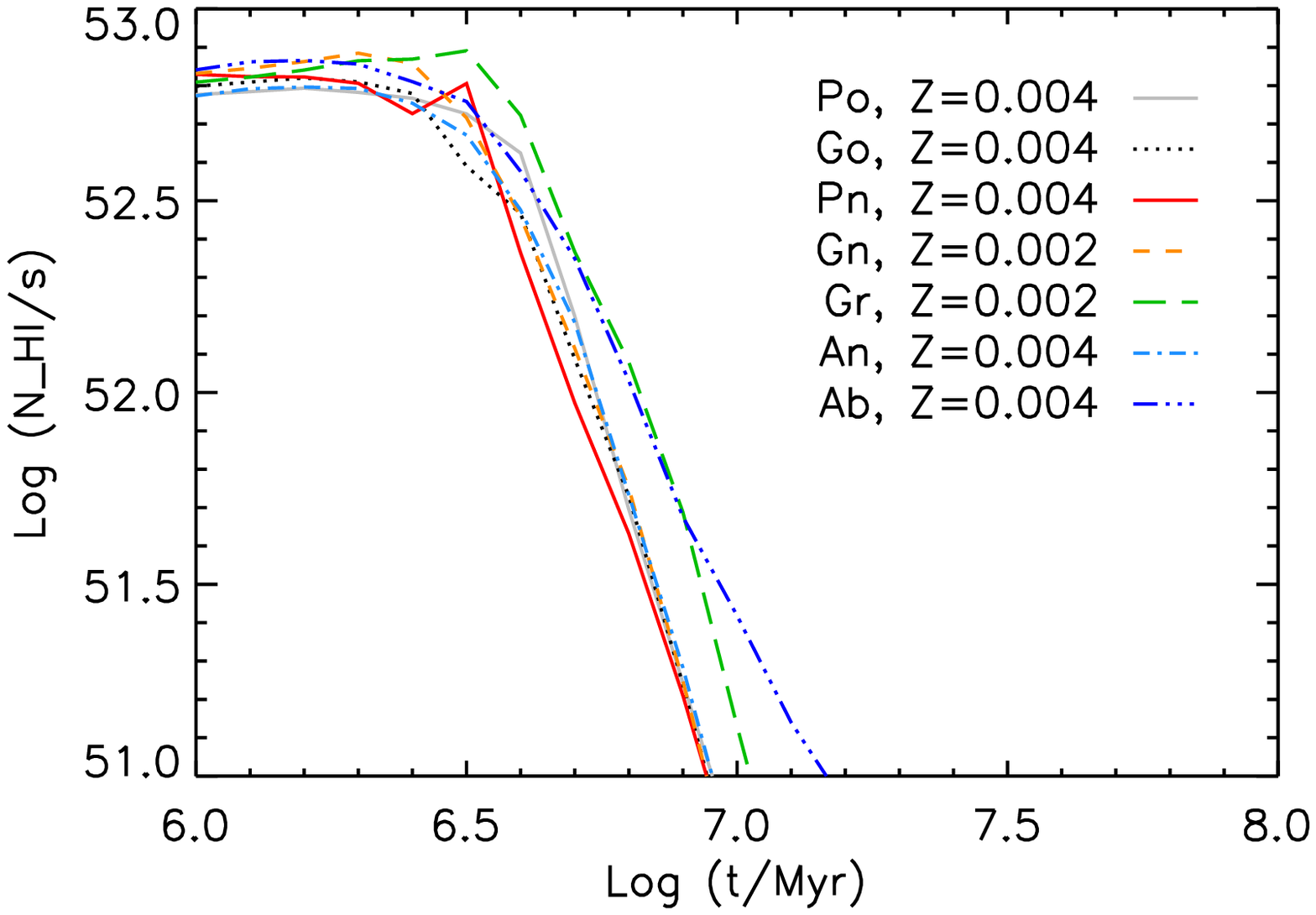}
\end{subfigure}
\begin{subfigure}
\centering
\includegraphics[width=0.99\columnwidth]{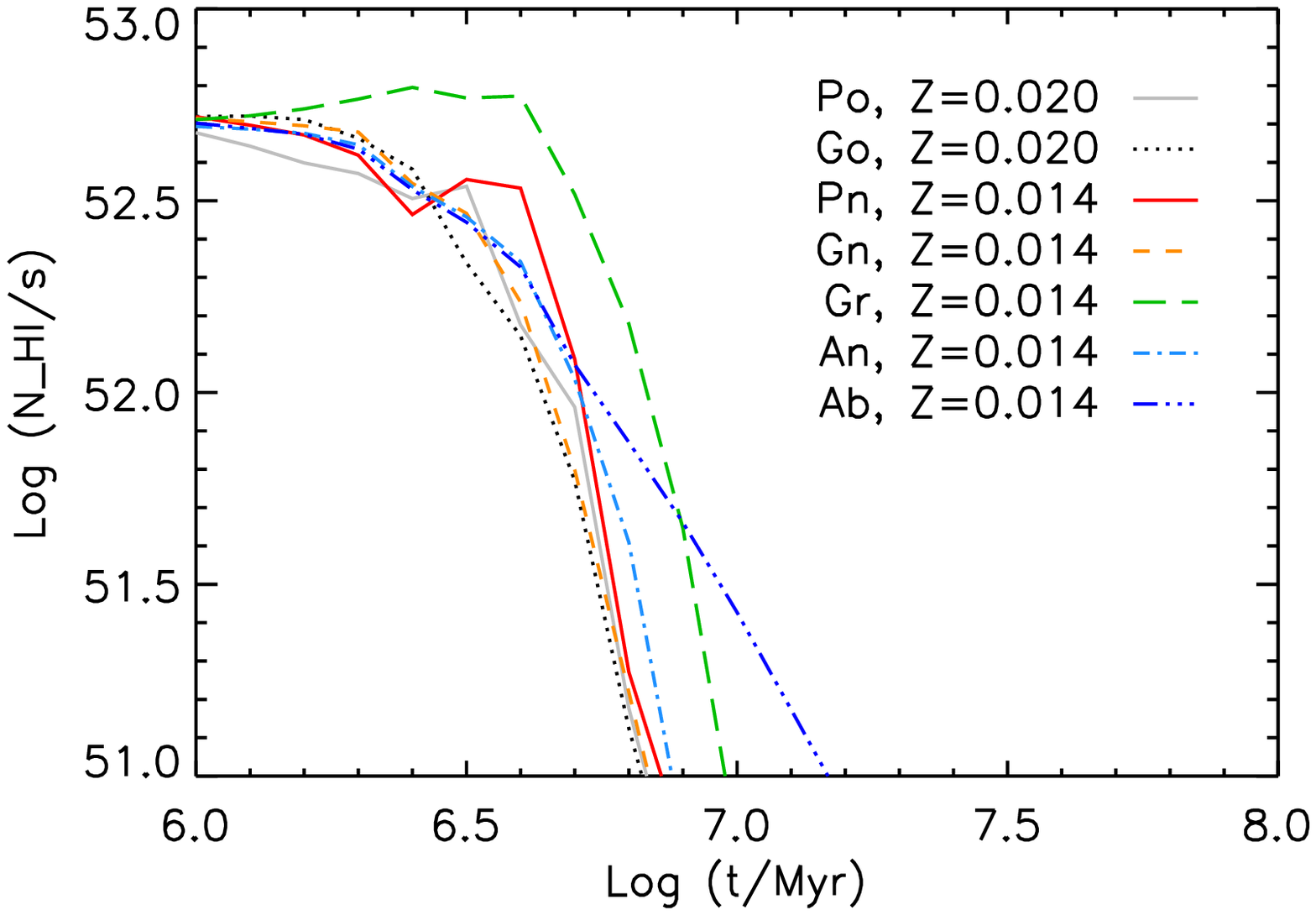}
\end{subfigure}
\begin{subfigure}
\centering
\includegraphics[width=0.99\columnwidth]{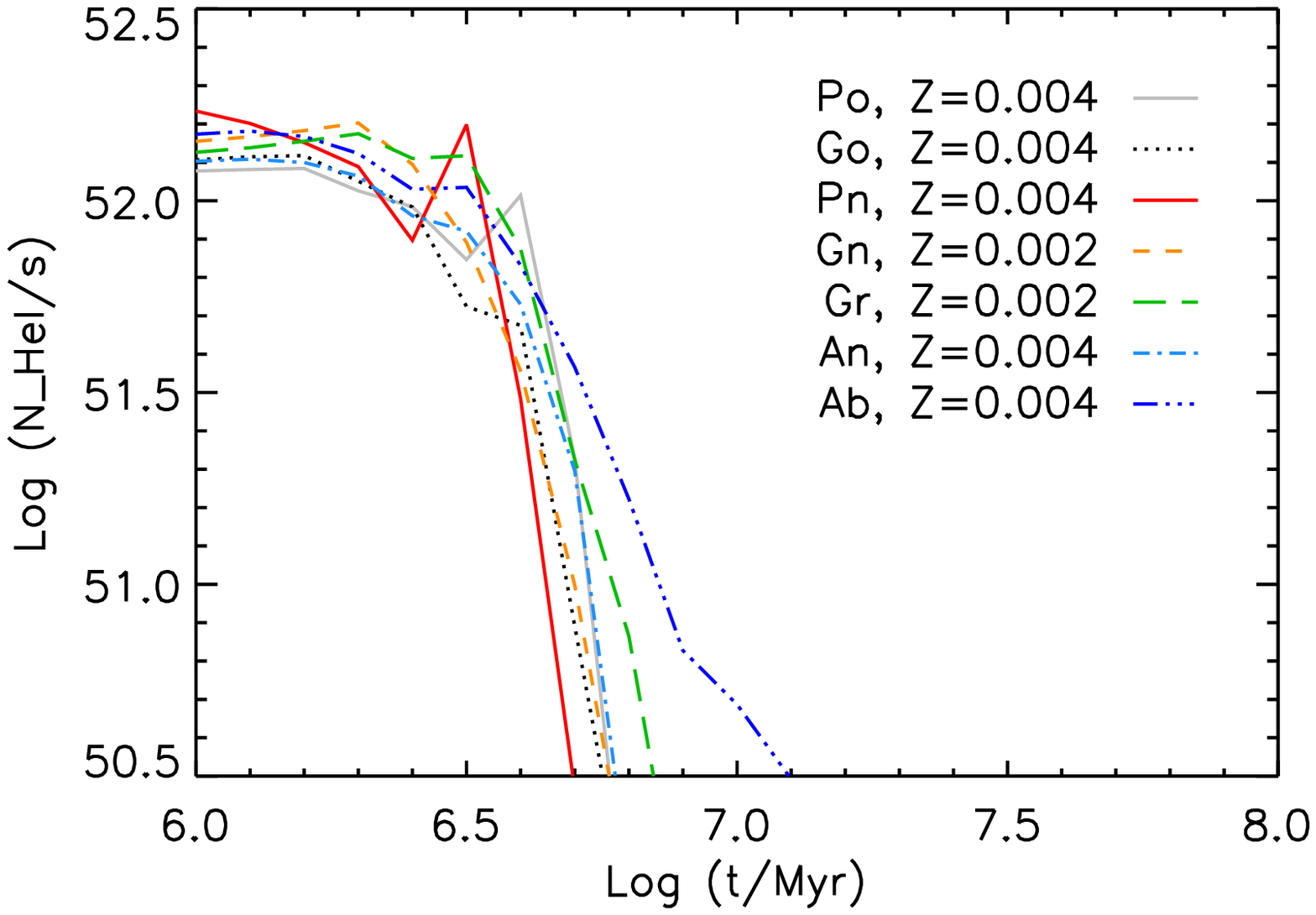}
\end{subfigure}
\begin{subfigure}
\centering
\includegraphics[width=0.99\columnwidth]{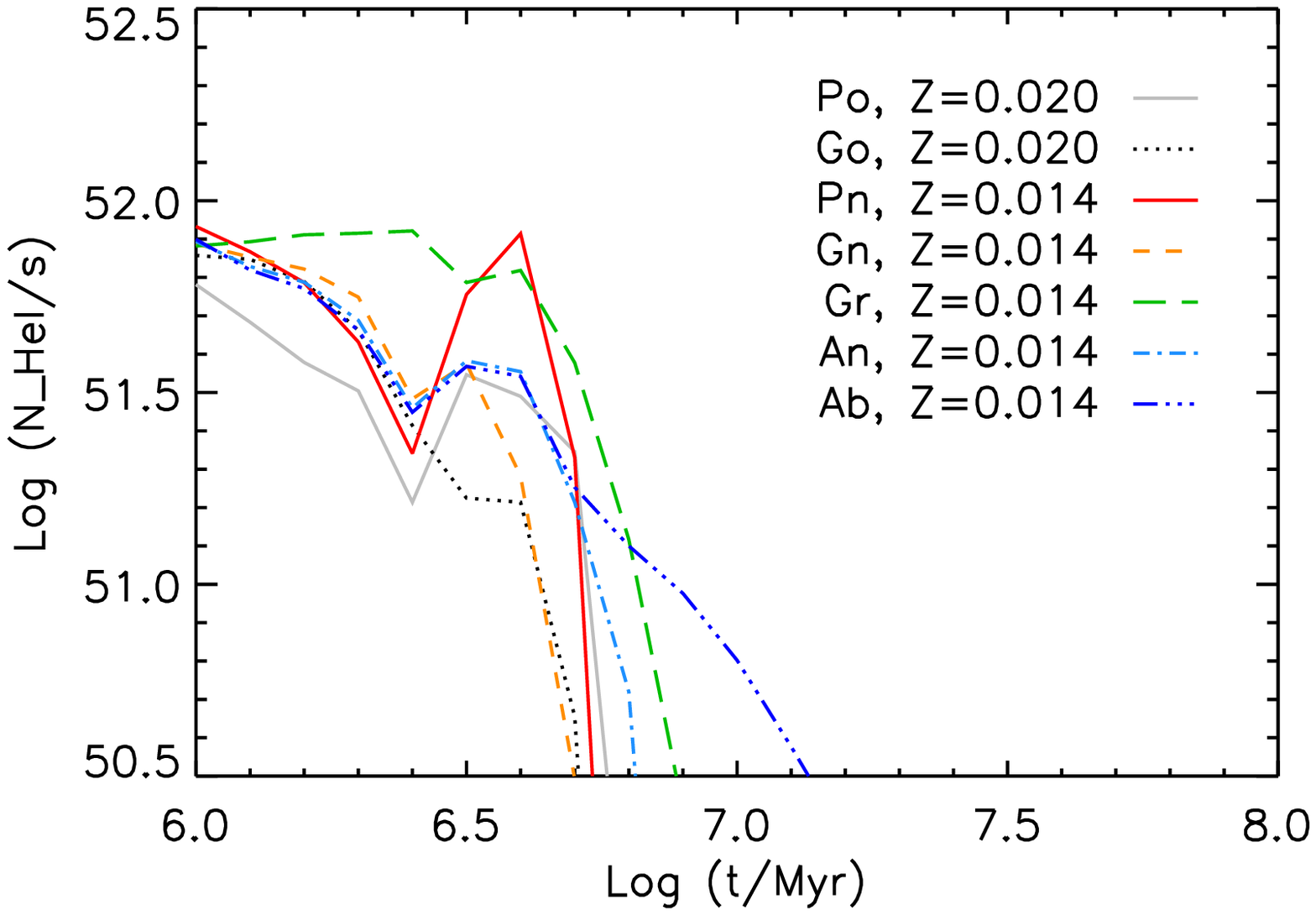}
\end{subfigure}
\begin{subfigure}
\centering
\includegraphics[width=0.99\columnwidth]{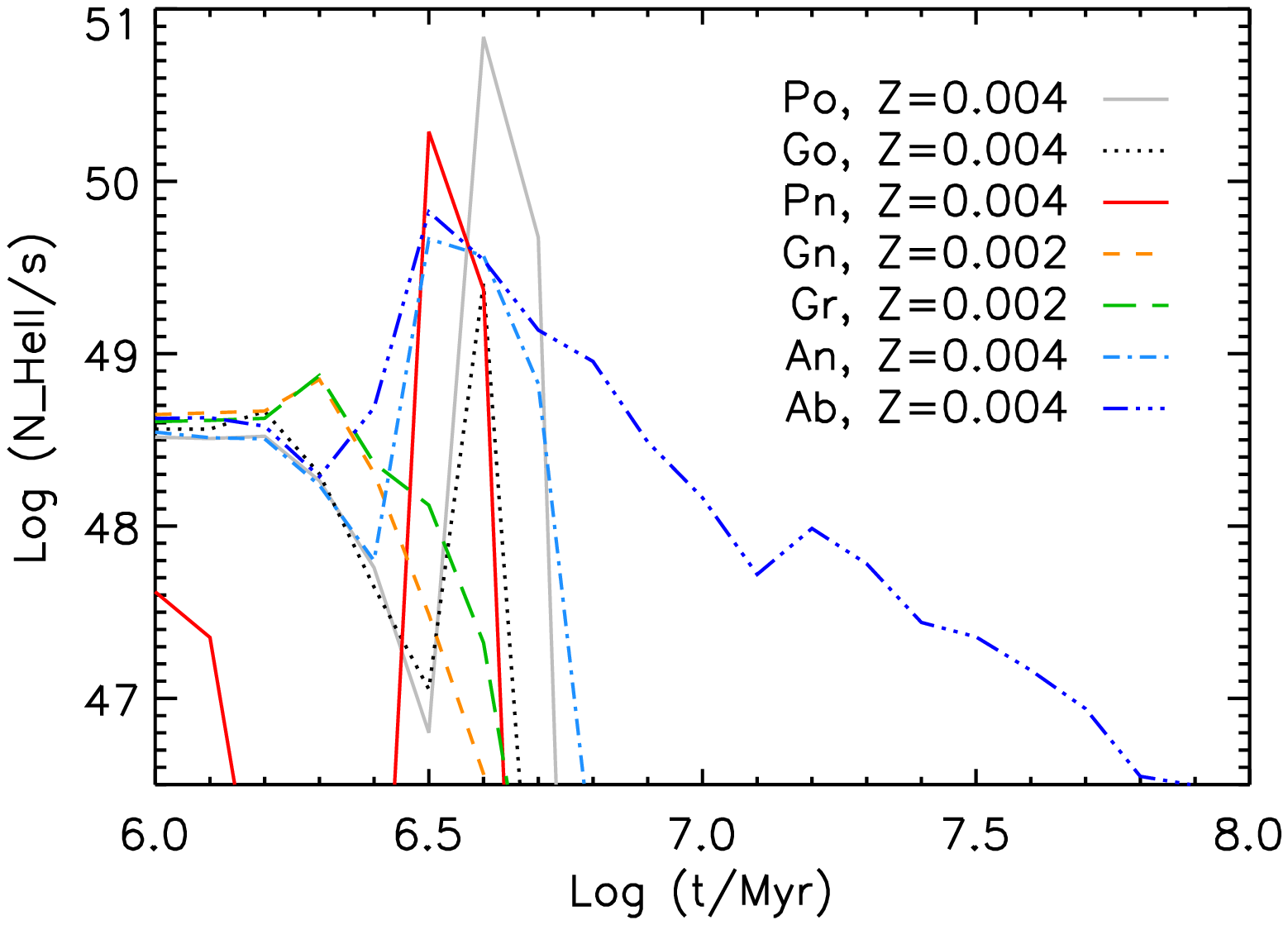}
\end{subfigure}
\begin{subfigure}
\centering
\includegraphics[width=0.99\columnwidth]{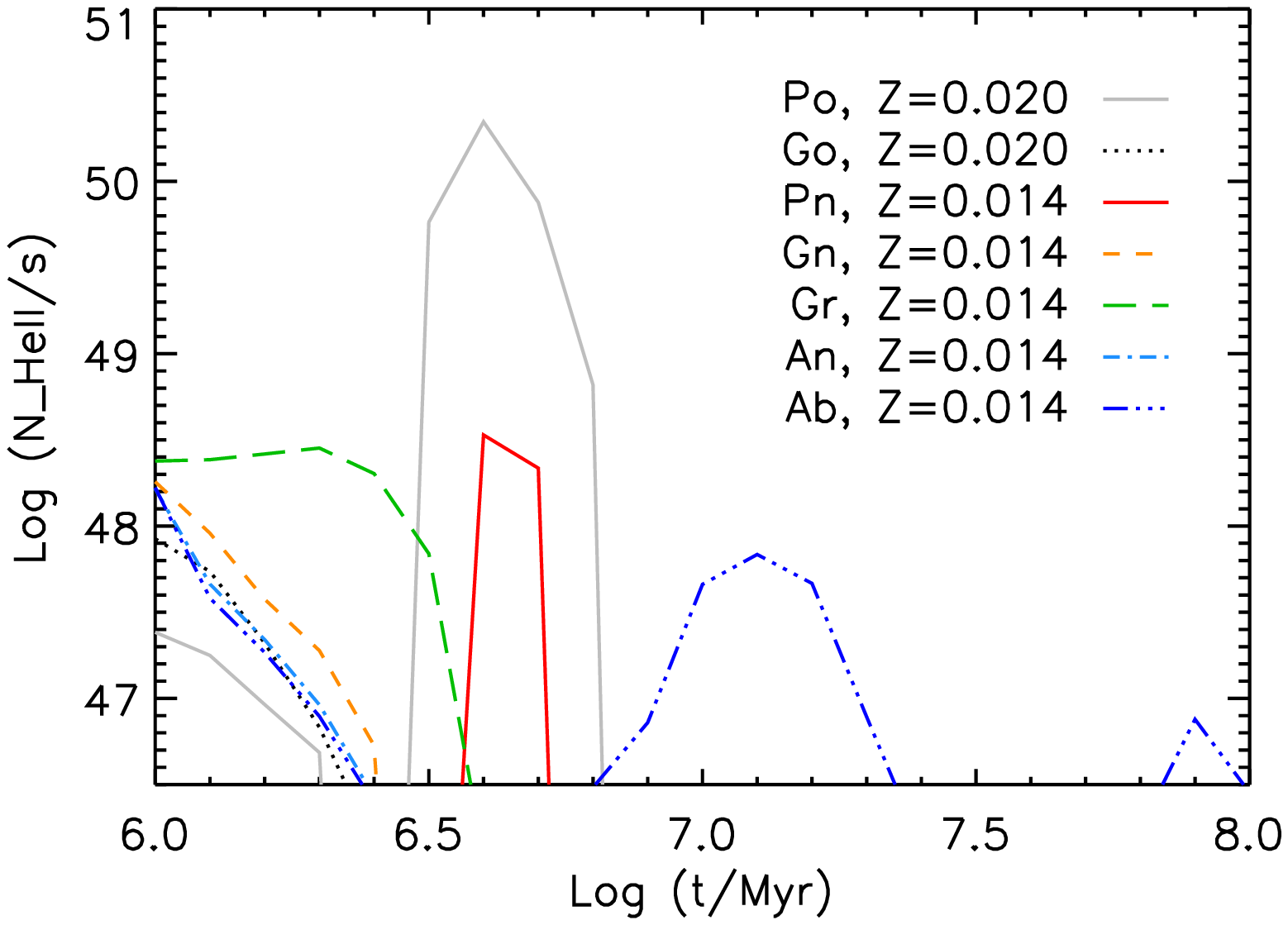}
\end{subfigure}
\caption{Evolution of the number of photons emitted per second in the H{\sc\,i}, He{\sc\,i}, and He{\sc\,ii} ionizing ranges (from top to bottom, respectively) by SSPs of $M_{\rm{cl}}=10^6\,$M$_\odot$. We show predictions from models that are based on different stellar evolution prescriptions and metallicities (lines of different styles, as specified in the legends). The low and high metallicity cases are shown on the left and right panels, respectively.}
\label{fig2}
\end{figure*}

Figure~\ref{fig2} shows the evolution of the H{\sc\,i}, He{\sc\,i}, and He{\sc\,ii} ionizing rates (number of photons per second) for different combinations of stellar evolution model and metallicity. We show results for SSPs of $10^6\,$M$_\odot$ in initial stellar mass. The H{\sc\,i} ionizing rates are higher at low compared to high metallicity because at low metallicity, O stars are hotter and their mass loss rates lower than at high metallicity. The lower mass loss rates make the winds less dense so there is less wind blanketing. The greater wind transparency means that more ionizing photons escape. Figure~\ref{fig2} also shows that the H{\sc\,i} ionizing rate of rotating models can be factors of a few larger than all other single-star models from 3 to $\sim$10 Myr, which was pointed out by \cite{lei14} when comparing Gr and Gn models.  Finally, Fig.~\ref{fig2}  shows that, while rotation and binaries both produce more ionizing photons, for Gr models, these are concentrated at a young age, while for Ab models, the ionizing flux is sustained to ages older than 10 Myr. This is due to three main processes: rejuvenation, mergers and envelope removal. The first two processes increase the mass of the secondary and primary star respectively, causing more massive stars to occur at later ages than expected in a single-star models. This is the dominant process leading to more hydrogen ionizing photons at late times in the binary-star models of Fig.~\ref{fig2}. The third process leads to the creation of helium stars, and low-luminosity WR stars at later ages than normally possible in single-star models.

%%%%%%%%%%%%%%%%%%%%%%%%%%%%%%
% Figure 2
%%%%%%%%%%%%%%%%%%%%%%%%%%%%%%

\subsection{Nebular component}

To compute the contribution of the ionized gas to the photometric filters used in this paper, we use version 13.03 of photoionization code {\tt{cloudy}} and adopt the nebular parameters of \cite{zac11}, which are the nebular parameters used by C15b and A16. The nebular parameters are: hydrogen density, $n_{\rmn{H}}=100\,\rmn{cm}^{-3}$; inner cloud radius, $R_{\rmn{in}}=100\,R_\odot(L/L_\odot)^{1/2}$, where $L$ is the bolometric luminosity of the model stellar population; gas filling factor, $f_{\rmn{fill}}=0.01$, meaning that the nebula's Str\"omgren radius, $R_{\rmn{S}}$, is given by $R_{\rmn{S}}^3=3\,Q_{\rmn{H}}/(4\,\pi\,n_{\rmn{H}}^2\,f_\rmn{fill}\,\alpha_B)$, where $Q_{\rmn{H}}$ is the number of ionizing photons per second and $\alpha_B$ the case-B hydrogen recombination coefficient (e.g., \citealt{cha01}); spherical ionized nebula, meaning that $\Delta\,r\sim\,R_{\rmn{S}}$, where $\Delta\,r$ is the the thickness of the nebula; and dust-free nebula with no depletion of elements in the gas onto dust grains. In our case, $Q_{\rmn{H}}$ is set by the spectral shape and total luminosity of the input SSP spectrum. For a description of how the above nebular parameters relate to the volume averaged ionization parameter, which is a quantity not used in the present work but often used to describe models of spherical ionized nebulae, we refer the reader to equation 3 of \cite{pan03}.

\subsection{Attenuation by dust}

In order to perform a meaningful comparison with the results of A16 and C15b, we do not include dust in the ionized gas (see previous section), but we do account for the effects of dust on the emergent spectrum, by means of either an extinction or an attenuation curve.  Extinction refers to the effect of a uniform dust screen in front of the stars, parametrized as $F(\lambda)_{out}=F(\lambda)_{model}10^{[-0.4*E(B-V)*k(\lambda)]}$, where $k(\lambda)$ is  the extinction curve \citep[e.g.][]{cal00}; while attenuation includes the effects of scattering on the absorption probability of photons in mixed  stars-gas geometries \citep[e.g.][]{wil11}. For each galaxy, we try the starburst attenuation law of Calzetti et al. (2000, we use a total to selective extinction $R_{\rmn{V}}=4.05$); and an extinction law based on the galaxy's metallicity. Specifically, we use the Milky Way (MW) extinction law of \cite{mat90} for clusters in NGC 1566, and the Small Magellanic Cloud (SMC) extinction law of \cite{gor03} for clusters in NGC 5253. In all cases, we assume equal attenuation of the stars and ionized gas. We consider $E(B-V)$ values in the range from 0 to 3, in steps of 0.01 mag. 

\subsection{Synthetic Magnitudes}

We obtain magnitudes in the Vega system, $m[z,t(z)]$, by convolving the model spectra at redshift $z$, $L_\lambda[\lambda(1+z)^{-1},t(z)]$, expressed in units of luminosity per unit wavelength, with the system filter throughputs, $R(\lambda)$. For each LEGUS/WFC3/UVIS filter or close ACS/HRC filter, the throughput curve was downloaded from the Space Telescope Science Institute website http://www.stsci.edu/$\sim$WFC3/UVIS/SystemThroughput/ or http://www.stsci.edu/hst/acs/analysis/throughputs, respectively. Using the Vega spectrum $C(\lambda)$ of \cite{boh07}, we write (e.g. \citealt{bc03})
\begin{equation}
m[z,t(z)]=-2.5\,\rmn{log}\,\frac{\int\limits_{-\infty}^\infty d\lambda\,\lambda\frac{L_\lambda[\lambda(1+z)^{-1},t(z)]}{(1+z)4\pi\,d_L^2(z)}\,R(\lambda)}{\int\limits_{-\infty}^\infty d\lambda\,\lambda\,C_\lambda(\lambda)R(\lambda)}.
\end{equation}
Dust attenuation is applied to the model spectrum prior to convolution with filter system throughput curves. At fixed metallicity, dust attenuation, and massive-star evolution flavor, the total number of models is 9331.

\subsection{Impact of Ionized Gas on Magnitudes and Colors}

\begin{figure*}
\centering
\begin{subfigure}
\centering
\includegraphics[width=0.99\columnwidth]{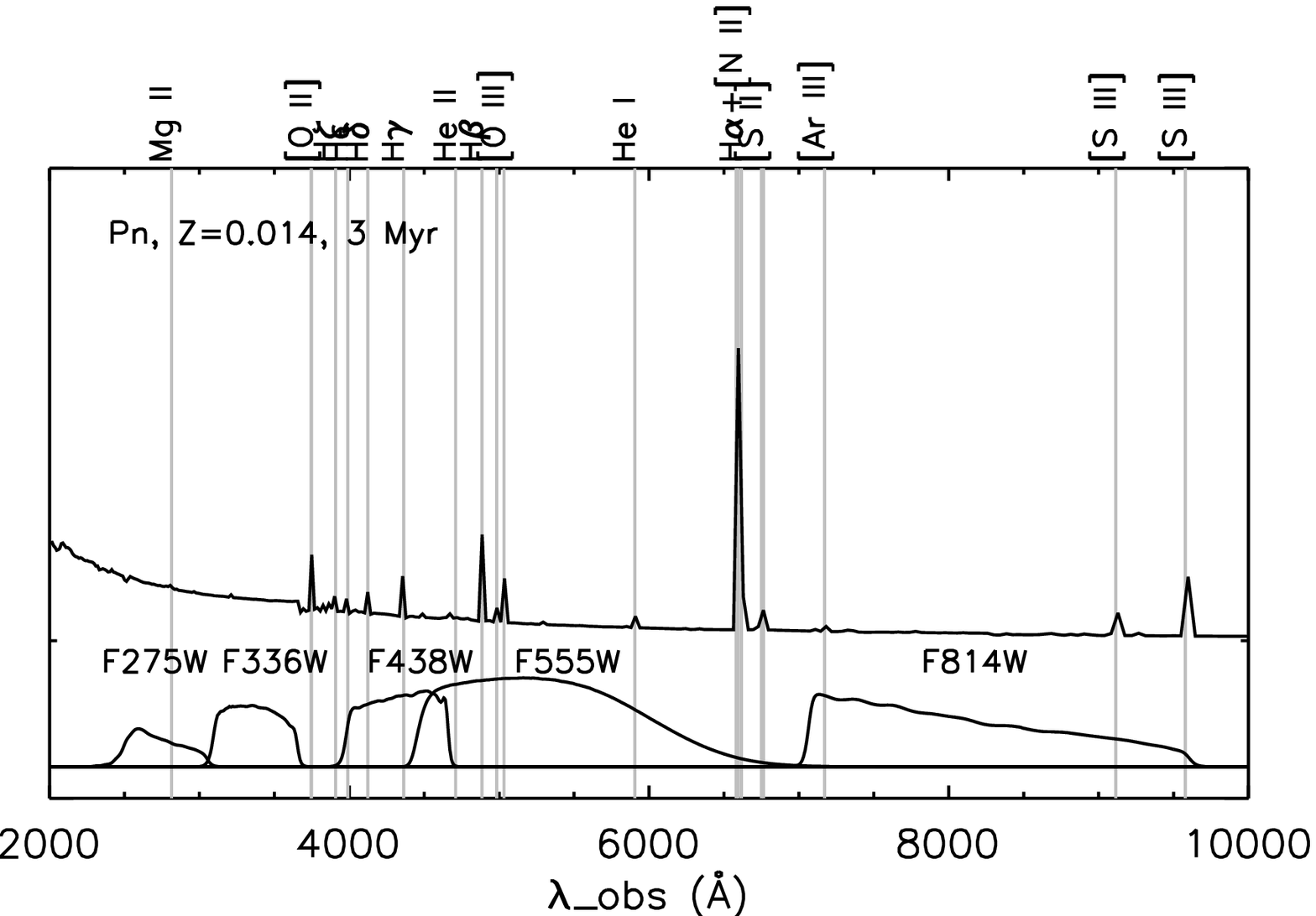}
\end{subfigure}
\begin{subfigure}
\centering
\includegraphics[width=0.99\columnwidth]{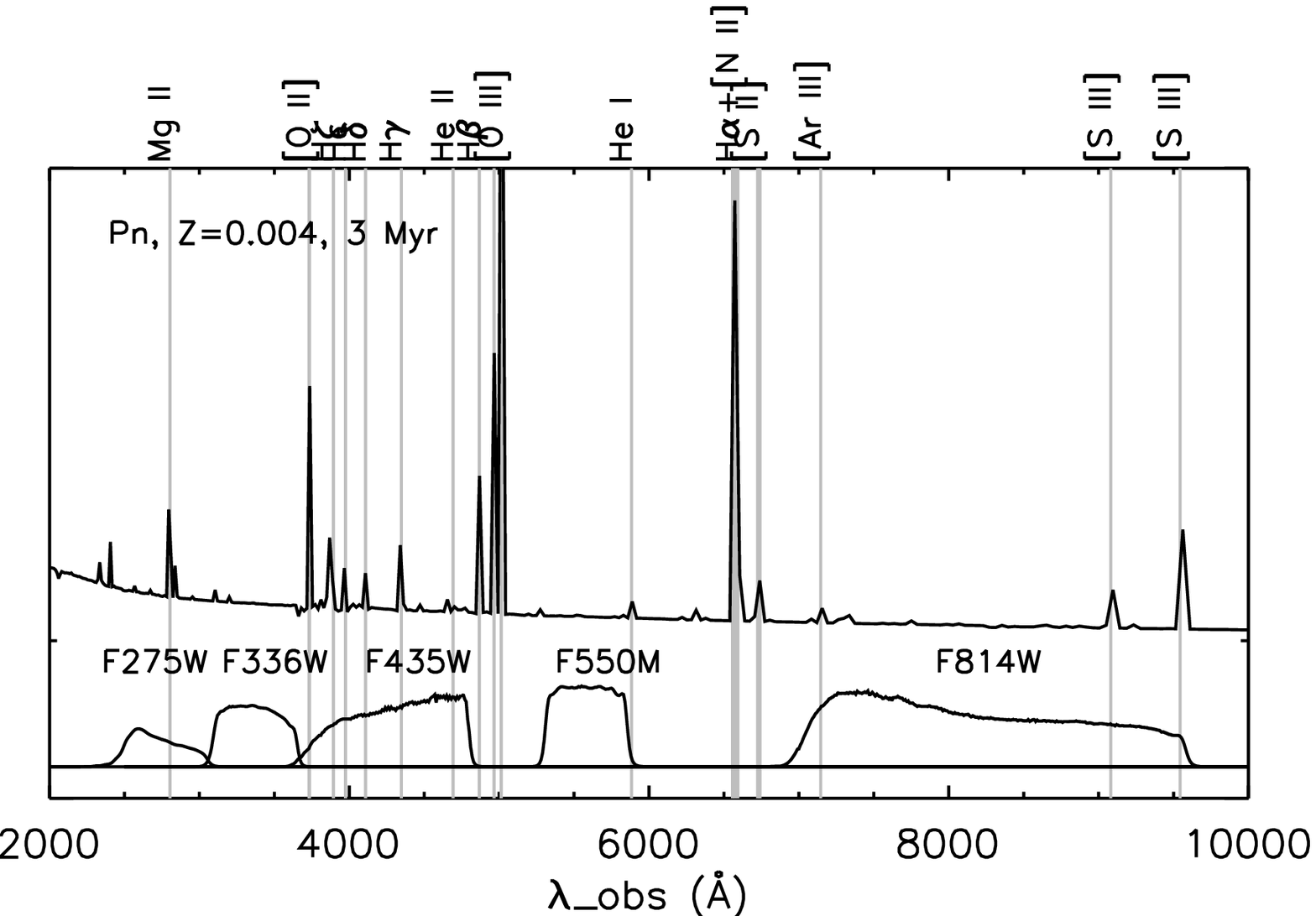}
\end{subfigure}
\caption{Left. Synthetic spectrum corresponding to an SSP of age 3 Myr and metallicity $Z=0.014$ at the redshift of NGC 1566 (see Section~\ref{sec:models} for more details on the models). The spectrum uses new Padova evolutionary tracks. Prominent lines in the spectrum are marked with vertical gray lines and labeled at the top. We overlay UVIS system throughputs. Right. Similar but metallicity is $Z=0.004$ and redshift is that of NGC 5253. We overlay UVIS and HRC system throughputs.}
\label{fig3}
\end{figure*}

Figure~\ref{fig3} shows the spectral features captured by the UVIS and HRC filters in the galaxies NGC 1566 (left panel, $Z=0.014$) and NGC 5253 (right panel, $Z=0.004$), as illustrated using 3\,Myr-old Pn models at the redshifts of the galaxies. The vertical axes of both panels use arbitrary units but identical scaling factors. Differences between the two spectra arise from the dependance of stellar evolution on metallicity. In particular, the ionizing rate at 3 Myr is higher at low compared to high metallicity due to hotter O stars and less dense winds. Note the relatively stronger collisionally-excited oxygen lines at low metallicity compared to high metallicity. This is because at low $Z$, the interstellar gas temperature is higher because of decreased cooling from metals \citep{cha01}. As can be seen in Fig.~\ref{fig3}, the use of F550M instead of F555W for NGC 5253 avoids contamination of broad-band photometry by strong H$\beta$ and [O{\sc\,iii}] lines at the redshift of this galaxy.

%%%%%%%%%%%%%%%%%%%%%%%%%%%%%%
% Figure 3
%%%%%%%%%%%%%%%%%%%%%%%%%%%%%%

\begin{figure*}
\begin{subfigure}
\centering
\includegraphics[width=0.89\columnwidth]{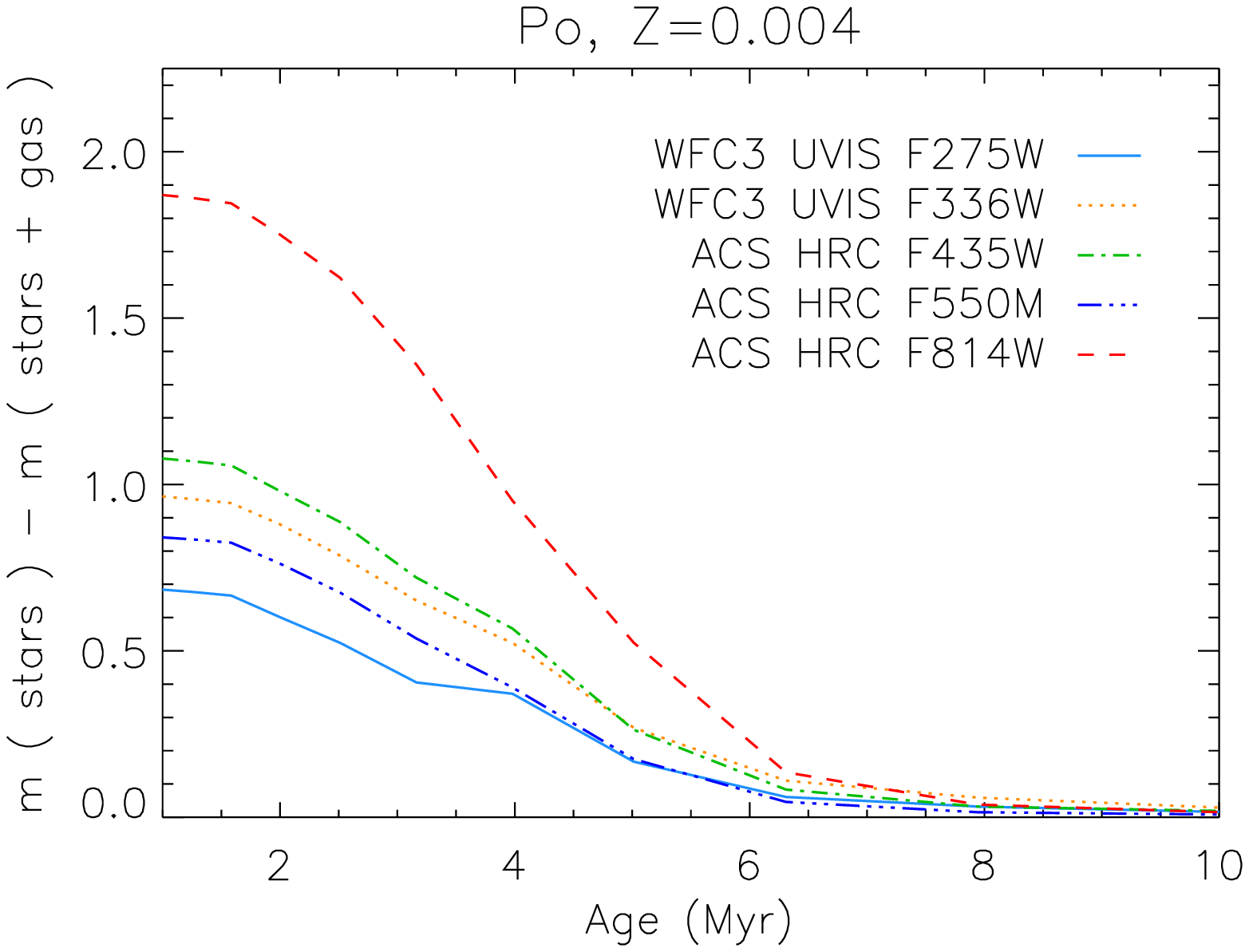}
\end{subfigure}
\begin{subfigure}
\centering
\includegraphics[width=0.89\columnwidth]{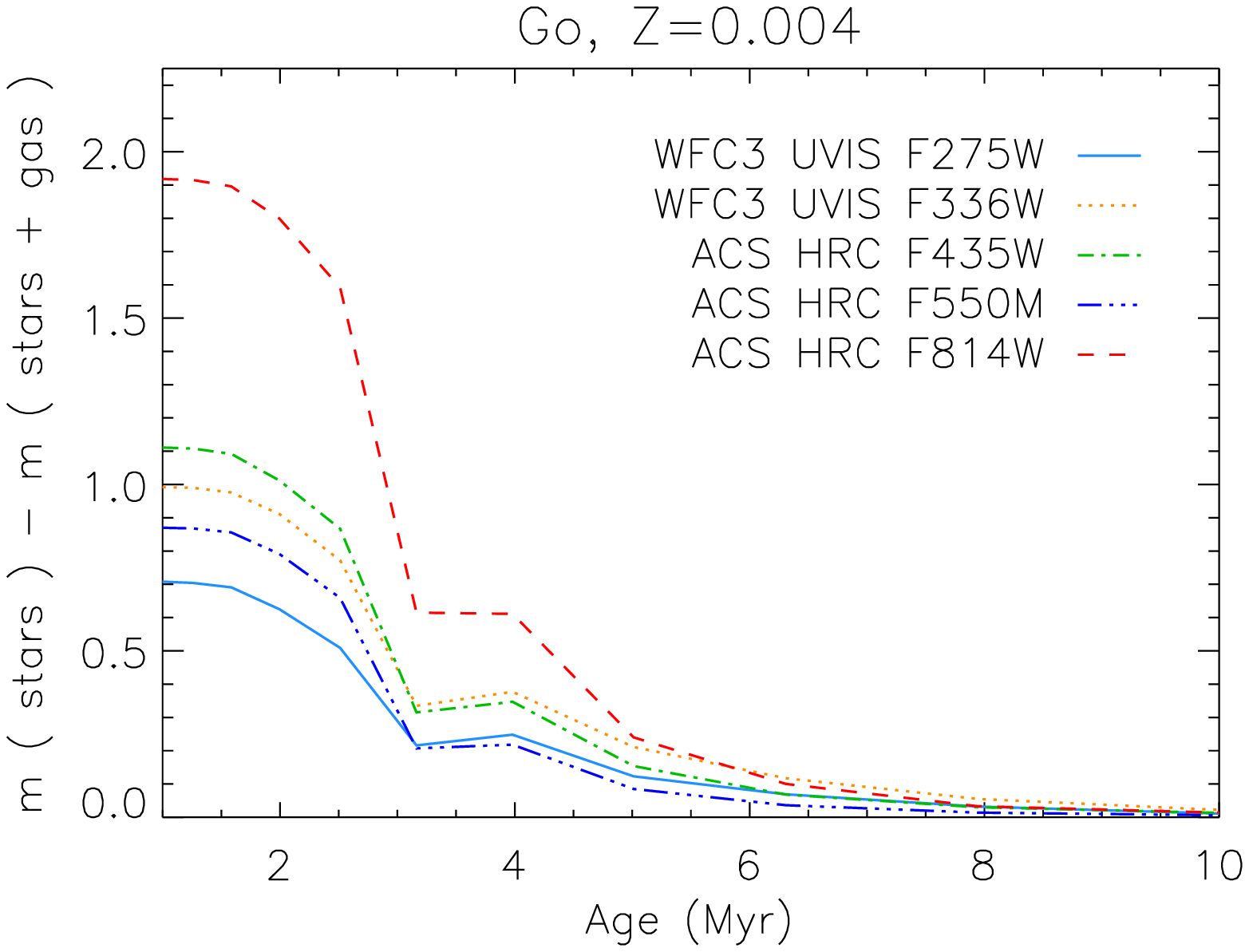}
\end{subfigure}
\begin{subfigure}
\centering
\includegraphics[width=0.89\columnwidth]{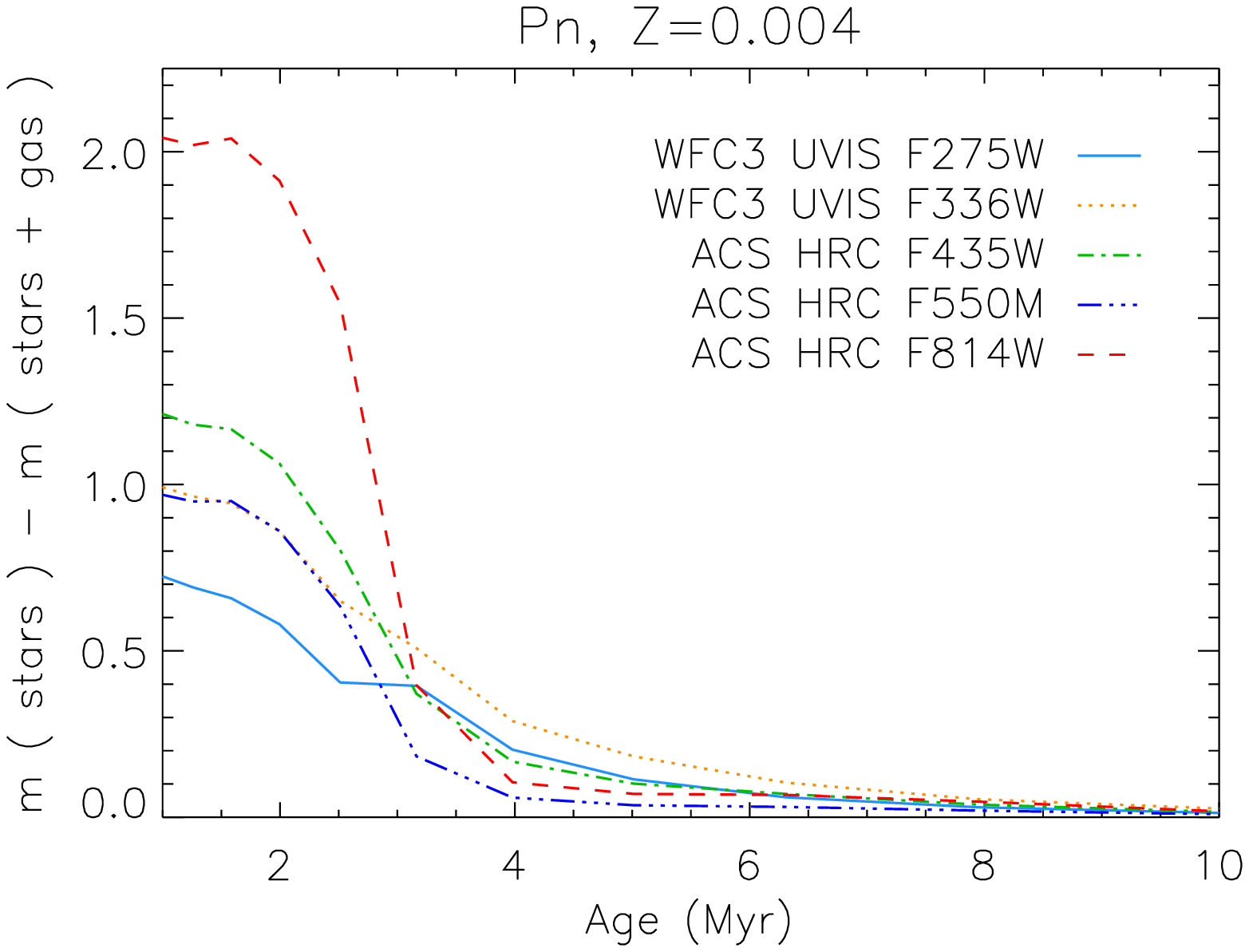}
\end{subfigure}
\begin{subfigure}
\centering
\includegraphics[width=0.89\columnwidth]{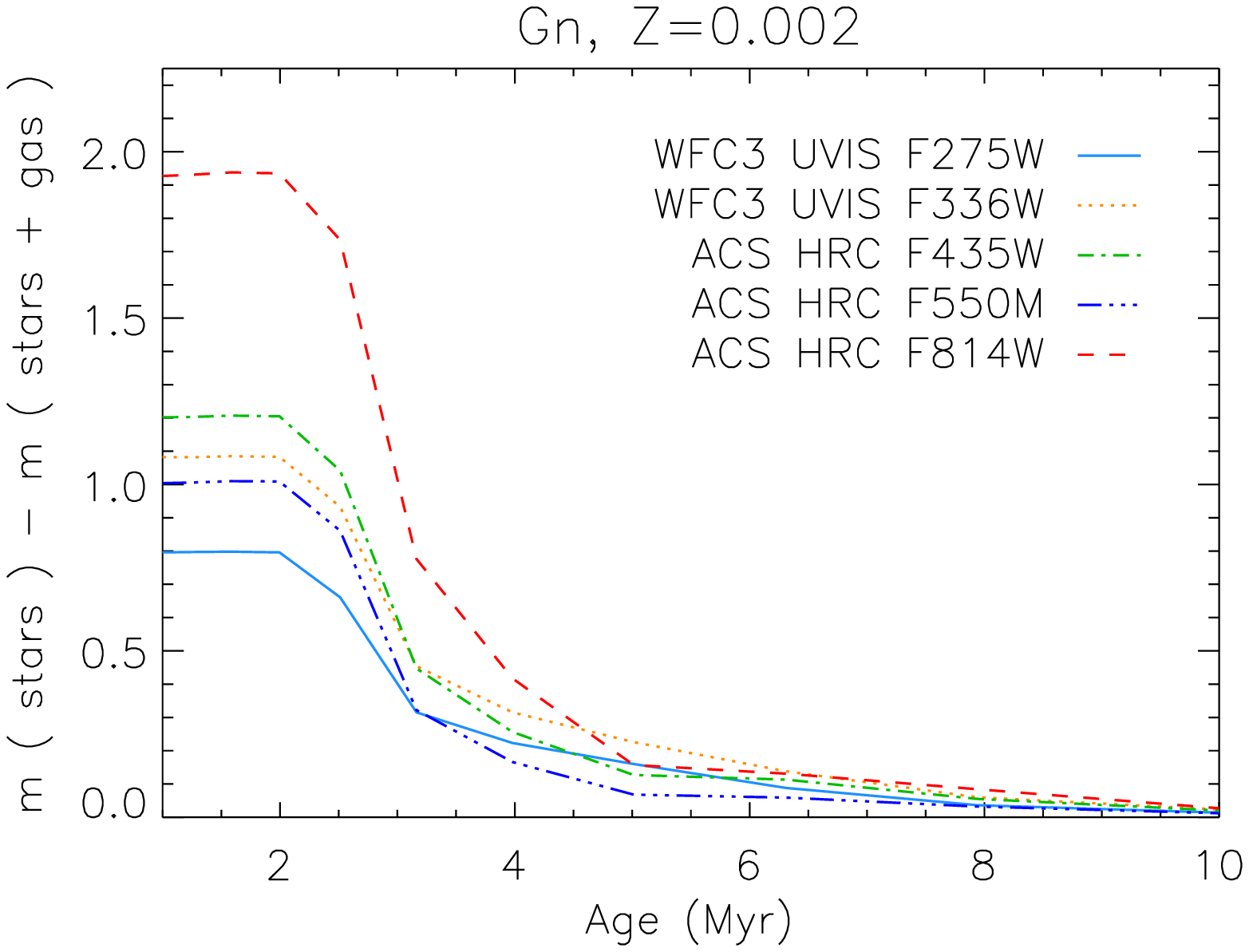}
\end{subfigure}
\begin{subfigure}
\centering
\includegraphics[width=0.89\columnwidth]{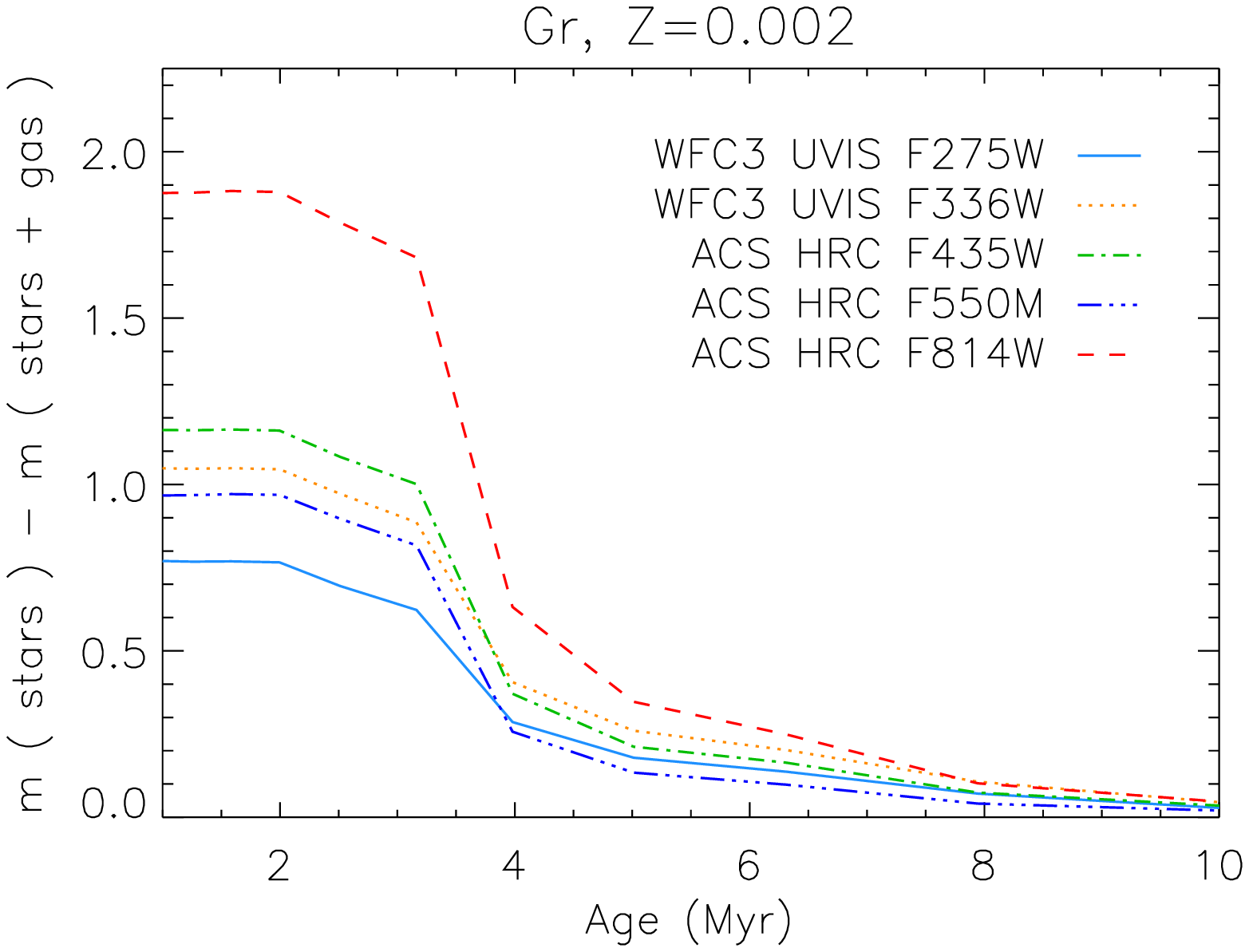}
\end{subfigure}
\begin{subfigure}
\centering
\includegraphics[width=0.89\columnwidth]{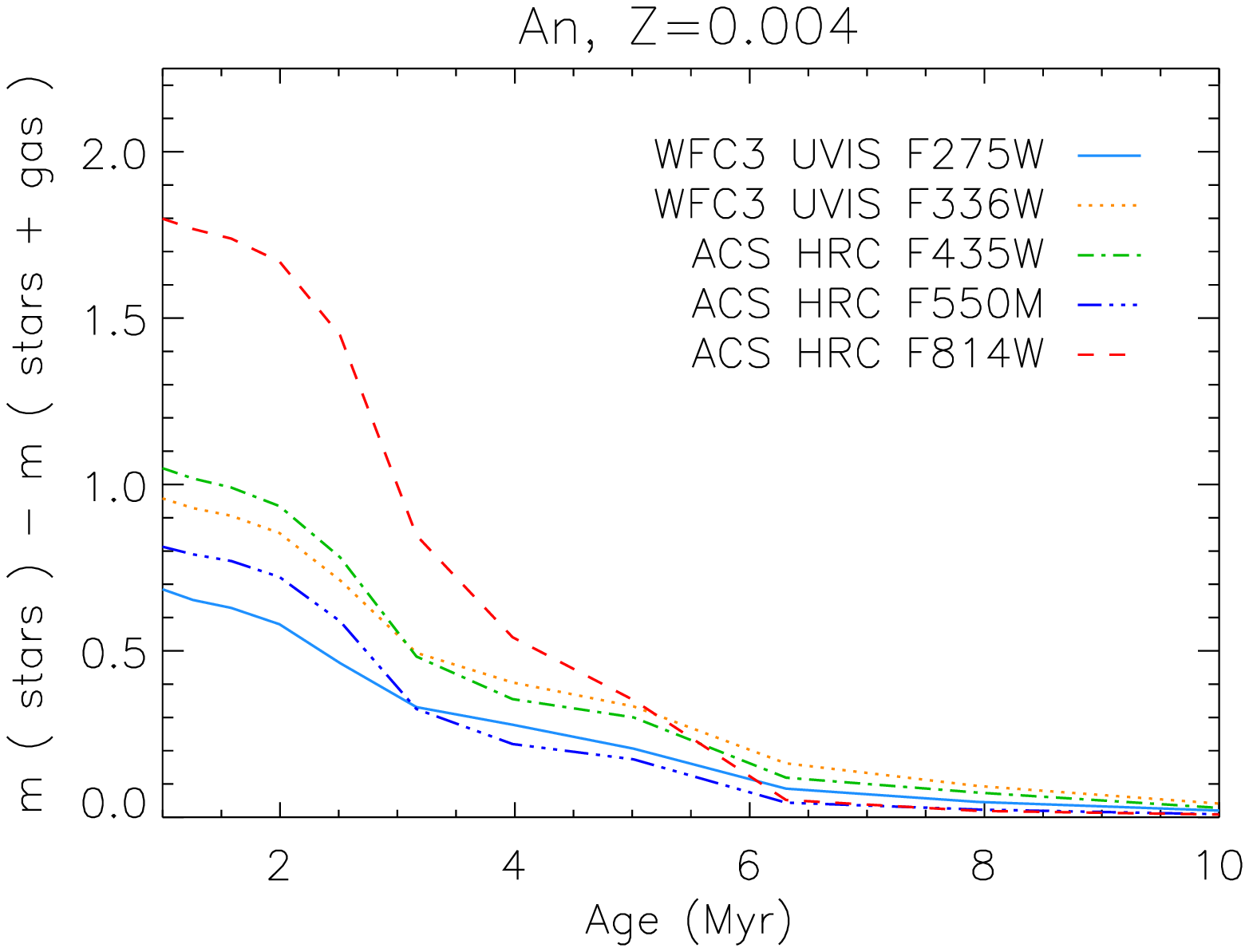}
\end{subfigure}
\begin{subfigure}
\centering
\includegraphics[width=0.89\columnwidth]{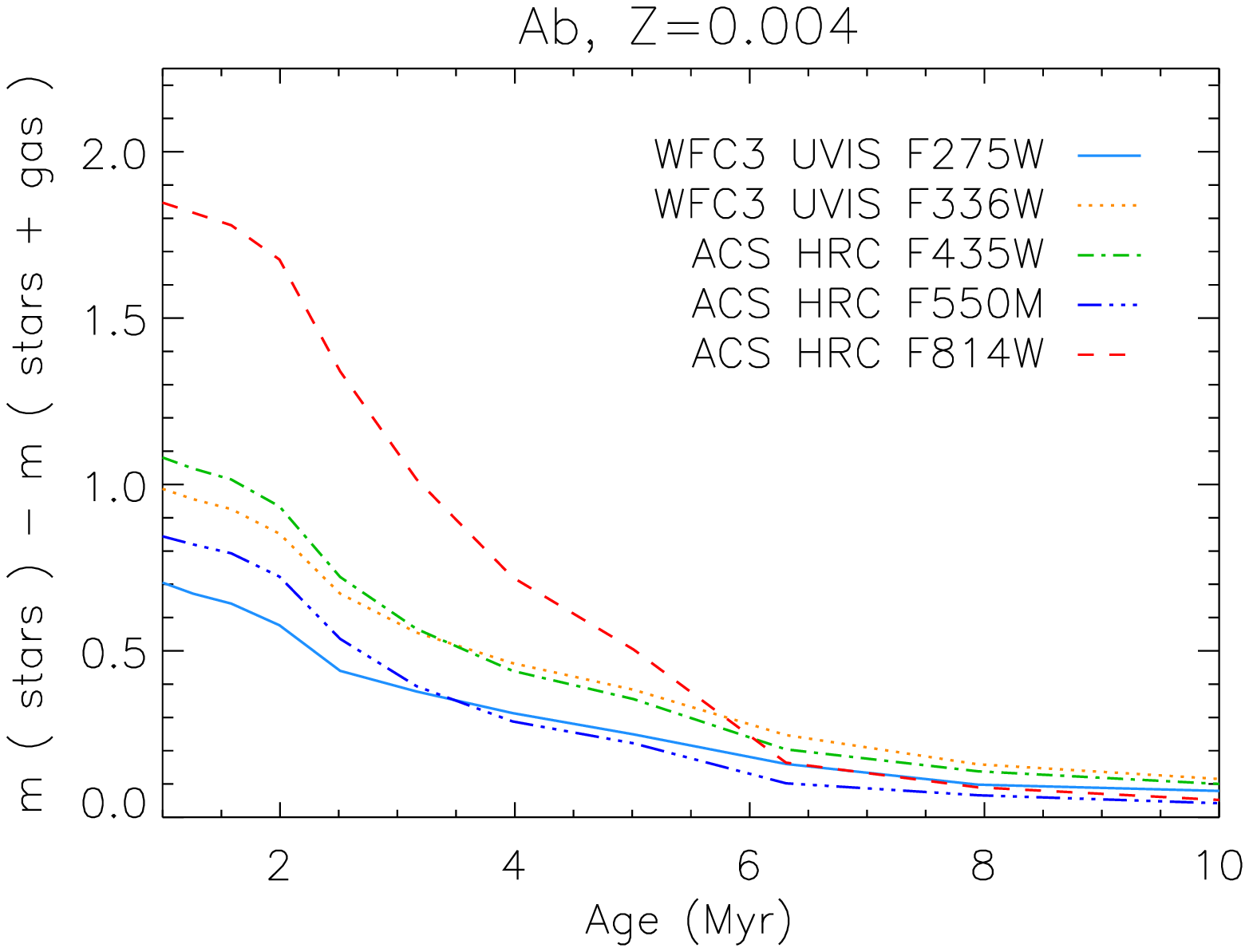}
\end{subfigure}
\caption{Magnitude difference as a function of age (from 1 to 10 Myr) between an SSP and an SSP+gas (both unattenuated). Curves of different styles show the different photometric bands used for NGC 5253. Each panel shows a different combination of model and metallicity, as indicated by the title. Note the different metallicities of the Gn and Gr models relative to that of the rest of the models.}
\label{fig4}
\end{figure*}

\begin{figure*}
\begin{subfigure}
\centering
\includegraphics[width=0.89\columnwidth]{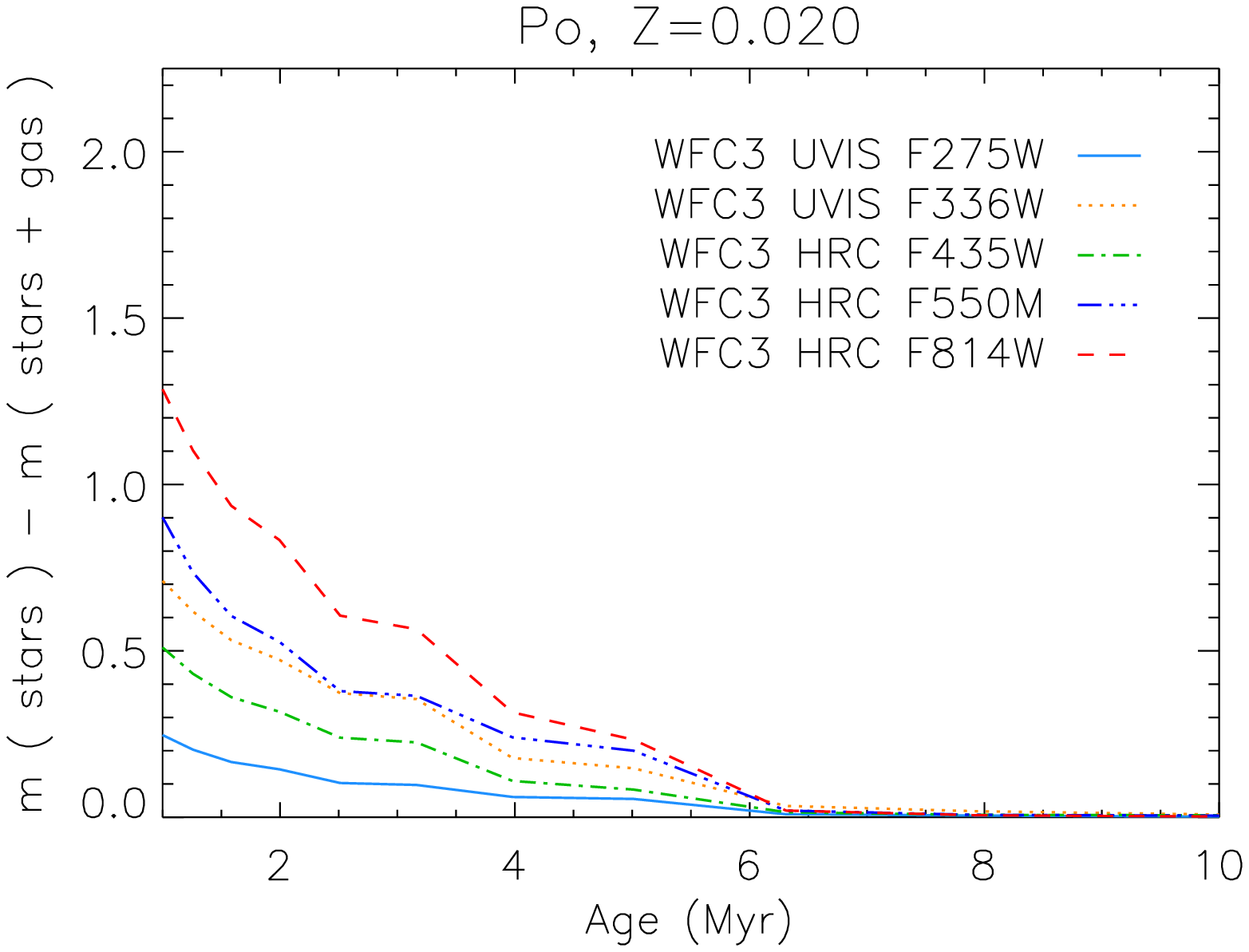}
\end{subfigure}
\begin{subfigure}
\centering
\includegraphics[width=0.89\columnwidth]{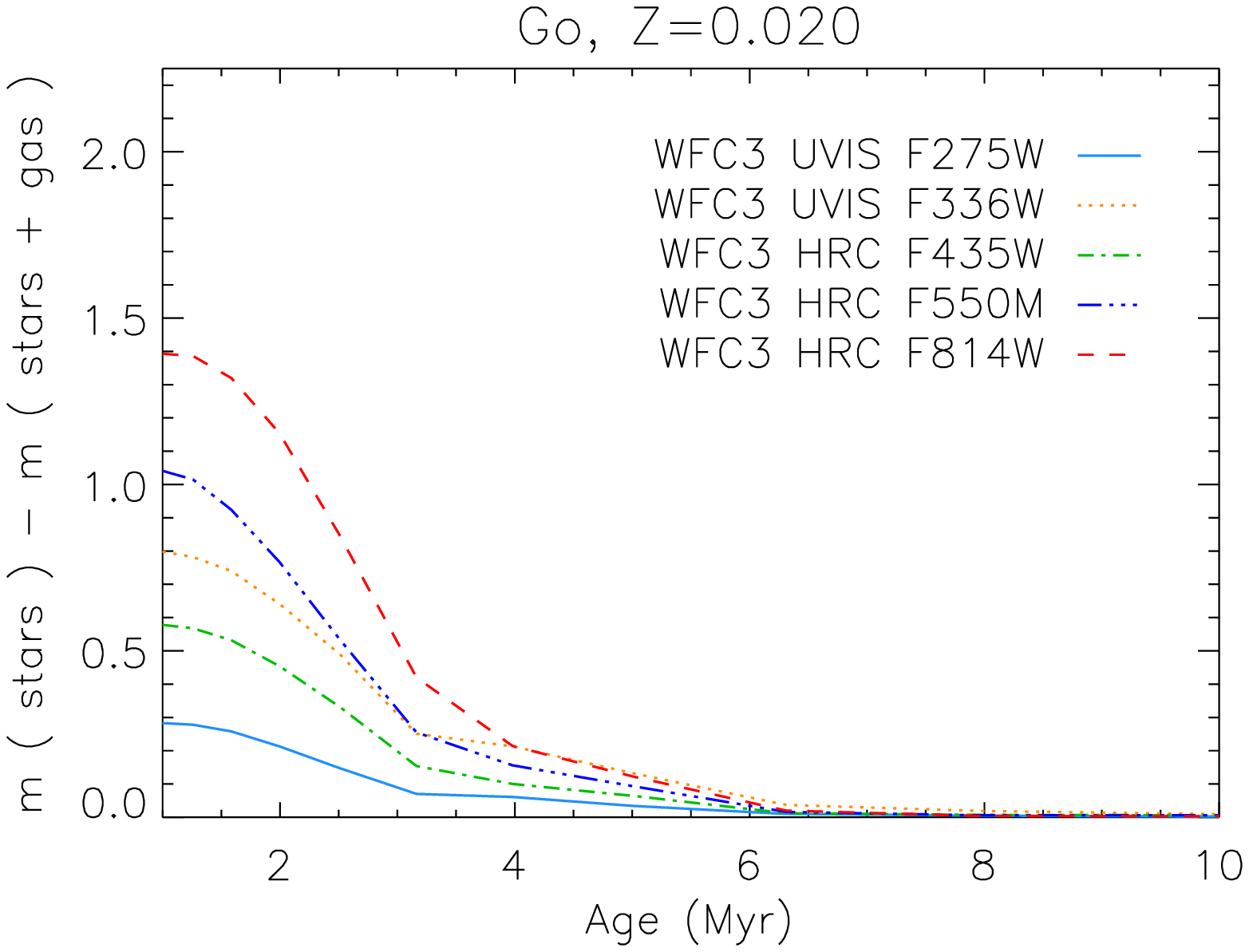}
\end{subfigure}
\begin{subfigure}
\centering
\includegraphics[width=0.89\columnwidth]{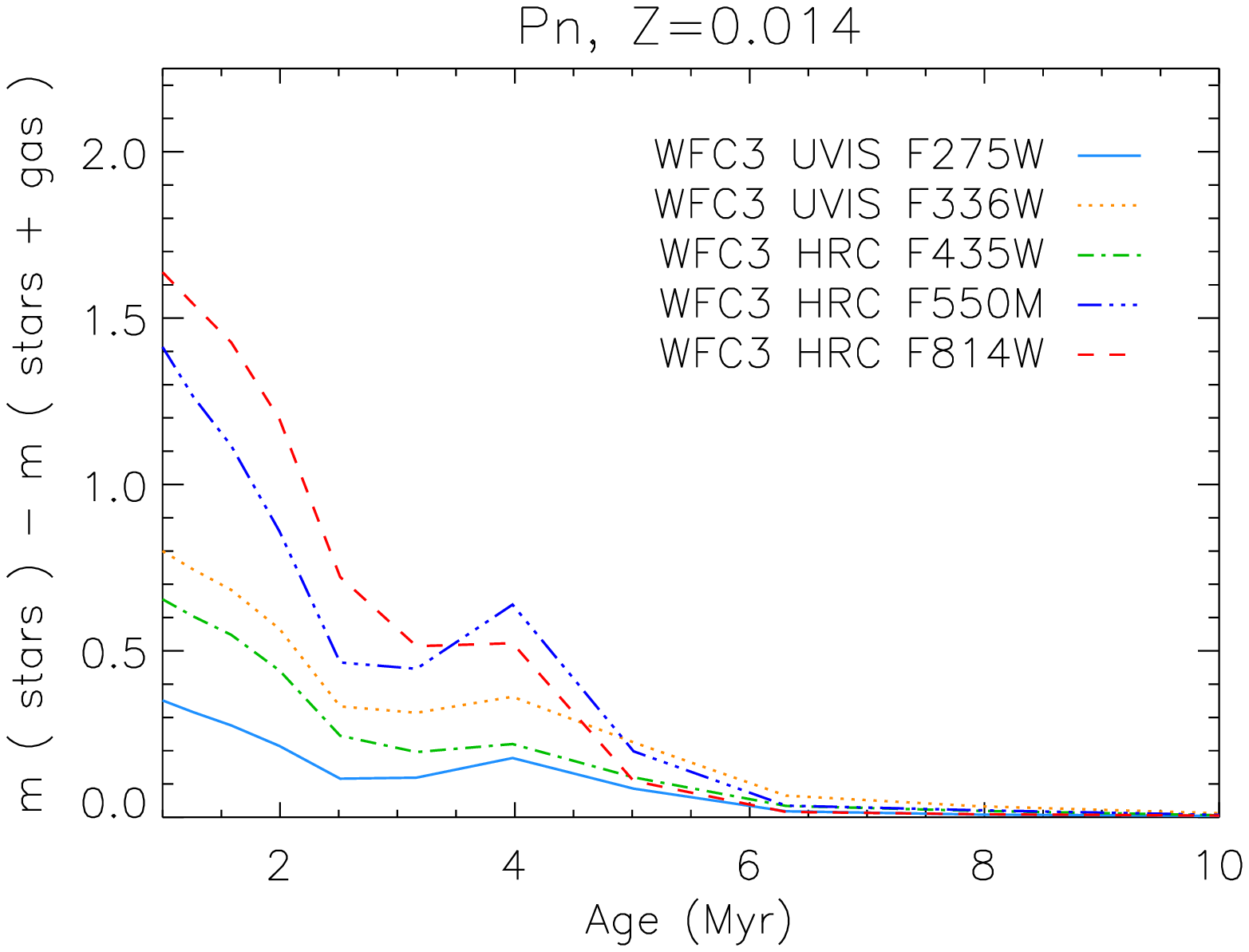}
\end{subfigure}
\begin{subfigure}
\centering
\includegraphics[width=0.89\columnwidth]{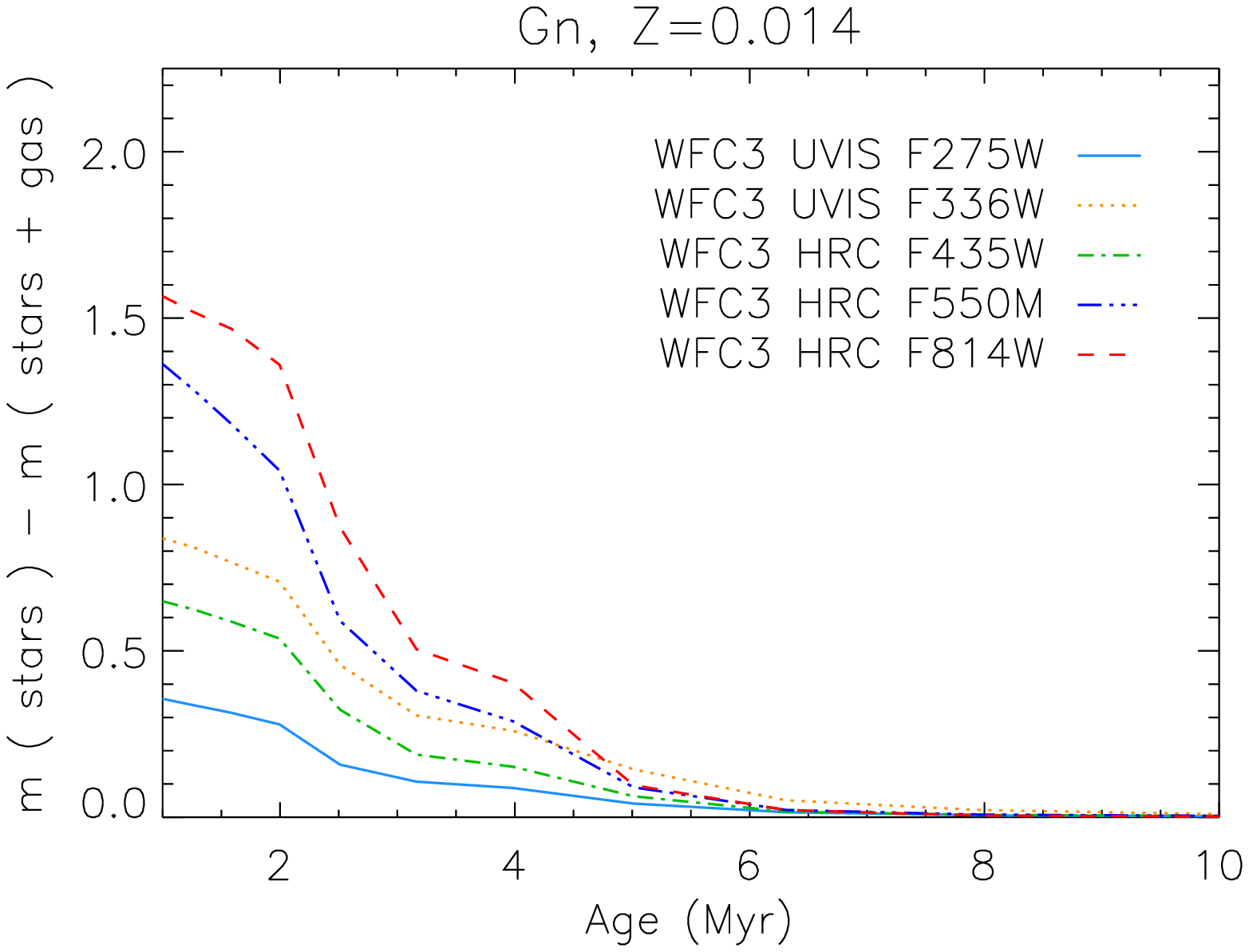}
\end{subfigure}
\begin{subfigure}
\centering
\includegraphics[width=0.89\columnwidth]{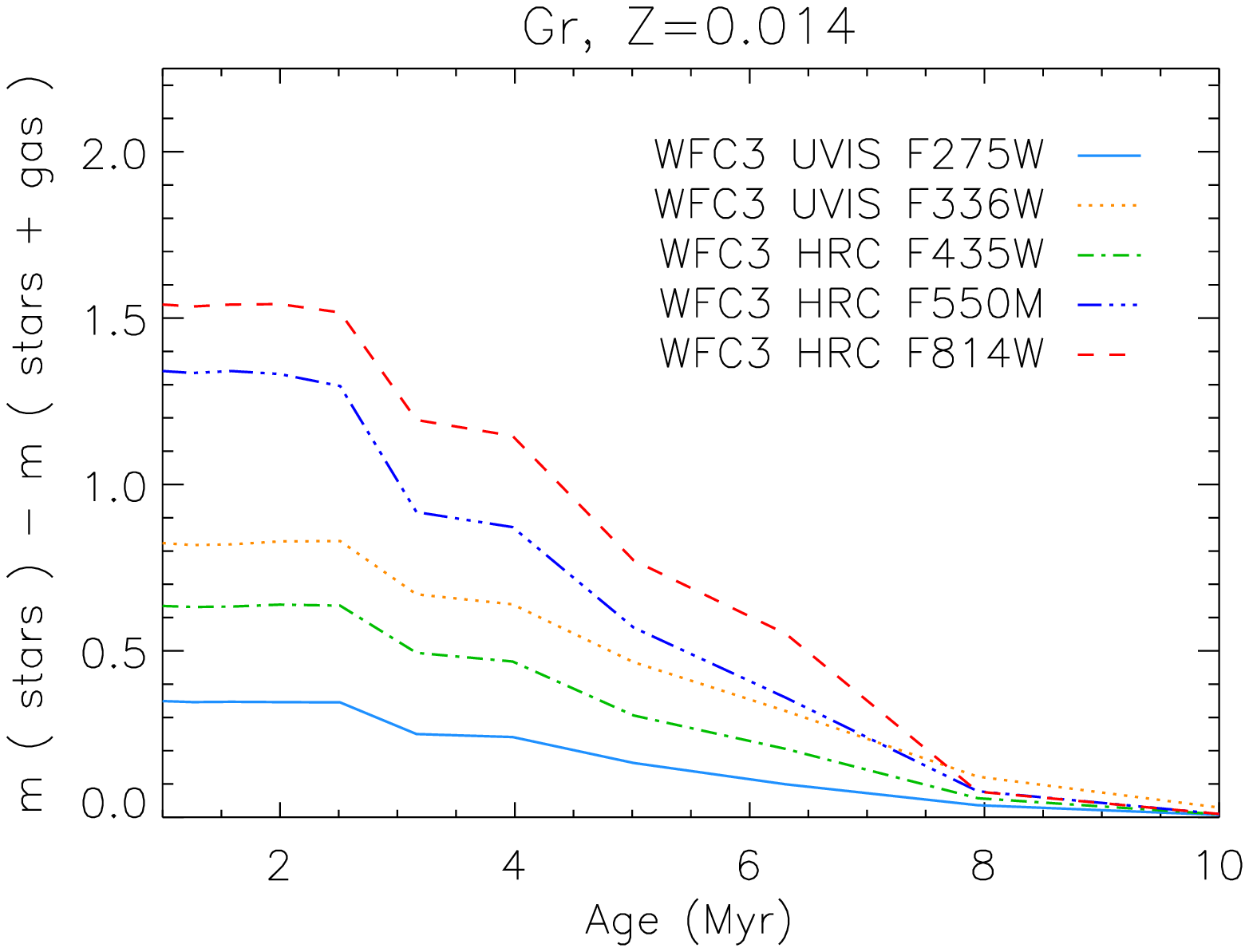}
\end{subfigure}
\begin{subfigure}
\centering
\includegraphics[width=0.89\columnwidth]{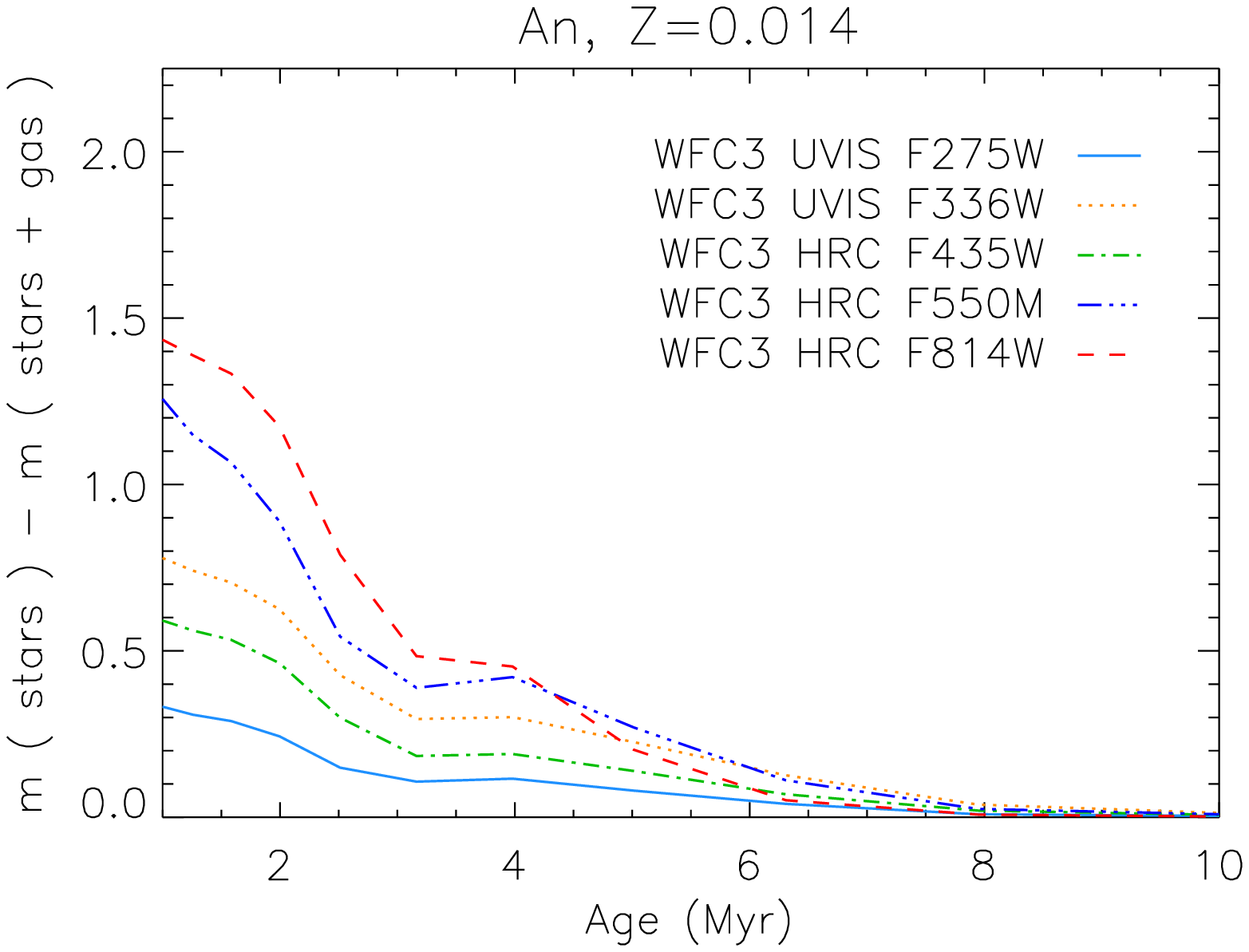}
\end{subfigure}
\begin{subfigure}
\centering
\includegraphics[width=0.89\columnwidth]{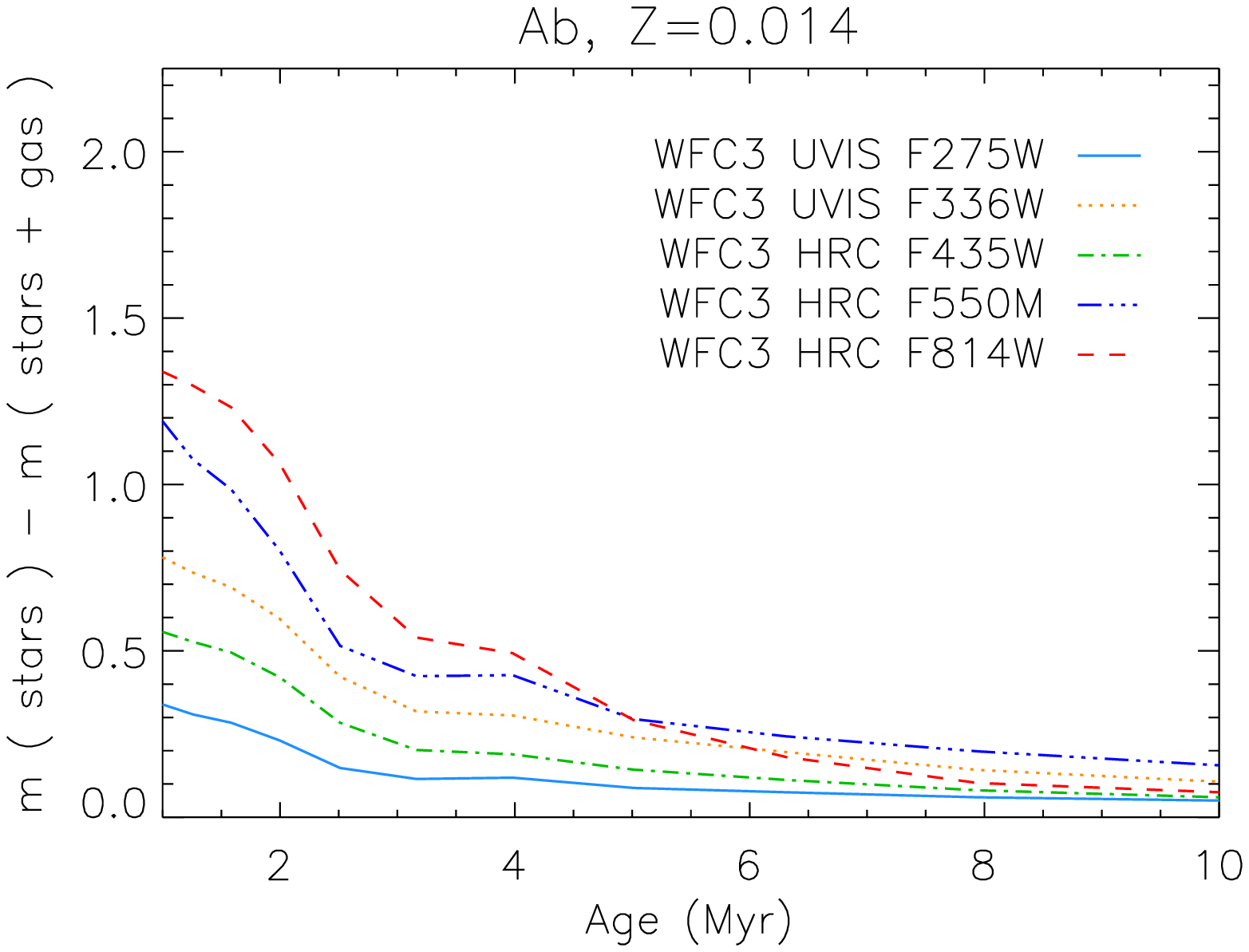}
\end{subfigure}
\caption{Similar to Fig. 4, but we show the photometric bands used for NGC 1566. Note the different metallicities of the Po and Go models relative to other models.}
\label{fig5}
\end{figure*}

As previously shown by, e.g., \cite{cha96}, \cite{zac01}, \cite{ber02}, \cite{and03}, \cite{ost03} and \cite{rei10}, the ionized gas associated with a young stellar population can provide a significant contribution to the total observed flux in a given filter through the contributions of both emission lines and the recombination continuum. Figs.~\ref{fig4} and~\ref{fig5} quantify the contribution of the nebular emission in the filters used to observe NGC 5253 and NGC 1566, respectively. They show the magnitude difference as a function of age between a pure stellar population and stars+ionized gas. The left panel of fig. 11 in \cite{rei10} shows Go, $Z=0.004$ predictions for WFC3 UVIS filters F547M and F814W. The latter predictions can be compared to our Go, $Z=0.004$ predictions for close ACS HRC filters F550M and F814W, which are plotted in the top-right panel of Fig.~\ref{fig4}. At 3 Myr, our predictions are higher by about 0.2 mag relative to \cite{rei10}. This presumably arises from differences in the models, detectors, and filters.

%%%%%%%%%%%%%%%%%%%%%%%%%%%%%%
% Figure 4
%%%%%%%%%%%%%%%%%%%%%%%%%%%%%%

%%%%%%%%%%%%%%%%%%%%%%%%%%%%%%
% Figure 5
%%%%%%%%%%%%%%%%%%%%%%%%%%%%%%

\begin{figure*}
\begin{subfigure}
\centering
\includegraphics[width=0.89\columnwidth]{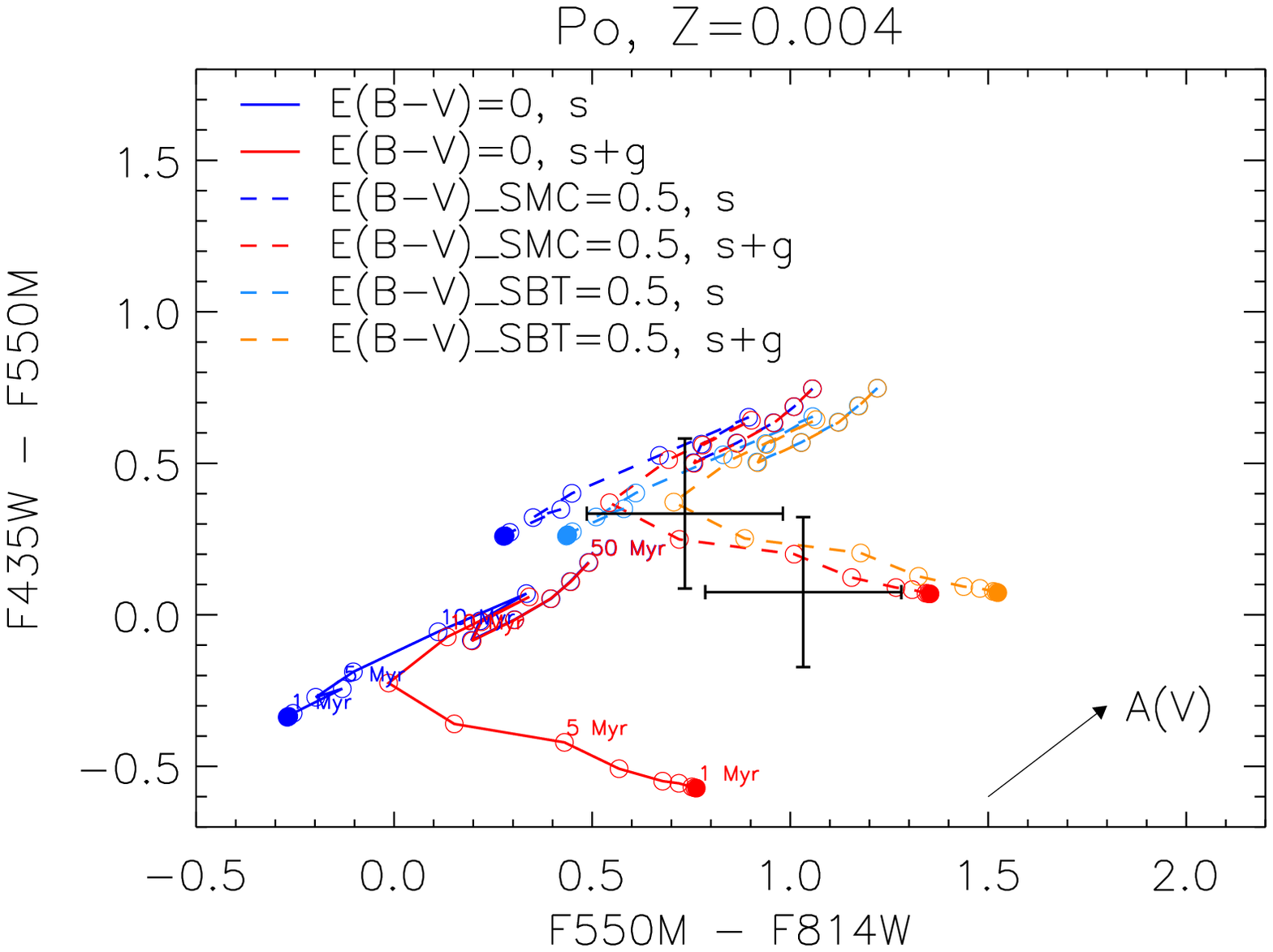}
\end{subfigure}
\begin{subfigure}
\centering
\includegraphics[width=0.89\columnwidth]{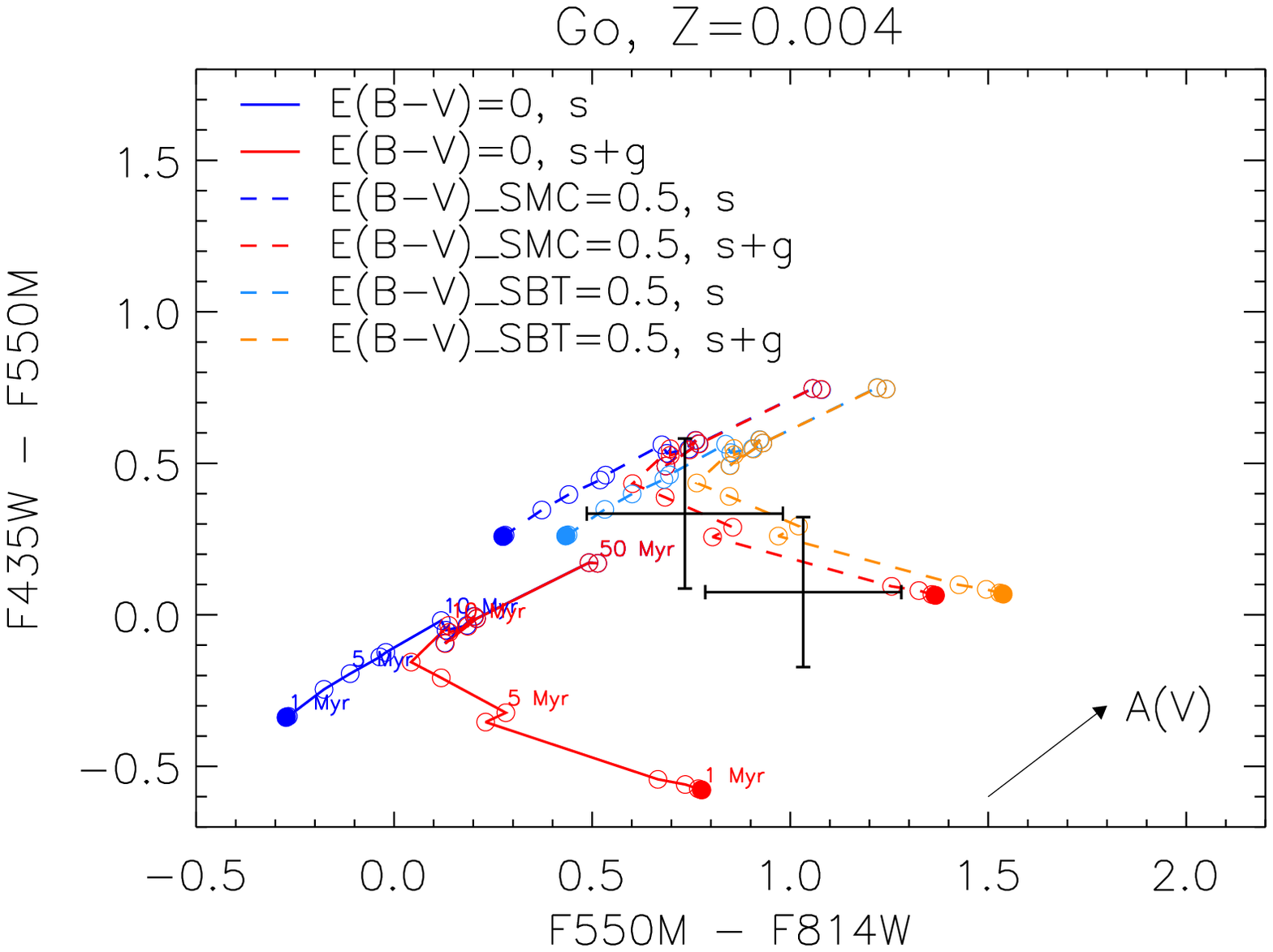}
\end{subfigure}
\begin{subfigure}
\centering
\includegraphics[width=0.89\columnwidth]{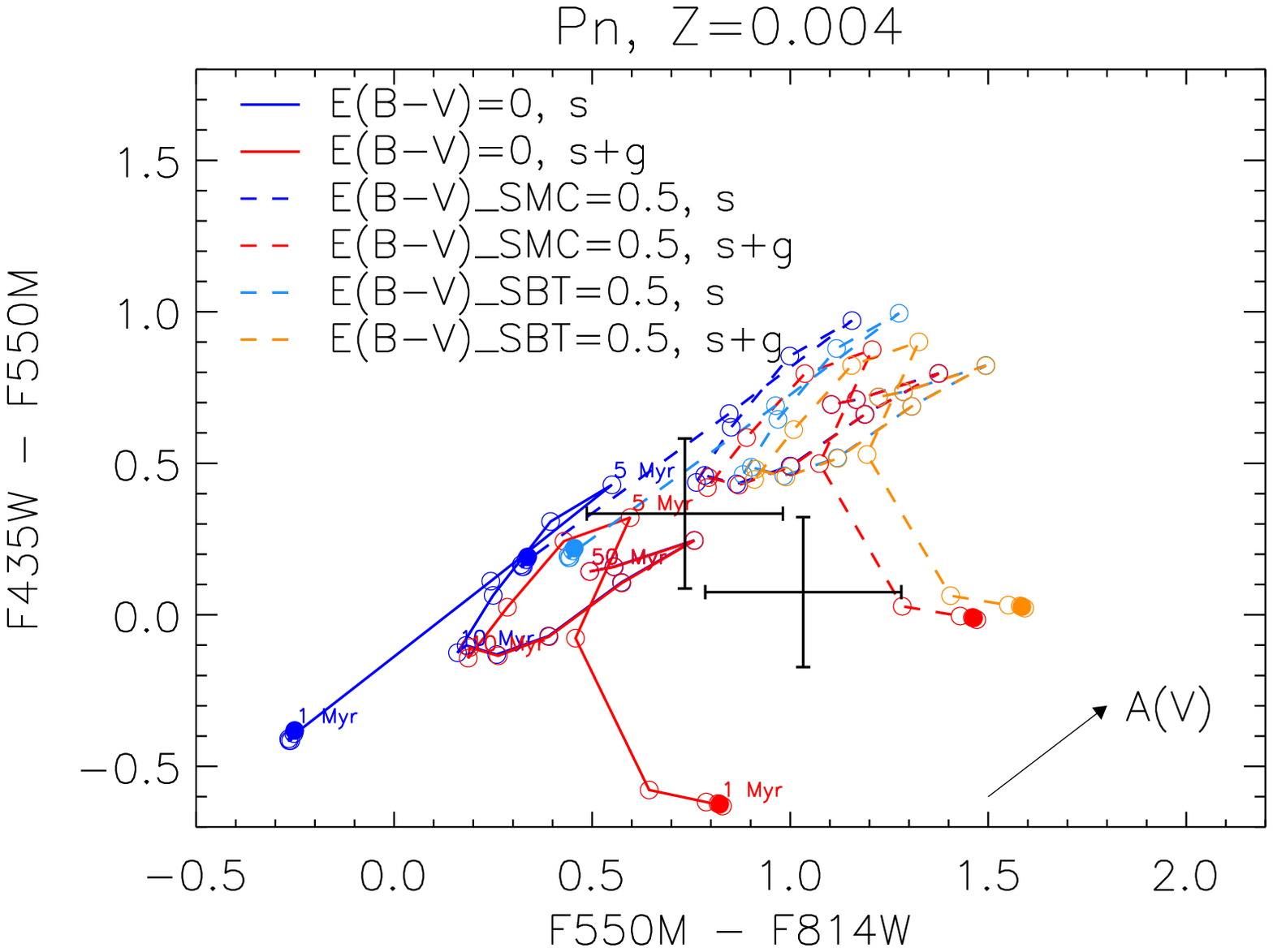}
\end{subfigure}
\begin{subfigure}
\centering
\includegraphics[width=0.89\columnwidth]{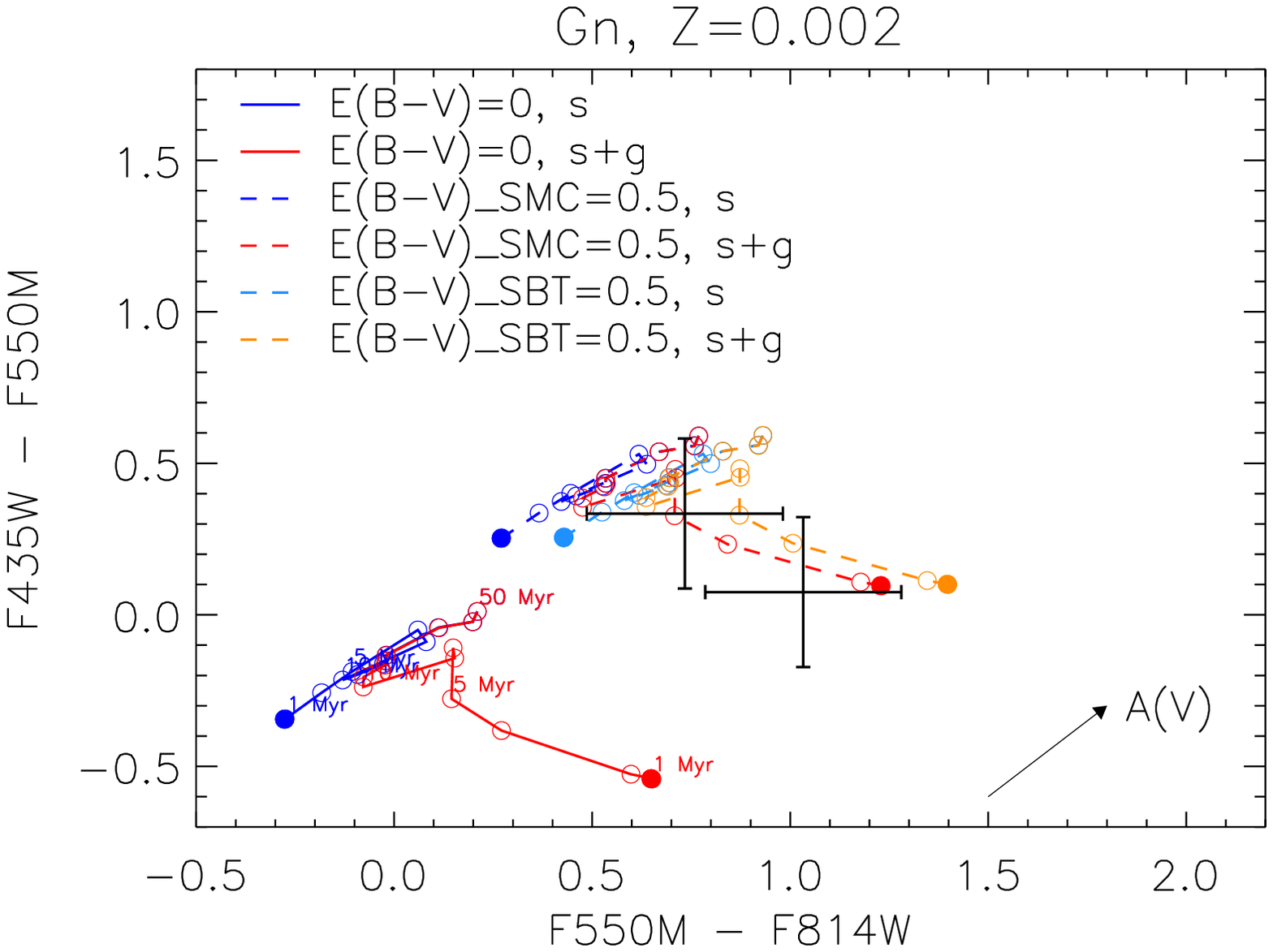}
\end{subfigure}
\begin{subfigure}
\centering
\includegraphics[width=0.89\columnwidth]{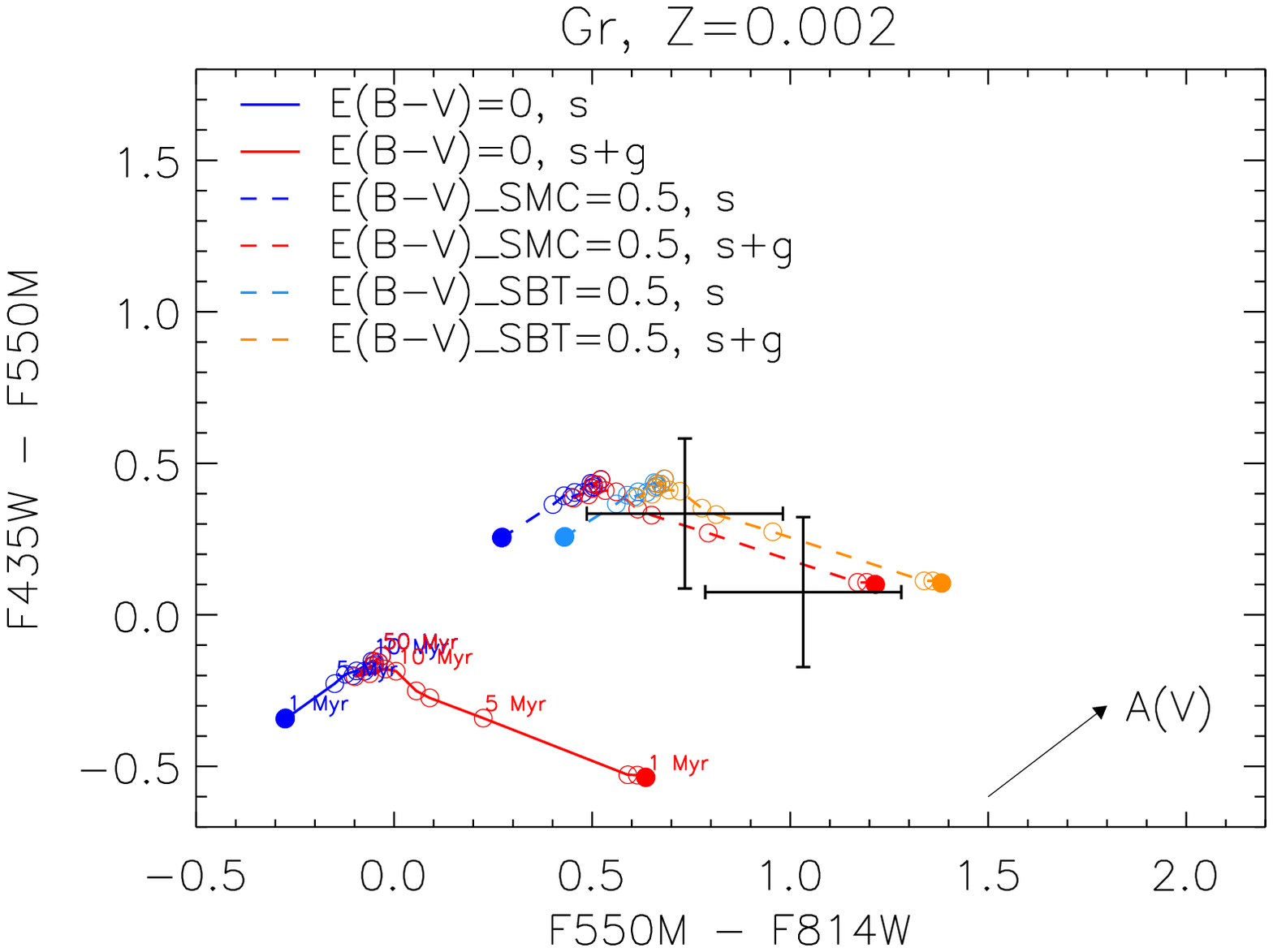}
\end{subfigure}
\begin{subfigure}
\centering
\includegraphics[width=0.89\columnwidth]{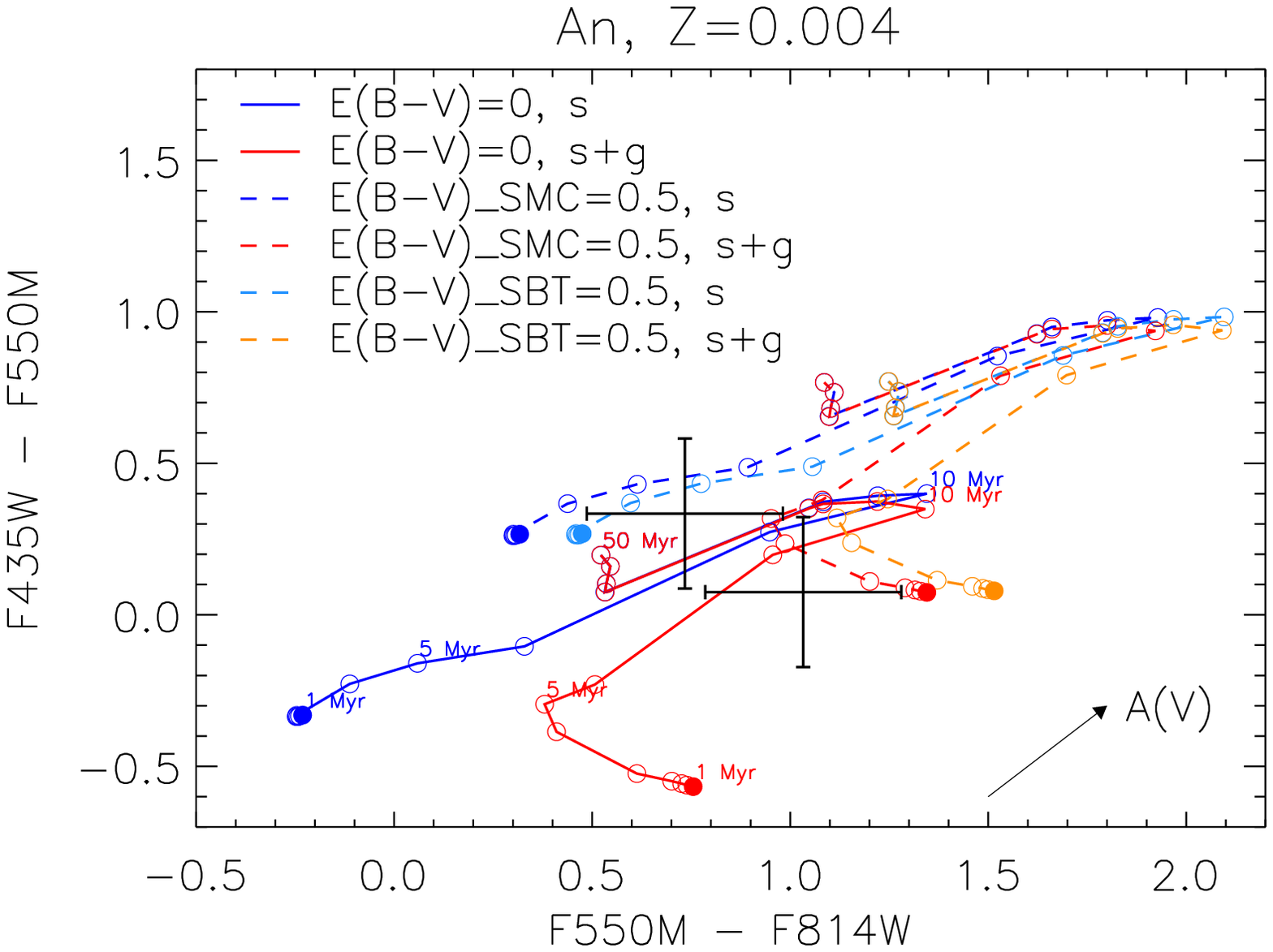}
\end{subfigure}
\begin{subfigure}
\centering
\includegraphics[width=0.89\columnwidth]{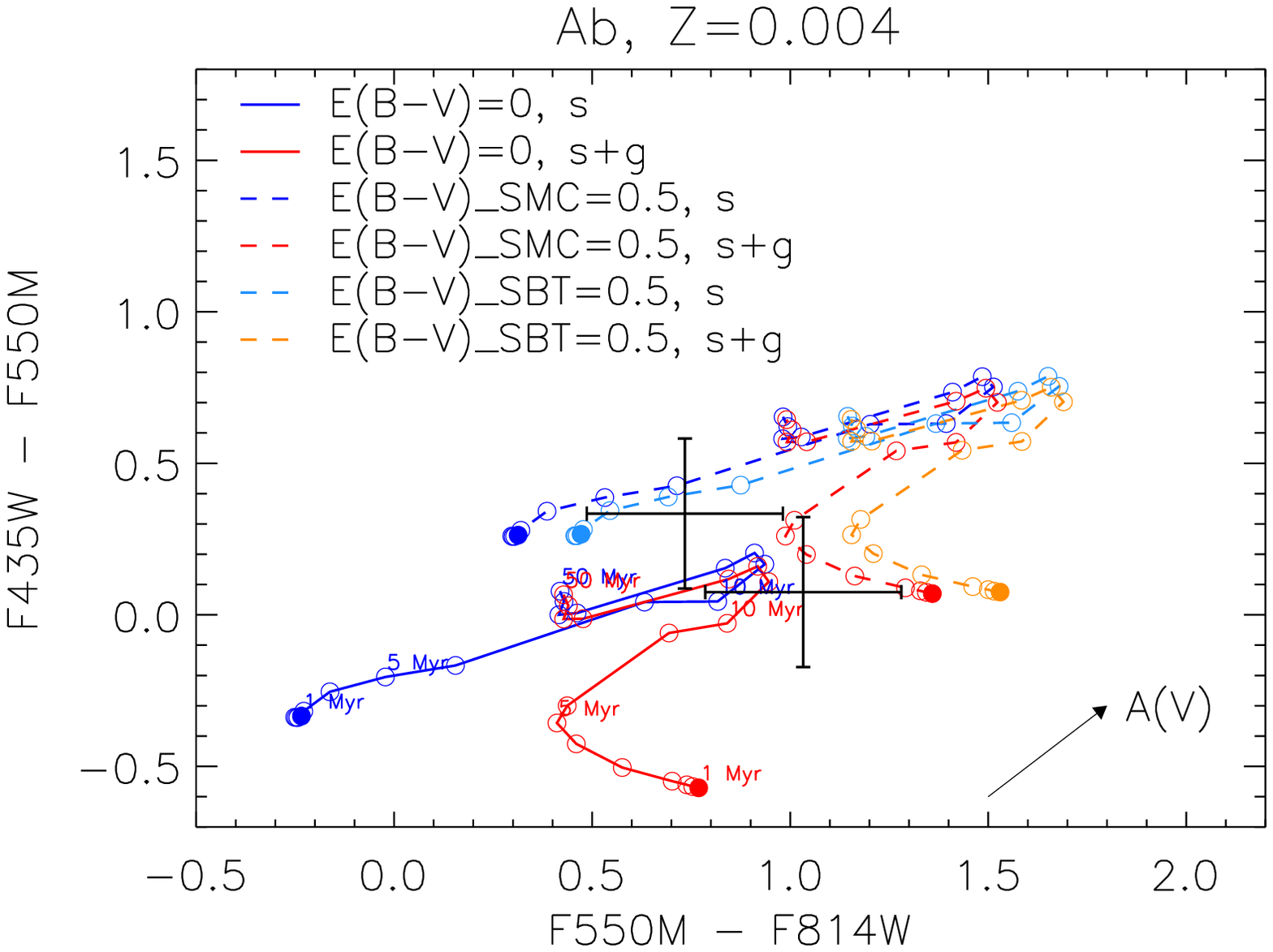}
\end{subfigure}
\caption{Color-color diagram based on the ACS HRC bands used for NGC 5253. We use unfilled circles to show the evolution from 1 Myr to 50 Myr in steps of $\Delta$log(t/yr) = 0.1 dex. We mark the position of t=1 Myr with a filled symbol. We use blue curves for pure SSPs and red curves for SSPs+ionized gas. We show unattenuated (solid curves) and attenuated (dashed curves) cases. For the latter, we use an arbitrary value of E(B-V)=0.5 mag and the SMC extinction law. The error bars correspond to the two clusters in NGC 5253. Each panel shows a different combination of model and metallicity, as indicated by the title. The metallicity of the Gn and Gr models is different relative to that of the rest of the models. The arrow indicates the direction of increasing extinction.}
\label{fig6}
\end{figure*}

\begin{figure*}
\begin{subfigure}
\centering
\includegraphics[width=0.89\columnwidth]{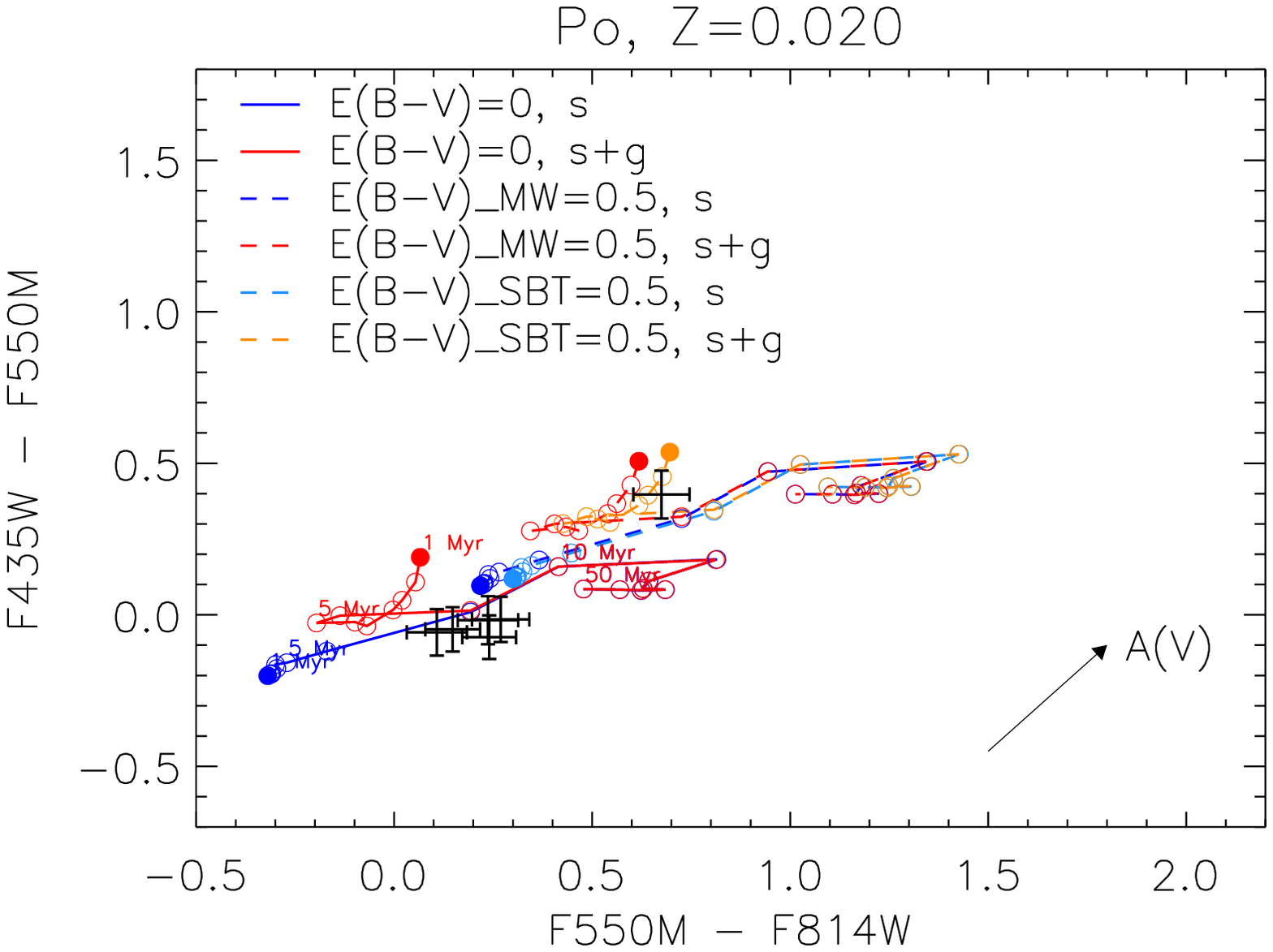}
\end{subfigure}
\begin{subfigure}
\centering
\includegraphics[width=0.89\columnwidth]{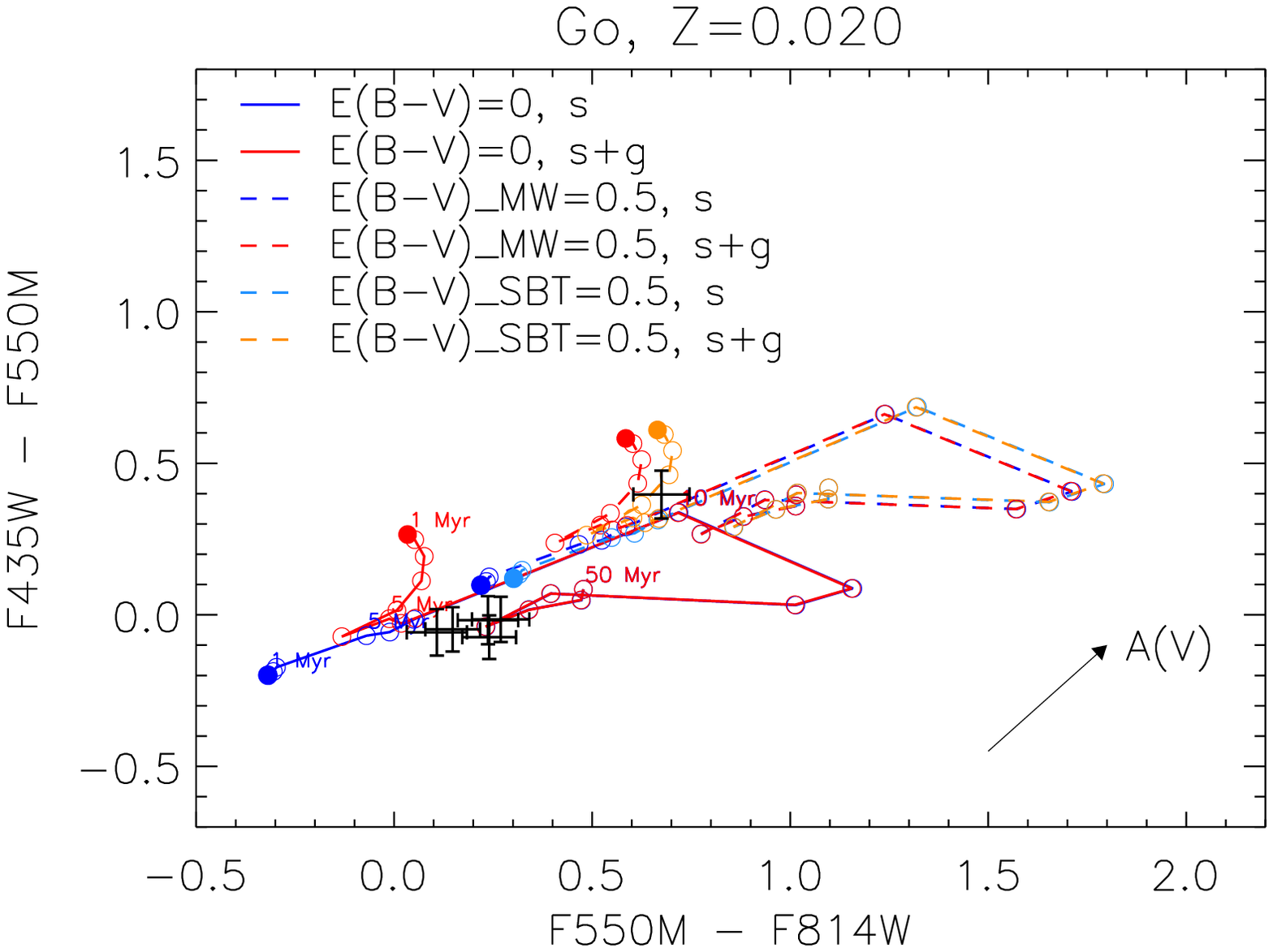}
\end{subfigure}
\begin{subfigure}
\centering
\includegraphics[width=0.89\columnwidth]{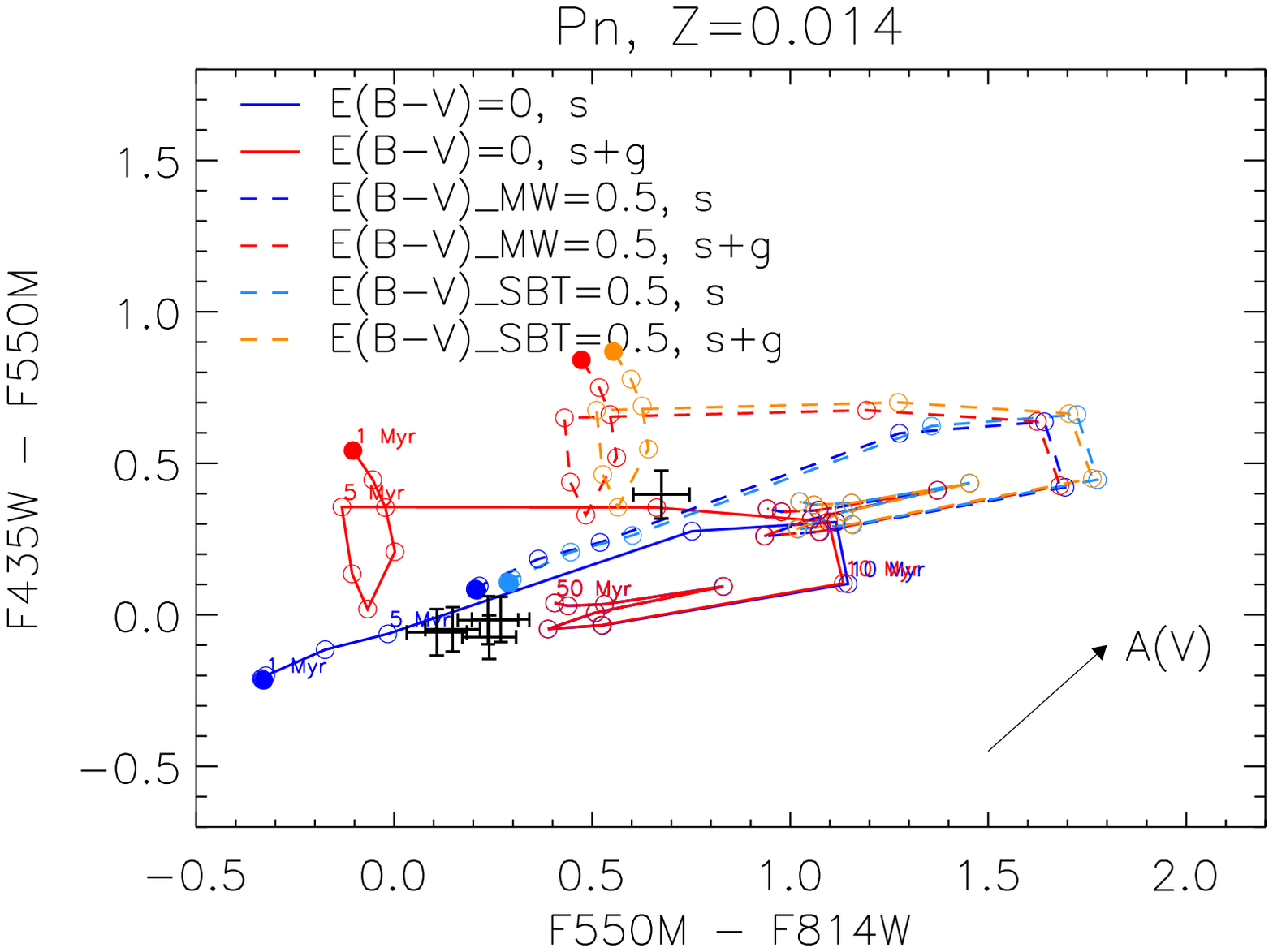}
\end{subfigure}
\begin{subfigure}
\centering
\includegraphics[width=0.89\columnwidth]{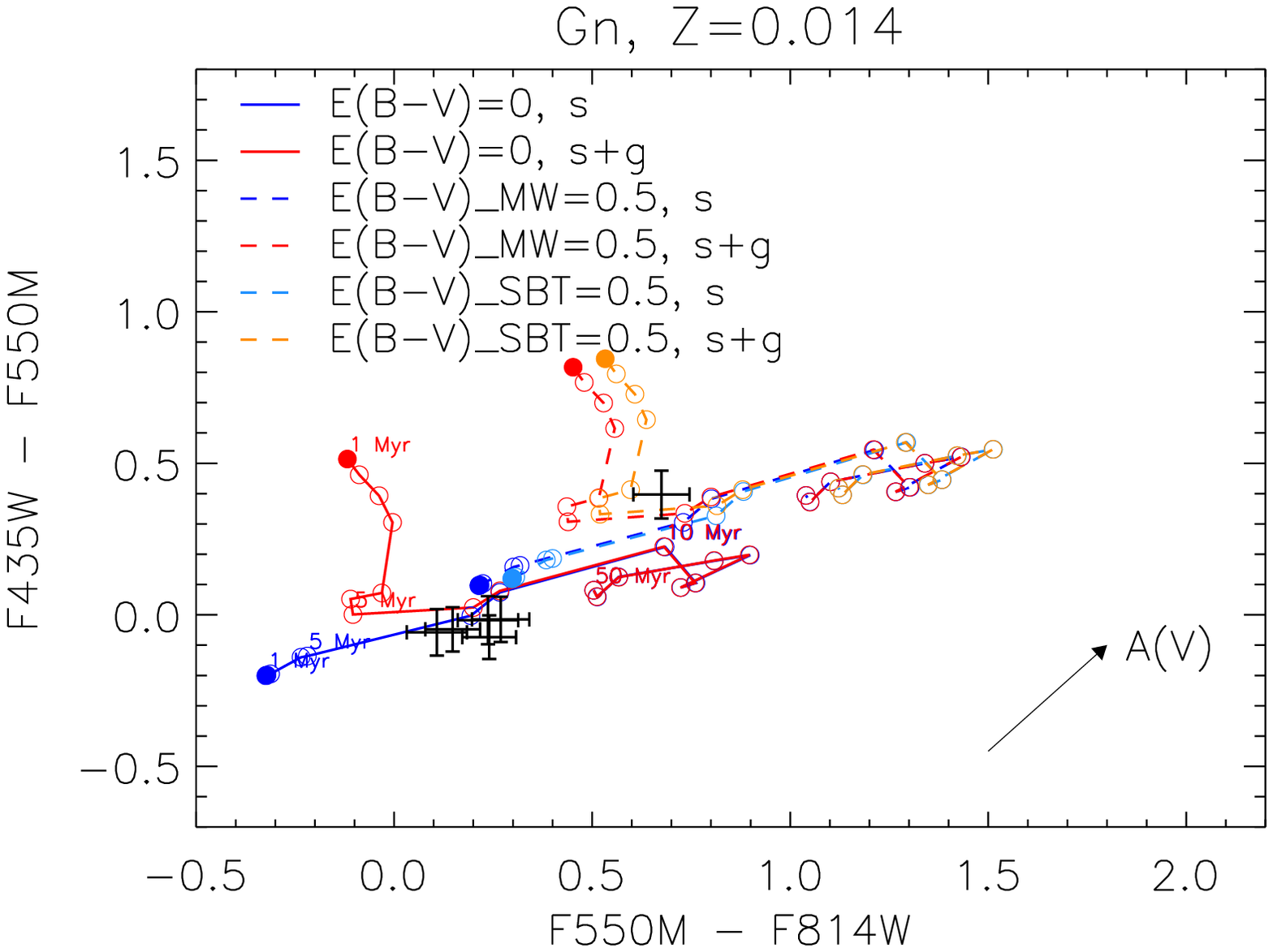}
\end{subfigure}
\begin{subfigure}
\centering
\includegraphics[width=0.89\columnwidth]{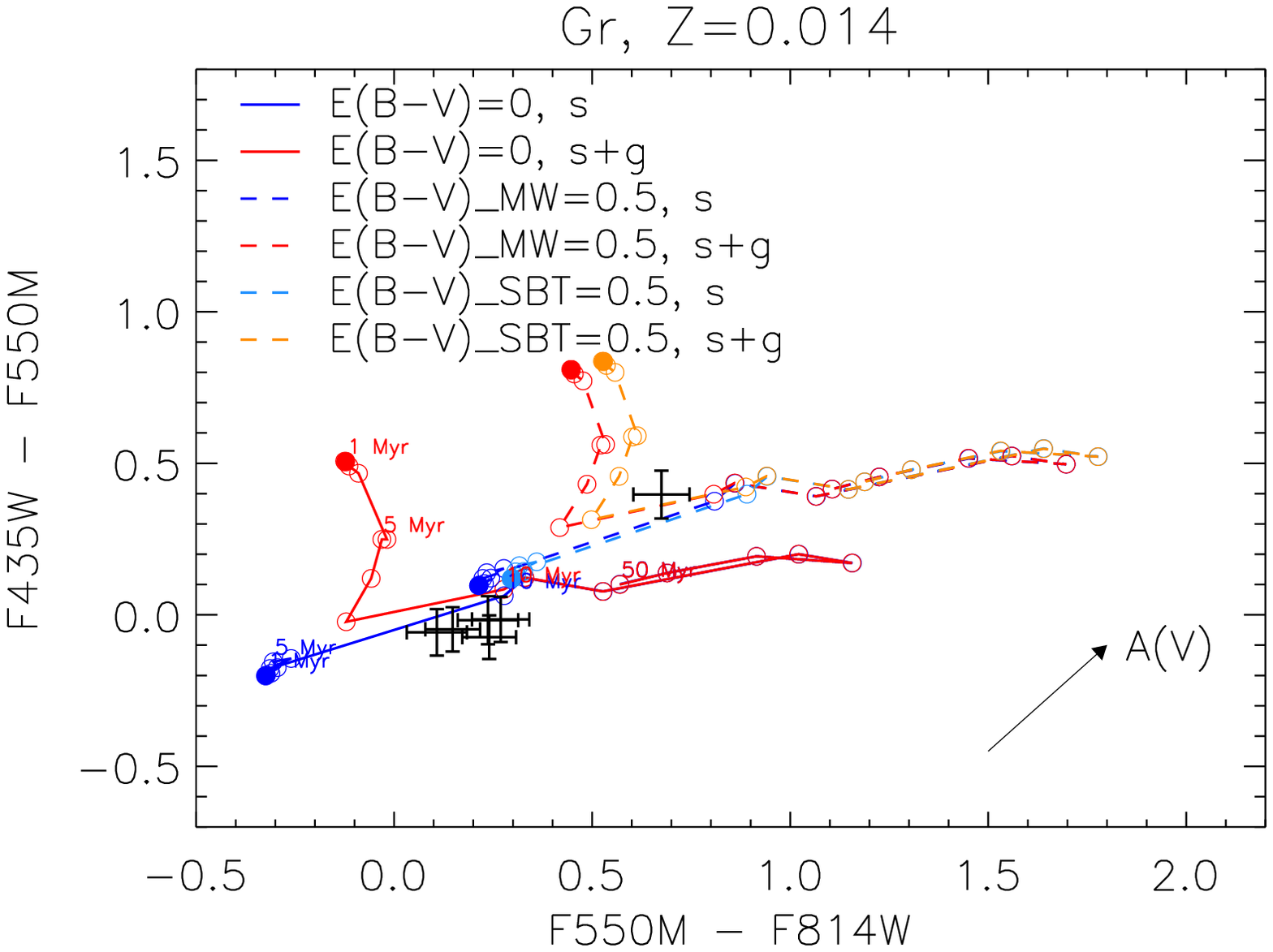}
\end{subfigure}
\begin{subfigure}
\centering
\includegraphics[width=0.89\columnwidth]{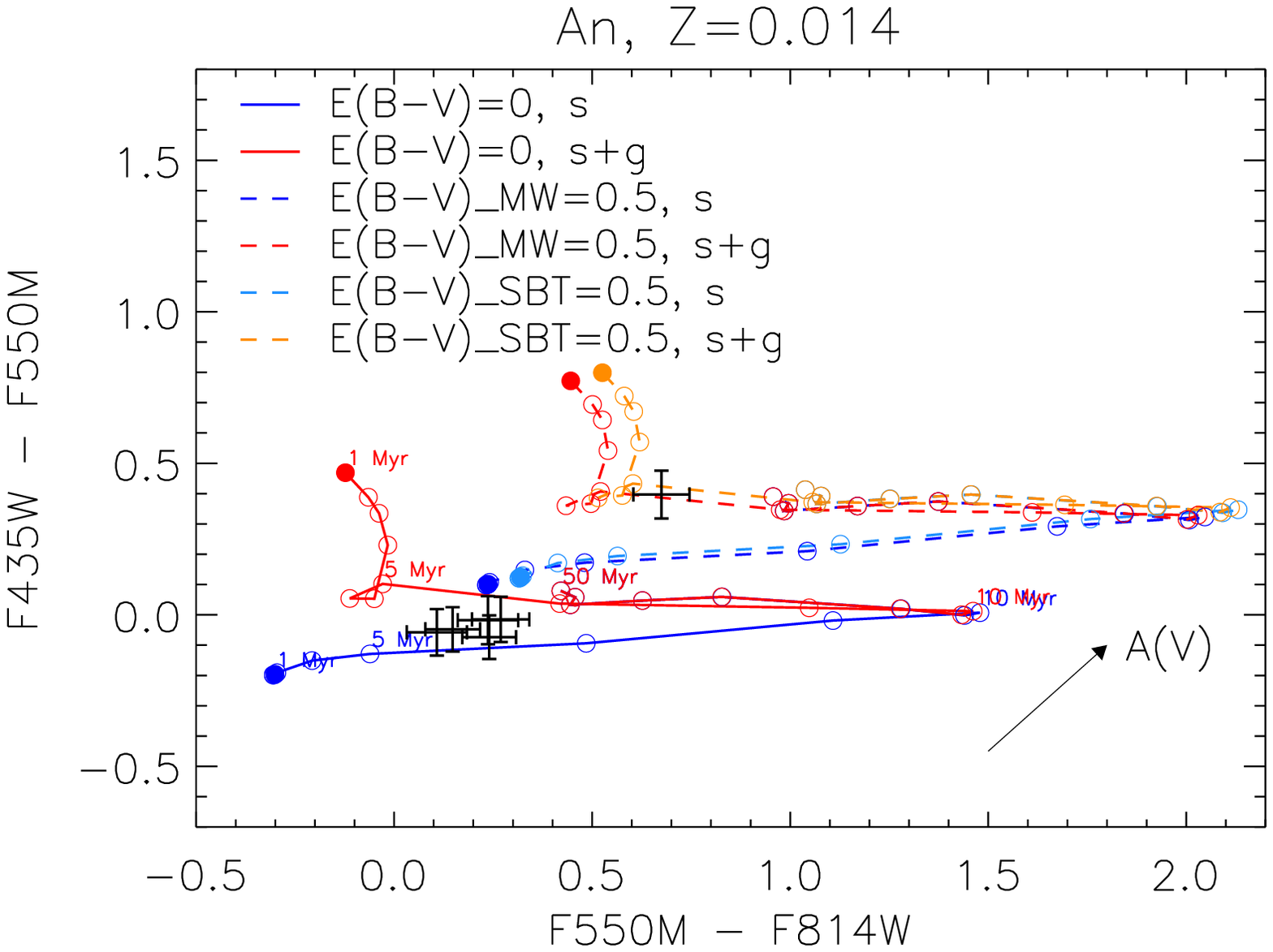}
\end{subfigure}
\begin{subfigure}
\centering
\includegraphics[width=0.89\columnwidth]{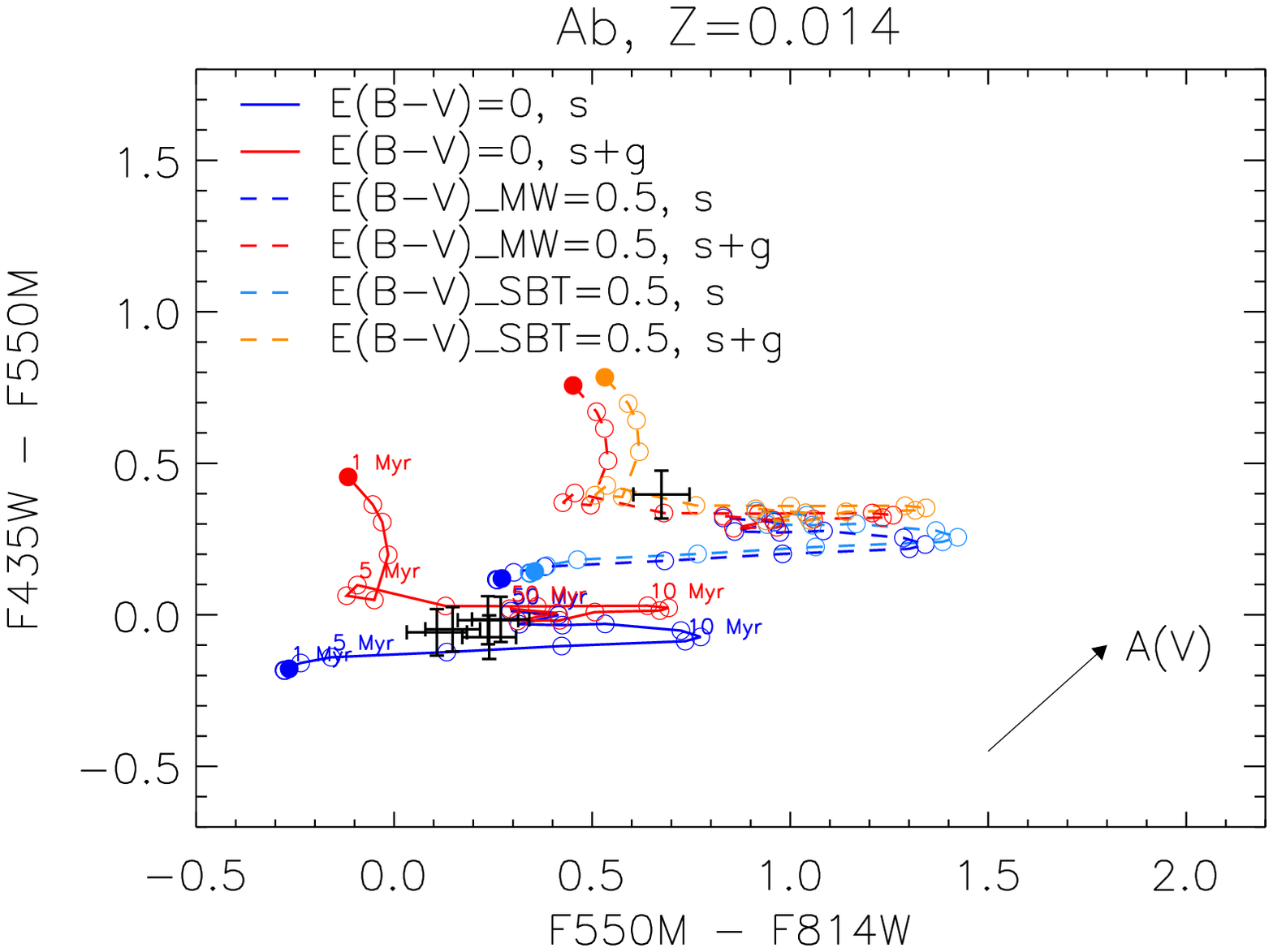}
\end{subfigure}
\caption{Similar to Fig. 6, but we show WFC3 UVIS bands, NGC 1566 data points, and we use the MW extinction law. The metallicity of the Po and Go models is different relative to that of other models.}
\label{fig7}
\end{figure*}

In Figs.~\ref{fig6} and~\ref{fig7} we show the impact of the ionized gas on colors for NGC 5253 and NGC 1566, respectively. We show model predictions for unattenuated and attenuated spectra, adopting $E(B-V)=0.5$ mag in the latter case. We compare results based on starburst and alternative attenuations, as indicated in the captions. We overlay the observations (symbols with error bars). The right panel of fig. 11 in \cite{rei10} can be compared with our Go predictions, which are plotted in the top-right panel of Fig.~\ref{fig6}. There is general agreement between the Go predictions presented in this work and in \cite{rei10}. There are clear differences from models based on different flavors of massive star evolution. For instance, the An models extend significantly further to the upper right of the color-color diagrams compared to other models in Fig.~\ref{fig6}. Differences in the color-color diagrams due to metallicity can be seen by comparing the same models in Fig.~\ref{fig6} and~\ref{fig7}. The diagrams show the importance of accounting for nebular emission at ages younger than 10 Myr for most models and even older ages for the binary models.

%%%%%%%%%%%%%%%%%%%%%%%%%%%%%%
% Figure 6
%%%%%%%%%%%%%%%%%%%%%%%%%%%%%%

%%%%%%%%%%%%%%%%%%%%%%%%%%%%%%
% Figure 7
%%%%%%%%%%%%%%%%%%%%%%%%%%%%%%

%%%%%%%%%%%%%%%%%%%%%%%%%%%%%%
%%% Method
%%%%%%%%%%%%%%%%%%%%%%%%%%%%%%

\section{Method}

We use Bayesian inference to constrain model parameters from the observations. For each cluster property, i.e., reddening [$E(B-V)$], mass ($M_{\rm{cl}}$) and age ($t$) we record two values, the best-fitting or minimum $\chi^2$ value, and the median of the posterior marginalized probability distribution function. We use flat priors in $E(B-V)$, log($M_{\rm{cl}}$) and log($t$). Our errors around the median correspond to the 16th and 84th percentiles of the probability density function. The posterior marginalized probability distribution functions are computed as follows. Let  $x_1=E(B-V)$,  $x_2=M_{\rm{cl}}$ and $x_3=t$ . The clusters are observed in five photometric bands. For a given cluster, let $\mathbf{y}_{\rmn{obs}}=(y_{\rmn{obs,1}},\,y_{\rmn{obs,2}},\,y_{\rmn{obs,3}},\,y_{\rmn{obs,4}},\,y_{\rmn{obs,5}})$ be the fluxes obtained from the observed reddening-uncorrected apparent magnitudes, and $\mathbf{\sigma}=(\sigma_1,\,\sigma_2,\,\sigma_3,\,\sigma_4,\,\sigma_5)$ the corresponding flux errors. In addition, let $\mathbf{y}_{\rmn{mod}}=(y_{\rmn{mod,1}},\,y_{\rmn{mod,2}},\,y_{\rmn{mod,3}},\,y_{\rmn{mod,4}},\,y_{\rmn{mod,5}})$ represent a set of fluxes obtained from a synthetic library of reddened magnitudes that mimic the redshift and distance of the observations. The synthetic fluxes correspond to the mass in living stars plus remnants at age t. We infer the marginal posterior probability distribution, $p(x_k|\mathbf{y}_{\rmn{obs}};\mathbf{\sigma})$ for the physical parameter $x_k$ given the observations and errors, using expression:

\begin{equation}
p(x_k|\mathbf{y}_{\rmn{obs}};\mathbf{\sigma})\propto\sum_{k'=1}^{n_{k'}}\rmn{exp}(\frac{-\chi_{x_1,\,x_2,\,x_3}^2}{2})\label{eq:pdf1}
\end{equation}
where $n_{k'}$ is the number of possible values for the physical parameter $x_{k'}$ ($k\ne\,k'$), for a fixed metallicity, prescription for attenuation by dust, and set of tracks; and $\chi_{x_1,\,x_2,\,x_3}^2$ is obtained from:
\begin{equation}
\chi_{x_1,x_2,x_3}^2=\sum_{i=1}^{5}\frac{(y_{\rmn{obs,i}}-A_{x_1,\,x_2,\,x_3}\,\cdot\,y_{\rmn{mod,i}})^2  }{ \sigma_\rmn{i}^2}
\end{equation}
where the sum is over the number of filters. The expression for $A_{x_1,\,x_2,\,x_3}$ is found by minimizing the difference between observed and model magnitudes:
\begin{equation}
\frac{\partial \chi_{x_1,x_2,x_3}^2}{\partial A_{x_1,x_2,x_3}}=0
\end{equation}
which yields
\begin{equation}
A_{x_1,\,x_2,\,x_3}=\frac{\sum_{i=1}^{5} y_{\rmn{obs,i}}\,\cdot\,y_{\rmn{mod,i}}  }{\sum_{i=1}^{5} y_{\rmn{mod,i}}^2}
\end{equation}
The stellar mass of the cluster is the product of $A_{x_1,\,x_2,\,x_3}$ and the predicted mass in living stars plus remnants at age $t$.\footnote{See Section~\ref{sec:m_init} for a discussion of how the adopted initial mass of the model stellar population affects the derived cluster properties).} The most probable values of  $x_1$, $x_2$, and $x_3$ are the peaks of the distributions given by equation (\ref{eq:pdf1}). The most typical values are the medians of these distributions. For  $x_1$ for example, the median of the PDF would be the value $x_{50}$, such that when computing the area under the PDF from the minimum value of $x_1$ to $x_{50}$, the area is 50 per cent of the total area. The 16th and 84th percentiles are computed in a similar way but using instead 16 per cent and 84 per cent of the total area.

%%%%%%%%%%%%%%%%%%%%%%%%%%%%%%
%%% Results
%%%%%%%%%%%%%%%%%%%%%%%%%%%%%%

\section{Results}

%%%%%%%%%%%%%%%%%%%%%%%%%%%%%%
%%% Fit quality
%%%%%%%%%%%%%%%%%%%%%%%%%%%%%%

\subsection{Spectral Energy Distribution Shapes}

\begin{figure*}
\begin{subfigure}
\centering
\includegraphics[width=0.50\columnwidth]{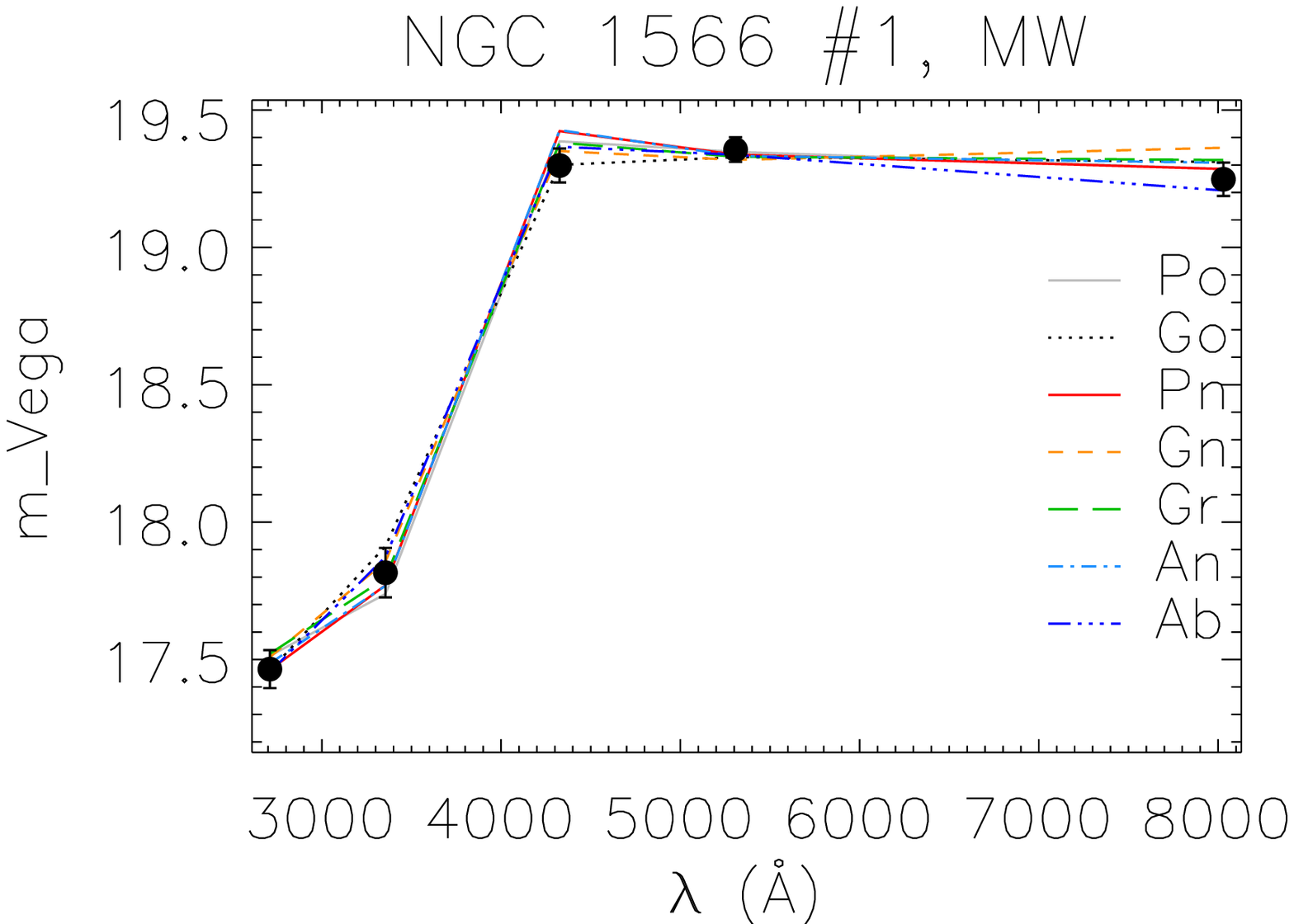}
\end{subfigure}
\begin{subfigure}
\centering
\includegraphics[width=0.50\columnwidth]{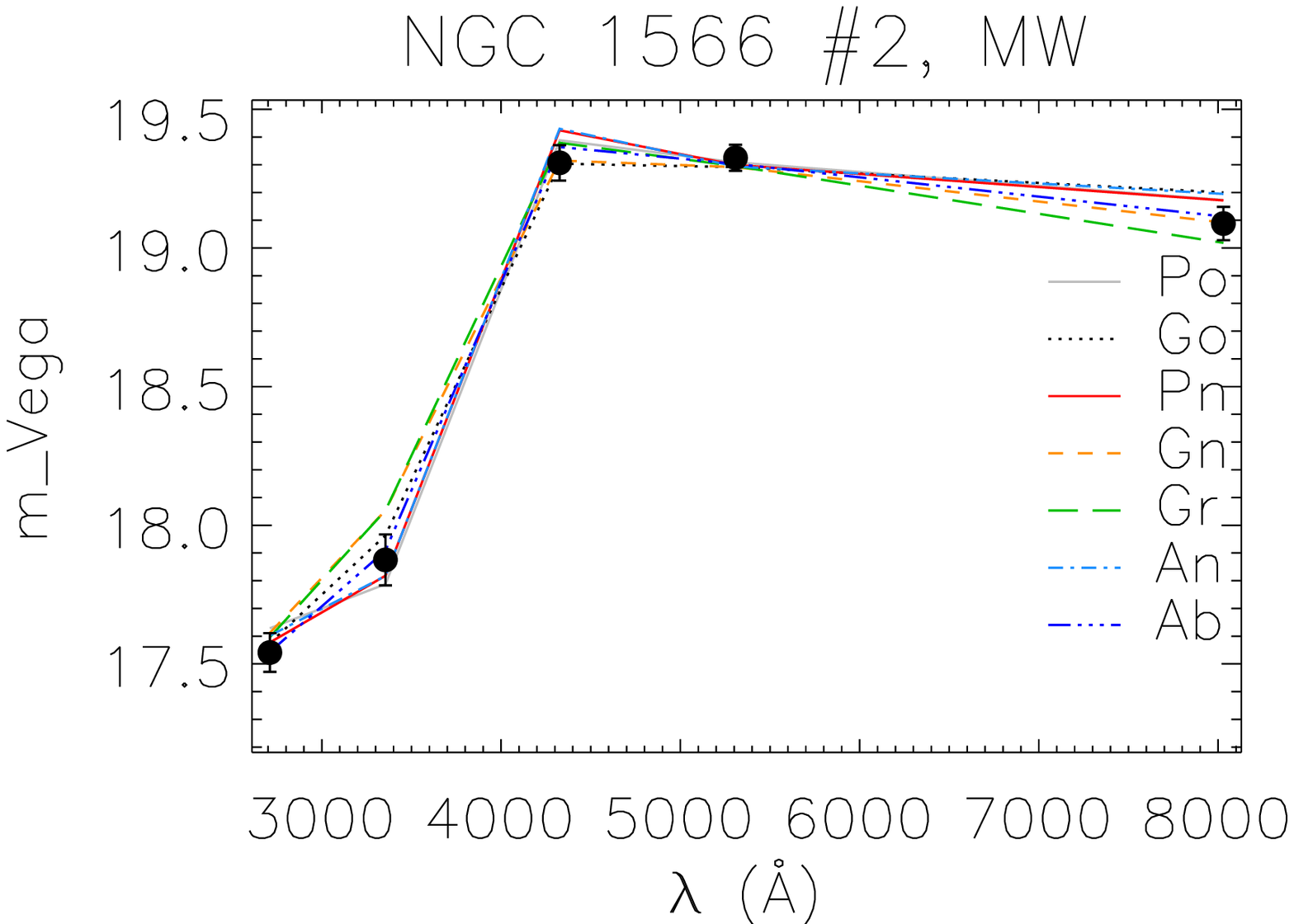}
\end{subfigure}
\begin{subfigure}
\centering
\includegraphics[width=0.50\columnwidth]{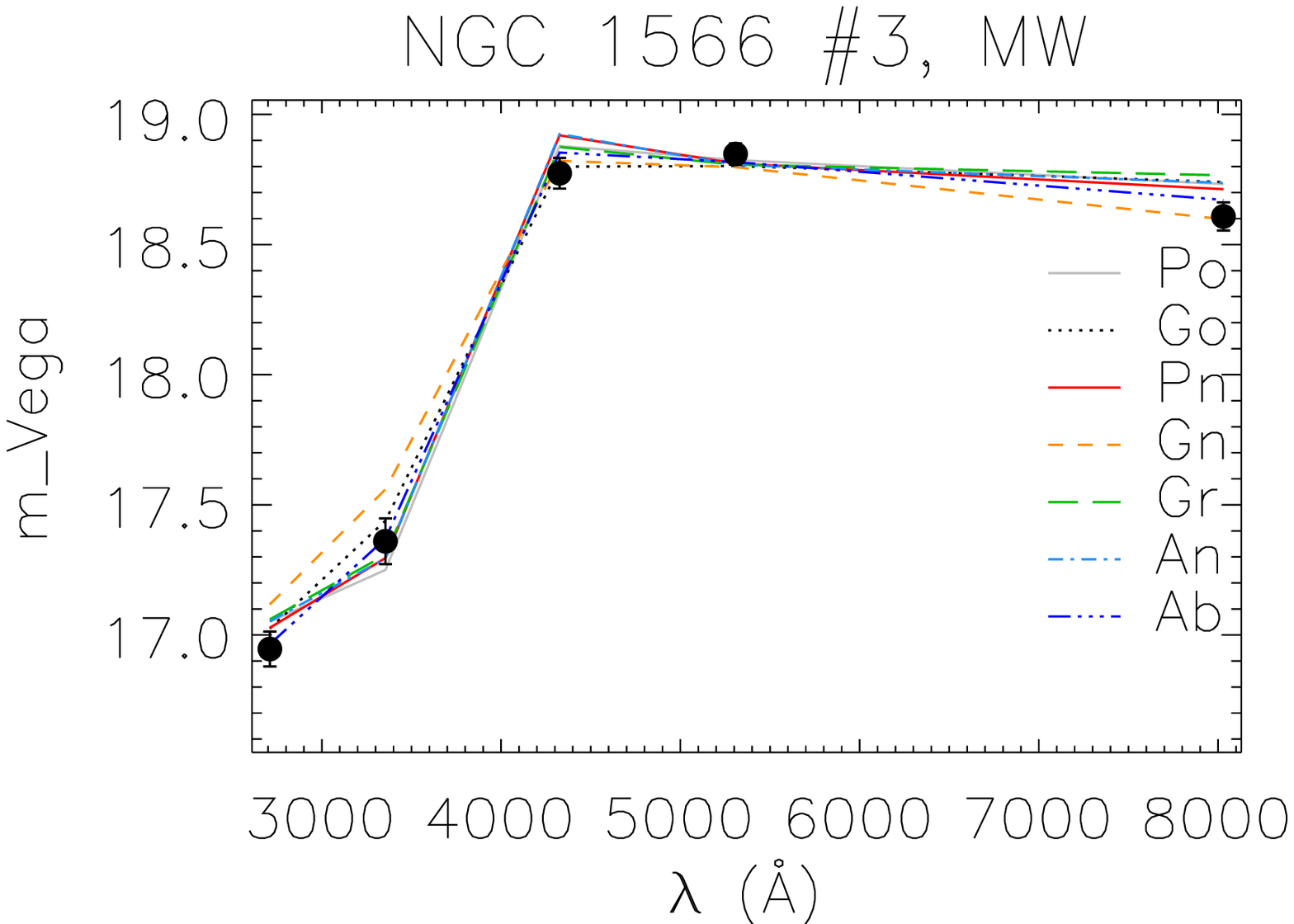}
\end{subfigure}
\begin{subfigure}
\centering
\includegraphics[width=0.50\columnwidth]{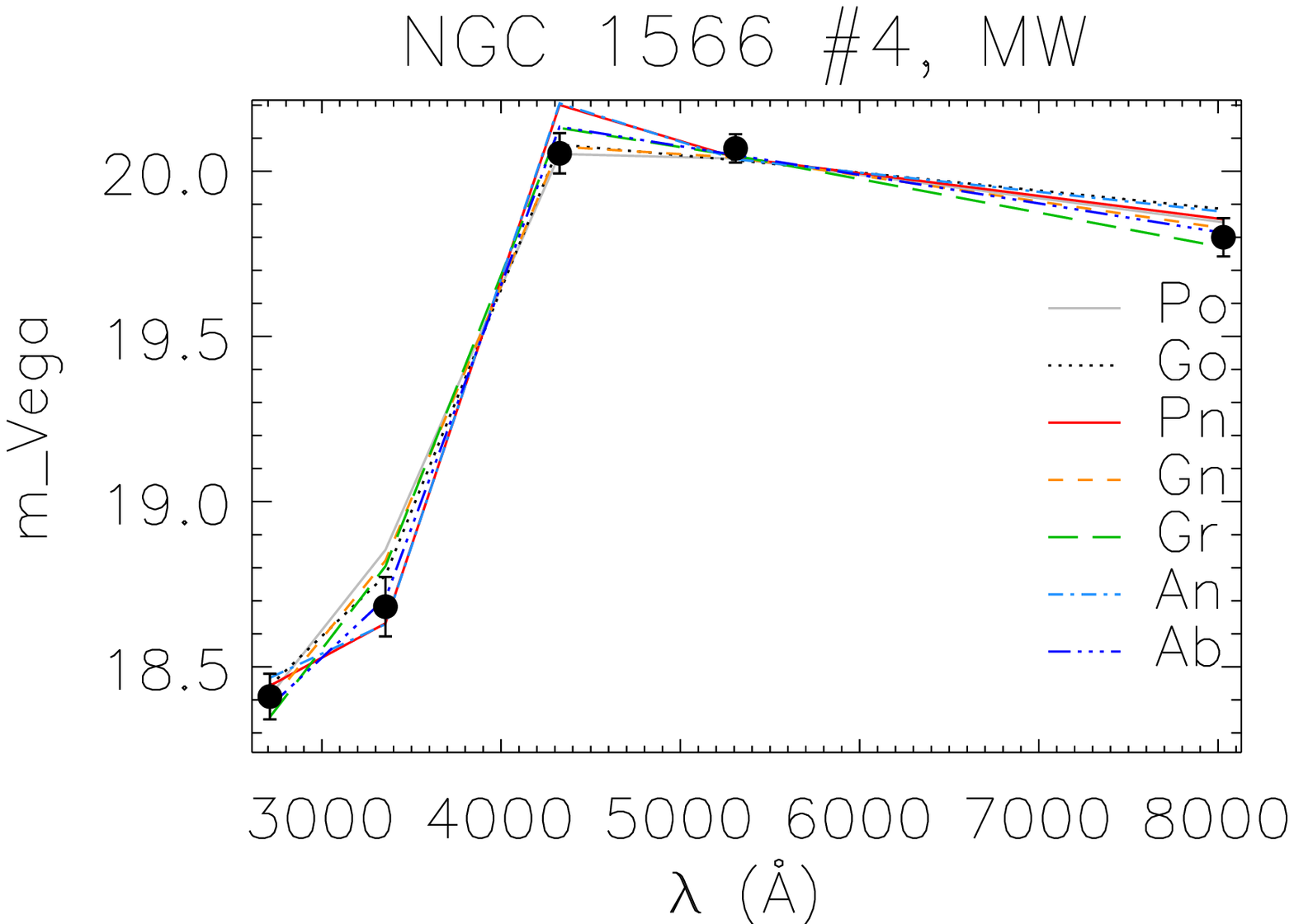}
\end{subfigure}
\begin{subfigure}
\centering
\includegraphics[width=0.50\columnwidth]{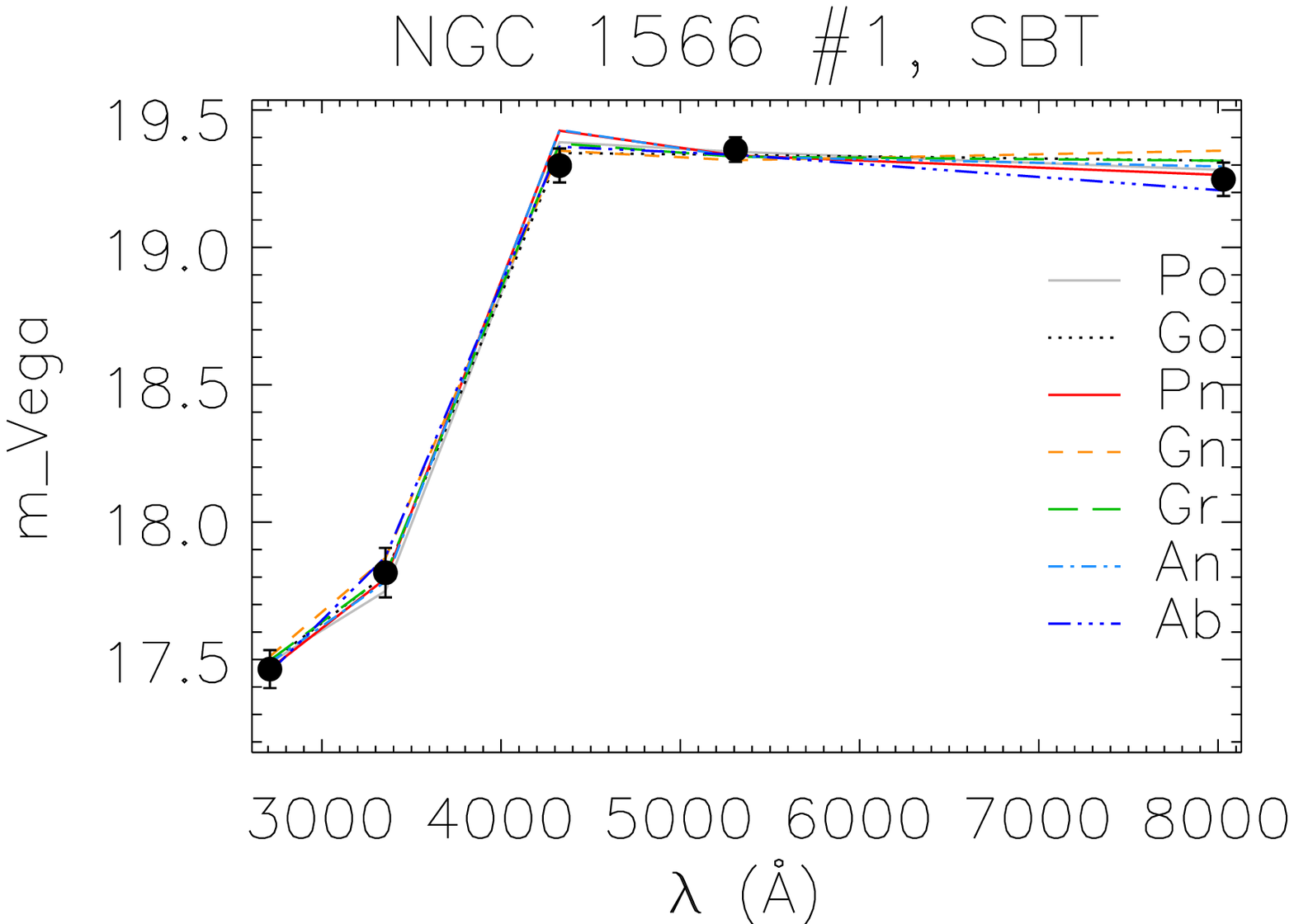}
\end{subfigure}
\begin{subfigure}
\centering
\includegraphics[width=0.50\columnwidth]{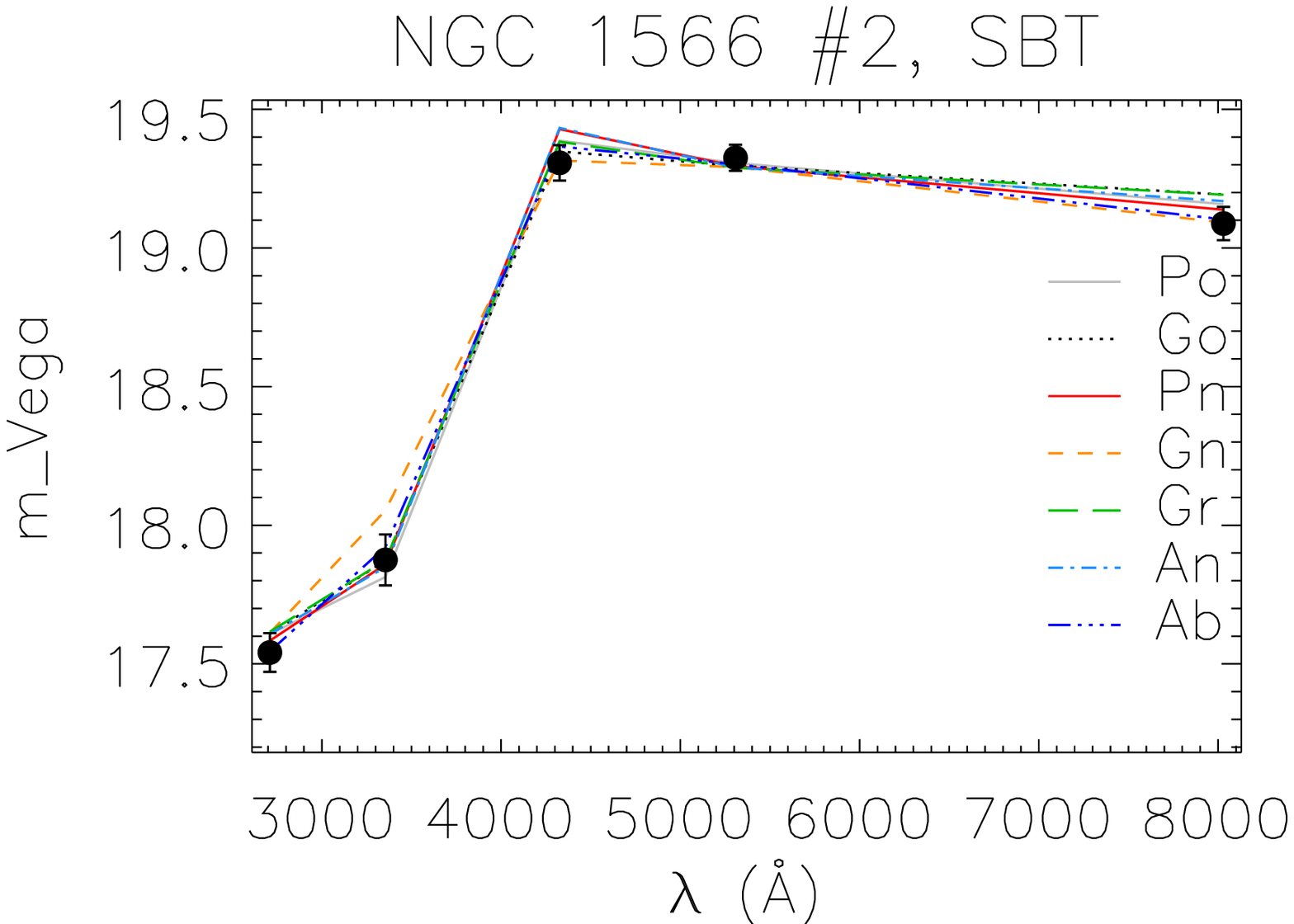}
\end{subfigure}
\begin{subfigure}
\centering
\includegraphics[width=0.50\columnwidth]{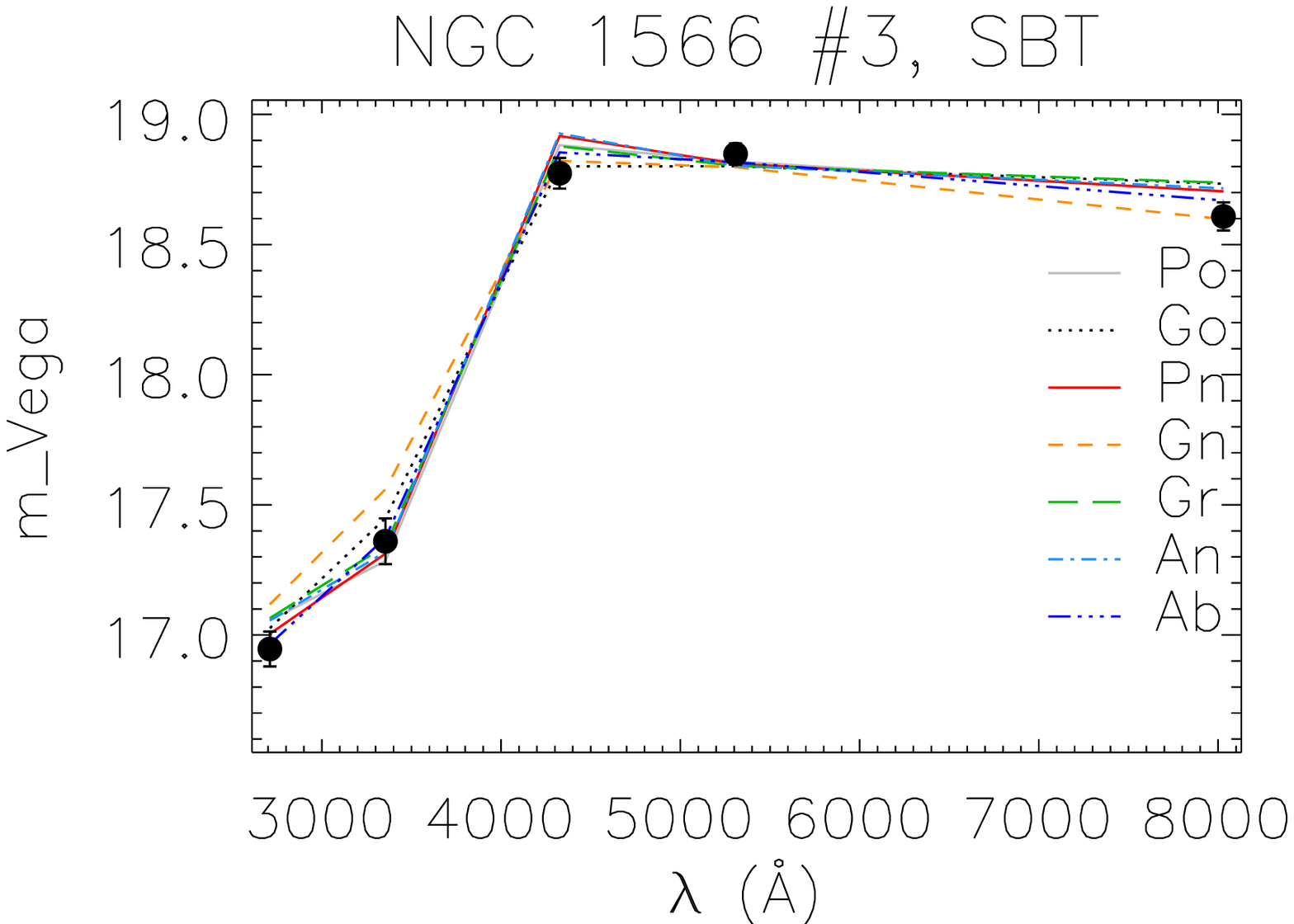}
\end{subfigure}
\begin{subfigure}
\centering
\includegraphics[width=0.50\columnwidth]{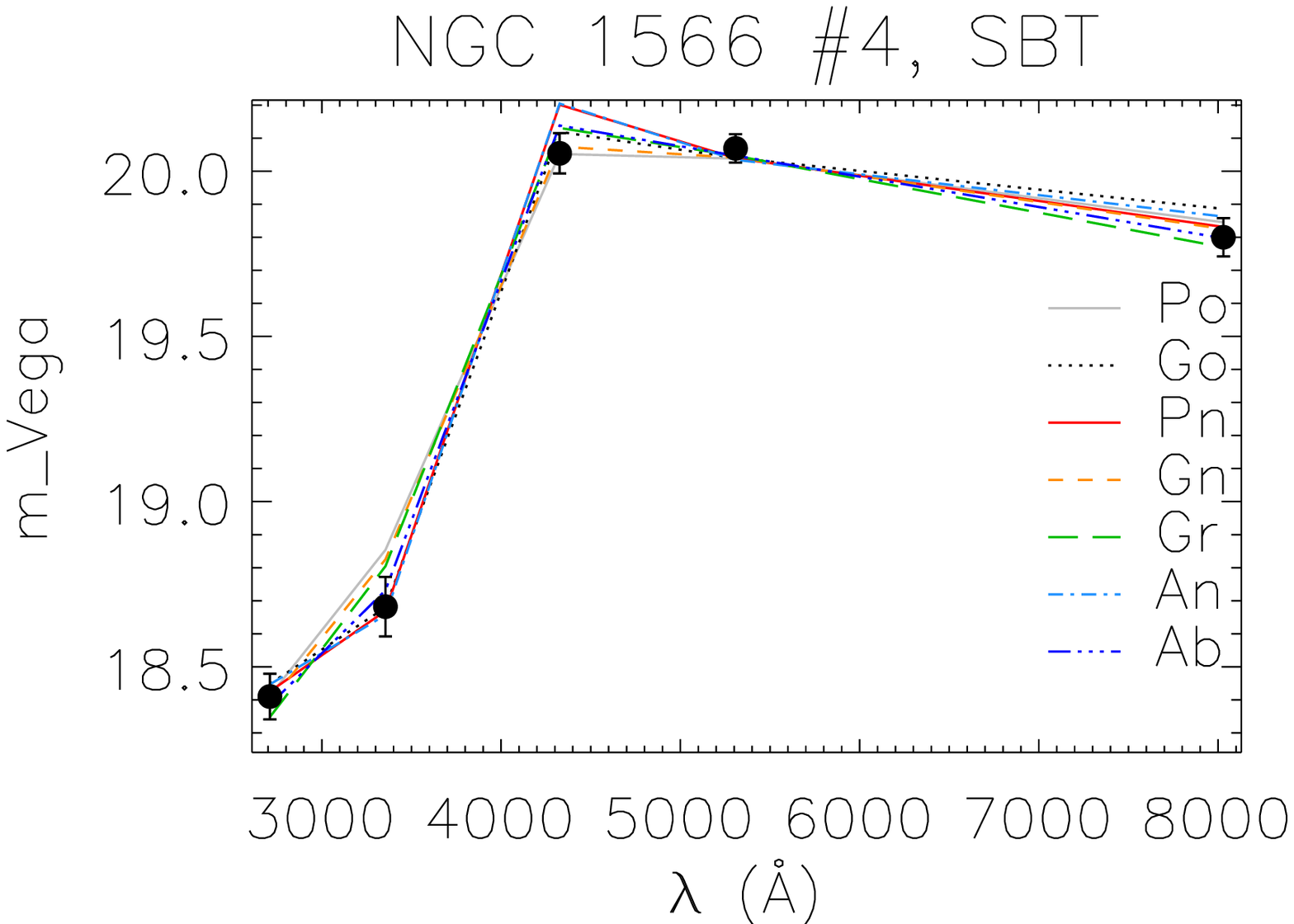}
\end{subfigure}
\begin{subfigure}
\centering
\includegraphics[width=0.50\columnwidth]{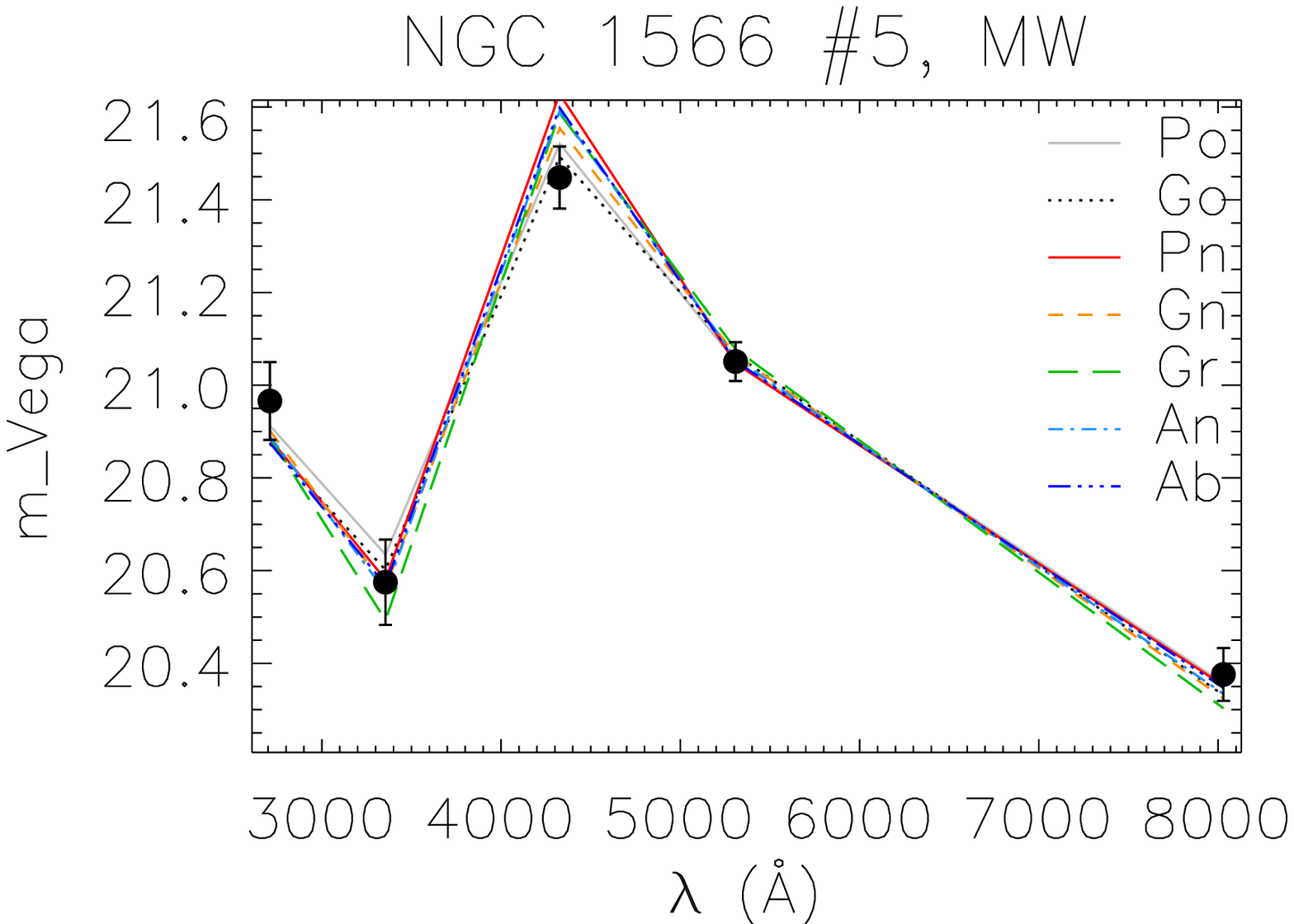}
\end{subfigure}
\begin{subfigure}
\centering
\includegraphics[width=0.50\columnwidth]{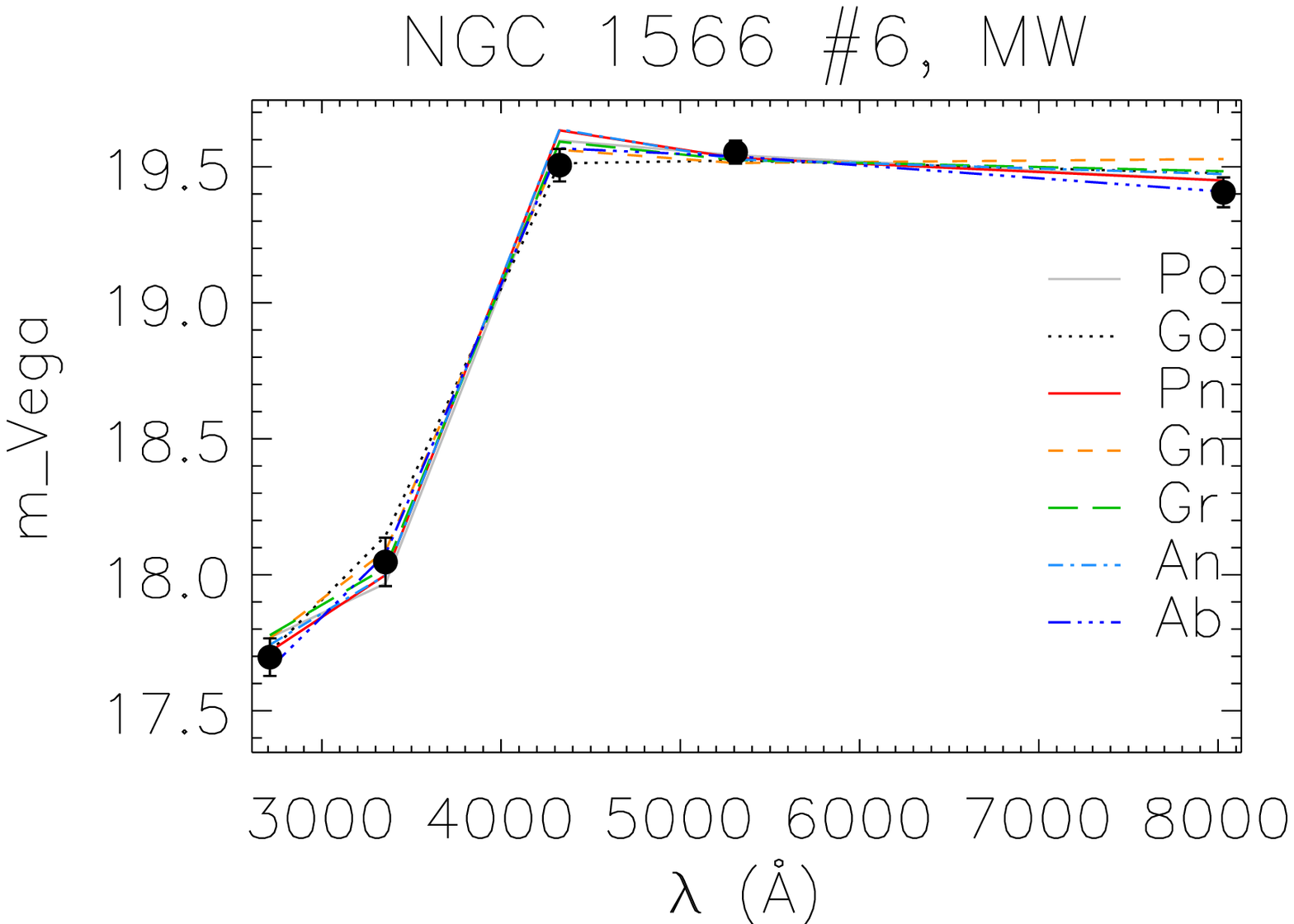}
\end{subfigure}
\begin{subfigure}
\centering
\includegraphics[width=0.50\columnwidth]{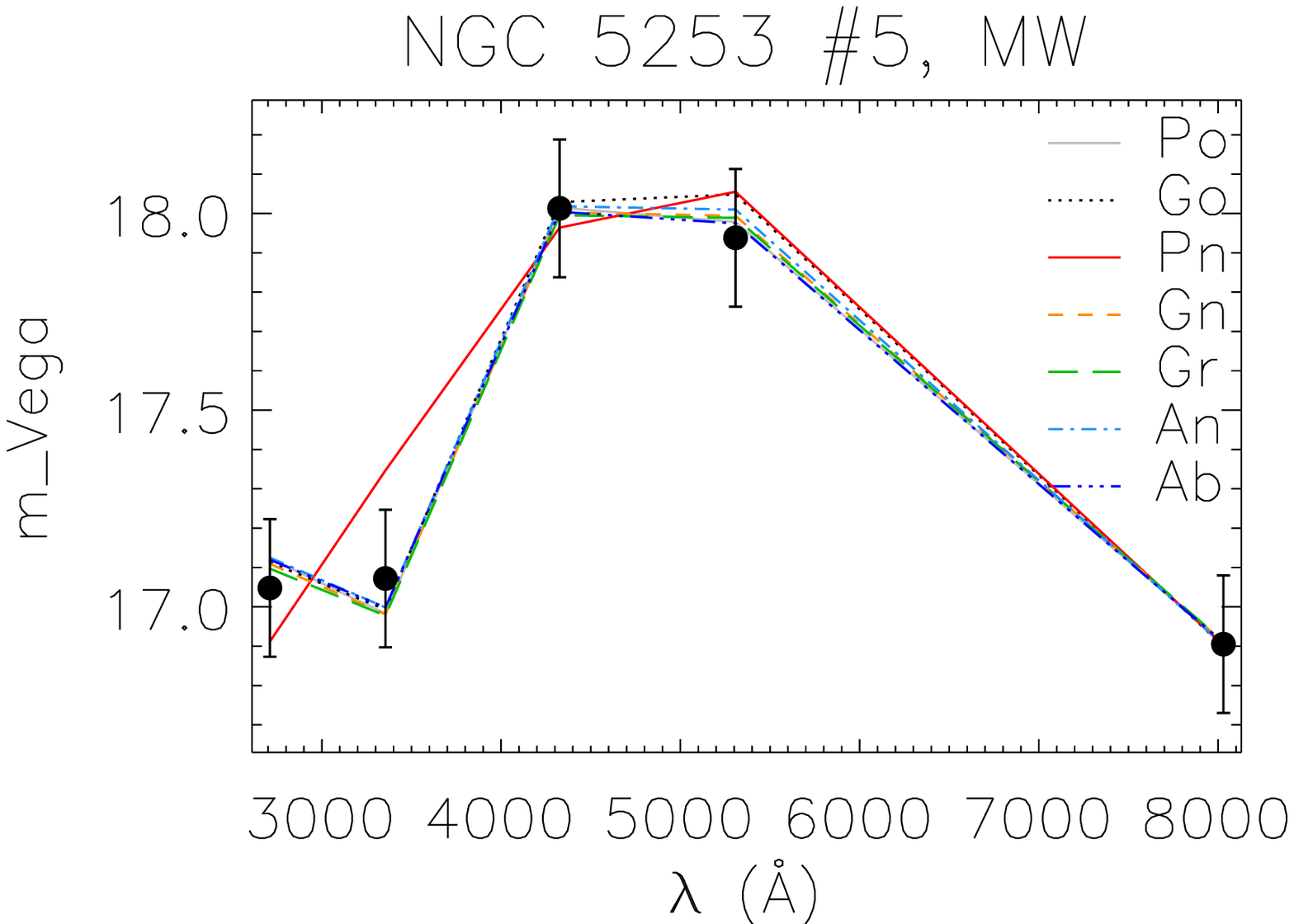}
\end{subfigure}
\begin{subfigure}
\centering
\includegraphics[width=0.50\columnwidth]{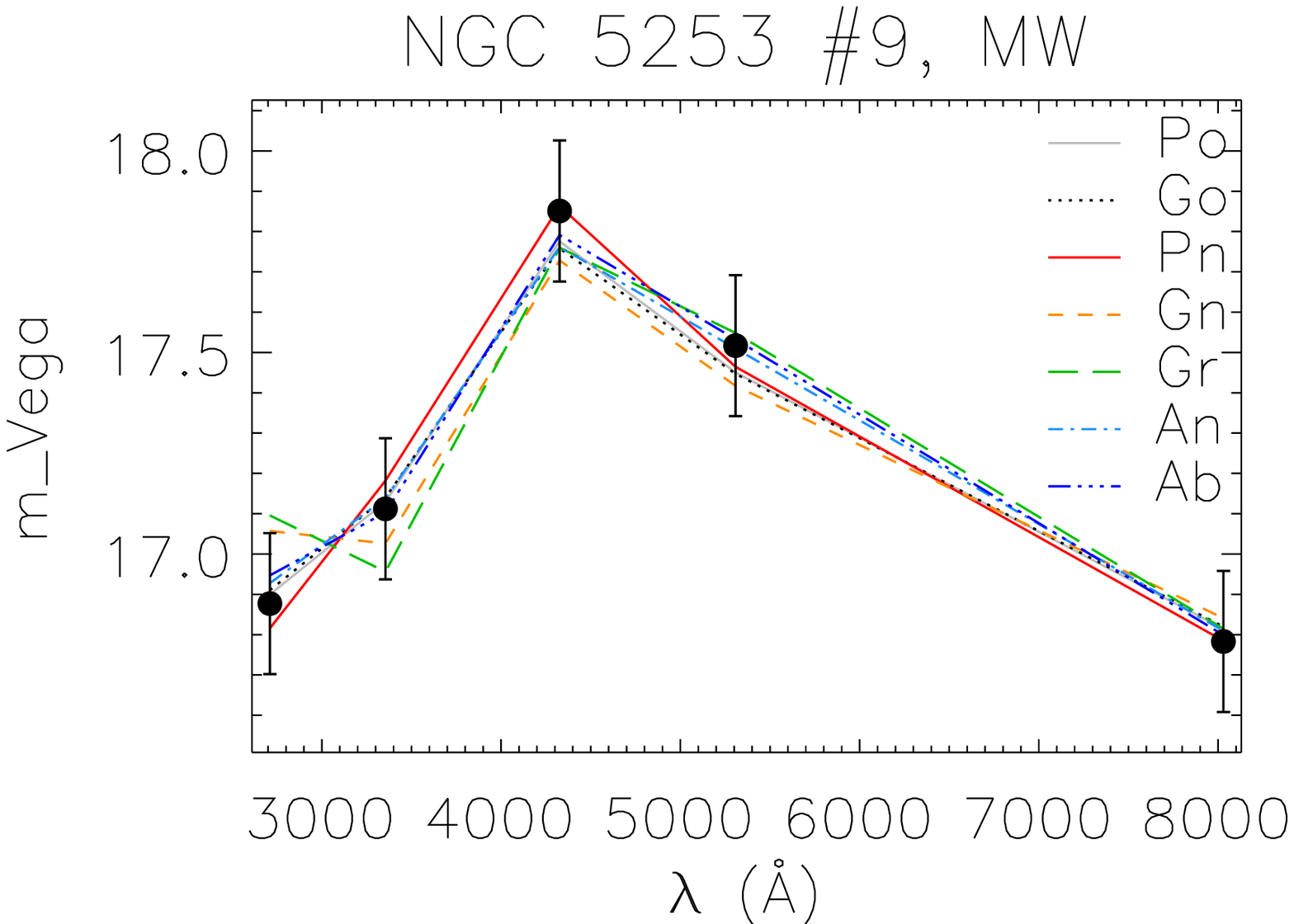}
\end{subfigure}
\begin{subfigure}
\centering
\includegraphics[width=0.50\columnwidth]{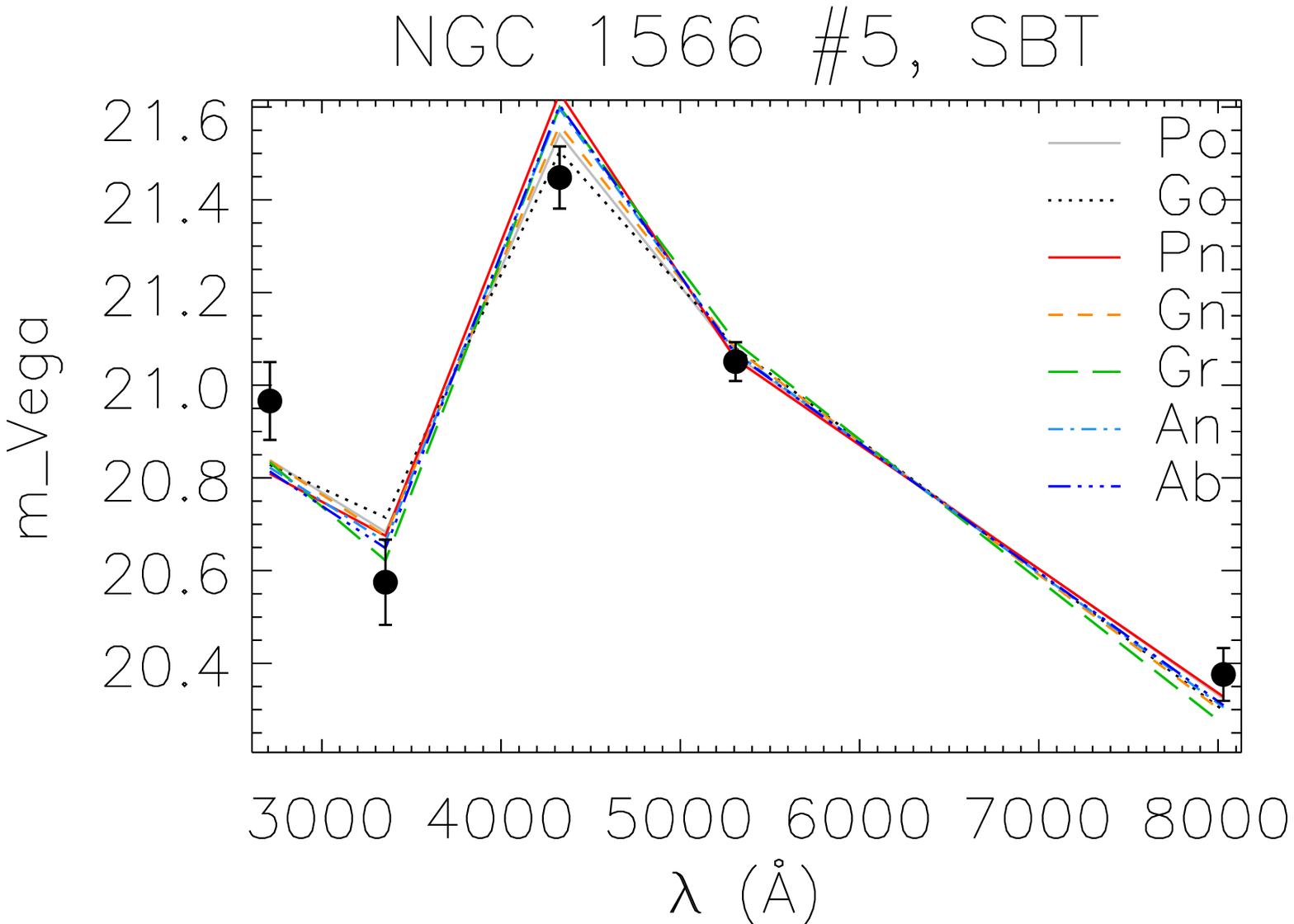}
\end{subfigure}
\begin{subfigure}
\centering
\includegraphics[width=0.50\columnwidth]{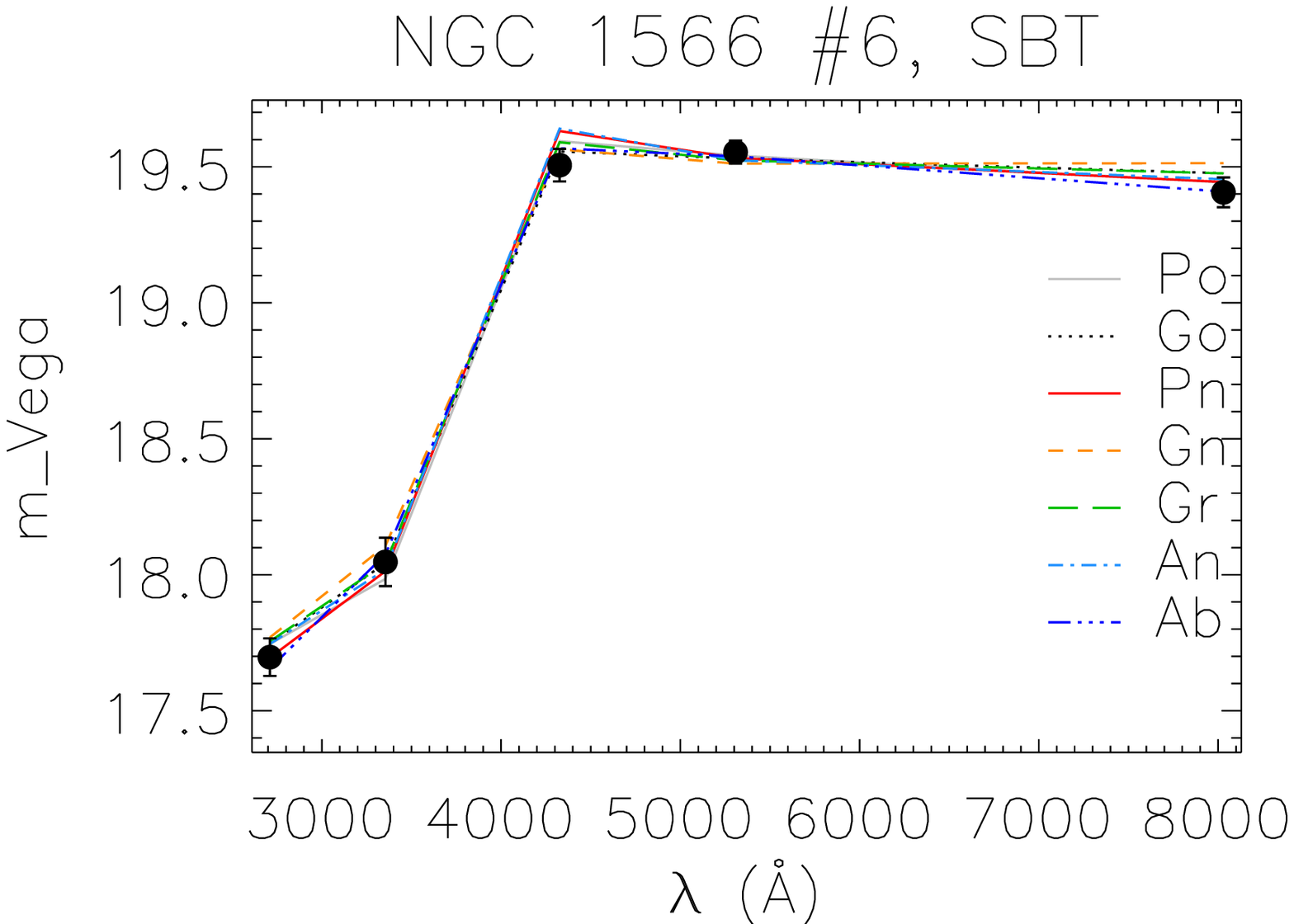}
\end{subfigure}
\begin{subfigure}
\centering
\includegraphics[width=0.50\columnwidth]{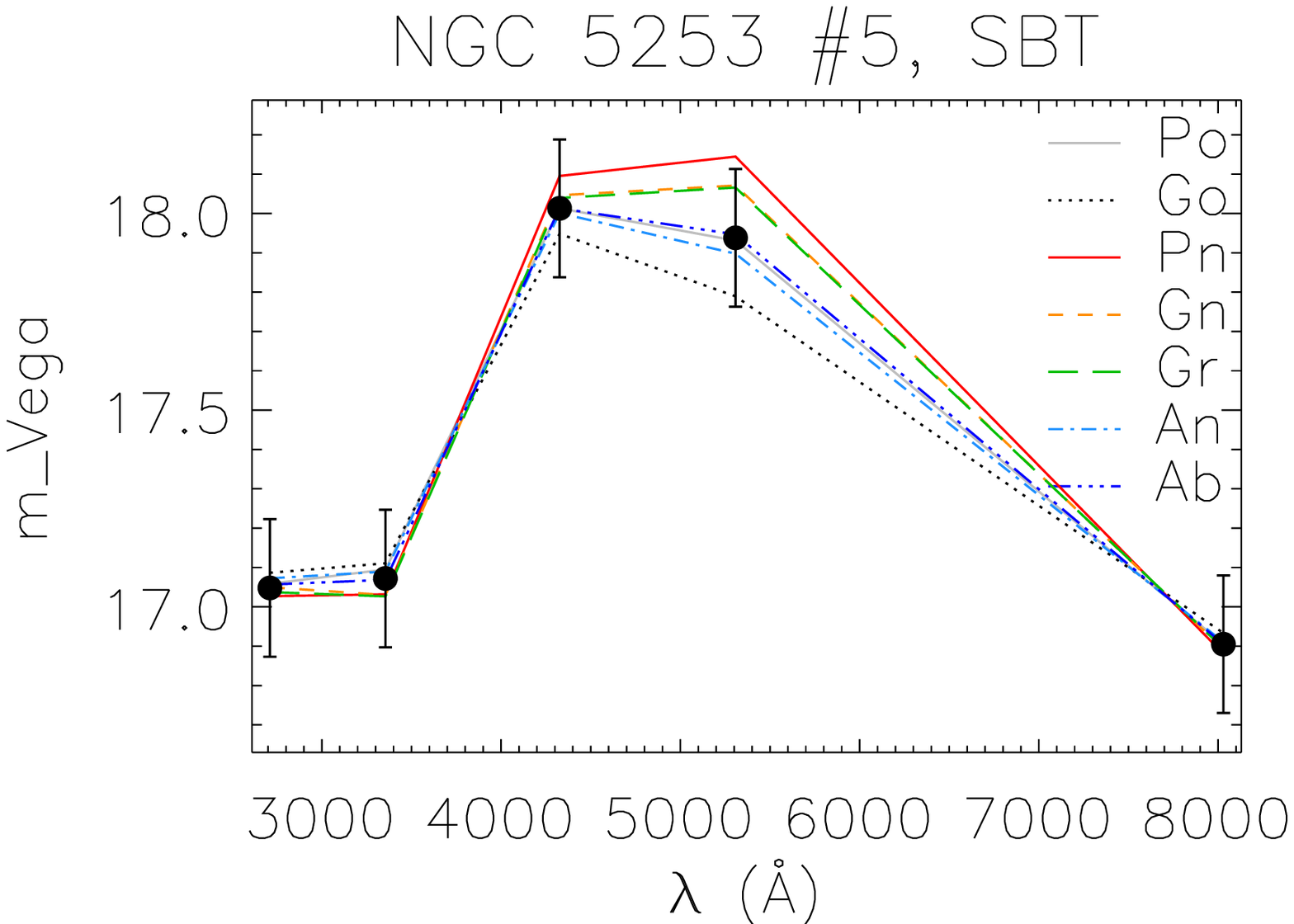}
\end{subfigure}
\begin{subfigure}
\centering
\includegraphics[width=0.50\columnwidth]{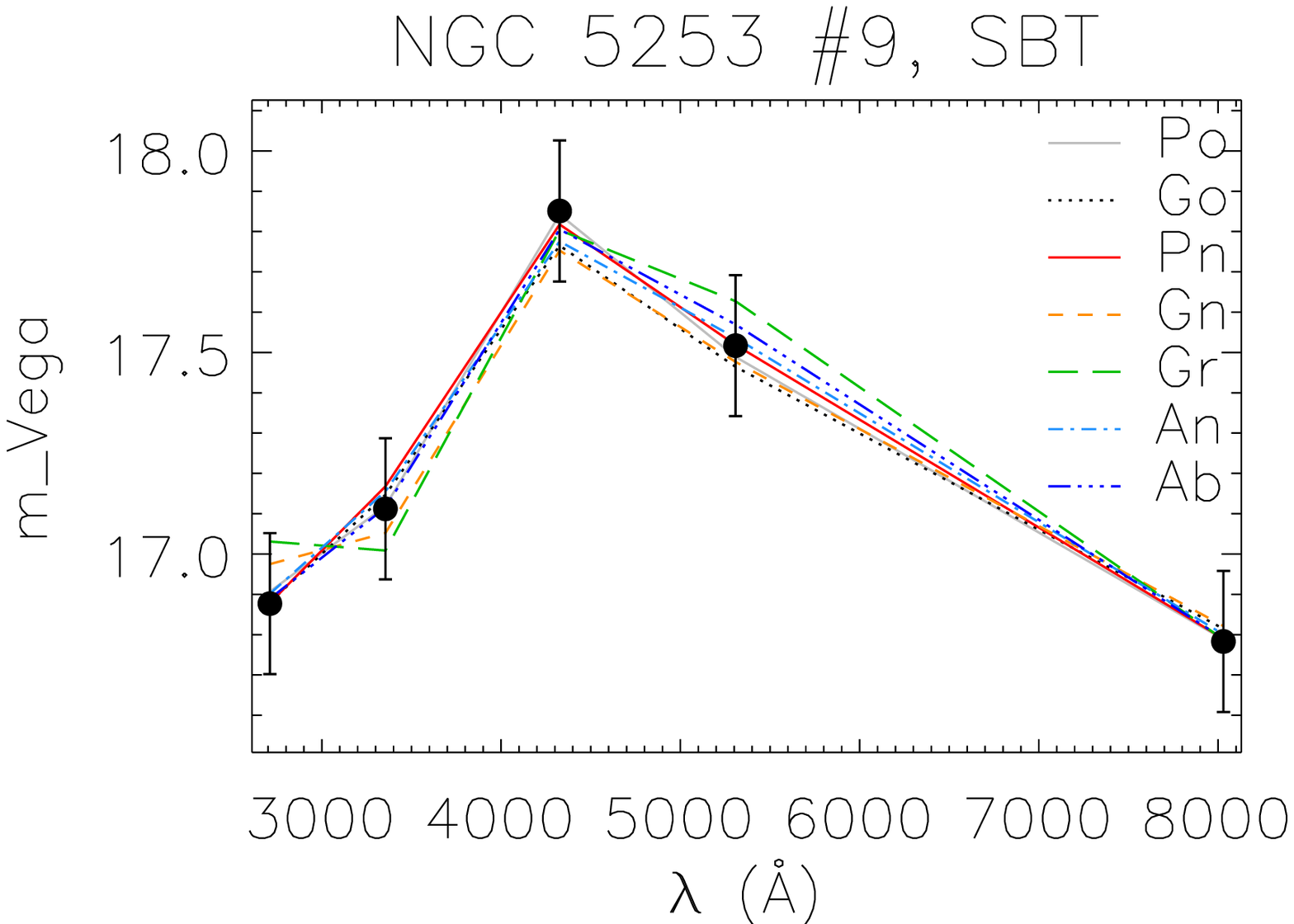}
\end{subfigure}

\caption{Comparison of observations (filled symbols with error bars) with seven different best-fit models (curves of different types, as indicated by the legend). Each panel corresponds to the cluster indicated in the title. The vertical axis gives apparent Vega magnitudes. The models shown in the 1st and 3rd rows use the MW (NGC 1566) or SMC (NGC 5253) extinction laws, while the models shown in the 2nd and 4th rows use the starburst attenuation law.}
\label{fig8}
\end{figure*}

In Fig.~\ref{fig8}, we compare the observed reddening-uncorrected apparent Vega magnitudes of the YMCs (black symbols with error bars) to 14 different best-fitting models, corresponding to the adopted 7 flavors of massive-star evolution (lines of different styles) and 2 prescriptions for attenuation by dust. Each panel corresponds to a different combination of cluster and dust attenuation. We use the MW (NGC 1566) or SMC (NGC 5253) extinction law for clusters in the first and third rows, and the starburst attenuation law for clusters in the second and fourth rows. Results obtained with two different laws are next to each other in the vertical direction. We find that all clusters in NGC 1566 have similar spectral energy distribution (SED) shapes, except for NGC 1566 \#5, which as shown below, is the most reddened cluster in this galaxy. On the other hand, the two clusters in NGC 5253 have different SED shapes relative to each other and clusters in NGC 1566. Differences in SED shapes of clusters in NGC 1566 and NGC 5253 are likely due to differences in cluster properties and the use of slightly different filter sets for the two galaxies. Overall, the models are successful in reproducing the data, as further quantified in the next section.

%%%%%%%%%%%%%%%%%%%%%%%%%%%%%%
%%% Figure 8
%%%%%%%%%%%%%%%%%%%%%%%%%%%%%%

\subsection{How well do the models fit the data?}\label{section:results1}

In this section, we discuss the performance of the different models in fitting the observations within the observational errors, explore if there is a preferred massive-star evolution flavor or prescription for attenuation by dust, and determine if there are bands which are better fitted than others. We answer these questions based on our analysis of best-fitting models.

\begin{figure*}
\centering
\includegraphics[width=2\columnwidth]{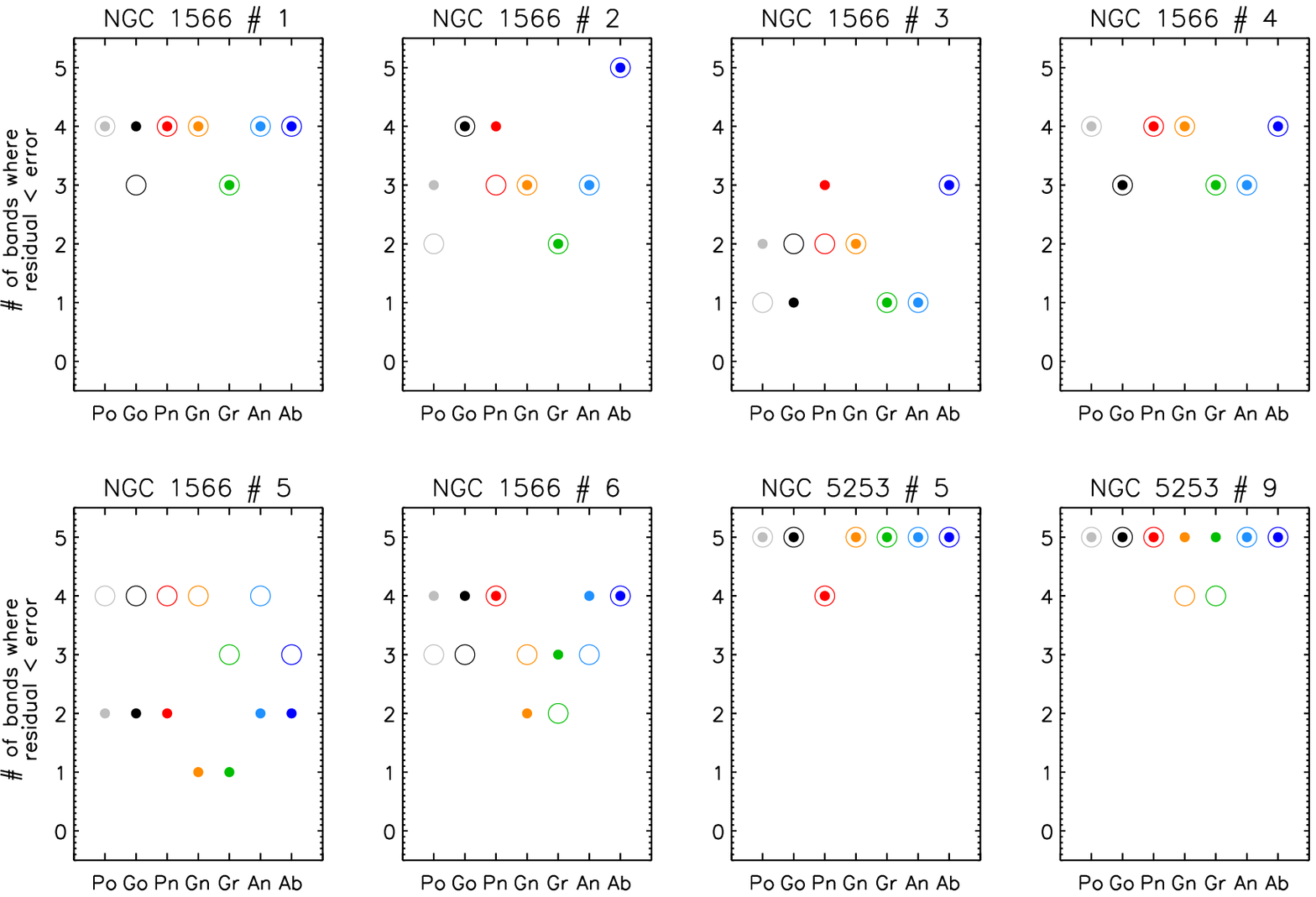}
\caption{Performance of different best-fitting models in fitting the most bands within the observational error. The models are indicated on the horizontal axis. The vertical axis gives the number of bands where the residual of the best-fit model magnitude is less than the observational error. We show results based on the starburst attenuation law (filled symbols) and MW (NGC 1566) or SMC (NGC 5253) extinction law (unfilled symbols). Each panel corresponds to a different cluster.}
\label{fig9}
\end{figure*}

In Fig.~\ref{fig9}, we quantify the performance of the different best-fitting models. The vertical axis of the Fig. gives the number of bands where the residual in the magnitude of the best-fitting model is within the observational error for the model indicated on the horizontal axis. We use filled symbols for results based on the starburst attenuation law and unfilled symbols for results based on the MW (NGC 1566) and SMC (NGC 5253) extinction laws. We show one cluster per panel. 

%%%%%%%%%%%%%%%%%%%%%%%%%%%%%%
% Figure 9
%%%%%%%%%%%%%%%%%%%%%%%%%%%%%%

Consider the total number of data points is the number of bands $\times$ the number of clusters, i.e., 40. Considering all data points we obtain, for each model and prescription for attenuation by dust, the following success rates of best-fitting models in fitting the observations within the observational errors: Po (72/70), Pn (75/75), Go (70/72), Gn (65/72), Gr (57/57), An (67/70), and Ab (80/82), where the first percentage uses the attenuation law and the second the extinction law. Thus, for both attenuation prescriptions, the Auckland binary models (Ab) are slightly more successful than the others and the Geneva rotating models (Gr) are the least successful. As previously mentioned, the Ab models employed in this work are standard models, i.e., we do not vary any parameters of these models to achieve a better fit. Confirmation of these results will require the use of a larger sample and further exploration of the model parameter space. In particular,  it is not clear that the above success rates will hold under different assumptions of metallicity and attenuation. Indeed, we remind the reader that in our analysis, we have fixed the metallicities of the two galaxies, because only two values of metallicity are available for all seven flavors of massive-star evolution. 

With regards to which dust attenuation prescription performs better, the current combination of LEGUS filters and limited sample size does not enable to answer this question conclusively. Which law performs better depends on the combination cluster and massive-star flavor. In several cases, the two prescriptions perform equally well. For NGC 1566 \#5 (the most reddened cluster in NGC 1566), the starburst law performs significantly worse than the MW law. Finally, in some cases, the starburst attenuation law performs better than the extinction law. Thus, answering this question would also benefit from a larger sample.

\begin{figure*}
\centering
\includegraphics[width=2.\columnwidth]{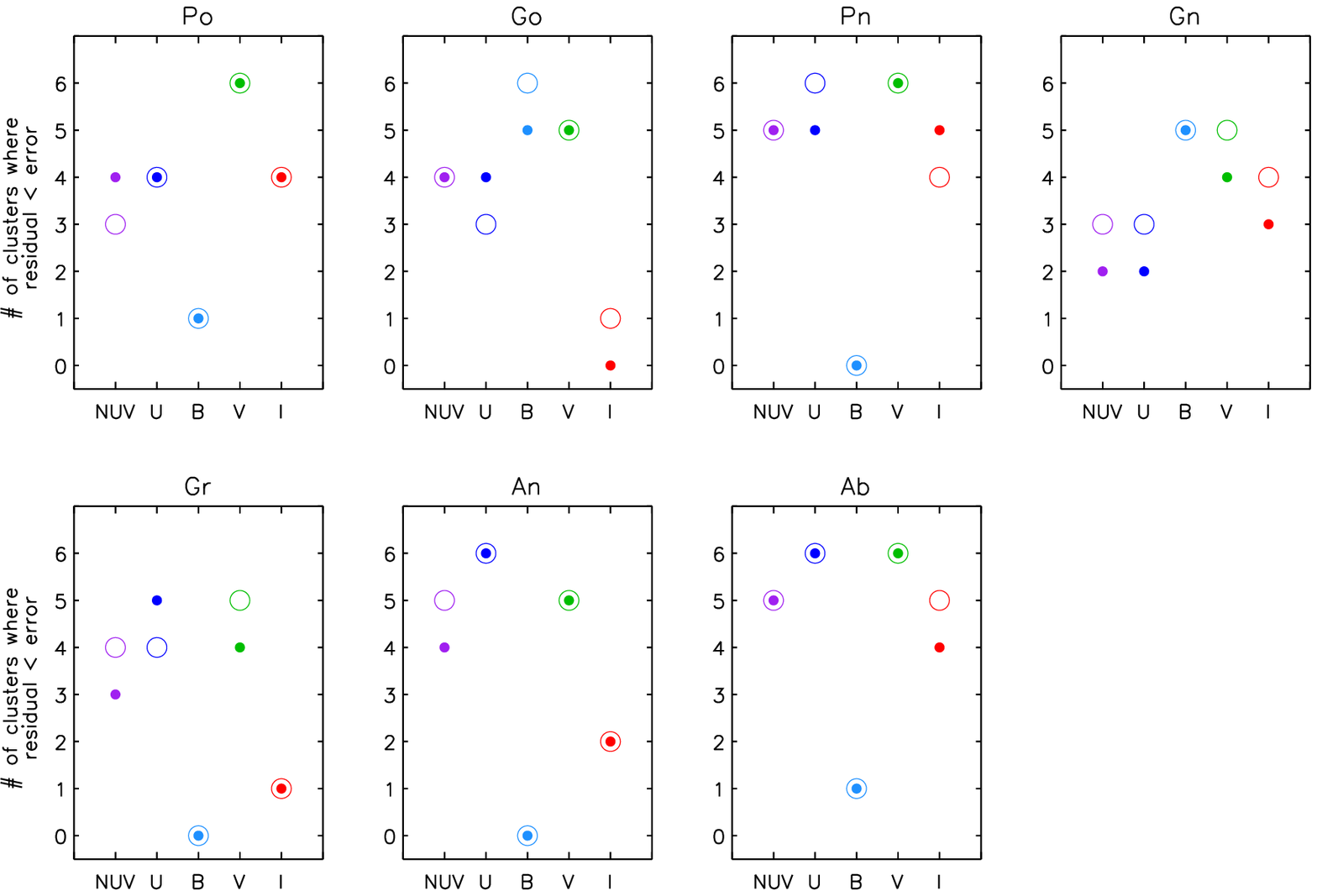}
\caption{Performance of different best-fitting models in fitting individual bands. The bands are indicated on the horizontal axis. The vertical axis gives the number of clusters where the residual of the best-fit model magnitude is less than the observational error. We show results based on the starburst attenuation law (filled symbols) and MW extinction law (unfilled symbols). Each panel corresponds to a different cluster. Since different sets of filters are used for the two galaxies, here we only include clusters in NGC 1566.}
\label{fig10}
\end{figure*}

The vertical axis of Fig.~\ref{fig10}, gives the number of clusters where the residual in the magnitude of the best-fit model is within the observational error of the band indicated on the horizontal axis. As in the previous Fig., we use filled symbols for results based on the starburst attenuation law and unfilled symbols for results based on the MW (NGC 1566) or SMC (NGC 5253) extinction law. We show one set of tracks per panel. The figure only includes NGC 1566 clusters, since a slightly different set of filters was used for NGC 5253. 

%%%%%%%%%%%%%%%%%%%%%%%%%%%%%%
% Figure 10
%%%%%%%%%%%%%%%%%%%%%%%%%%%%%%

Consider that the total number of data points is the number of massive-star evolution flavors $\times$ the number of clusters, i.e., 42. Considering all data points we obtain, for each LEGUS band and prescription for attenuation by dust, the following success rates of best-fitting models in fitting the observations within the observational errors: NUV (64/69), U (76/76), B (29/31), V (86/90), and I (45/50). Thus, overall, the V band is fitted the best, as expected because this band has the lowest photometric errors; and the B band is fitted the worse. This band has larger photometric errors. For NGC 5253, most bands are fitted within the observational errors which are larger than for NGC 1566. 

%%%%%%%%%%%%%%%%%%%%%%%%%%%%%%
%%% PDFs
%%%%%%%%%%%%%%%%%%%%%%%%%%%%%%

\subsection{How well constrained are the cluster properties?}

\begin{figure*}
\begin{subfigure}
\centering
\includegraphics[width=0.89\columnwidth]{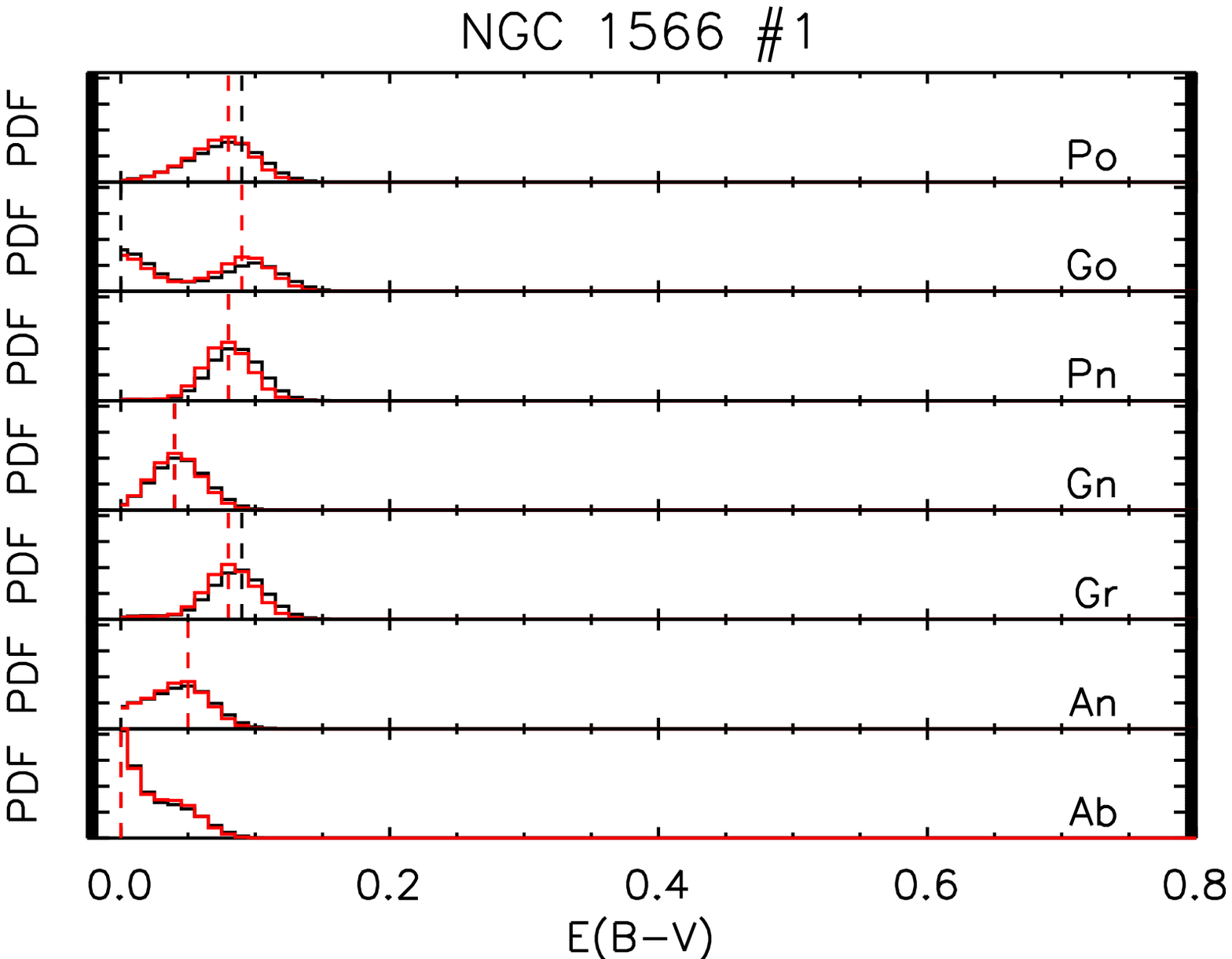}
\end{subfigure}
\begin{subfigure}
\centering
\includegraphics[width=0.89\columnwidth]{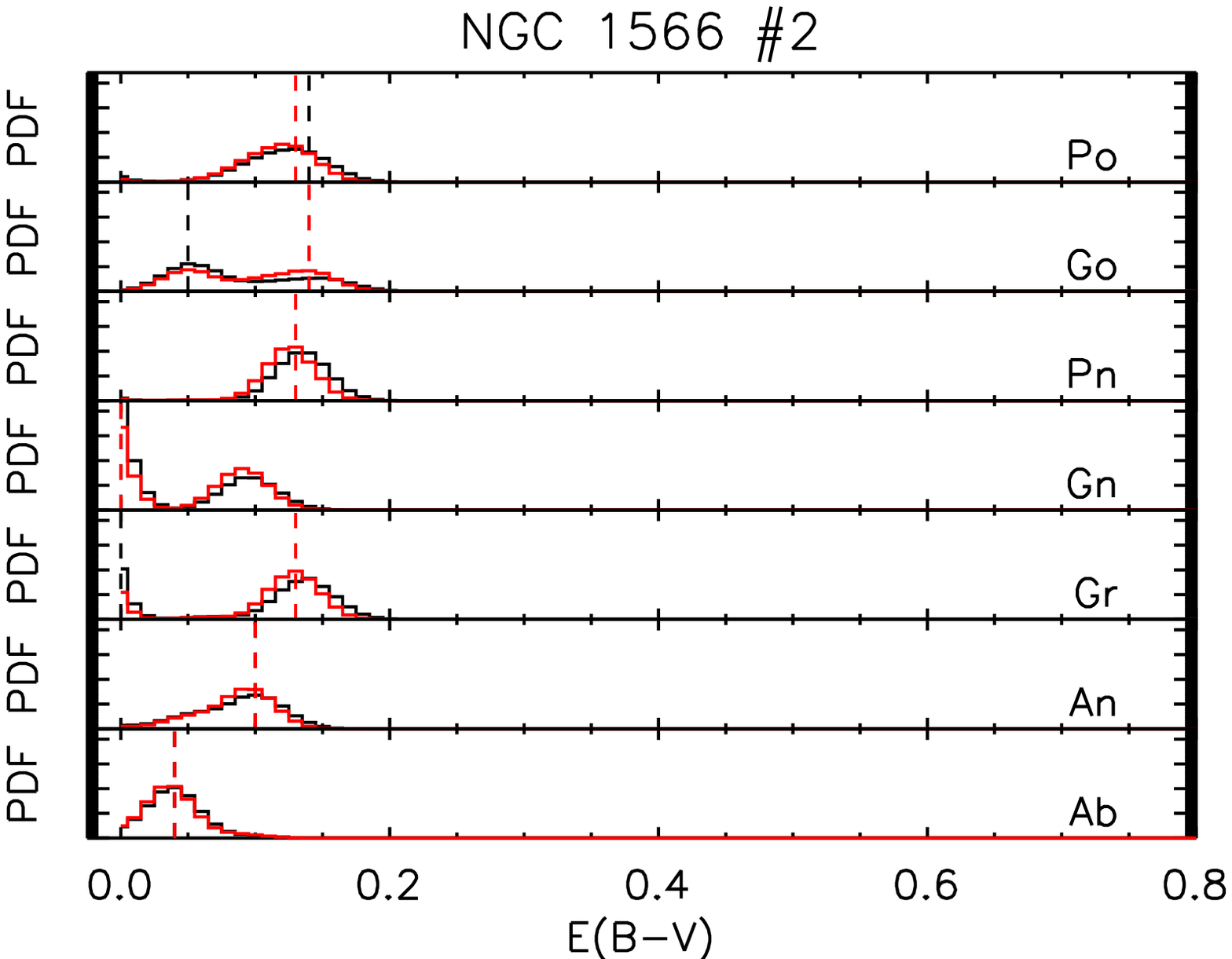}
\end{subfigure}
\begin{subfigure}
\centering
\includegraphics[width=0.89\columnwidth]{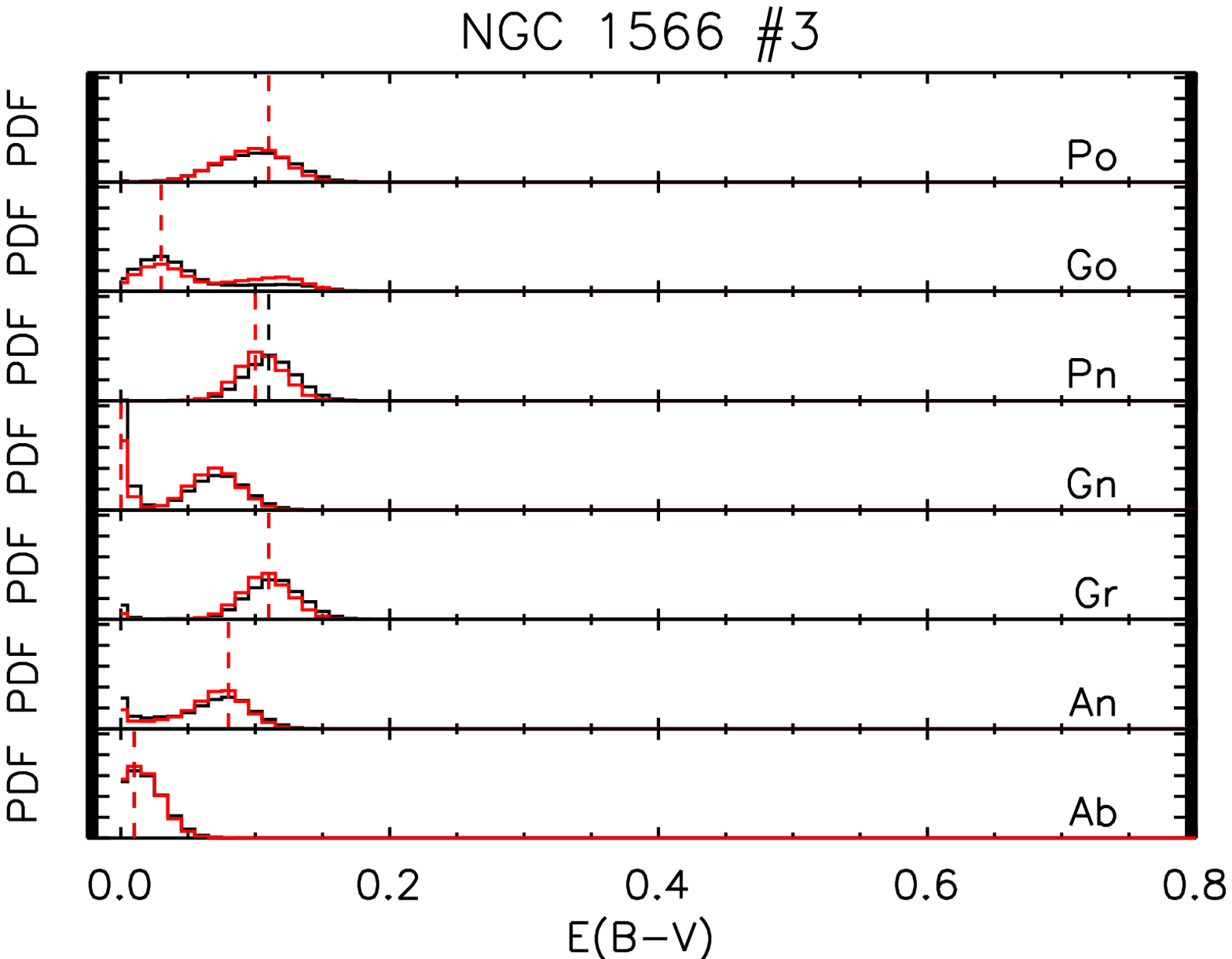}
\end{subfigure}
\begin{subfigure}
\centering
\includegraphics[width=0.89\columnwidth]{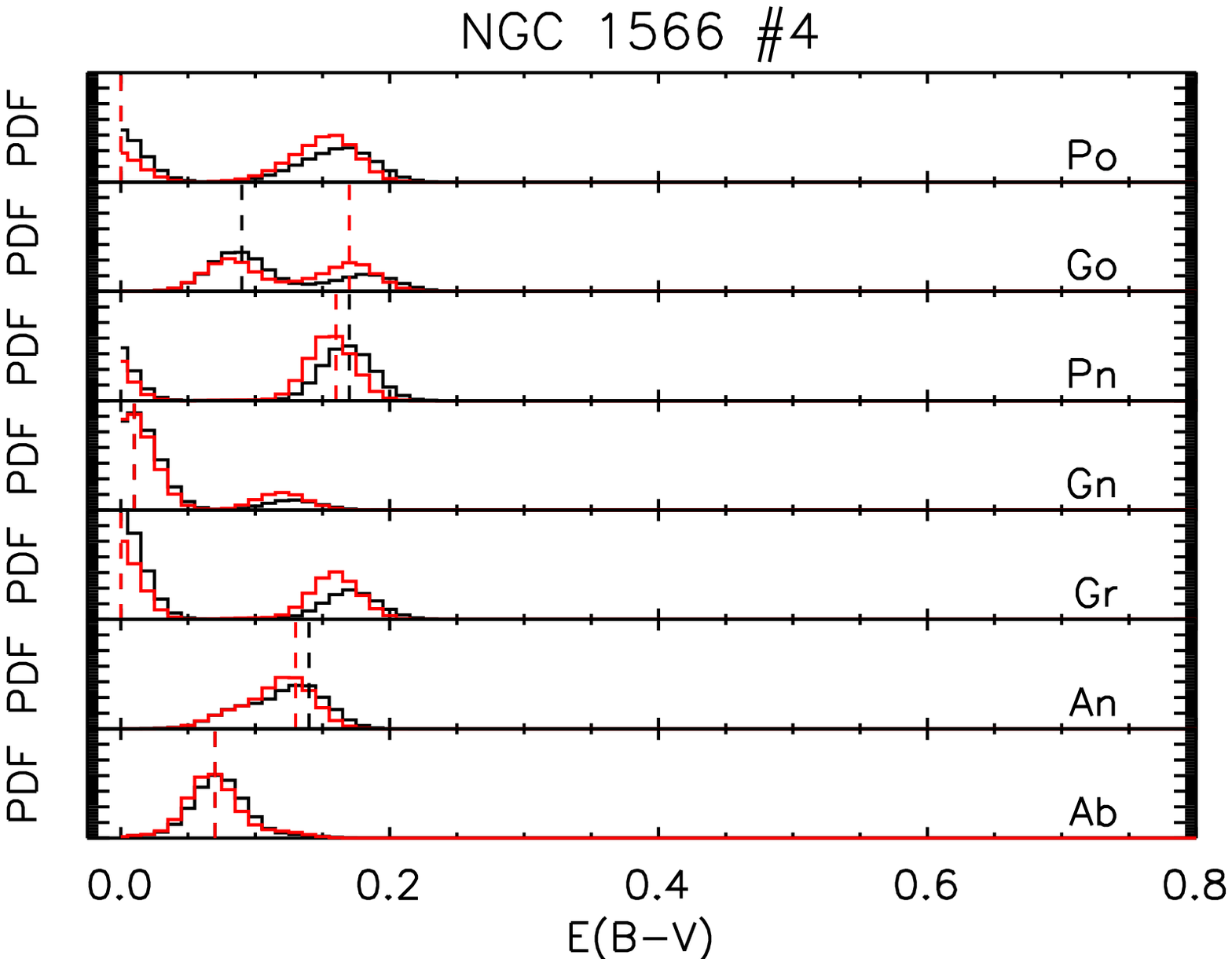}
\end{subfigure}
\begin{subfigure}
\centering
\includegraphics[width=0.89\columnwidth]{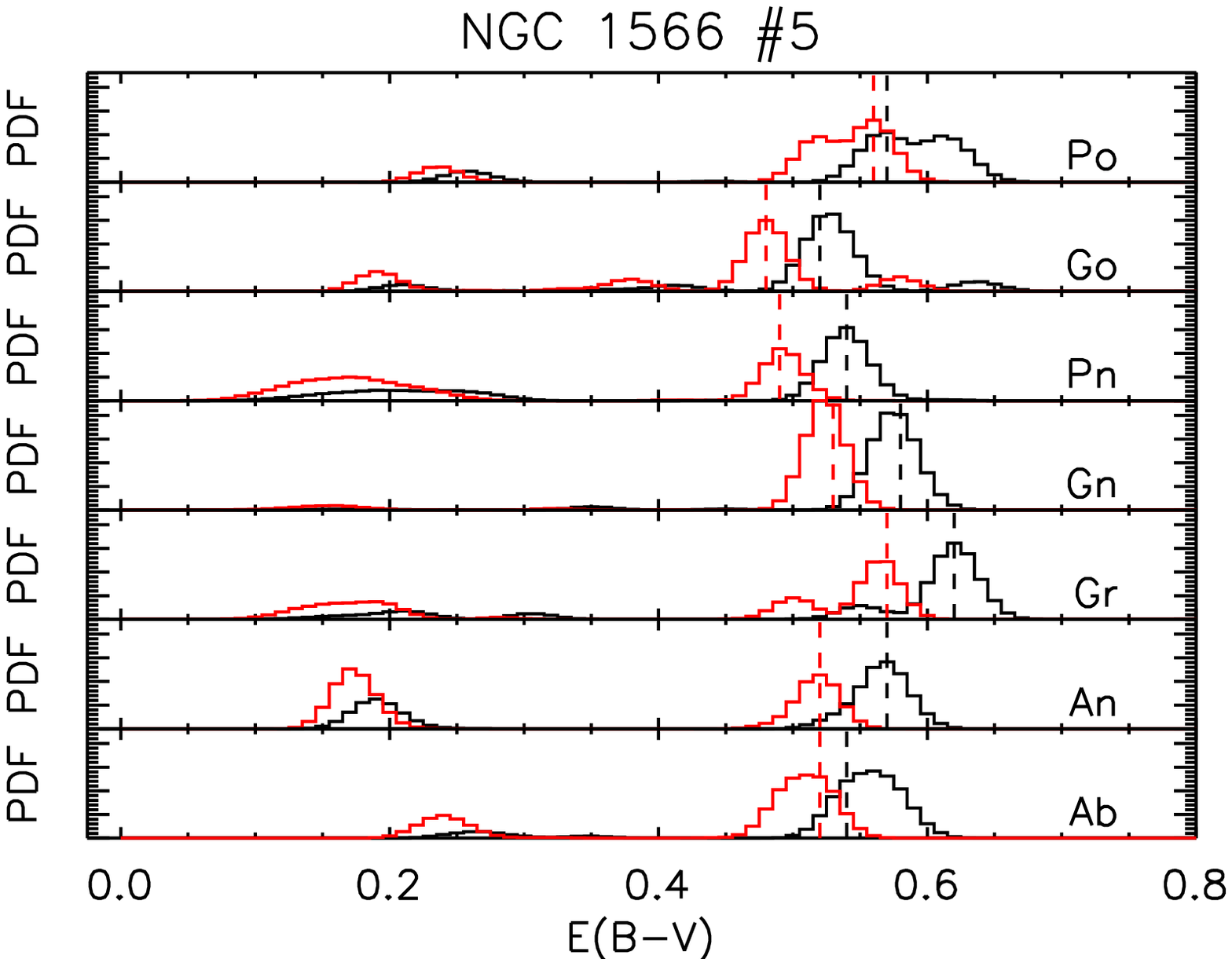}
\end{subfigure}
\begin{subfigure}
\centering
\includegraphics[width=0.89\columnwidth]{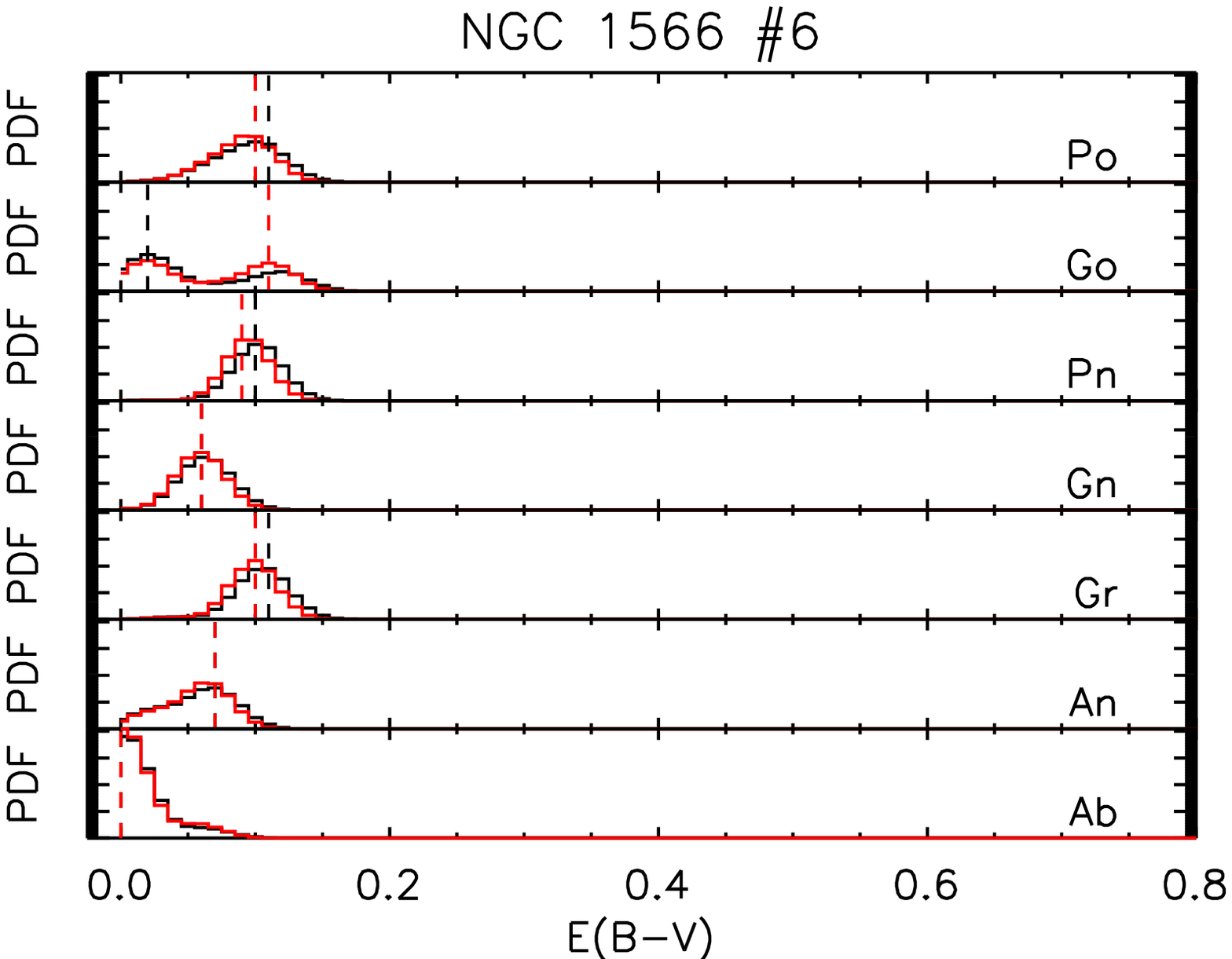}
\end{subfigure}
\begin{subfigure}
\centering
\includegraphics[width=0.89\columnwidth]{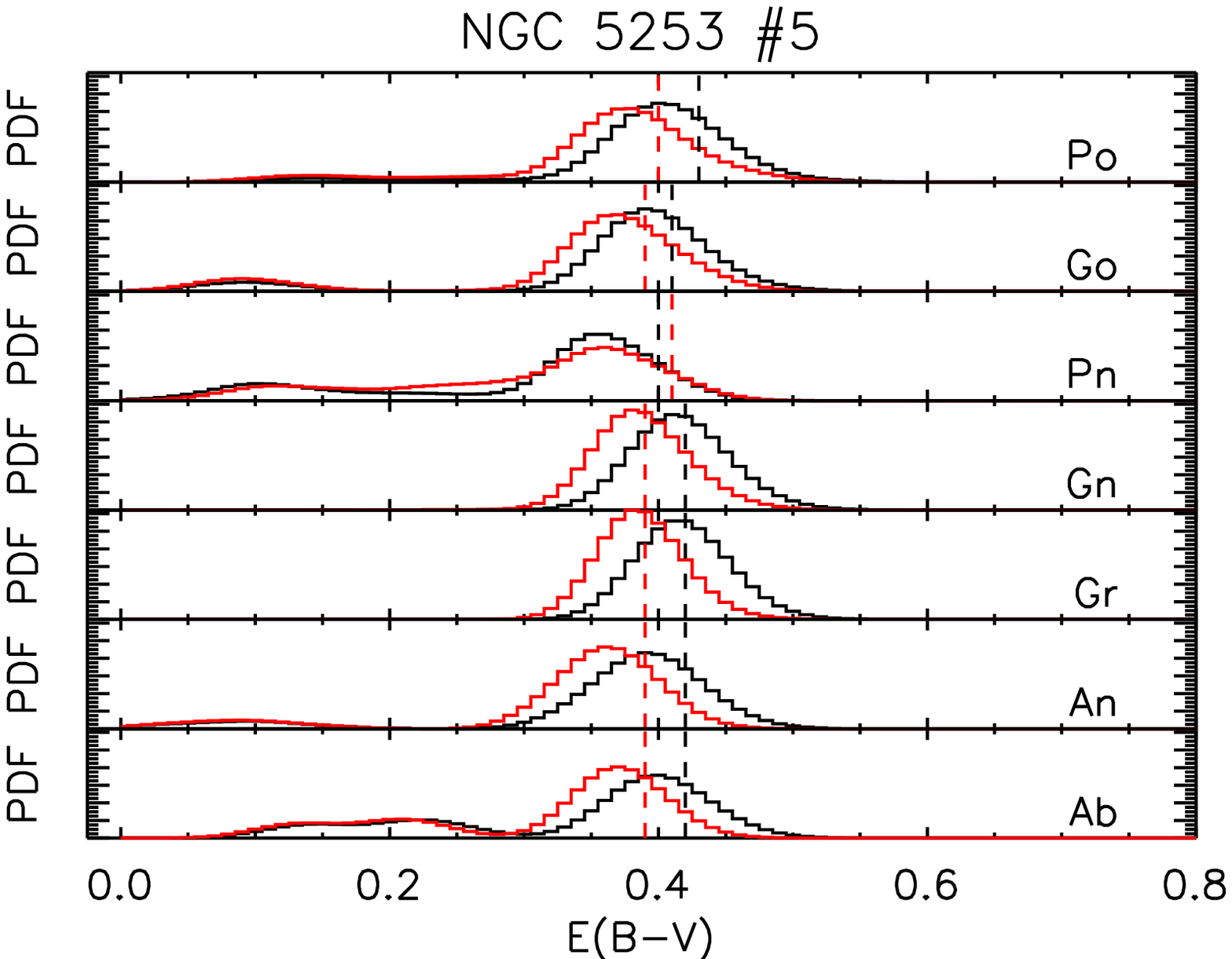}
\end{subfigure}
\begin{subfigure}
\centering
\includegraphics[width=0.89\columnwidth]{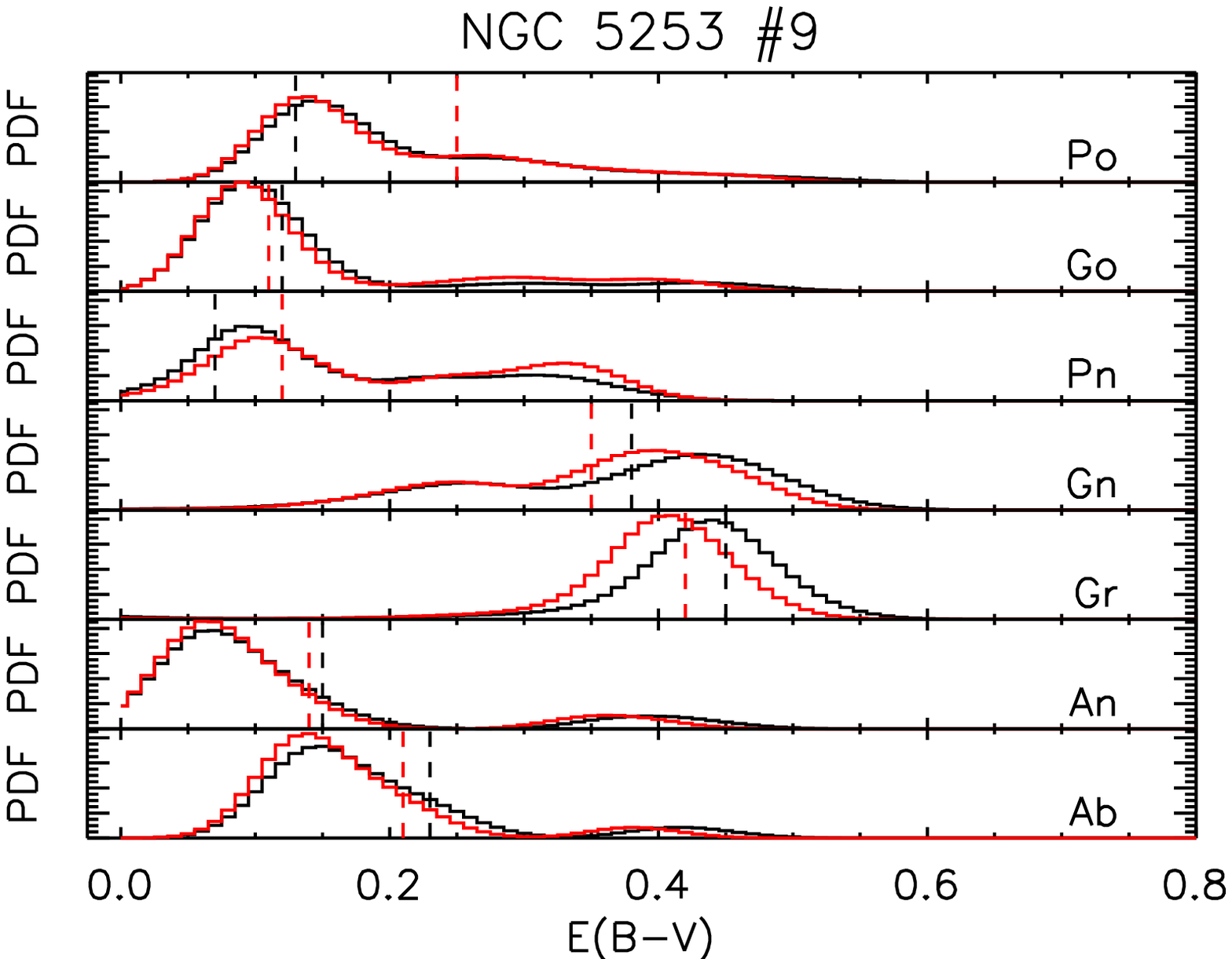}
\end{subfigure}
\caption{$E(B-V)$ posterior marginalized PDFs (solid lines) and best-fitting solutions (dashed vertical lines) obtained with the models given on the bottom right of each sub-panel and with the prescriptions for dust attenuation given by the color (black=MW [NGC 1566] or SMC [NGC 5253] extinction law; red=starburst attenuation law). The title of each panel gives the cluster ID. The PDFs are normalized to the area of the distribution and for a given cluster use the same vertical-axis scale for each sub-panel.  We use $\Delta$$E(B-V)=0.01$ mag.}
\label{fig11}
\end{figure*}

\begin{figure*}
\begin{subfigure}
\centering
\includegraphics[width=0.89\columnwidth]{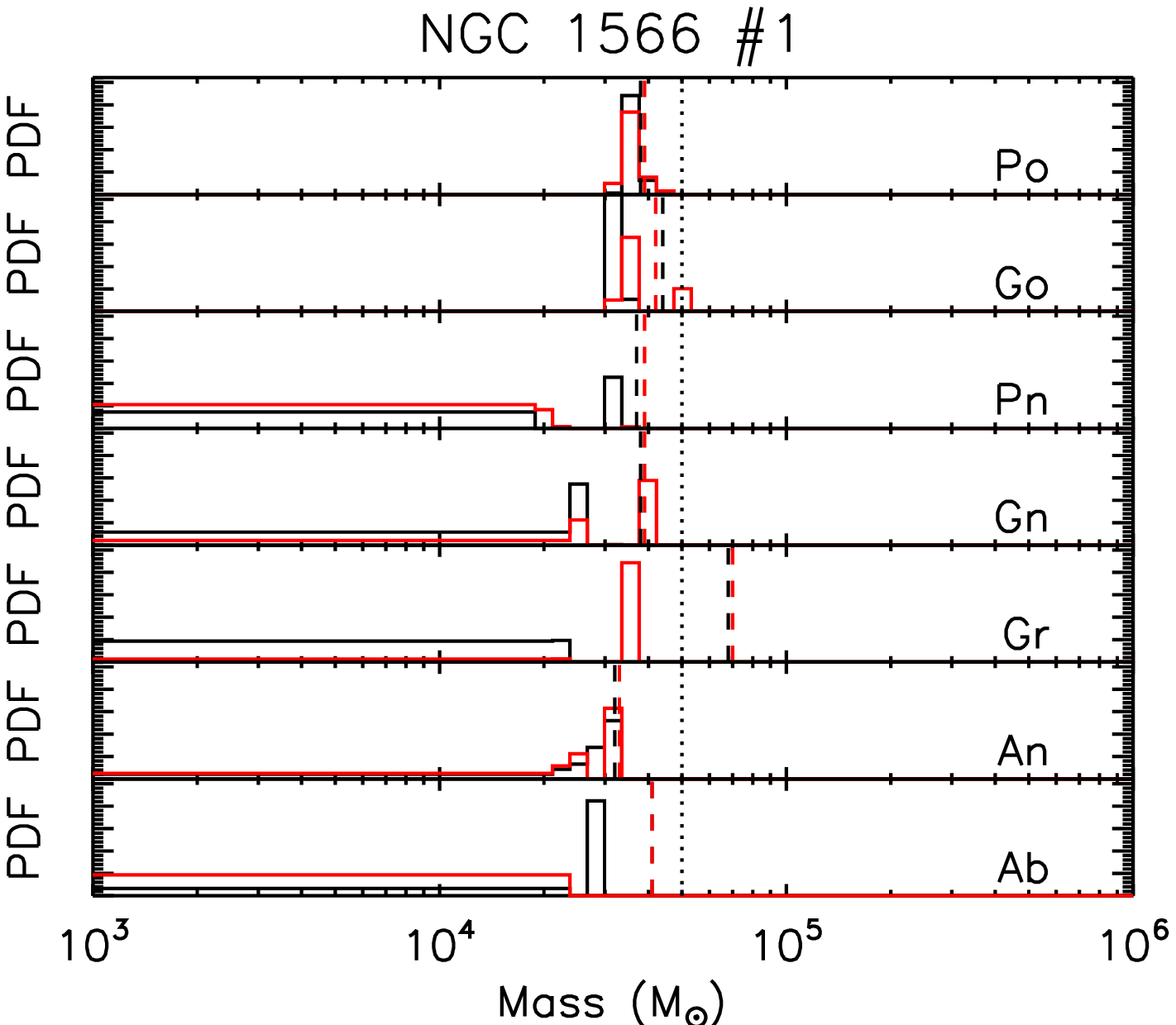}
\end{subfigure}
\begin{subfigure}
\centering
\includegraphics[width=0.89\columnwidth]{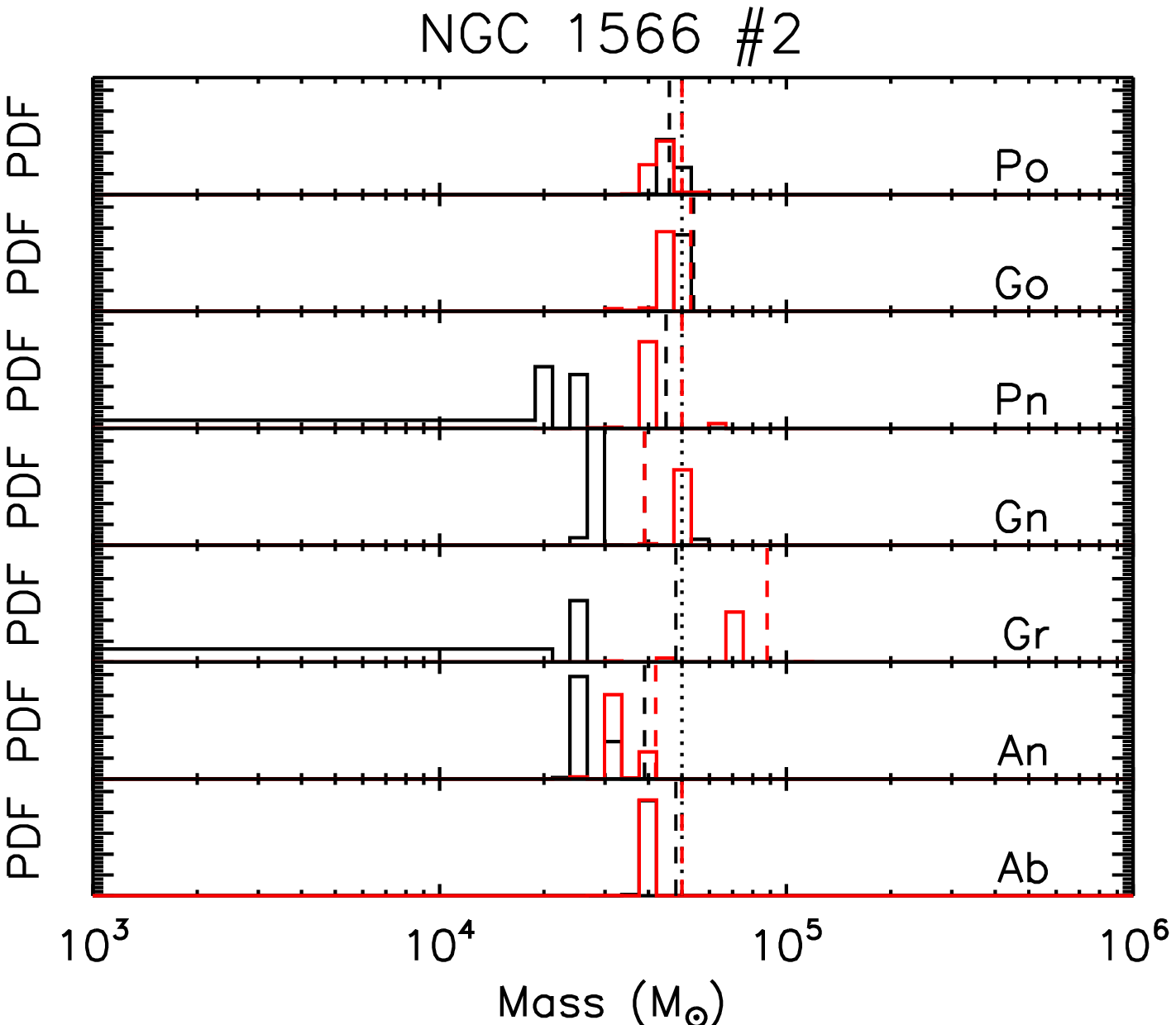}
\end{subfigure}
\begin{subfigure}
\centering
\includegraphics[width=0.89\columnwidth]{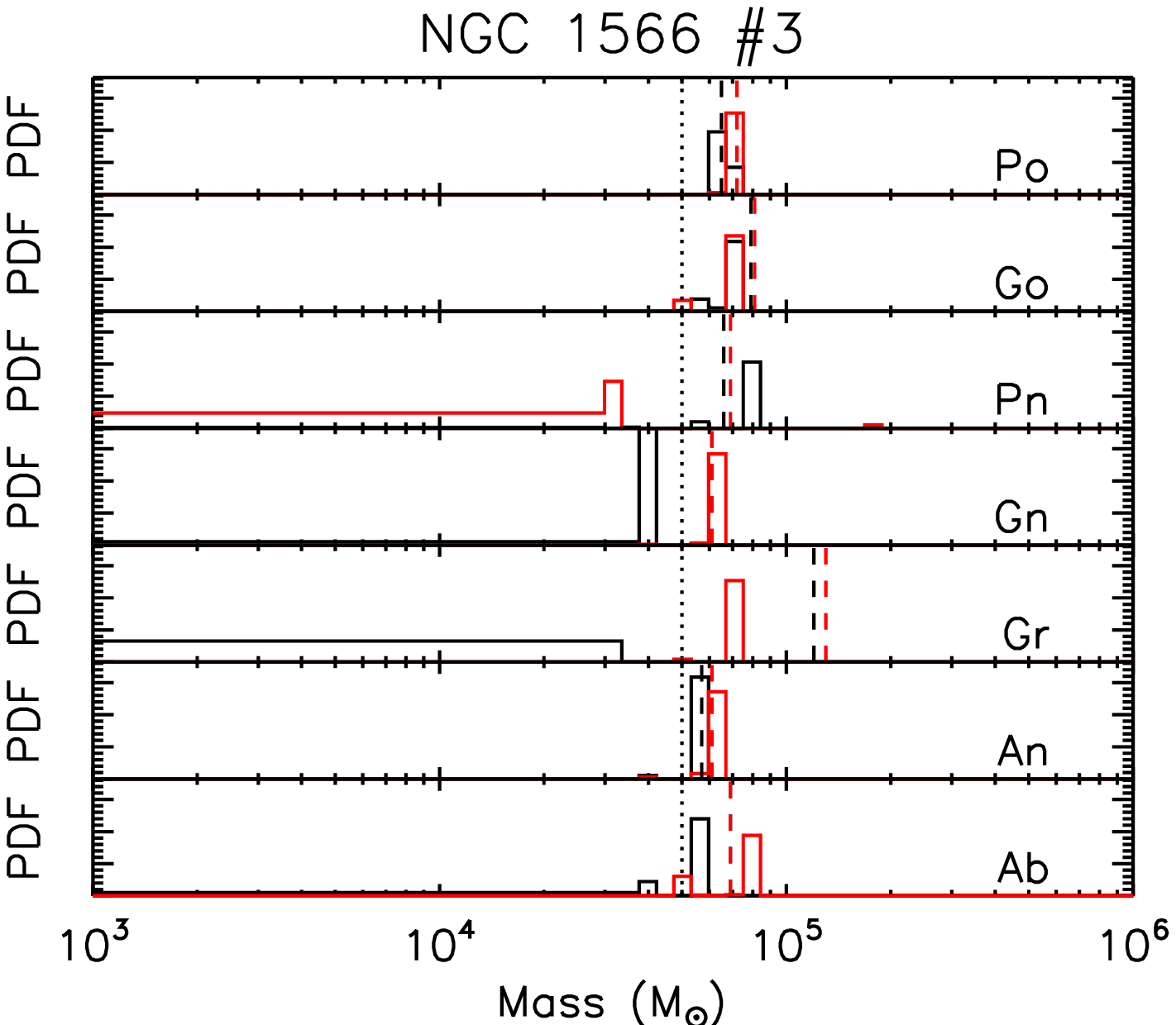}
\end{subfigure}
\begin{subfigure}
\centering
\includegraphics[width=0.89\columnwidth]{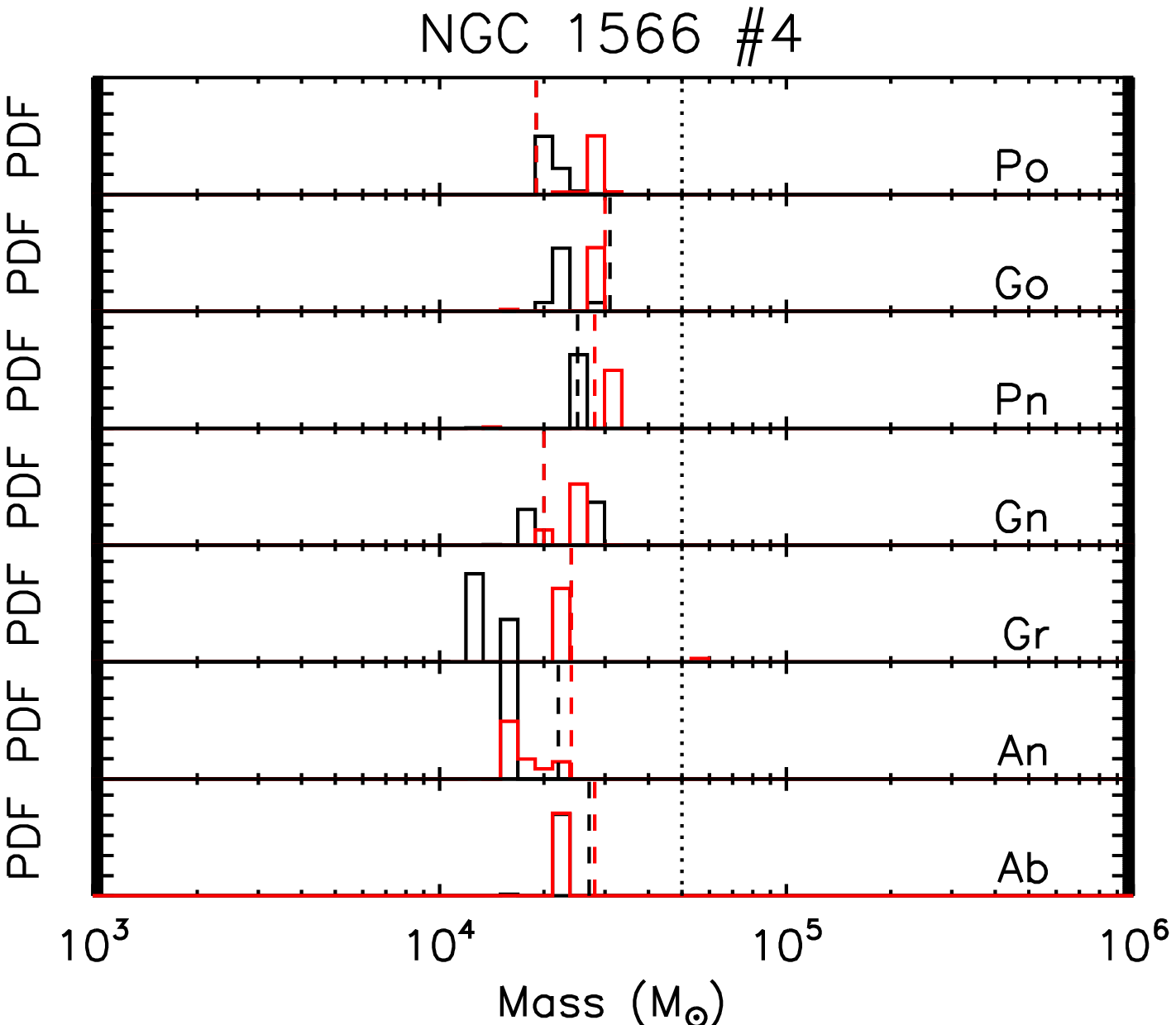}
\end{subfigure}
\begin{subfigure}
\centering
\includegraphics[width=0.89\columnwidth]{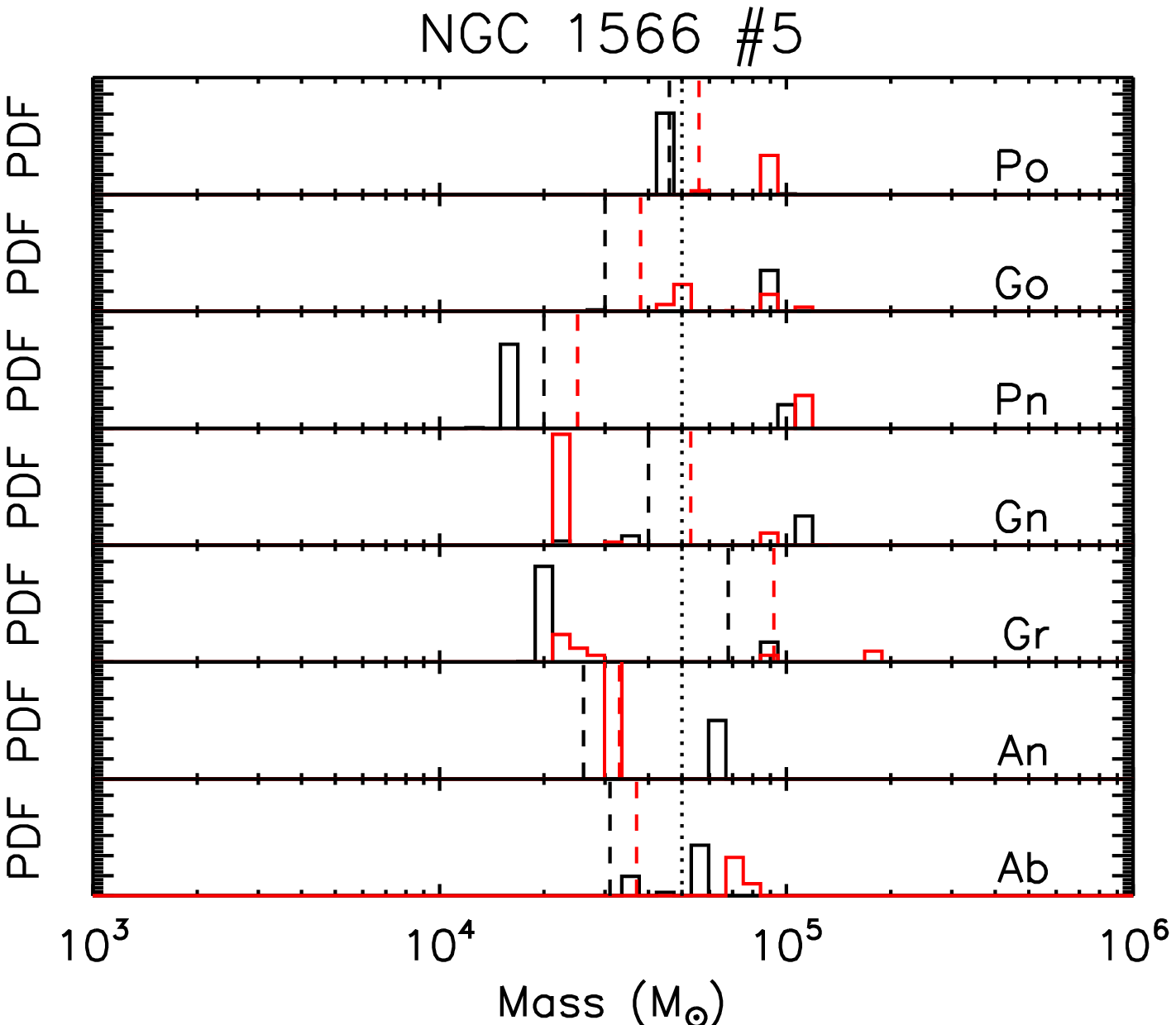}
\end{subfigure}
\begin{subfigure}
\centering
\includegraphics[width=0.89\columnwidth]{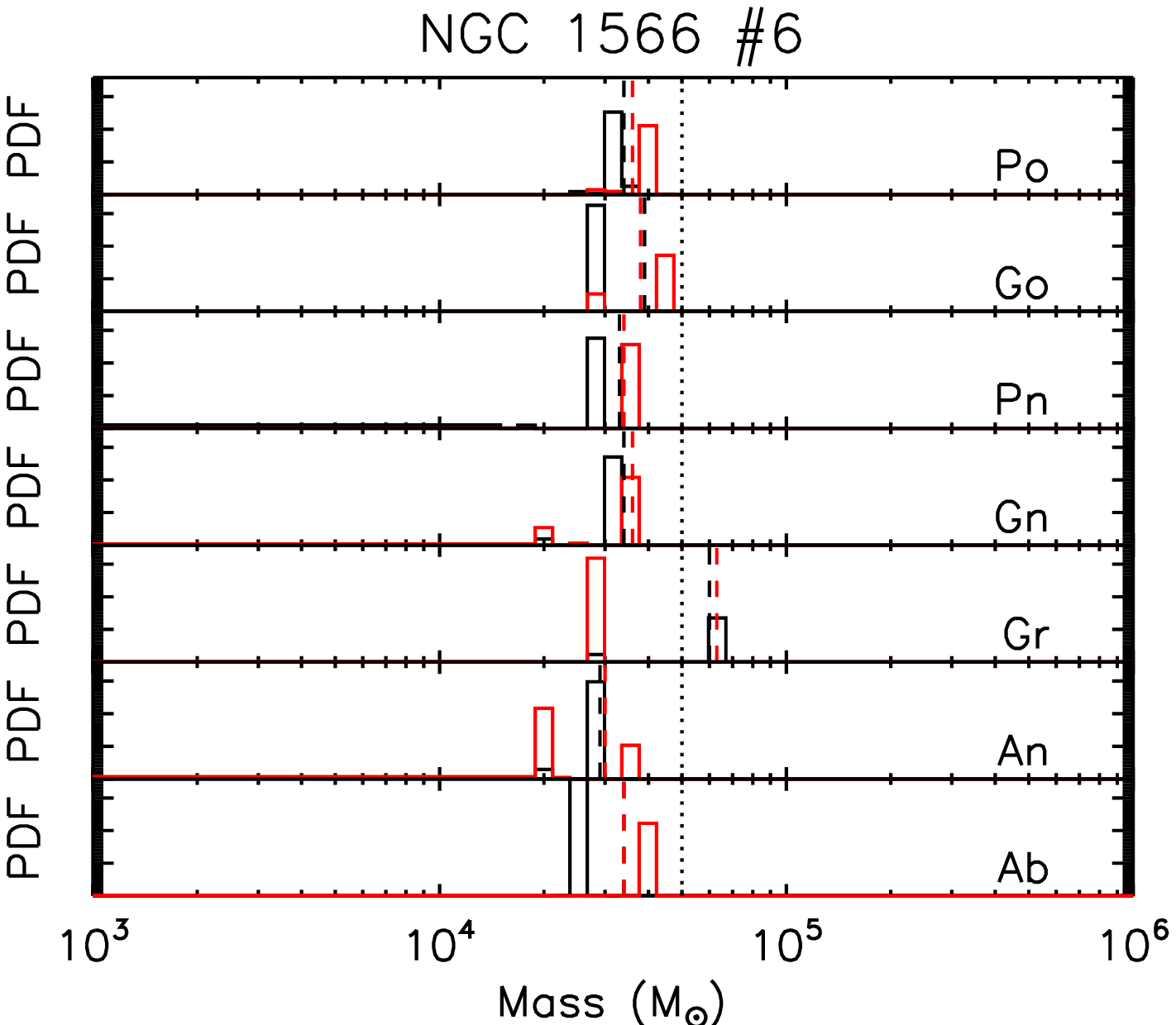}
\end{subfigure}
\begin{subfigure}
\centering
\includegraphics[width=0.89\columnwidth]{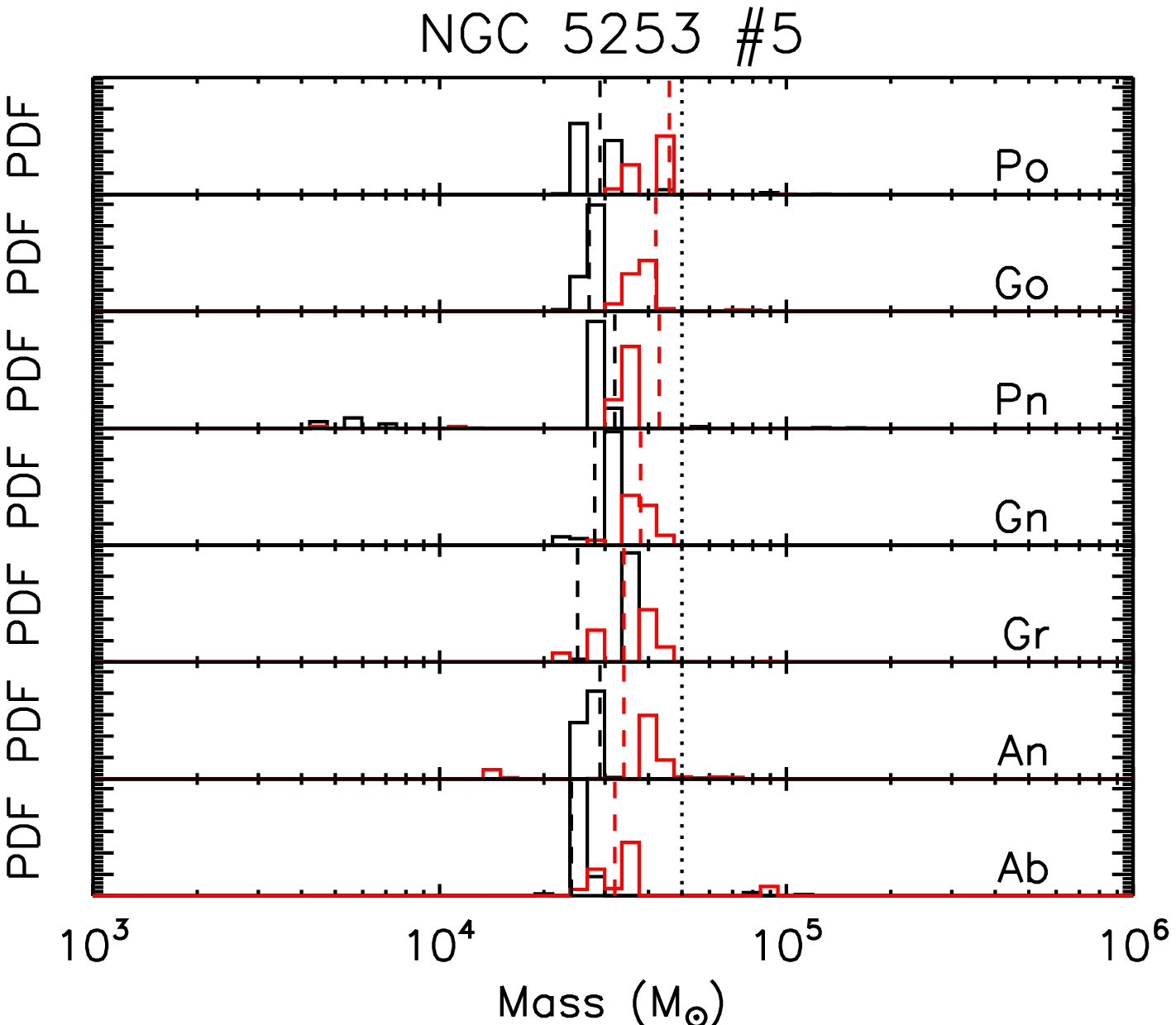}
\end{subfigure}
\begin{subfigure}
\centering
\includegraphics[width=0.89\columnwidth]{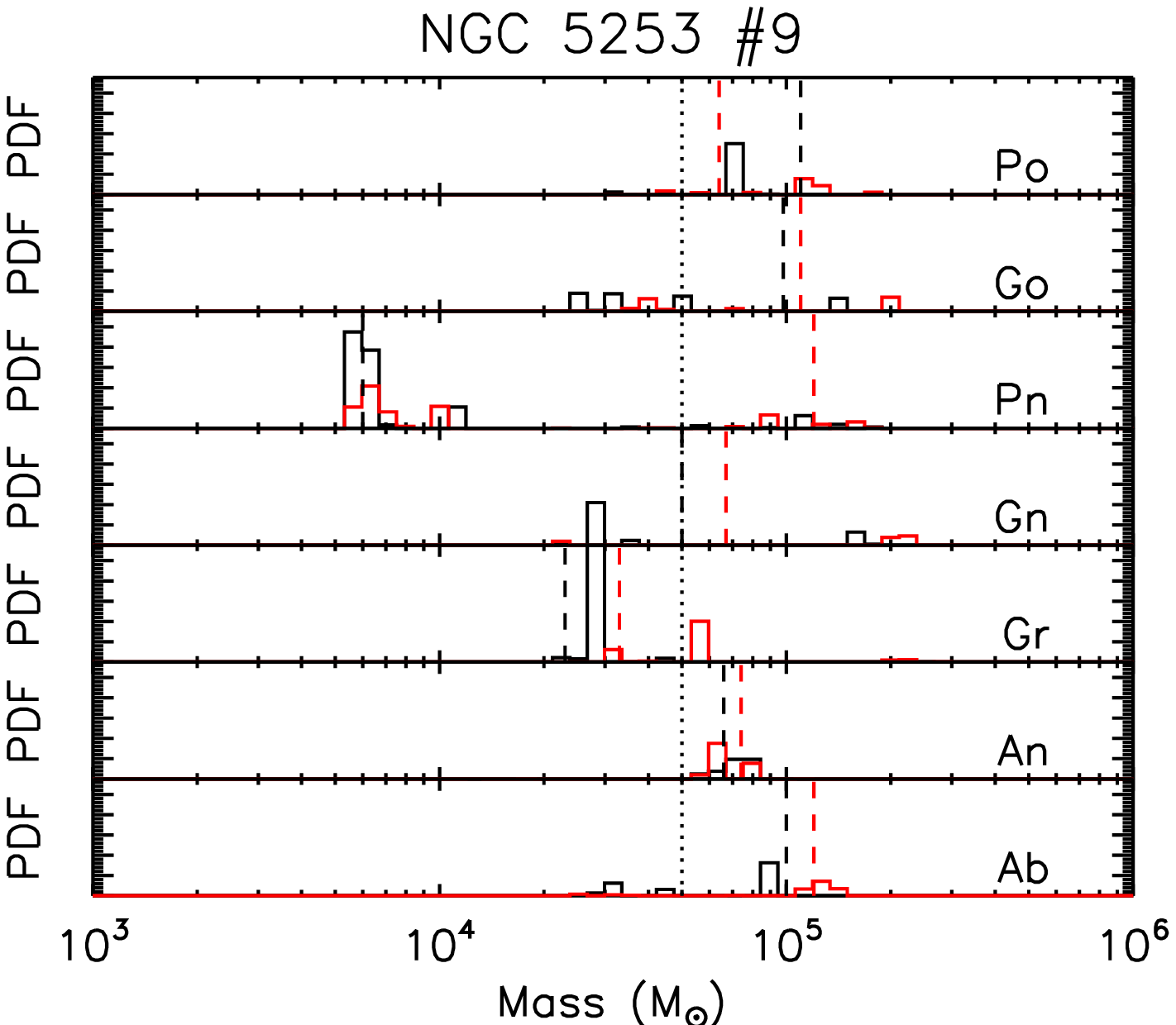}
\end{subfigure}
\caption{Similar to Fig.~\ref{fig11} but we show results for the cluster masses. We use $\Delta$log($M_{cl}/M_\odot$)=0.05 dex. The dotted vertical line marks the position of $5E4$ M$_\odot$ position.}
\label{fig12}
\end{figure*}

\begin{figure*}
\begin{subfigure}
\centering
\includegraphics[width=0.89\columnwidth]{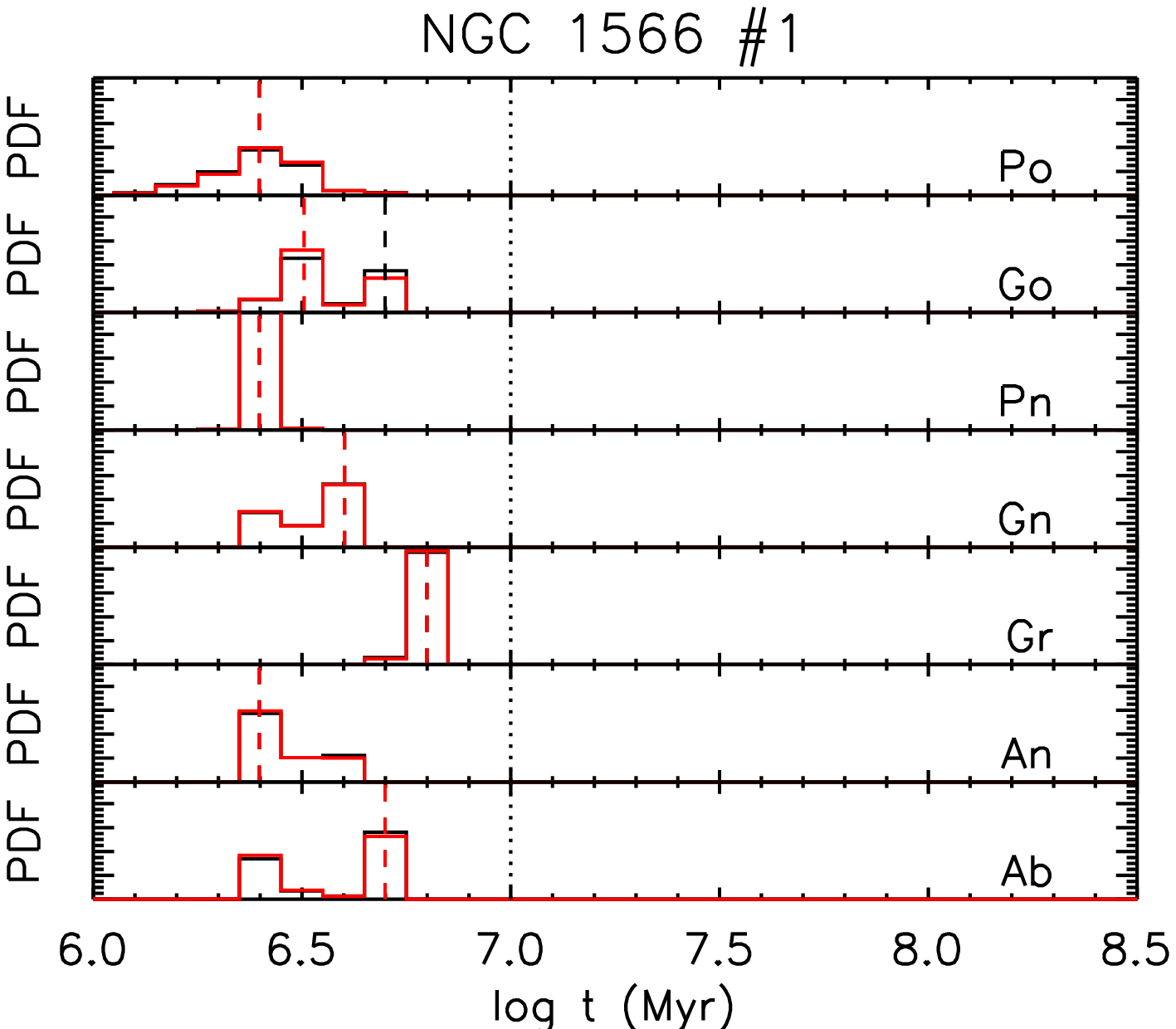}
\end{subfigure}
\begin{subfigure}
\centering
\includegraphics[width=0.89\columnwidth]{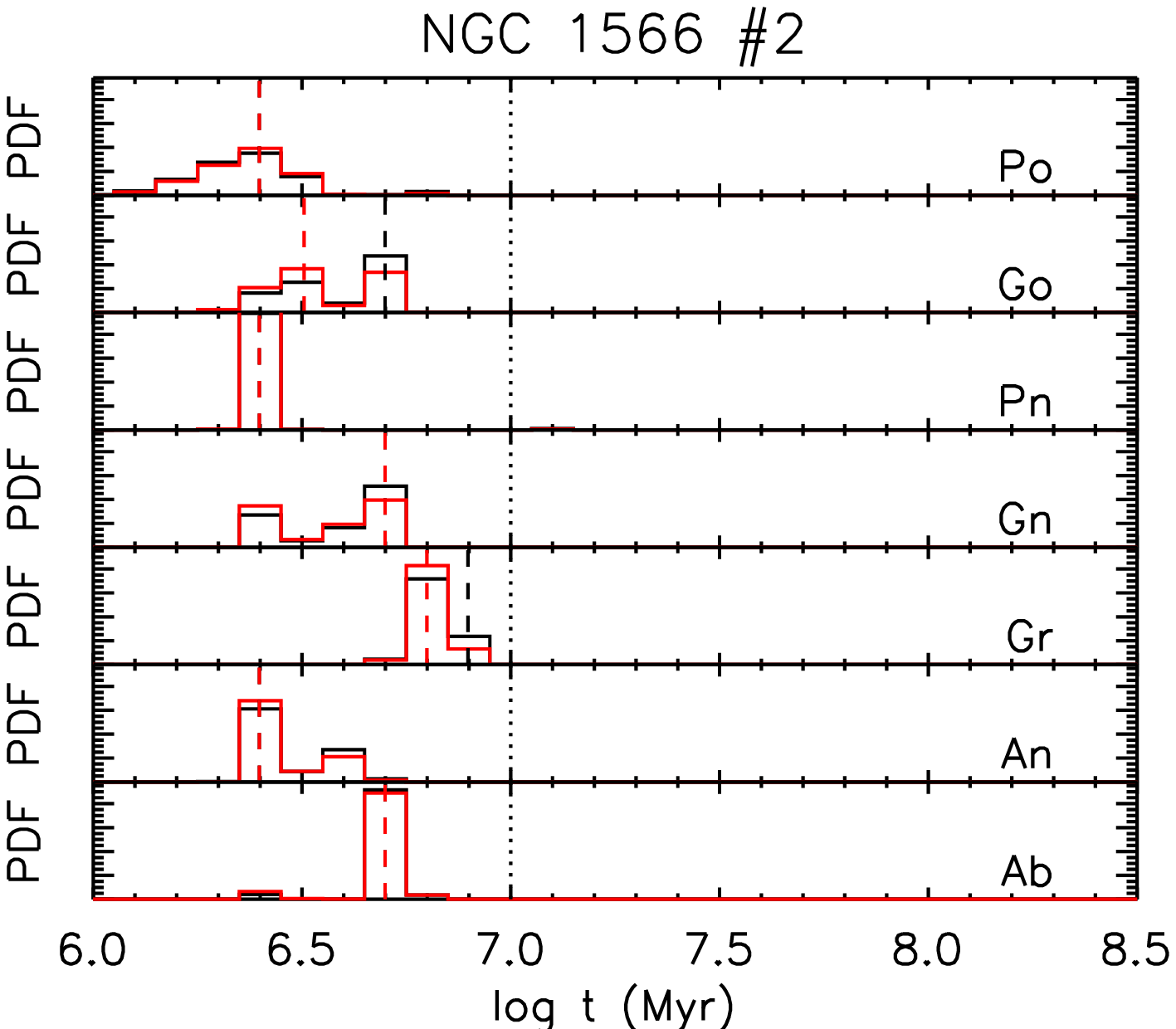}
\end{subfigure}
\begin{subfigure}
\centering
\includegraphics[width=0.89\columnwidth]{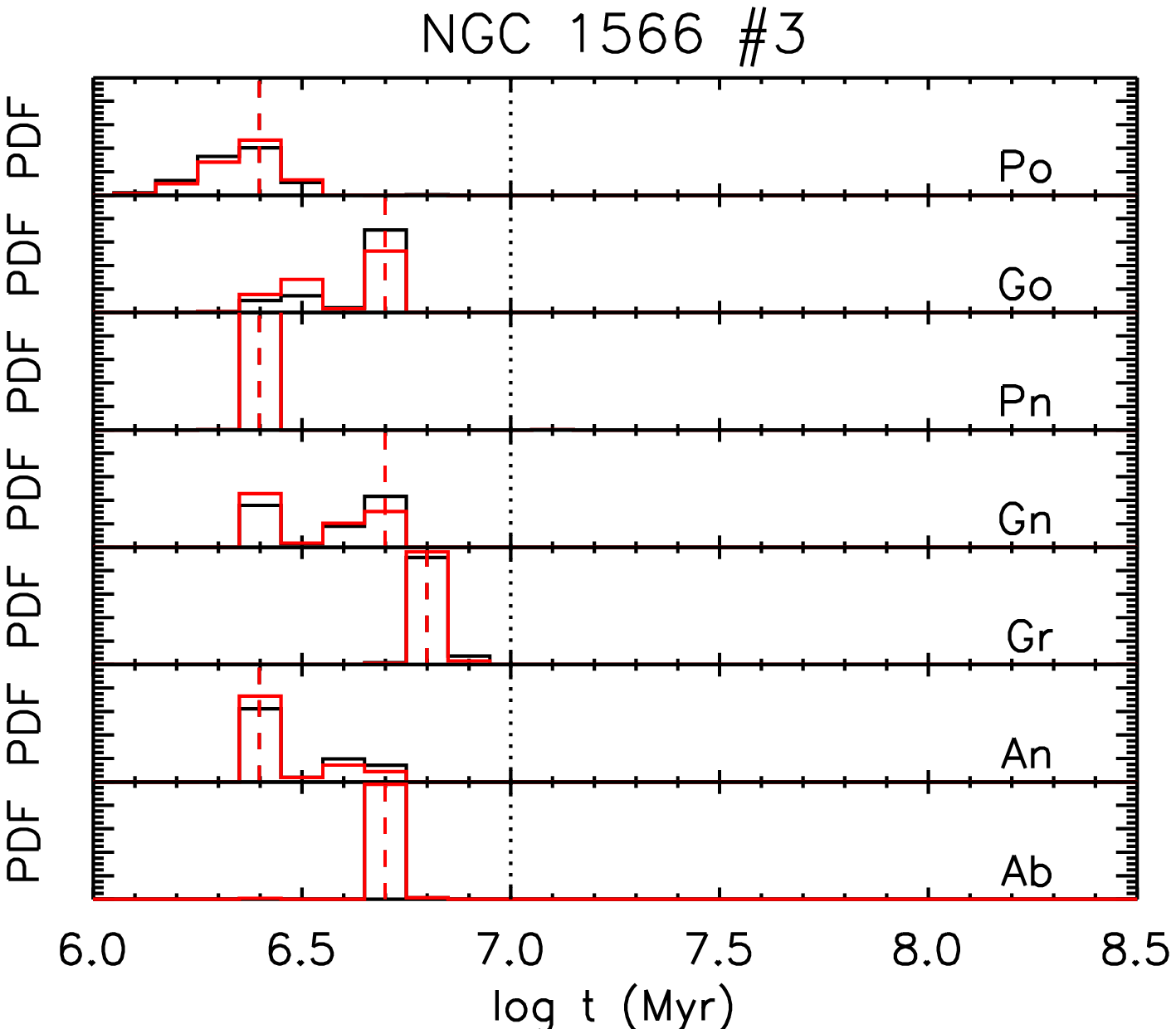}
\end{subfigure}
\begin{subfigure}
\centering
\includegraphics[width=0.89\columnwidth]{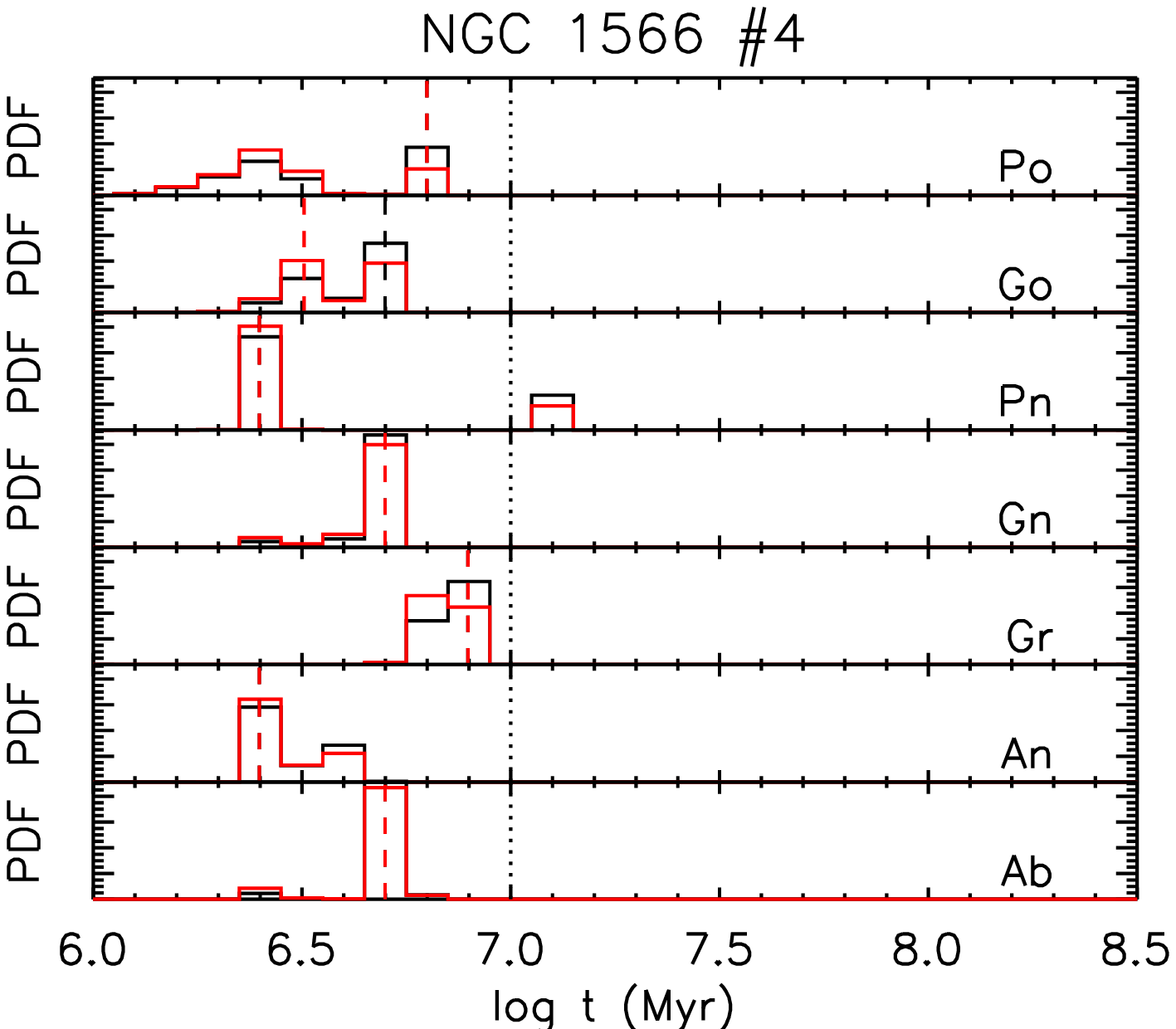}
\end{subfigure}
\begin{subfigure}
\centering
\includegraphics[width=0.89\columnwidth]{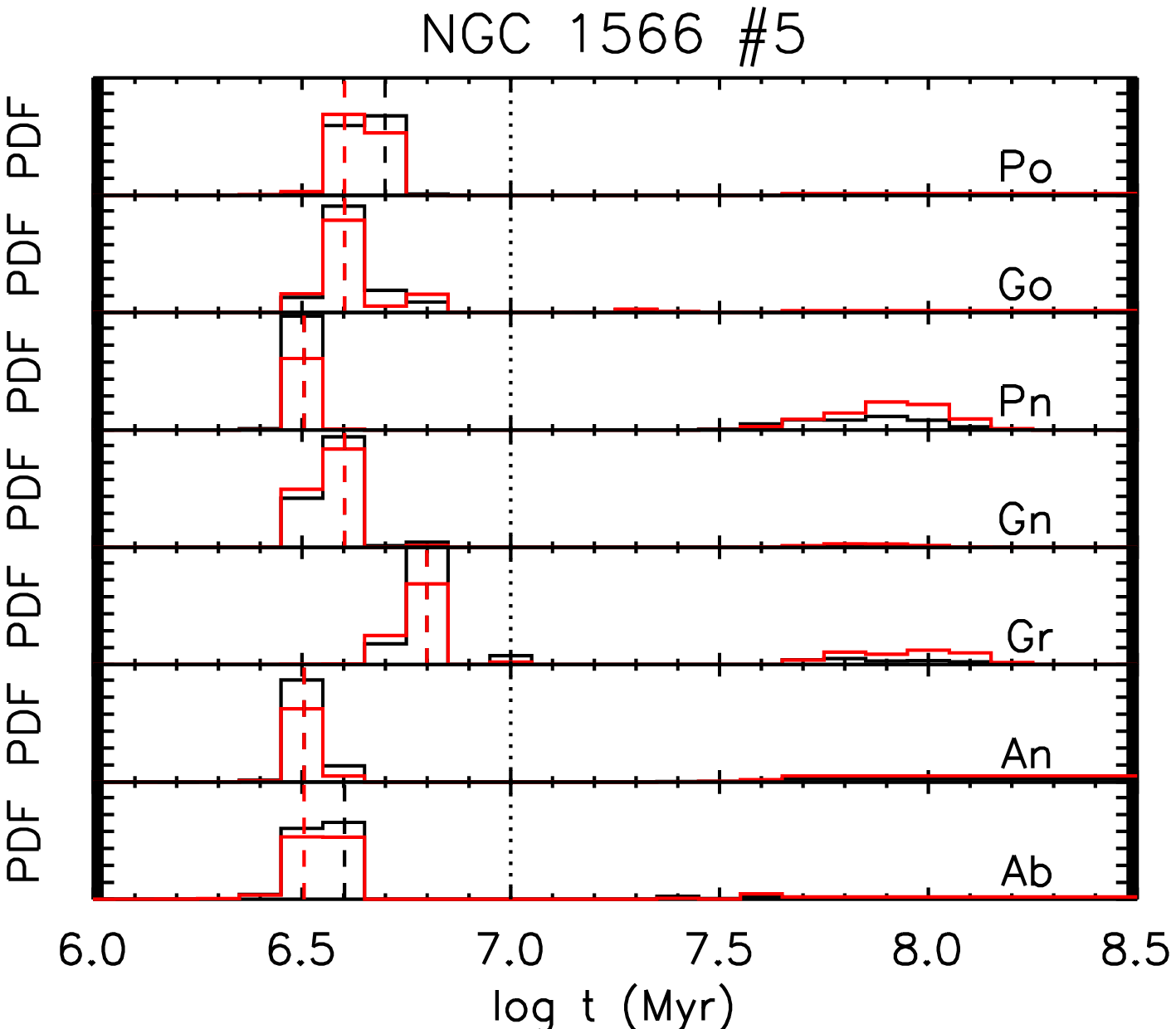}
\end{subfigure}
\begin{subfigure}
\centering
\includegraphics[width=0.89\columnwidth]{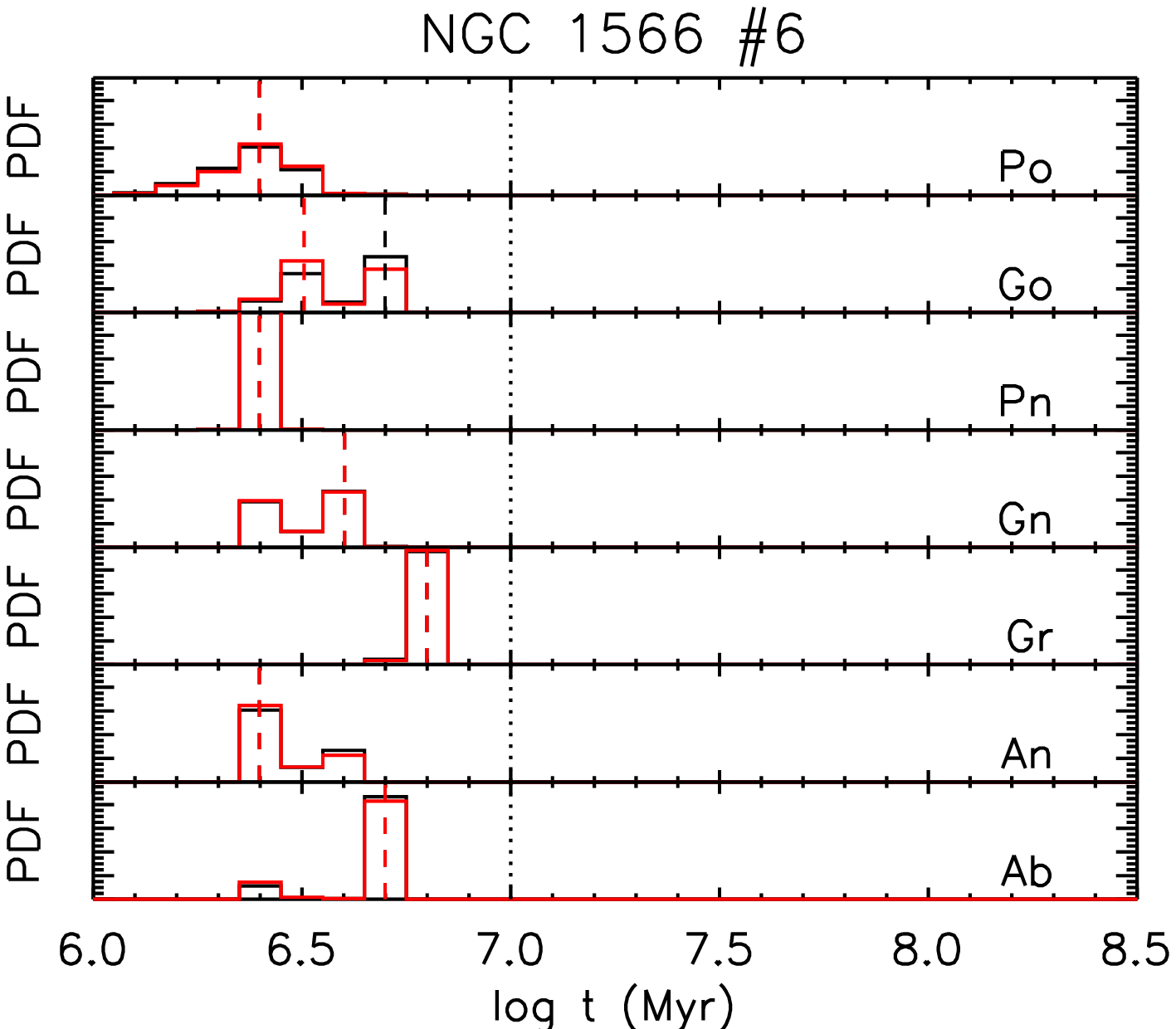}
\end{subfigure}
\begin{subfigure}
\centering
\includegraphics[width=0.89\columnwidth]{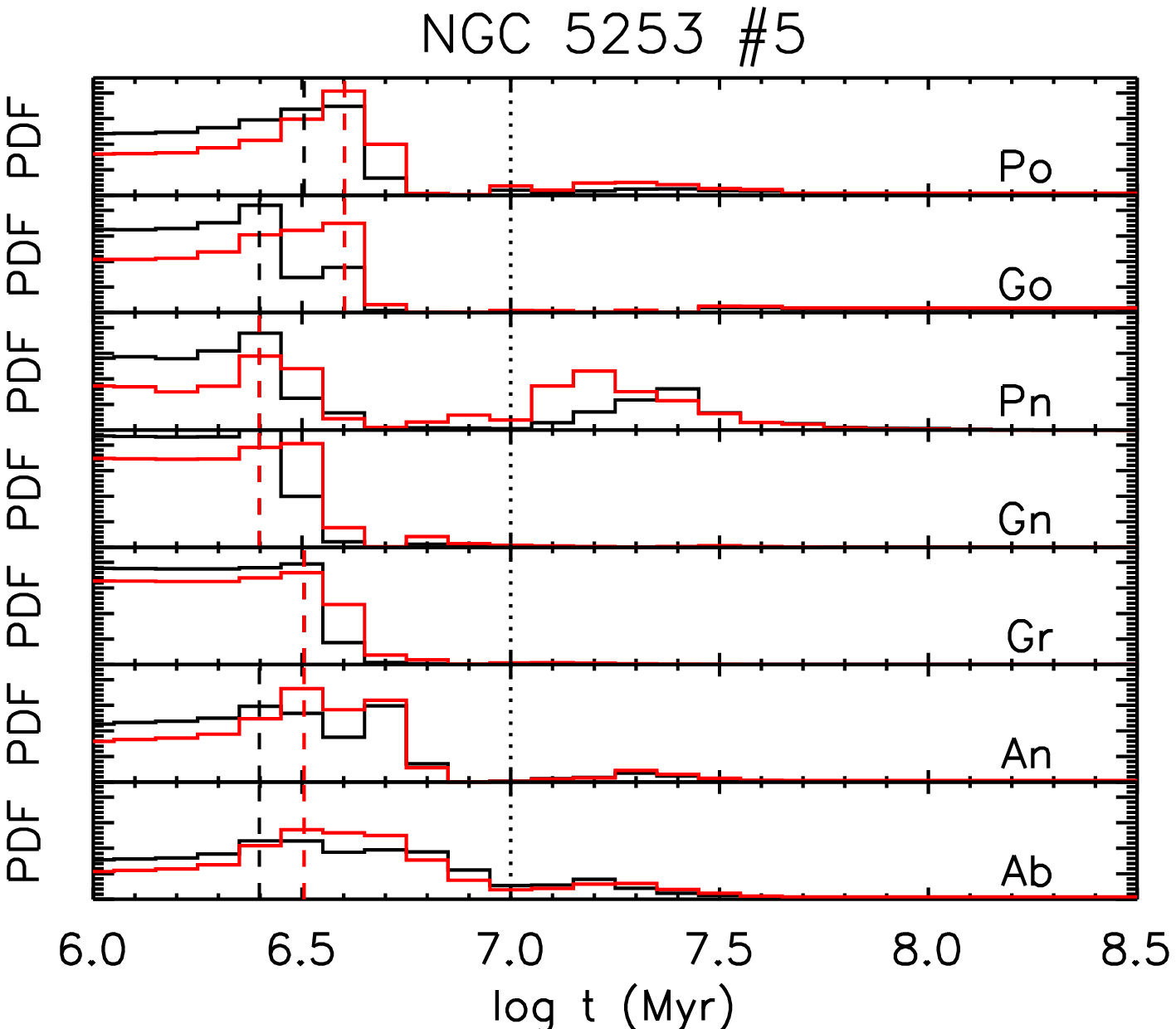}
\end{subfigure}
\begin{subfigure}
\centering
\includegraphics[width=0.89\columnwidth]{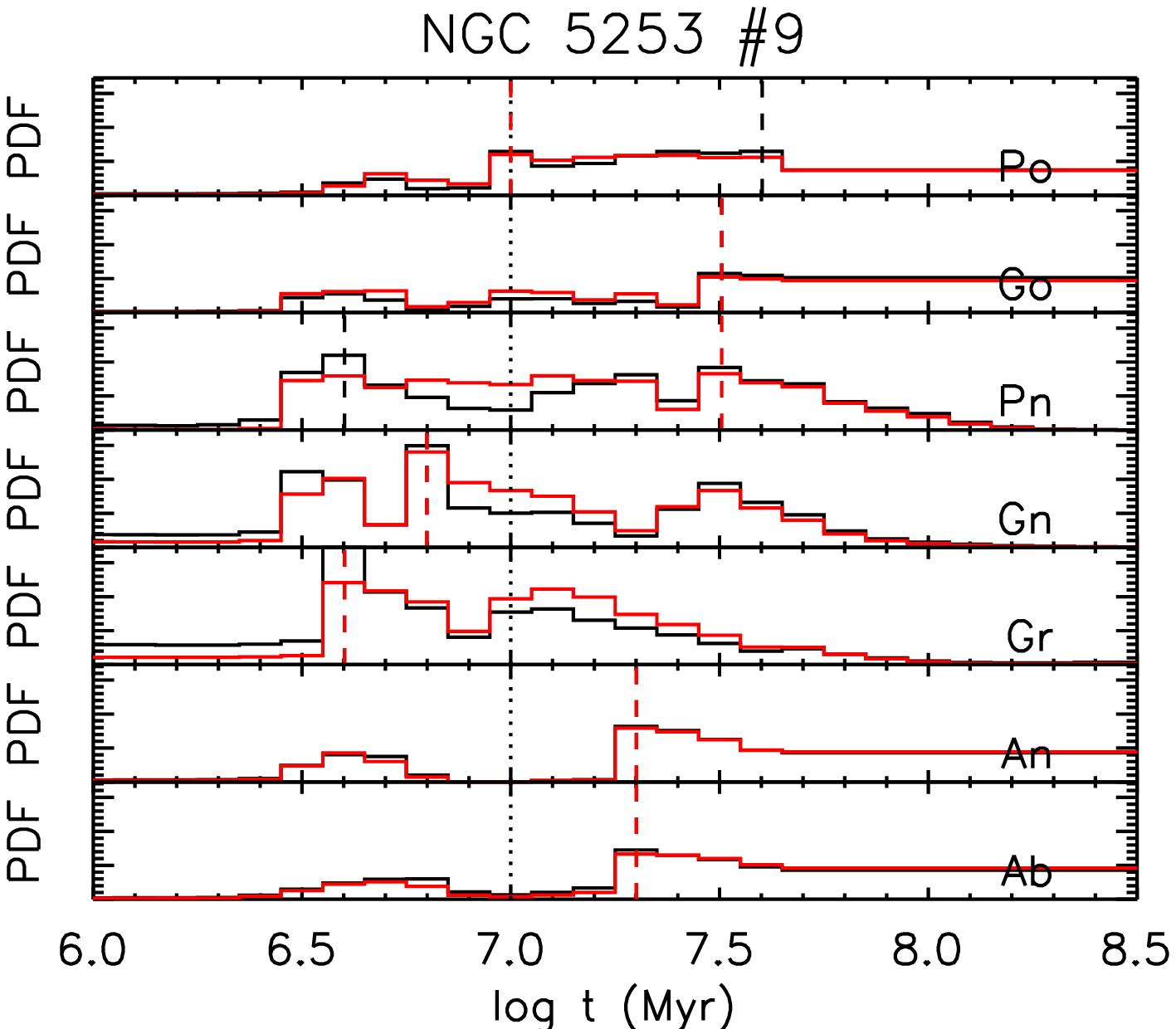}
\end{subfigure}
\caption{Similar to Fig.~\ref{fig11} but we show results for the cluster ages. We use $\Delta$log($t$/yr)=0.1 dex. The dotted vertical line marks the position of 10 Myr.}
\label{fig13}
\end{figure*}

In order to assess if the cluster properties are well constrained by LEGUS five-band photometry and the different models, we look at the posterior marginalized PDFs of $E(B-V)$, mass, and age of the clusters, which are plotted in Fig.~\ref{fig11}-\ref{fig13}, respectively. Curves of different colors were obtained with different prescriptions for the dust attenuation, as indicated by the captions. Each panel corresponds to a different cluster and each sub-panel to a model based on a different flavor of massive-star evolution. Narrow single-peaked PDFs indicate that a given property is well constrained, while broad PDFs and/or PDFs with multiple peaks indicate that the properties are not as well constrained. Additionally, if the peaks of the PDFs line up vertically, then there is little scatter in properties derived with different models. Note the broader $E(B-V)$ and age PDFs of the two clusters in NGC 5253 (shown in the bottom rows of Fig.~\ref{fig11} and~\ref{fig13}, respectively) relative to clusters in NGC 1566. This is due to the larger photometric errors of clusters in NGC 5253. The dashed vertical lines indicate the corresponding best-fitting solutions. Since these come from a combination of the PDFs of different independent variables, they do not necessarily coincide with the peaks of the posterior marginalized PDFs for all the variables. We proceed to look at each property individually.

For $E(B-V)$, Fig.~\ref{fig11} shows that: 1) in general, the color-excess is well constrained; 2) in general, models based on different stellar evolutionary tracks are in good agreement with each other;  3) the PDFs obtained with the starburst and alternative laws are very similar, except for the most reddened clusters in the sample (NGC 1566 \#5 and NGC 5253 \#5), for which the starburst law yields that are shifted towards lower $E(B-V)$ values; 4) most clusters in our sample have low attenuation by dust; and 5) for the oldest cluster in our sample (NGC 5253  \#9, see Fig.~\ref{fig13}), Gn and Gr models yield larger $E(B-V)$ values relative to other models, which could be because of the lower metallicity of these two models ($Z=0.002$), relative to the rest of models ($Z=0.004$).

%%%%%%%%%%%%%%%%%%%%%%%%%%%%%%
% Figure 11
%%%%%%%%%%%%%%%%%%%%%%%%%%%%%%

For the cluster masses, Fig.~\ref{fig12} shows that: 1) masses are mostly between $10^4-10^5\,$M$_\odot$, except for NGC 5253 \#9, which has a broad range of solutions; 2) masses obtained with the starburst attenuation law appear to yield overall higher masses relative to those obtained with the extinction law (more on this below); 3) for Po and Go models, in general, the PDF masses obtained in this work are slightly below our lower mass selection limit of $\ge5\times10^4$ M$\odot$, likely because C15b and A16 use a different mass binning and compute best-fitting masses instead of masses derived from the median of the PDF. Our best-fitting masses are below our median of PDF masses. However, for two clusters, we do obtain Po and Go median of PDF masses that are $\ge5\times10^4$ M$\odot$.

%%%%%%%%%%%%%%%%%%%%%%%%%%%%%%
% Figure 12
%%%%%%%%%%%%%%%%%%%%%%%%%%%%%%

Finally, for the cluster ages, Fig.~\ref{fig13} shows that all models agree that the majority of clusters are young, i.e., $<10\,$Myr, with NGC 5253-5 and NGC 5253-9 being the youngest and oldest clusters, respectively.

%%%%%%%%%%%%%%%%%%%%%%%%%%%%%%
% Figure 13
%%%%%%%%%%%%%%%%%%%%%%%%%%%%%%

\subsection{How different are properties derived with different models?}

\begin{figure*}
\begin{subfigure}
\centering
\includegraphics[width=1.59\columnwidth]{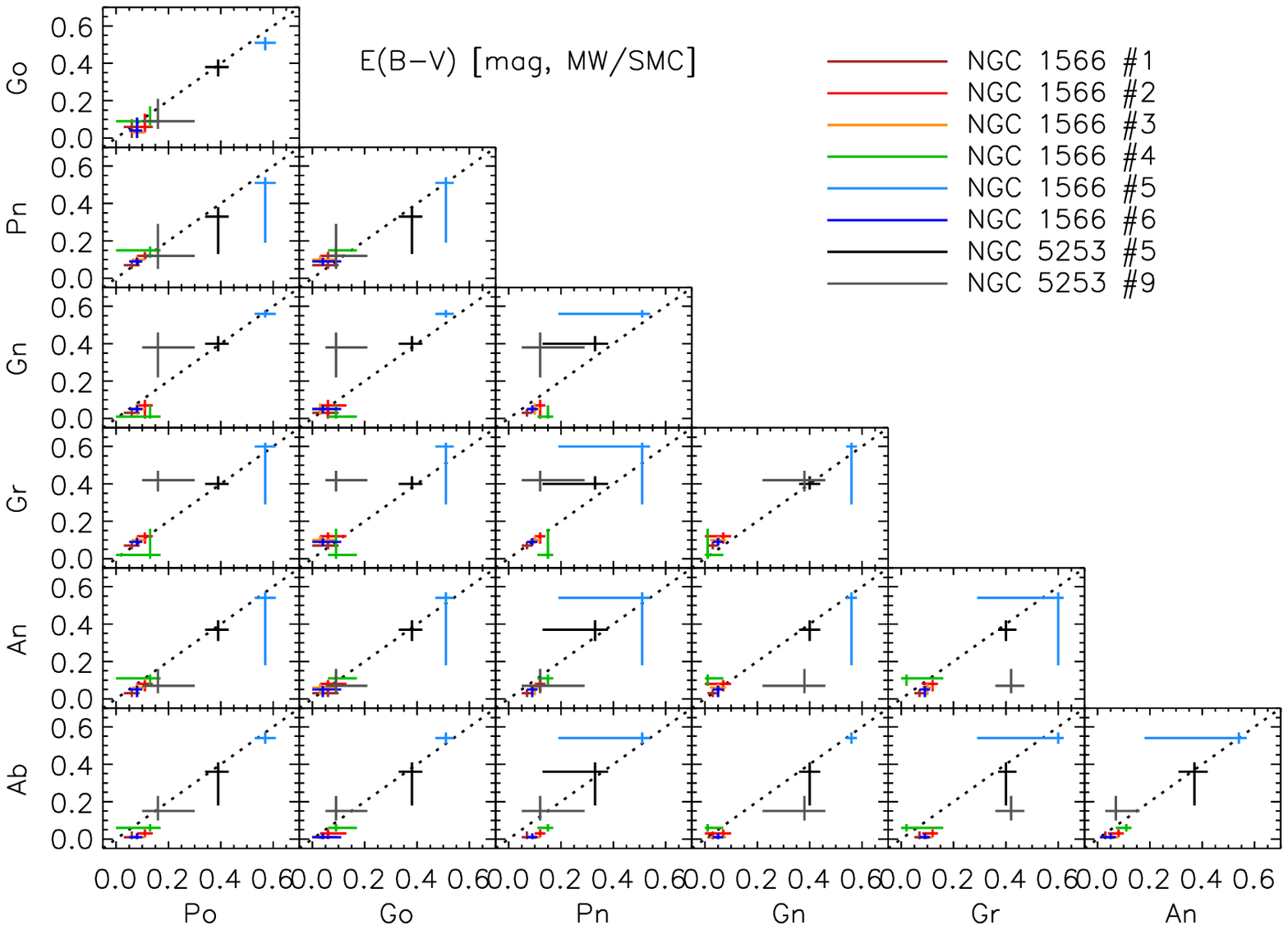}
\end{subfigure}
\begin{subfigure}
\centering
\includegraphics[width=1.59\columnwidth]{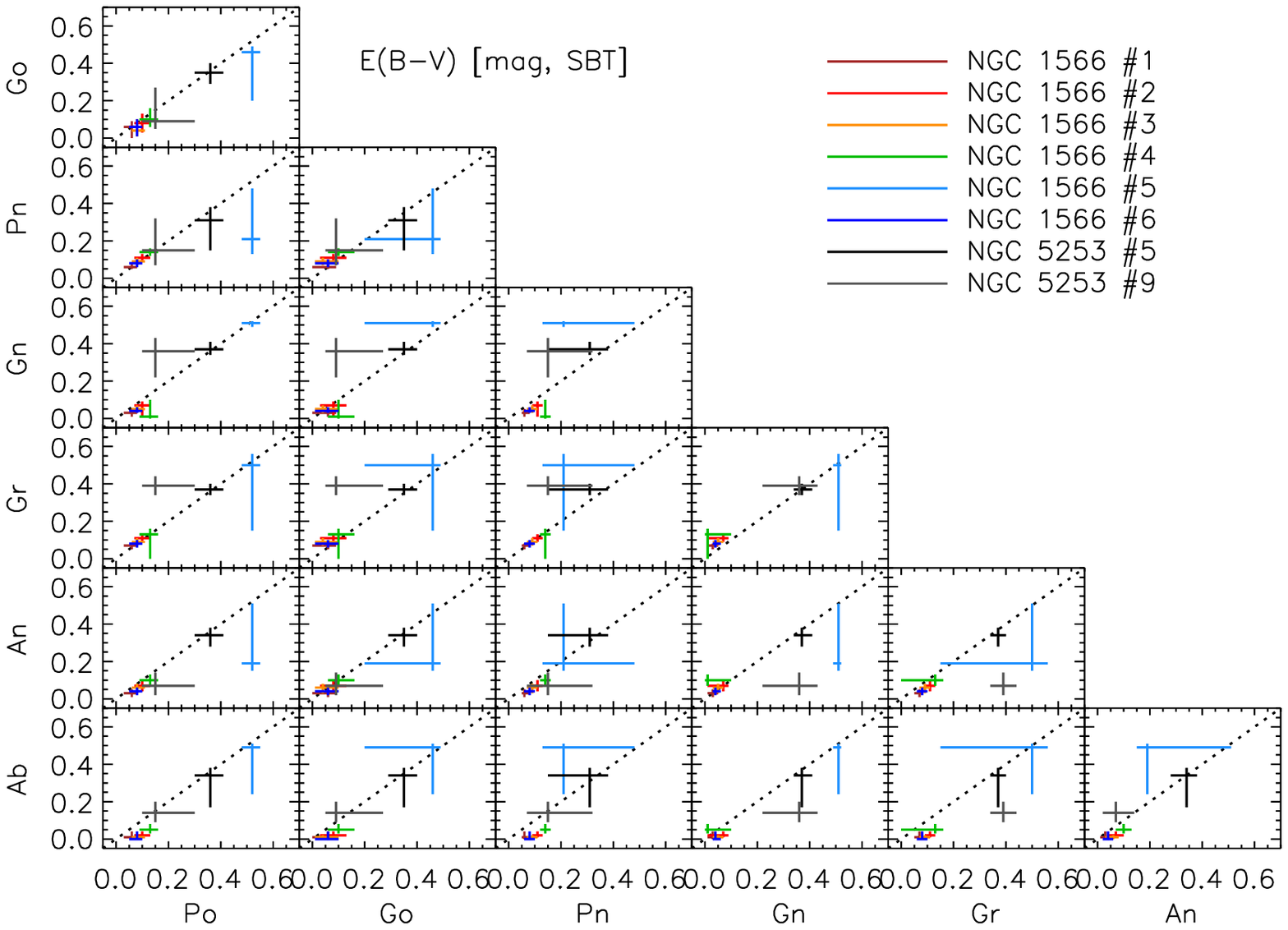}
\end{subfigure}
\caption{One to one comparison of median $E(B-V)$ values obtained seven different models. The symbols with error-bars correspond to the clusters in the legend. The models use the MW (NGC 1566) or SMC (NGC 5253) extinction laws.}
\label{fig14}
\end{figure*}

\begin{figure*}
\begin{subfigure}
\centering
\includegraphics[width=1.59\columnwidth]{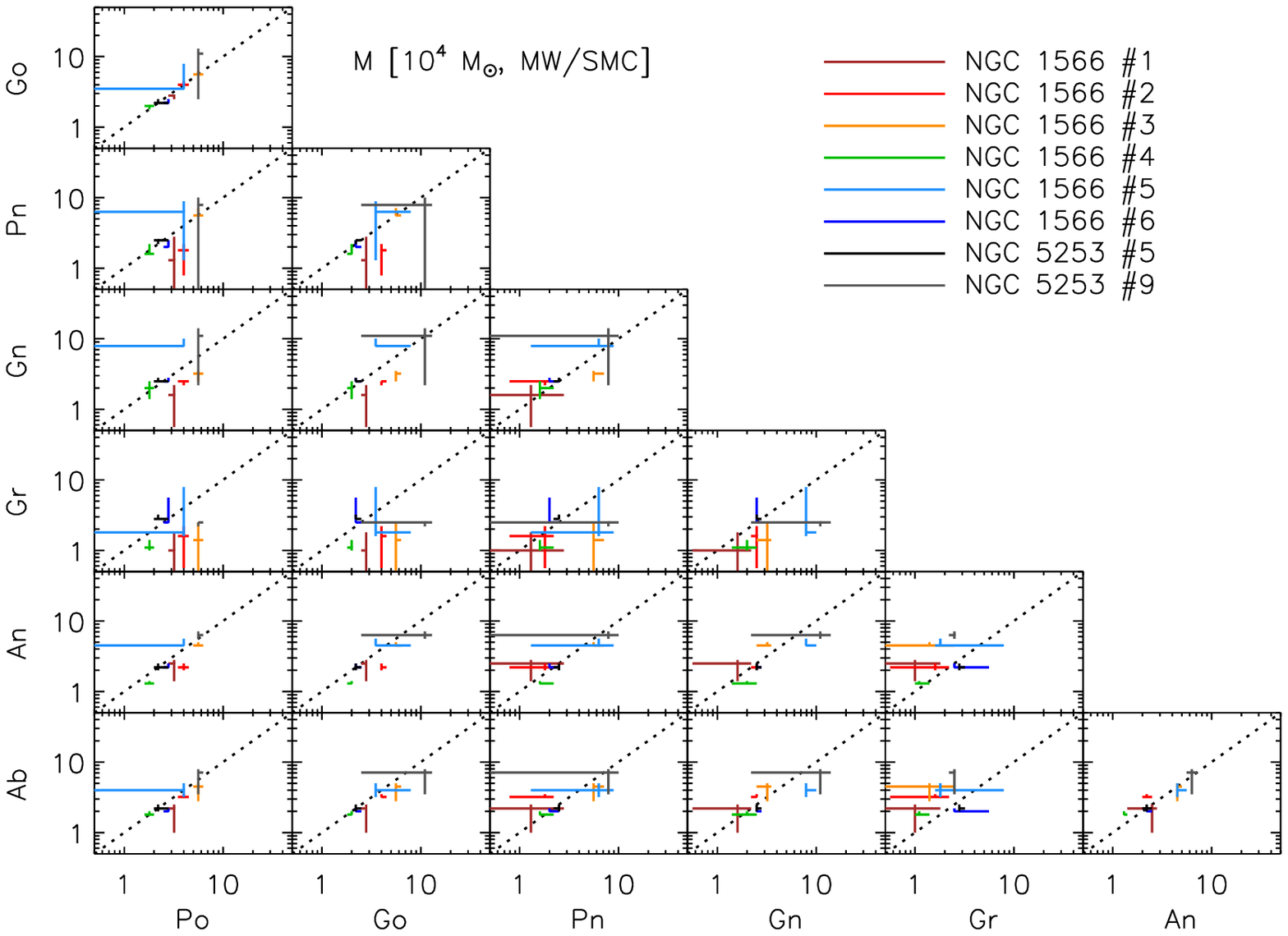}
\end{subfigure}
\begin{subfigure}
\centering
\includegraphics[width=1.59\columnwidth]{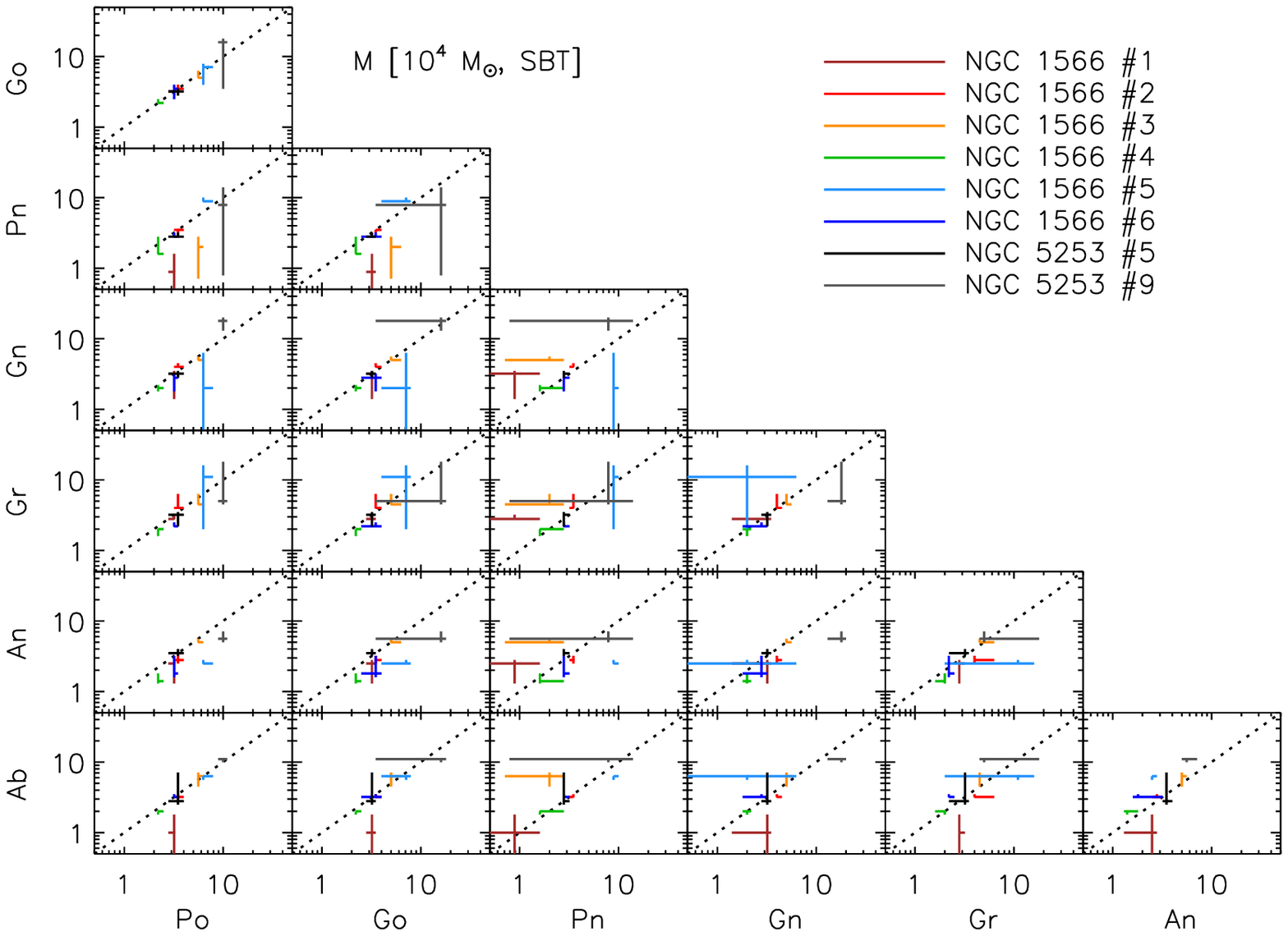}
\end{subfigure}
\caption{Similar to Fig.~\ref{fig14} but we compare the masses of the clusters.}
\label{fig15}
\end{figure*}

\begin{figure*}
\begin{subfigure}
\centering
\includegraphics[width=1.59\columnwidth]{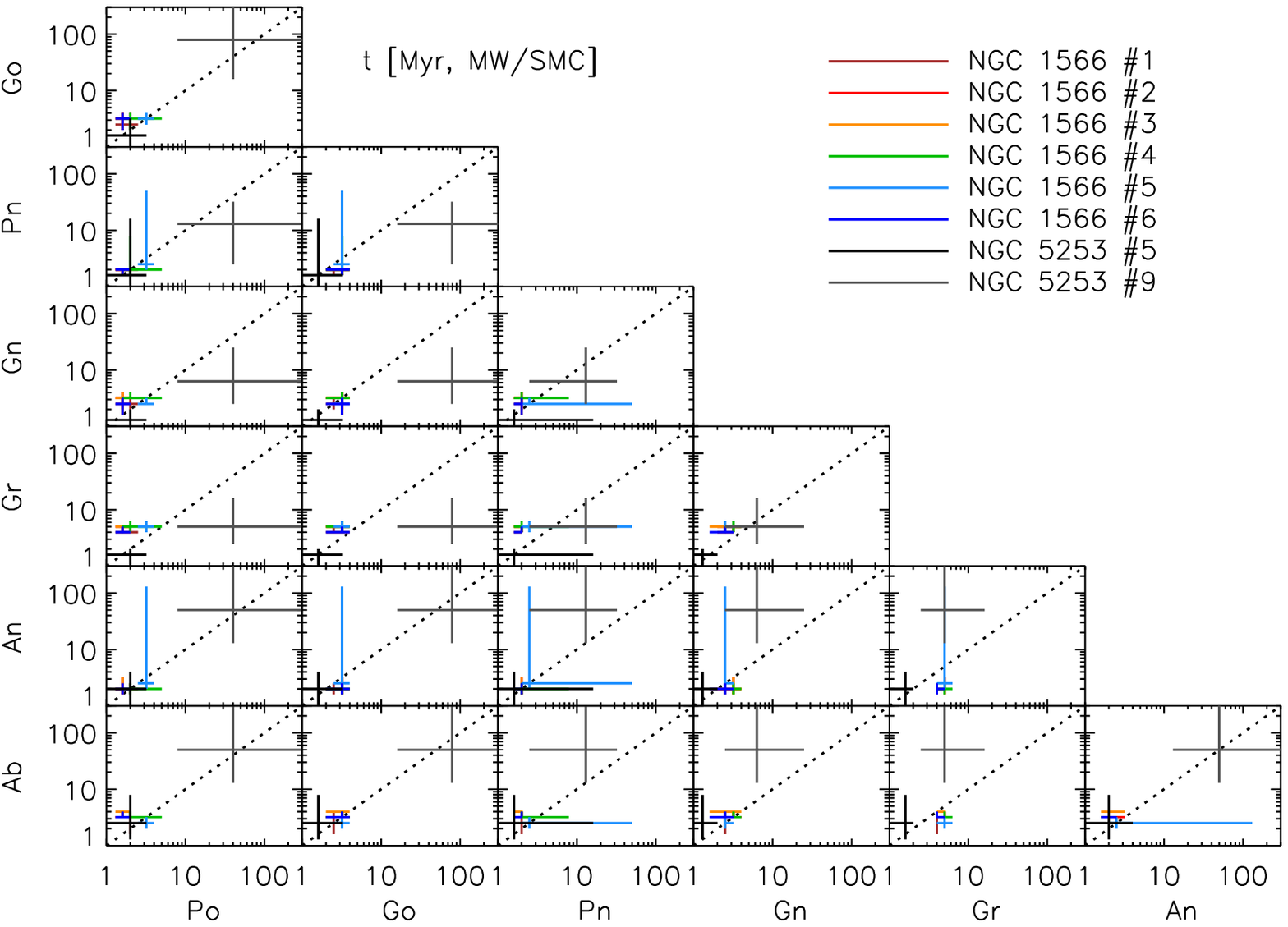}
\end{subfigure}
\begin{subfigure}
\centering
\includegraphics[width=1.59\columnwidth]{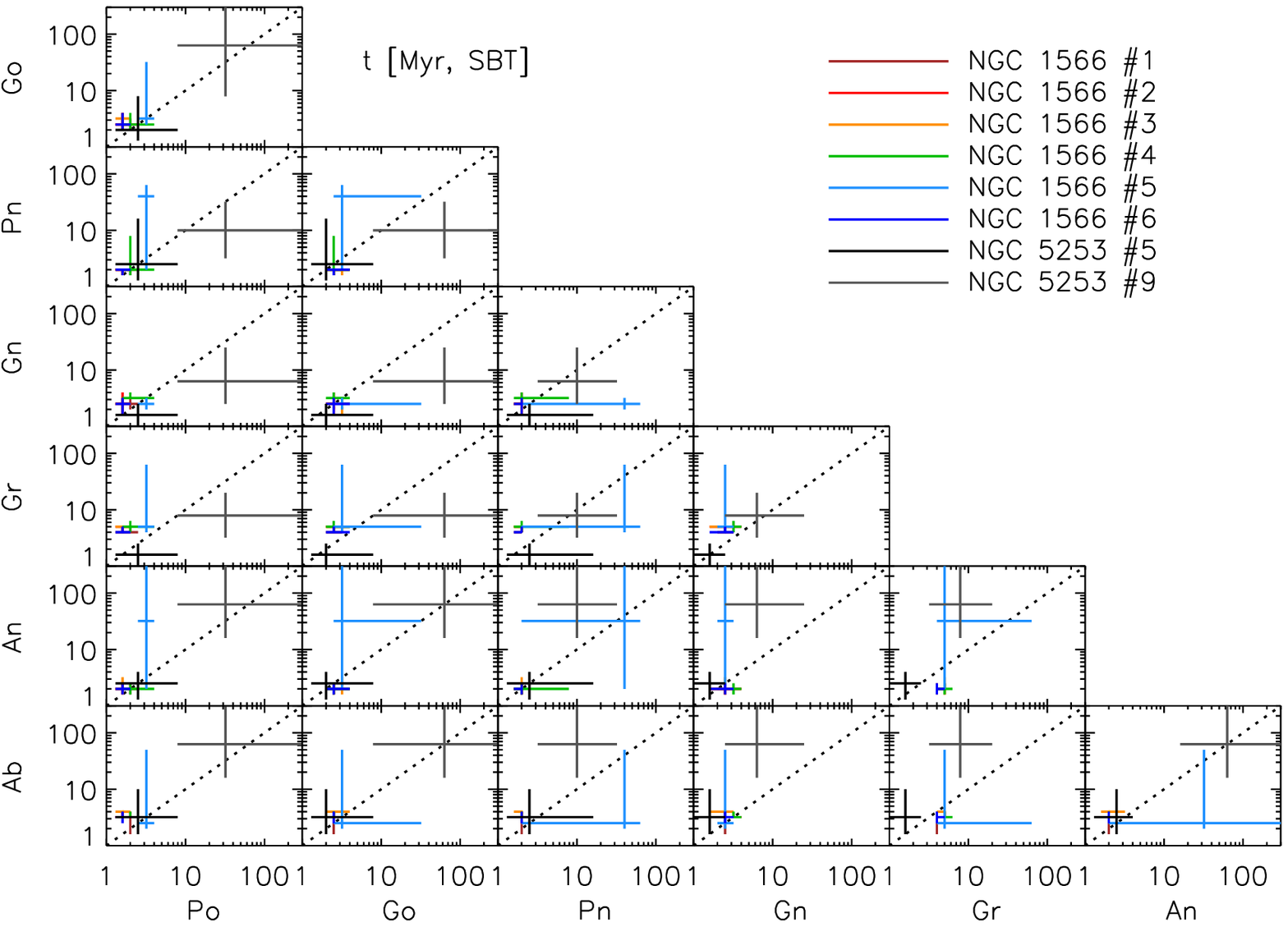}
\end{subfigure}
\caption{Similar to Fig.~\ref{fig14} but we compare the ages of the clusters.}
\label{fig16}
\end{figure*}

Figures~\ref{fig14} to~\ref{fig16} show one-to-one correspondence plots between pairs of models for the median values of $E(B-V)$, mass, and age, respectively. We use error bars (16th and 84th percentiles of the PDF) of different colors for the different clusters, as indicated in the legend. We show results obtained with the MW (NGC 1566) and SMC (NGC 5253) extinction law in the top panel and with the starburst attenuation law in the bottom panel. The dotted lines mark the positions where the properties obtained with the two models are equal.

For $E(B-V)$, Fig.~\ref{fig14} shows that: 1) the different models are in agreement with each other within the errors, with a few exceptions, e.g., Gn and Gr models yield higher $E(B-V)$ values for NGC 5253 \#9 (in gray), as previously mentioned; but most models agree that NGC 1566 \#5 (in light blue) is the most reddened cluster followed by NGC 5253 \#5 (in black); and 2) for the most reddened clusters (light blue and black symbols), the starburst law yields lower $E(B-V)$ values.

%%%%%%%%%%%%%%%%%%%%%%%%%%%%%%
% Figure 14
%%%%%%%%%%%%%%%%%%%%%%%%%%%%%%

For the mass, Fig.~\ref{fig15} shows the following. 1) The Po, Go, Pn, An, and Ab masses are in better agreement within the errors when using the extinction law than when using the starburst attenuation law. 2) Conversely, the Gn and Gr masses are in better agreement with the rest of masses (within the errors) when using the starburst attenuation law. 3) According to all but the Gr models, cluster NGC 5253 \#9 is the most massive. 

%%%%%%%%%%%%%%%%%%%%%%%%%%%%%%
% Figure 15
%%%%%%%%%%%%%%%%%%%%%%%%%%%%%%

Finally, Fig.~\ref{fig16} shows that: 1) most models agree that all clusters are younger than 10 Myr and that NGC 5253 \#9 (the most massive cluster in our sample) is the oldest; 2) for clusters with ages $<10$ Myr, Gr and Ab models yield systematically larger ages than the rest of models; and 3) for cluster NGC 1566 \#5 (one of the most reddened clusters in our sample), when using the Pn and An models there is a large discrepancy in the ages obtained with the starburst law ($>30$ Myr) and MW law (2.5 Myr) (corresponding PDFs show that the age is not well constrained).

%%%%%%%%%%%%%%%%%%%%%%%%%%%%%%
% Figure 16
%%%%%%%%%%%%%%%%%%%%%%%%%%%%%%

The values of properties obtained with the different models, along with their error bars are provided in Tables~\ref{tab6} to~\ref{tab8} for $E(B-V)$, mass, and age, respectively. Each table shows two sets of values, obtained with the starburst attenuation law and MW (NGC 1566) or SMC (NGC 5253) extinction law, as indicated in the first column after the cluster ID. For each property, the range of values spanned by our sample at fixed attenuation/extinction law can be obtained from the two rows of the table that give the minimum and maximum values in the column. The last three columns in each table give the median of the values obtained with all models, the corresponding standard deviation, and the difference or ratio (depending on the property) between the maximum and minimum values obtained with the different models. The median of standard deviations of the different models, i.e., the typical scatter between models for a given cluster and property, is given in the second to last column and last row of each table (value after the comma). The median of the maximum difference between models is given in the last column and last row of each table (value after the comma).

When considering all present models, we find that clusters span ranges of $E(B-V)=0.05\pm0.02$ to $0.54\pm0.13$ mag; $M_{\rm{cl}}=1.8\pm0.3$ to $10\pm5\times10^4\,M_\odot$; and $t=1.6\pm0.4$ to $40\pm27.7$ Myr; and that the typical scatter in properties derived with different models, for a given cluster is $\sigma[E(B-V)]\sim0.03$ mag, $\sigma({\rmn{mass}})=\sim10^4\,M_\odot$, and $\sigma({\rmn{age}})\sim1$ Myr. In addition, typical maximum differences in properties derived with different models, for a given cluster, are $\pm0.09$ mag in $E(B-V)$, a factor of 2.8 in mass, and a factor of 2.5 in age. 

%%%%%%%%%%%%%%%%%%%%%%%%%%%%%%
% Table 6
%%%%%%%%%%%%%%%%%%%%%%%%%%%%%%

\begin{table*} 
\centering
\caption{$E(B-V)$ values obtained with models based on different massive-star tracks and dust extinction/attenuation laws.}
 \label{tab6}
\begin{threeparttable}
\begin{tabular}{lllllllllll}
\hline
ID\tnote{a}&Po\tnote{b}&Go\tnote{b}&Pn\tnote{b}&Gn\tnote{b}&Gr\tnote{b}&An\tnote{b}&Ab\tnote{b}&median\tnote{c}&$\sigma$\tnote{d}&$\Delta$\tnote{e}\\
\hfill&(mag)&(mag)&(mag)&(mag)&(mag)&(mag)&(mag)&(mag)&(mag)&(mag)\\
\hline
 1566-1-mwy&0.06$^{+0.09}_{-0.03}$&0.06$^{+0.10}_{-0.00}$&0.07$^{+0.09}_{-0.05}$&0.03$^{+0.05}_{-0.01}$&0.07$^{+0.09}_{-0.05}$&0.03$^{+0.05}_{-0.01}$&0.01$^{+0.04}_{-0.00}$&0.06&0.02&0.06\\
 1566-2-mwy&0.11$^{+0.14}_{-0.08}$&0.06$^{+0.13}_{-0.03}$&0.12$^{+0.14}_{-0.10}$&0.07$^{+0.10}_{-0.00}$&0.12$^{+0.14}_{-0.08}$&0.08$^{+0.10}_{-0.04}$&0.03$^{+0.05}_{-0.01}$&0.08&0.03&0.09\\
 1566-3-mwy&0.08$^{+0.11}_{-0.06}$&0.03$^{+0.09}_{-0.00}$&0.10$^{+0.12}_{-0.08}$&0.05$^{+0.08}_{-0.02}$&0.10$^{+0.12}_{-0.08}$&0.06$^{+0.08}_{-0.02}$&0.01$^{+0.02}_{-0.00}$&0.06&0.03&0.09\\
 1566-4-mwy&0.13$^{+0.17}_{-0.00}$&0.09$^{+0.17}_{-0.06}$&0.15$^{+0.17}_{-0.11}$&0.01$^{+0.07}_{-0.00}$&0.02$^{+0.16}_{-0.00}$&0.11$^{+0.13}_{-0.07}$&0.06$^{+0.08}_{-0.04}$&0.09&0.05&0.14\\
 1566-5-mwy&0.57$^{+0.61}_{-0.53}$&0.51$^{+0.54}_{-0.47}$&0.51$^{+0.54}_{-0.19}$&0.56$^{+0.58}_{-0.54}$&0.60$^{+0.62}_{-0.29}$&0.54$^{+0.57}_{-0.18}$&0.54$^{+0.57}_{-0.51}$&0.54&0.03&0.09\\
 1566-6-mwy&0.08$^{+0.10}_{-0.05}$&0.04$^{+0.11}_{-0.00}$&0.09$^{+0.11}_{-0.07}$&0.05$^{+0.07}_{-0.03}$&0.09$^{+0.11}_{-0.07}$&0.05$^{+0.07}_{-0.01}$&0.01$^{+0.03}_{-0.00}$&0.05&0.03&0.08\\
 5253-5-smc&0.36$^{+0.41}_{-0.30}$&0.35$^{+0.40}_{-0.29}$&0.31$^{+0.38}_{-0.15}$&0.37$^{+0.41}_{-0.34}$&0.37$^{+0.40}_{-0.34}$&0.34$^{+0.38}_{-0.28}$&0.34$^{+0.38}_{-0.17}$&0.35&0.02&0.06\\
 5253-9-smc&0.15$^{+0.30}_{-0.10}$&0.09$^{+0.27}_{-0.05}$&0.15$^{+0.32}_{-0.07}$&0.36$^{+0.43}_{-0.22}$&0.39$^{+0.44}_{-0.34}$&0.07$^{+0.14}_{-0.02}$&0.14$^{+0.20}_{-0.09}$&0.15&0.13&0.32\\
  min\tnote{f}&0.06&0.03&0.07&0.01&0.02&0.03&0.01&0.05&0.02&0.06\\
  max\tnote{g}&0.57&0.51&0.51&0.56&0.60&0.54&0.54&0.54&0.13&0.32\\
 1566-1-sbt&0.06$^{+0.08}_{-0.03}$&0.06$^{+0.09}_{-0.00}$&0.06$^{+0.08}_{-0.05}$&0.03$^{+0.05}_{-0.01}$&0.07$^{+0.08}_{-0.05}$&0.03$^{+0.05}_{-0.01}$&0.01$^{+0.04}_{-0.00}$&0.06&0.02&0.06\\
 1566-2-sbt&0.10$^{+0.13}_{-0.07}$&0.08$^{+0.13}_{-0.03}$&0.11$^{+0.13}_{-0.09}$&0.07$^{+0.09}_{-0.01}$&0.11$^{+0.13}_{-0.09}$&0.07$^{+0.10}_{-0.04}$&0.02$^{+0.04}_{-0.01}$&0.08&0.03&0.09\\
 1566-3-sbt&0.08$^{+0.11}_{-0.06}$&0.04$^{+0.10}_{-0.01}$&0.09$^{+0.11}_{-0.07}$&0.05$^{+0.07}_{-0.03}$&0.09$^{+0.11}_{-0.07}$&0.06$^{+0.08}_{-0.03}$&0.01$^{+0.02}_{-0.00}$&0.06&0.03&0.08\\
 1566-4-sbt&0.13$^{+0.16}_{-0.09}$&0.10$^{+0.16}_{-0.06}$&0.14$^{+0.16}_{-0.12}$&0.01$^{+0.10}_{-0.00}$&0.13$^{+0.16}_{-0.00}$&0.10$^{+0.13}_{-0.07}$&0.05$^{+0.08}_{-0.03}$&0.10&0.05&0.13\\
 1566-5-sbt&0.52$^{+0.55}_{-0.48}$&0.46$^{+0.49}_{-0.20}$&0.21$^{+0.48}_{-0.13}$&0.51$^{+0.52}_{-0.49}$&0.50$^{+0.56}_{-0.15}$&0.19$^{+0.51}_{-0.15}$&0.49$^{+0.51}_{-0.24}$&0.49&0.15&0.33\\
 1566-6-sbt&0.08$^{+0.10}_{-0.05}$&0.06$^{+0.10}_{-0.01}$&0.08$^{+0.10}_{-0.06}$&0.04$^{+0.06}_{-0.03}$&0.08$^{+0.10}_{-0.06}$&0.04$^{+0.06}_{-0.02}$&0.00$^{+0.04}_{-0.00}$&0.06&0.03&0.08\\
 5253-5-sbt&0.39$^{+0.43}_{-0.34}$&0.38$^{+0.42}_{-0.33}$&0.33$^{+0.38}_{-0.13}$&0.40$^{+0.44}_{-0.36}$&0.40$^{+0.44}_{-0.37}$&0.37$^{+0.42}_{-0.31}$&0.36$^{+0.41}_{-0.18}$&0.38&0.03&0.07\\
 5253-9-sbt&0.16$^{+0.30}_{-0.10}$&0.09$^{+0.21}_{-0.05}$&0.12$^{+0.29}_{-0.05}$&0.38$^{+0.46}_{-0.22}$&0.42$^{+0.47}_{-0.36}$&0.07$^{+0.16}_{-0.03}$&0.15$^{+0.23}_{-0.10}$&0.15&0.14&0.35\\
  min\tnote{f}&0.06&0.04&0.06&0.01&0.07&0.03&0.00&0.06&0.02&0.06\\
  max\tnote{g}&0.52&0.46&0.33&0.51&0.50&0.37&0.49&0.49&0.15, 0.03&0.35, 0.09 \\
\hline
\end{tabular}
\begin{tablenotes}
\item [a] Cluster ID and extinction/attenuation law. The first four digits are the NGC number of the galaxy. The extinction/attenuation law is given by the last characters. mwy=Milky Way. smc=SMC. sbt=starburst.
\item [b] Median of the $E(B-V)$  PDF.
\item [c] Median of the $E(B-V)$ values in the row.
\item [d] Standard deviation of the $E(B-V)$ values in the row. The value after the comma in the last row of this column is the median of the values in the column (excluding min/max rows).
\item [e] Difference between maximum and minimum $E(B-V)$ values in the row. The value after the comma in the last row of this column is the median of the values in the column (excluding min/max rows).
\item [f] Minimum of eight values in the column for a fixed dust extinction/attenuation law. 
\item [g] Maximum of eight values in the column for a fixed dust extinction/attenuation law. For the last column, the second value is the median of the 16 values in the column. 
\end{tablenotes}
\end{threeparttable} 
\end{table*} 

%%%%%%%%%%%%%%%%%%%%%%%%%%%%%%
% Table 6
%%%%%%%%%%%%%%%%%%%%%%%%%%%%%%

\begin{table*} 
\centering
\caption{Cluster masses obtained with models based on different massive-star tracks and dust extinction/attenuation laws.}
 \label{tab7}
\begin{threeparttable}
\begin{tabular}{llllllllllll}
\hline
ID\tnote{a}&Po\tnote{b}&Go\tnote{b}&Pn\tnote{b}&Gn\tnote{b}&Gr\tnote{b}&An\tnote{b}&Ab\tnote{b}&median\tnote{c}&$\sigma$\tnote{d}&$\rho$\tnote{e}\\
\hfill&($10^4\,M_\odot$)&($10^4\,M_\odot$)&($10^4\,M_\odot$)&($10^4\,M_\odot$)&($10^4\,M_\odot$)&($10^4\,M_\odot$)&($10^4\,M_\odot$)&($10^4\,M_\odot$)&($10^4\,M_\odot$)&\hfill\\
\hline
 1566-1-mwy& 3.2$^{+ 3.2}_{- 2.8}$& 2.8$^{+ 2.8}_{- 2.5}$& 1.3$^{+ 2.8}_{- 0.5}$& 1.6$^{+ 2.2}_{- 0.6}$& 1.0$^{+ 1.8}_{- 0.4}$& 2.5$^{+ 2.8}_{- 1.4}$& 2.2$^{+ 2.5}_{- 1.0}$& 2.2& 0.8& 3.2\\
 1566-2-mwy& 4.0$^{+ 4.5}_{- 3.5}$& 4.0$^{+ 4.5}_{- 4.0}$& 1.8$^{+ 2.2}_{- 0.8}$& 2.5$^{+ 2.5}_{- 2.2}$& 1.6$^{+ 2.2}_{- 0.6}$& 2.2$^{+ 2.5}_{- 2.0}$& 3.2$^{+ 3.5}_{- 3.2}$& 2.5& 1.0& 2.5\\
 1566-3-mwy& 5.6$^{+ 6.3}_{- 5.0}$& 5.6$^{+ 6.3}_{- 5.6}$& 5.6$^{+ 7.1}_{- 5.6}$& 3.2$^{+ 3.5}_{- 2.5}$& 1.4$^{+ 2.5}_{- 0.5}$& 4.5$^{+ 5.0}_{- 4.5}$& 4.5$^{+ 5.0}_{- 2.8}$& 4.5& 1.6& 4.0\\
 1566-4-mwy& 1.8$^{+ 2.0}_{- 1.6}$& 2.0$^{+ 2.0}_{- 1.8}$& 1.6$^{+ 2.2}_{- 1.6}$& 2.0$^{+ 2.5}_{- 1.4}$& 1.1$^{+ 1.4}_{- 1.0}$& 1.3$^{+ 1.4}_{- 1.3}$& 1.8$^{+ 2.0}_{- 1.8}$& 1.8& 0.3& 1.8\\
 1566-5-mwy& 4.0$^{+ 4.0}_{- 0.1}$& 3.5$^{+ 7.9}_{- 3.5}$& 6.3$^{+ 8.9}_{- 1.3}$& 7.9$^{+10.0}_{- 7.9}$& 1.8$^{+ 7.9}_{- 1.6}$& 4.5$^{+ 5.6}_{- 4.5}$& 4.0$^{+ 5.0}_{- 3.2}$& 4.0& 2.0& 4.4\\
 1566-6-mwy& 2.8$^{+ 2.8}_{- 2.5}$& 2.2$^{+ 2.5}_{- 2.2}$& 2.0$^{+ 2.5}_{- 2.0}$& 2.5$^{+ 2.8}_{- 2.5}$& 2.5$^{+ 5.6}_{- 2.5}$& 2.2$^{+ 2.5}_{- 2.2}$& 2.0$^{+ 2.2}_{- 2.0}$& 2.2& 0.3& 1.4\\
 5253-5-smc& 3.5$^{+ 4.0}_{- 2.8}$& 3.2$^{+ 3.5}_{- 2.8}$& 2.8$^{+ 3.2}_{- 2.8}$& 3.2$^{+ 3.5}_{- 2.8}$& 3.2$^{+ 3.5}_{- 2.2}$& 3.5$^{+ 4.0}_{- 3.2}$& 2.8$^{+ 7.1}_{- 2.5}$& 3.2& 0.3& 1.2\\
 5253-9-smc&10.0$^{+11.0}_{- 8.9}$&16.0$^{+18.0}_{- 3.5}$& 7.9$^{+14.0}_{- 0.8}$&18.0$^{+20.0}_{-13.0}$& 5.0$^{+18.0}_{- 4.5}$& 5.6$^{+ 7.1}_{- 5.0}$&11.0$^{+11.0}_{-10.0}$&10.0& 5.0& 3.6\\
  min\tnote{f}& 1.8& 2.0& 1.3& 1.6& 1.0& 1.3& 1.8& 1.8& 0.3& 1.2\\
  max\tnote{g}&10.0&16.0& 7.9&18.0& 5.0& 5.6&11.0&10.0& 5.0& 4.4\\
 1566-1-sbt& 3.2$^{+ 3.2}_{- 2.8}$& 3.2$^{+ 3.5}_{- 2.8}$& 0.9$^{+ 1.6}_{- 0.3}$& 3.2$^{+ 3.5}_{- 1.4}$& 2.8$^{+ 3.2}_{- 2.8}$& 2.5$^{+ 2.8}_{- 1.3}$& 1.0$^{+ 1.8}_{- 0.4}$& 2.8& 1.0& 3.6\\
 1566-2-sbt& 3.5$^{+ 4.0}_{- 3.2}$& 3.5$^{+ 4.0}_{- 3.5}$& 3.5$^{+ 3.5}_{- 3.2}$& 4.0$^{+ 4.5}_{- 4.0}$& 4.0$^{+ 6.3}_{- 4.0}$& 2.8$^{+ 3.2}_{- 2.5}$& 3.2$^{+ 3.5}_{- 3.2}$& 3.5& 0.4& 1.4\\
 1566-3-sbt& 5.6$^{+ 6.3}_{- 5.6}$& 5.0$^{+ 6.3}_{- 5.0}$& 2.0$^{+ 2.8}_{- 0.7}$& 5.0$^{+ 5.6}_{- 5.0}$& 4.5$^{+ 6.3}_{- 4.5}$& 5.0$^{+ 5.6}_{- 5.0}$& 6.3$^{+ 7.1}_{- 4.5}$& 5.0& 1.3& 3.1\\
 1566-4-sbt& 2.2$^{+ 2.5}_{- 2.2}$& 2.2$^{+ 2.5}_{- 2.2}$& 1.6$^{+ 2.8}_{- 1.6}$& 2.0$^{+ 2.2}_{- 1.8}$& 2.0$^{+ 2.0}_{- 1.6}$& 1.4$^{+ 1.8}_{- 1.3}$& 2.0$^{+ 2.0}_{- 1.8}$& 2.0& 0.3& 1.6\\
 1566-5-sbt& 6.3$^{+ 7.9}_{- 6.3}$& 7.1$^{+ 7.9}_{- 4.0}$& 8.9$^{+10.0}_{- 8.9}$& 2.0$^{+ 6.3}_{- 0.1}$&11.0$^{+16.0}_{- 2.0}$& 2.5$^{+ 2.8}_{- 2.5}$& 6.3$^{+ 6.3}_{- 5.6}$& 6.3& 3.2& 5.5\\
 1566-6-sbt& 3.2$^{+ 3.5}_{- 3.2}$& 3.5$^{+ 4.0}_{- 2.5}$& 2.8$^{+ 3.2}_{- 2.8}$& 2.8$^{+ 3.2}_{- 1.8}$& 2.2$^{+ 2.5}_{- 2.2}$& 1.8$^{+ 3.2}_{- 1.6}$& 3.2$^{+ 3.5}_{- 3.2}$& 2.8& 0.6& 1.9\\
 5253-5-sbt& 2.2$^{+ 2.8}_{- 2.0}$& 2.2$^{+ 2.5}_{- 2.2}$& 2.5$^{+ 2.5}_{- 2.2}$& 2.5$^{+ 2.8}_{- 2.5}$& 2.8$^{+ 3.2}_{- 2.8}$& 2.2$^{+ 2.5}_{- 2.0}$& 2.2$^{+ 2.5}_{- 2.0}$& 2.2& 0.2& 1.3\\
 5253-9-sbt& 5.6$^{+ 6.3}_{- 5.6}$&11.0$^{+13.0}_{- 2.5}$& 7.9$^{+10.0}_{- 0.5}$&11.0$^{+14.0}_{- 2.2}$& 2.5$^{+ 2.5}_{- 2.2}$& 6.3$^{+ 7.1}_{- 5.6}$& 7.1$^{+ 7.9}_{- 3.5}$& 7.1& 3.0& 4.4\\
  min\tnote{f}& 2.2& 2.2& 0.9& 2.0& 2.0& 1.4& 1.0& 2.0& 0.2& 1.3\\
  max\tnote{g}& 6.3&11.0& 8.9&11.0&11.0& 6.3& 7.1& 7.1& 3.2, 0.9& 5.5, 2.8\\
\hline
\end{tabular}
\begin{tablenotes}
\item [a] Cluster ID and extinction/attenuation law. The first four digits are the NGC number of the galaxy. The extinction/attenuation law is given by the last characters. mwy=Milky Way. smc=SMC. sbt=starburst.
\item [b] Median of the mass PDF.
\item [c] Median of the masses in the row.
\item [d] Standard deviation of the masses in the row. The value after the comma in the last row of this column is the median of the values in the column (excluding min/max rows).
\item [e] Ratio between the maximum and minimum masses in the row. The value after the comma in the last row of this column is the median of the values in the column (excluding min/max rows).
\item [f] Minimum of eight values in the column for a fixed dust extinction/attenuation law. 
\item [g] Maximum of eight values in the column for a fixed dust extinction/attenuation law. For the last column, the second value is the median of the 16 values in the column. 
\end{tablenotes}
\end{threeparttable} 
\end{table*} 

%%%%%%%%%%%%%%%%%%%%%%%%%%%%%%
% Table 7
%%%%%%%%%%%%%%%%%%%%%%%%%%%%%%

\begin{table*} 
\centering
\caption{Cluster ages obtained with models based on different massive-star tracks and dust extinction/attenuation laws.}
 \label{tab8}
\begin{threeparttable}
\begin{tabular}{lllllllllllll}
\hline
ID\tnote{a}&Po\tnote{b}&Go\tnote{b}&Pn\tnote{b}&Gn\tnote{b}&Gr\tnote{b}&An\tnote{b}&Ab\tnote{b}&median\tnote{c}&$\sigma$\tnote{d}&$\rho$\tnote{e}\\
\hfill&(Myr)&(Myr)&(Myr)&(Myr)&(Myr)&(Myr)&(Myr)&(Myr)&(Myr)&\hfill\\
\hline
1566-1-mwy&2.0$^{+2.5}_{-1.3}$&2.5$^{+4.0}_{-2.0}$&2.0$^{+2.0}_{-1.6}$&2.5$^{+3.2}_{-2.0}$&4.0$^{+5.0}_{-4.0}$&2.0$^{+2.5}_{-1.6}$&3.2$^{+4.0}_{-1.6}$&2.5&0.8&2.0\\
1566-2-mwy&1.6$^{+2.0}_{-1.3}$&3.2$^{+4.0}_{-2.0}$&2.0$^{+2.0}_{-1.6}$&3.2$^{+4.0}_{-2.0}$&5.0$^{+5.0}_{-4.0}$&2.0$^{+3.2}_{-1.6}$&3.2$^{+4.0}_{-3.2}$&3.2&1.2&3.1\\
1566-3-mwy&1.6$^{+2.0}_{-1.3}$&3.2$^{+4.0}_{-2.0}$&2.0$^{+2.0}_{-1.6}$&3.2$^{+4.0}_{-1.6}$&5.0$^{+5.0}_{-4.0}$&2.0$^{+3.2}_{-1.6}$&4.0$^{+4.0}_{-3.2}$&3.2&1.2&3.1\\
1566-4-mwy&2.0$^{+5.0}_{-1.6}$&3.2$^{+4.0}_{-2.0}$&2.0$^{+7.9}_{-1.6}$&3.2$^{+4.0}_{-3.2}$&5.0$^{+6.3}_{-4.0}$&2.0$^{+2.5}_{-1.6}$&3.2$^{+4.0}_{-3.2}$&3.2&1.1&2.5\\
1566-5-mwy&3.2$^{+4.0}_{-2.5}$&3.2$^{+4.0}_{-2.5}$&2.5$^{+50.0}_{-2.0}$&2.5$^{+3.2}_{-2.5}$&5.0$^{+6.3}_{-4.0}$&2.5$^{+130.0}_{-2.0}$&2.5$^{+3.2}_{-2.0}$&2.5&0.9&2.0\\
1566-6-mwy&1.6$^{+2.0}_{-1.3}$&3.2$^{+4.0}_{-2.0}$&2.0$^{+2.0}_{-1.6}$&2.5$^{+3.2}_{-1.6}$&4.0$^{+5.0}_{-4.0}$&2.0$^{+2.5}_{-1.6}$&3.2$^{+4.0}_{-3.2}$&2.5&0.9&2.5\\
5253-5-smc&2.5$^{+7.9}_{-1.3}$&2.0$^{+7.9}_{-1.3}$&2.5$^{+16.0}_{-1.3}$&1.6$^{+2.5}_{-1.0}$&1.6$^{+2.5}_{-1.0}$&2.5$^{+4.0}_{-1.3}$&3.2$^{+10.0}_{-1.6}$&2.5&0.6&2.0\\
5253-9-smc&32.0$^{+320.0}_{-7.9}$&63.0$^{+400.0}_{-7.9}$&10.0$^{+32.0}_{-3.2}$&6.3$^{+25.0}_{-2.5}$&7.9$^{+20.0}_{-3.2}$&63.0$^{+320.0}_{-16.0}$&63.0$^{+400.0}_{-16.0}$&32.0&27.5&10.0\\
min\tnote{f}&1.6&2.0&2.0&1.6&1.6&2.0&2.5&2.5&0.6&2.0\\
max\tnote{g}&32.0&63.0&10.0&6.3&7.9&63.0&63.0&32.0&27.5&10.0\\
1566-1-sbt&2.0$^{+2.5}_{-1.3}$&2.5$^{+3.2}_{-2.0}$&2.0$^{+2.0}_{-1.6}$&2.5$^{+3.2}_{-2.0}$&4.0$^{+5.0}_{-4.0}$&2.0$^{+2.5}_{-1.6}$&3.2$^{+4.0}_{-1.6}$&2.5&0.8&2.0\\
1566-2-sbt&1.6$^{+2.0}_{-1.3}$&2.5$^{+4.0}_{-2.0}$&2.0$^{+2.0}_{-1.6}$&2.5$^{+4.0}_{-1.6}$&5.0$^{+5.0}_{-4.0}$&2.0$^{+2.5}_{-1.6}$&3.2$^{+4.0}_{-3.2}$&2.5&1.1&3.1\\
1566-3-sbt&1.6$^{+2.0}_{-1.3}$&3.2$^{+4.0}_{-2.0}$&2.0$^{+2.0}_{-1.6}$&2.5$^{+3.2}_{-1.6}$&5.0$^{+5.0}_{-4.0}$&2.0$^{+3.2}_{-1.6}$&4.0$^{+4.0}_{-3.2}$&2.5&1.2&3.1\\
1566-4-sbt&2.0$^{+4.0}_{-1.6}$&2.5$^{+4.0}_{-2.0}$&2.0$^{+7.9}_{-1.6}$&3.2$^{+4.0}_{-3.2}$&5.0$^{+6.3}_{-4.0}$&2.0$^{+2.5}_{-1.6}$&3.2$^{+4.0}_{-3.2}$&2.5&1.1&2.5\\
1566-5-sbt&3.2$^{+4.0}_{-2.5}$&3.2$^{+32.0}_{-2.5}$&40.0$^{+63.0}_{-2.0}$&2.5$^{+3.2}_{-2.0}$&5.0$^{+63.0}_{-4.0}$&32.0$^{+320.0}_{-2.0}$&2.5$^{+50.0}_{-2.0}$&3.2&16.2&16.0\\
1566-6-sbt&1.6$^{+2.0}_{-1.3}$&2.5$^{+4.0}_{-2.0}$&2.0$^{+2.0}_{-1.6}$&2.5$^{+3.2}_{-1.6}$&4.0$^{+5.0}_{-4.0}$&2.0$^{+2.5}_{-1.6}$&3.2$^{+4.0}_{-2.5}$&2.5&0.8&2.5\\
5253-5-sbt&2.0$^{+3.2}_{-1.0}$&1.6$^{+3.2}_{-1.0}$&1.6$^{+16.0}_{-1.0}$&1.3$^{+2.0}_{-1.0}$&1.6$^{+2.0}_{-1.0}$&2.0$^{+4.0}_{-1.0}$&2.5$^{+7.9}_{-1.3}$&1.6&0.4&1.9\\
5253-9-sbt&40.0$^{+320.0}_{-7.9}$&79.0$^{+400.0}_{-16.0}$&13.0$^{+32.0}_{-2.5}$&6.3$^{+25.0}_{-2.5}$&5.0$^{+16.0}_{-2.5}$&50.0$^{+320.0}_{-13.0}$&50.0$^{+320.0}_{-13.0}$&40.0&27.7&15.8\\
min\tnote{f}&1.6&1.6&1.6&1.3&1.6&2.0&2.5&1.6&0.4&1.9\\
max\tnote{g}&40.0&79.0&40.0&6.3&5.0&50.0&50.0&40.0&27.7, 1.1&16.0, 2.5\\
\hline
\end{tabular}
\begin{tablenotes}
\item [a] Cluster ID and extinction/attenuation law. The first four digits are the NGC number of the galaxy. The extinction/attenuation law is given by the last characters. mwy=Milky Way. smc=SMC. sbt=starburst.
\item [b] Median of the age PDF.
\item [c] Median of the ages in the row.
\item [d] Standard deviation of ages in the row. The value after the comma in the last row of this column is the median of the values in the column (excluding min/max rows).
\item [e] Ratio between the maximum and minimum ages in the row. The value after the comma in the last row of this column is the median of the values in the column (excluding min/max rows).
\item [f] Minimum of eight values in the column for a fixed dust extinction/attenuation law. 
\item [g] Maximum of eight values in the column for a fixed dust extinction/attenuation law. For the last column, the second value is the median of the 16 values in the column. 
\end{tablenotes}
\end{threeparttable} 
\end{table*} 

%%%%%%%%%%%%%%%%%%%%%%%%%%%%%
%%% Discussion
%%%%%%%%%%%%%%%%%%%%%%%%%%%%%

\section{Discussion}\label{sec:discussion}

\subsection{Implications of model assumptions.}

\subsubsection{Is an SSP a reasonable assumption?}

An SSP is a reasonable first-order approximation for the clusters in our sample. At the adopted distances of NGC 1566 and NGC 5253 (see Table~\ref{tab1}), the radii of the apertures used for the photometry are ~10 pc and 2 pc, respectively. For reference, the mass and radius of the most massive YMC in the Local Group (NGC 2070 in the Large Magellanic Cloud) are $8.7\times10^4\,M_\odot$ and $>20$ pc, respectively (see fig. 1 of \citealt{cig15}). The stars in NGC 2070 have an age spread of 7 Myr, peaking at 3 Myr. This excludes the central concentration of the cluster, which has a half-light radius of 1 pc and is known as R136.  Although it is possible that for clusters in NGC 1566, multiple ages are captured by the aperture, any assumption on the star formation other than SSP would be somewhat arbitrary. In addition, the regions are not sufficiently large to assume continuous star formation.

\subsubsection{Pre-main sequence stars}

In this work, we do not include pre-MS stars, i.e., we start the SSP models with all stars sampling the full IMF on the main sequence (MS) at age t=0. According to Karttunen et al. (2007), the typical contraction time is ~50 Myr for a 1 solar mass star and ~200 Myr for a half a solar mass star. Thus, it could be that the lower-mass pre-MS stars do not reach the MS before the most massive stars have left the MS. Implementing this phase in spectral evolution modeling is not straightforward as it requires a prescription for their spectra, and protostars are still enshrouded in dust at least a fraction of their lives. In clusters that are deeply embedded and very young, massive young stellar objects are expected to increase the intensity at near-IR wavelengths due to circumstellar IR-excess emission. This could be the case of cluster \#5 in NGC 1566 and both clusters in NGC 5253, since they all show higher extinction and deep red SEDs. The rest of NGC 1566 clusters are UV- and U-bright, i.e., not enshrouded. Thus, for these clusters, our assumption of no pre-MS stars is safe.

\subsubsection{Upper mass limit of the IMF}

 The {\tt{yggdrasil}} models, which are used as templates in this work, adopt a canonical value of 100 $M_\odot$ for the upper mass limit of the IMF. While the value of the upper mass limit of stars is presently uncertain \citep{sch14}, there is a great deal of evidence of the existence of stars with initial masses well above 100 $M_\odot$. A number of close binaries have component initial masses above 100 $M_\odot$ (\citealt{sch08, sch09}). \cite{cro10} and Crowther et al. (submitted) propose initial masses of up to 300 $M_\odot$ for several stars in the LMC based on their luminosities. Finally, \cite{wof14} suggest that very massive stars are present in superstar cluster NGC 3125-A1, based on a study of its strong He II 1640 \AA~emission and O V $\lambda$1371 absorption. Such high mass stars are expected to be short-lived $<3$ Myr. Their inclusion in our study is not crucial, since we are only interested in relative differences in cluster properties obtained with different models. In addition, the mass loss rates of very massive stars, which determine how long they would stay "very massive" are very uncertain. Finally, models for very massive stars are only available with the Pn, An, and Ab tracks. 
 
\subsubsection{Initial stellar mass of the models}\label{sec:m_init}

Typical derived cluster masses are between one and two orders of magnitude less than those of the original SSP models, even after accounting for the recycled fraction in the models. Thus, the cluster luminosities used as input for photoionization calculations are significantly larger than the expected masses of the observed clusters, which could influence the properties of the ionized nebulae (e.g., ionization parameter) and cluster properties. In order to check the effect of changing the initial stellar mass of the models, we computed posterior marginalized PDFs and best-fitting properties of NGC 1566 \#3, using the MW extinction law and a library of Pn models with an initial mass in stars of $5\times10^4\,M_\odot$, which is closer to the corresponding mass reported in Table~\ref{tab7}. Next, we give the properties obtained for this cluster using initial stellar masses of $10^6\,M_\odot$ and $5\times10^4\,M_\odot$, where the value in parenthesis corresponds to the lower mass. For the best-fitting properties we obtain: $E(B-V)=0.011$ mag (0.12 mag), mass\,=6.5\,$\times10^4\,M_\odot$ ($6.9\times10^4\,M_\odot$), age\,=2.5\,$\times10^6$ Myr ($2.5\times10^6$ Myr), i.e., no significant change in properties. The corresponding medians of the posterior marginalized PDFs are: $E(B-V)=0.10^{+0.12}_{-0.08}$ mag ($0.10^{+0.12}_{-0.08}$ mag), mass\,=\,$5.6^{+7.1}_{-5.6}\times10^4\,M_\odot$ ($2.8^{+3.5}_{-1.1}\times10^4\,M_\odot$), age\,=2.0$^{+2.0}_{-1.6}$\,$\times10^6$ Myr ($2.0^{+2.0}_{-1.6}\times10^6$ Myr), i.e., the median properties derived using the two initial SSP masses are in agreement with each other within the errors given by the 16th and 84th percentiles. We note that the fluxes hitting the model nebulae are the same for any initial stellar mass because of the relation between the adopted radius of the inner face of the cloud and the SSP luminosity (section 4.2 above).

\subsubsection{Nebular emission}

Figure 5 in \cite{ost03} shows that different prescriptions for the ionized gas can have dramatically different effects. In the latter example, this is due to the [O III] line flux entering the F555W filter. Ionized nebulae are bigger than star clusters, have evolving sizes, and have a diversity of densities and geometries which impact their sizes. There is no simple relation between cluster age and nebula's size. These issues pose challenges for modelers but also for observers. Indeed, there is no safe way of doing the photometry that guarantees a homogeneous inclusion of the ionized gas component. A method used by some observers to overcome the challenge of dealing with the ionized gas is to select filters that are not strongly contaminated by nebular emission, or to use narrow band images, e.g., in H$\alpha$ and H$\beta$ to remove the gas spectrum and its reddening from broader band filters. However, this is not always possible. Hence, further testing of the nebular parameter space (filling factor, ionization parameter, geometry, etc.) is necessary, in order to understand the impact of these parameters on spectral synthesis models. 

\subsubsection{Metallicity and attenuation by dust}

Ideally, metallicity and attenuation due to dust should be left as free parameters. In our case, we chose not to leave metallicity as a free parameter because Geneva tracks are only available at two metallicities and we want to include these widely known models in our comparison. In the present analysis, we find that the prescription for attenuation by dust that works best depends on the combination cluster +  model flavor. Testing models where the dust is directly included in the gas and the metals in the gas are depleted accordingly is also required to understand the impact of different approaches in accounting for the dust attenuation.

\subsection{YMCs in NGC 1566 and environment}

It is interesting to note that the properties of the YMCs in NGC 1566 which are listed in Tables~\ref{tab4} to~\ref{tab6} may be related to their environment. Gouliermis et al. (in preparation) show that five of the YMCs in NGC 1566 are located inside the boundaries of star-forming stellar structures, i.e., large loose complexes identified at significance of 1 and 2 sigma above the background surface density of OB-type stars. The five clusters themselves are identified to coincide with more compact stellar concentrations at significance levels of 5 and 7 sigmas (sigma being the standard deviation of the stellar surface density). The only cluster not found in any complex (marginally on the borders of one and located at the edge of a spiral arm) is cluster \#5, which is the faintest in the sample. Interestingly enough, cluster \#5 is the only cluster in the NGC1566 sample with an SED dropping at V and I wavebands. It also has the higher extinction as found in this paper, which possibly makes the case for the remaining clusters having lower extinction due to their location inside radiative bright stellar complexes. The existence of a rich blue stellar population around the clusters has also implications on any age-spread and multiple ages that may be captured by the aperture. This positional correlation of the YMCs with stellar complexes suggests that they are probably products of hierarchical star formation, representing the centers of active nested star formation within larger stellar constellations. For more details on the effect of environment on the properties of young star clusters in NGC 1566 see Gouliermis et al. (in preparation).

\subsection{Impact of results on cluster age and mass functions}

Statistically driven studies of young star clusters (YSCs) rely on SED techniques and model assumptions like those explored in this work. To date many key aspects of YSC properties remain unanswered, like the rate at which cluster form (e.g. \citealt{ada15}, \citealt{cha15}) and dissolve (e.g. \citealt{sil14}, \citealt{cha14}), and the shape of the cluster mass function (e.g. Larsen 2009).   Therefore, it is crucial to understand degeneracies and uncertainties when inferring cluster properties. Overall, we see (Fig. 14 to 16) that the filter combination we have access to is not allowing us to put strong constraints on the best spectral synthesis model. Derived ages, extinctions, and masses are sensitive to the choice of model and the applied extinction law. We also see that the quality of the fit improves when models including interacting binary stars are used. However when we compare derived ages, extinctions, and masses obtained with the newest Ab models to those obtained with the oldest models, Po and Go, which are widely used in the literature, the differences are typically around 0.1~dex in age and mass, and within the uncertainties in extinction (although maximum differences can reach up to 0.4~dex on a case-by-case basis). It is challenging to extrapolate the impact of the use of these different models to statistical studies of cluster populations. In general, we observe differences and larger uncertainties in derived cluster properties for objects which have higher extinction. To probe the impact of the different models on the derived cluster population properties, as the age and mass distributions, we propose to extend the current study to a larger sample of YMCs, using the whole population of YMC detected in the LEGUS galaxies. 
 
\subsection{NGC 5253 \#5 and Feedback.}

Along with cluster NGC 5253 \#11, cluster NGC 5253 \#5 is of great interest for reasons discussed in detail in C15b. These clusters drive much of the energetics in this galaxy. In Table~\ref{tab9}, we compare the $E(B-V)$, mass, and age values of this cluster obtained by C15b and this work. C15b use a total to selective extinction $R_{\rmn{V}}=4.05$ (same as in this paper) and obtain a slightly higher reddening and higher mass. This is because of the different approaches for calculating cluster properties in this paper and C15b. In particular, C15b use differential attenuation for stellar continuum and nebular ionized gas, such that the stellar continuum is subject to half the attenuation of the nebular gas. This boosts up the stellar continuum (and mass). The ages that we derive with both prescriptions for dust attenuation are in agreement, within the uncertainties, with the very young age obtained by C15b, and with the age obtained by Smith et al. (in preparation), i.e., 1--2 Myr, based on archival UV spectroscopic data. 

%%%%%%%%%%%%%%%%%%%%%%%%%%%%%%
% Table 9
%%%%%%%%%%%%%%%%%%%%%%%%%%%%%%

\begin{table*} 
\caption{NGC 5253 \#5.}
\label{tab9}
\begin{threeparttable}[b]
\begin{tabular}{lccc}
\hline
Source &  $E(B-V)$ & Mass & Age \\
\hfill & mag & $10^4\,M_\odot$ & Myr \\
\hline
C15b\tnote{a} & $0.46^{+0.04}_{-0.04}$ & $7.46^{+0.20}_{-0.27}$ &  $1\pm1$ \\
This work - SMC\tnote{b} & $0.35\pm0.02$ &  $3.2\pm0.3$ &  $2.5\pm0.6$ \\
This work - SBT\tnote{c}  & $0.38\pm0.03$ & $2.2\pm0.2$ & $1.6\pm0.4$ \\
\hline
\end{tabular}
\begin{tablenotes}
\item [a]  Values from table 5 of C15b.
\item [b] Median of models discussed in this work using the SMC law.
\item [c] Median of models discussed in this work using the SBT law.
\end{tablenotes}
\end{threeparttable} 
\end{table*} 
 
%%%%%%%%%%%%%%%%%%%%%%%%%%%%%
%%% Summary and conclusions
%%%%%%%%%%%%%%%%%%%%%%%%%%%%%

\section{Summary and conclusions}

We undertake a comprehensive comparative test of spectral synthesis models of massive-star populations which are based on seven different flavors of massive-star evolution. For this purpose, we use near-ultraviolet to near-infrared photometry of eight young massive clusters in galaxies, NGC 1566 and NGC 5253 (Fig.~\ref{fig1} and Table~\ref{tab1}), which were selected to mitigate the effect of the stochastic sampling of the stellar IMF and ensure the presence of massive stars in the clusters. The observations are from \textit{HST}'s Treasury program LEGUS and the archive (Tables~\ref{tab2} and ~\ref{tab3}). The models are for SSPs of well-sampled IMFs and include older and state-of-the-art massive-star evolution models accounting for updated input stellar physics, rotation, and interacting binaries, from independent groups in Padova, Geneva, and Auckland. Our libraries of models account for stellar and nebular emission and use two different prescriptions for attenuation by dust. For a homogeneous comparison with the cluster properties used to select our sample, we adopt the nebular parameters of \cite{zac11}.

In our Bayesian fitting approach, metallicity is not a free parameter because only two metallicities are available for all seven flavors of massive-star evolution included in this work. The adopted massive-star evolution models match the published gas-phase metallicities of the two galaxies in our sample. From Bayesian fits to the observations, we find cluster reddenings, masses and ages in the ranges $(0.05\pm0.02-0.54\pm0.13)$ mag, $(1.8\pm0.3-10\pm5)\times10^4\,M_\odot$ and $(1.6\pm0.4-40\pm27.7)$\,Myr (median and standard deviation of values from all models, Tables~\ref{tab6}-\ref{tab8}). These properties are characterised by typically small standard deviations derived for individual clusters using different models ($\sim0.03\,$mag, $\sim10^4\,M_\odot$ and $\sim1\,$Myr, last row and second to last column of Tables~\ref{tab6}-\ref{tab8}), although maximum differences are typically 0.09 mag in $E(B-V)$, a factor of 2.8 in mass, and a factor of 2.5 in age (last row and last column of Tables~\ref{tab6}-\ref{tab8}). 

In terms of best fit, the observations are slightly better reproduced by models with interacting binaries as implemented in {\tt{bpass}} and least well reproduced by models with single rotating stars as implemented in {\tt{starburst99}} (Fig.~\ref{fig8} and~\ref{fig9}). This could be because the Geneva rotating models are not intensively calibrated on observations of individual stars. The fact that binary models slightly better match observations compared to single star models could be because they better reflect the observed complexity of stellar populations. We do not adjust the parameters of binary models to achieve a better fit.

The available combination of LEGUS filters and our limited sample size does not allow us to put strong constraints on the best prescription for attenuation by dust. Which prescription is more successful in reproducing the observations depends on the combination cluster + model flavor (Fig.~\ref{fig9}).

For NGC 1566, the B and I bands are the least-well fitted (Fig.~\ref{fig10}), while the V band, which has the lowest observational errors, is the best fitted. For NGC 5253 most bands are well fitted due to larger observational errors (Fig.~\ref{fig8}). 

We check that the adopted initial mass in stars of the models does not strongly affect the derived cluster properties. This is done for one cluster (NGC 1566 \#3). We find that the median properties of the cluster obtained with initial model masses of $5\times10^4$ and $1\times10^6\,M_\odot$ are in agreement within the errors given by the 16th and 84th percentiles.
 
For NGC 1566, we find that the properties of clusters may be related to the environment (Section 7.2).

In order to assess the impact of the different model flavors on studies of cluster populations, we propose to extend the current study to the entire LEGUS sample of YMCs  (Section 7.3).

Our study provides a first quantitative estimate of the accuracies and uncertainties of the most recent spectral evolution models of massive-star populations, demonstrates the good progress of models in fitting high-quality observations, and highlights the needs for using a larger cluster sample and testing more extensively the model parameter space.

\section*{Acknowledgments}
AW and SC acknowledge support from the ERC via an Advanced Grant under grant agreement no. 321323-NEOGAL. SdM acknowledges support through a Marie Sklodowska-Curie  Reintegration Fellowship, H2020-MSCA-IF-2014, project id 661502. EZ acknowledges research funding from the Swedish Research Council (project 2011-5349). We are very thankful to the following scientists for answering questions relevant to this paper: C. Leitherer ({\tt{starburst99}} models); P. Crowther (age spread in cluster NGC 2070); W.-R. Hamann ({\tt{PoWR}} models and H-burning WR stars); B. Elmegreen and M. Krumholz (pre-MS stars). Finally, we want to thank the referee for carefully reading this manuscript and providing ideas that enhanced the scientific outcome of the paper.

\bsp

\label{lastpage}

\end{document}